\documentclass[twoside,openright,12pt]{thesis}

\ifpdf
    \pdfcatalog { /PageMode (/UseOutlines)
                  /OpenAction (fitbh)  }
\fi

\title{Modeling pre-biotic self-organization: The chemical dynamics of autocatalytic networks}

\ifpdf
  \author{Varun Giri}
  \collegeordept{Department of Physics and Astrophysics}
  \university{University of Delhi}
\else
  \author{Varun Giri}
  \collegeordept{Department of Physics and Astrophysics}
  \university{University of Delhi}
\fi

\degree{Doctor of Philosophy}
\degreedate{December 2011}

\newcommand{\ie}{{\it i.e.}}

\def\signature#1#2{\parbox[b]{1in}{\smash{#1}\vskip12pt} \hfill \parbox[t]{2.5in}{\shortstack{\vrule width 2.5in height 0.4pt\\\\\small#2}}}

\begin{document}

\renewcommand\baselinestretch{1.2}
\baselineskip=18pt plus1pt
\pagestyle{empty}

\maketitle

\chapter*{}
\thispagestyle{empty}
\begin{quote}
\begin{center}
{\Large \bf \textsf{Declaration}}
\end{center}
\vspace*{5ex} \noindent {This thesis describes work done by the candidate during his tenure as Ph.D. student at the Department of Physics and Astrophysics, University of Delhi, Delhi, India under the supervision of Sanjay Jain. The work reported in this thesis is original and it has not been submitted earlier for any degree to any university.}

\par
\vspace*{16ex}\noindent \signature{Candidate:}{Varun Giri}
\par
\vspace*{10ex}\noindent \signature{Supervisor:}{Sanjay Jain}
\par
\vspace*{10ex}\noindent \signature{Head of the Department:}{R. P. Tandon}
\end{quote}

\cleardoublepage
\section*{List of Publications}
\subsubsection*{In International Refereed Journals}
\begin{enumerate}
    \item The origin of large molecules in primordial autocatalytic reaction networks,\\ {\bf Varun Giri} and Sanjay Jain,\\ Accepted for publication in {\it PLoS ONE} (2011)\\ doi: 10.1371/journal.pone.0029546.
    \item A universal power law and proportionate change process characterize the evolution of metabolic networks,\\ Shalini Singh, Areejit Samal, {\bf Varun Giri}, Sandeep Krishna, N. Raghuram and Sanjay Jain,\\ {\it Eur. Phys. J. B}, {\bf 57:75-80} (2007).
    \item Low degree metabolites explain essential reactions and enhance modularity in biological networks,\\ Areejit Samal, Shalini Singh, {\bf Varun Giri}, Sandeep Krishna, N. Raghuram and Sanjay Jain,\\ {\it BMC Bioinformatics}, {\bf 7:118} (2006).
\end{enumerate}
\noindent The work reported in this thesis is based on publication 1 mentioned above.

\subsubsection*{In Conference Proceedings}
\begin{enumerate}
    \item Modelling stock market: Expectation bubbles and crashes,\\ {\bf Varun Giri} and Areejit Samal,\\ Student paper for IMSc Complex Systems School, January 2-27 2006, held at the Institute of Mathematical Sciences, Chennai, India in association with Santa Fe Institute, New Mexico, USA.
\end{enumerate}

\newpage
\section*{Important Talks and Oral Presentations}
\begin{enumerate}
    \item Presented a talk titled ``Dynamics of autocatalytic reaction networks and the origin of life'' at Sixth Dynamics Day Delhi, held at Sri Venkateswara College, University of Delhi on 27th November, 2010.
    \item Presented a talk titled ``Modeling the chemical dynamics of autocatalytic networks in a pre-biotic scenario'' at
        \begin{enumerate}
            \item Complex Structures in Biology and Cognition Group, Max Planck Institute for Mathematics in the Sciences, Leipzig on 1st October, 2010,
            \item Herbstseminar Bioinformatik 2010, Vysoka Lipa (Decin), Czech organized by Institut f\"{u}r Informatik, University of Leipzig, Leipzig on 8th October, 2010,
            \item Potsdam Institute for Climate Impact Research, Potsdam on 12th October, 2010,
            \item Biologische Experimental Physik Group, University of Saarlandes, Saarbrucken on 13th October, 2010.
        \end{enumerate}
    \item Presented a talk titled ``Using network topology to optimize molecular production in an artificial chemistry model'' at Application of Control Theory and Optimization Techniques in Biochemical Pathways, satellite symposium of the International Congress of Mathematics, held at Hyderabad International Convention Centre (HICC), Hyderabad, India, from 16th to 18th August, 2010.
    \item Presented a talk titled ``Modeling the effects of autocatalytic networks on pre-biotic organization'' at International Workshop on Chemical Evolution and Origin of Life, organized by Department of Chemistry, Indian Institute of Technology Roorkee, held at IIT-R, Roorkee from 5th to 7th March, 2010.
    \item Presented a talk titled ``Bistability in autocatalytic reaction networks'' at Dynamics Day Delhi, held at Department of Physics and Astrophysics, University of Delhi, Delhi on 15th November, 2008.
    \item Invited talk with hands-on-session titled ``A practical guide to graph-theoretic analysis of large scale biological networks'' at Workshop on knowledge discovery in Life Sciences: Tools \& Techniques in Bioinformatics, January 29 to February 2, 2007, held at Bioinformatics Centre, University of Pune, India. This talk was presented with Areejit Samal.
\end{enumerate}

\section*{Poster Presentations}
\begin{enumerate}
    \item Presented a poster titled ``Modeling the chemical dynamics of autocatalytic networks in a pre-biotic scenario'' at Multi-scale dynamics and evolvability of biological networks, organized by Max Planck Society (MPG) and the Centre National de la Recherche Scientifique (CNRS), held at Max Planck Institute for Mathematics in the Sciences, Leipzig from 4th to 6th October, 2010.
    \item Coauthored poster titled ``Classifying reactions and locating evolutionary hotspots in metabolic networks'' at International Conference on Bioinformatics, held at Hotel Ashok, New Delhi, India, from 18th to 20th December, 2006.
    \item Presented a poster titled ``Modeling stock markets: Formation of bubbles and crashes'' at International Workshop on Econophysics of Stock Markets and Minority Games (Econophysics II), held at Saha Institute of Nuclear Physics (SINP), Kolkata, India from 14th to 17th February, 2006.
    \item Presented a poster titled ``Bow-Tie decomposition of metabolic networks'' at Statphys - Kolkata V: Complex Networks: Structure, Function and Processes, held at S. N. Bose National Centre for Basic Sciences, Kolkata, India from 27th to 30th June, 2004.
\end{enumerate}

\begin{dedication}
`Life'
\end{dedication}

\newpage
\thispagestyle{empty} \mbox{}

\frontmatter
\pagestyle{fancy}
\begin{abstract}
Earth was formed about 4.54 billion years ago. In less than a billion years the first life forms appeared on a still very `young' earth. Even the simplest cells today are extremely complicated and require several components to work coherently. It is hard to imagine how such complexity, as seen in life, would have arisen at the origin of life. Several approaches, experimental as well as theoretical, are being used to explain the origin(s) of life, but a comprehensive explanation still eludes us.

Miller's experiment in 1953 was one of the first successful attempts to show that some of the precursors of life, amino acids, can be produced in simulated pre-biotic environments. Recent experiments have also presented a possible way of producing ribonucleotides in prebiotically plausible conditions. It has been thus far not possible to show how these precursors can combine or self-organize to form large molecules that are needed for life to function. Large molecules such as proteins and nucleic acids are crucial for life. The production of these large molecules, as we know it today, requires good catalysts, and the only good catalysts we know that can accomplish this task consist of large molecules. There exists no other natural process we know of that produces large molecules and that does not itself use large molecules. Thus the origin of large molecules is a `chicken and egg problem' in chemistry. We expect that the answer to the question lies in the processes that occurred before life originated.

In this thesis we present a mechanism, based on autocatalytic sets (ACSs), that is a possible solution to this problem. We present a mathematical model describing the population dynamics of molecules in an artificial but prebiotically plausible chemistry in a well stirred chemical reactor. In the model large molecules can in principle be produced by successive ligations of pairs of smaller molecules. The smallest molecules in the chemistry compose the `food set', whose concentrations are considered to be buffered. The chemistry contains a large number of spontaneous reactions of which a small subset could be catalyzed by molecules produced in the chemistry with varying catalytic strengths. Normally the concentrations of large molecules in such a scenario are very small, diminishing exponentially with their size. ACSs, if present in the catalytic network, can focus the resources of the system into a sparse set of molecules. ACSs can produce a bistability in the population dynamics and, in particular, steady states wherein the ACS molecules dominate the population, \ie, have higher concentrations compared to the rest of molecules in the chemistry (background).

In this thesis we attempt to address two main questions: First, under what circumstances do molecules belonging to the ACSs dominate over the background, and second, starting from an initial condition that does not contain good catalysts, can a sparse set of large molecules (containing several tens or a few hundred monomers) that are good catalysts arise and be maintained in the reactor at concentrations significantly above the background? We find that the mere existence of ACSs in the chemistry at the level of network topology is not sufficient to guarantee their domination at the population level (as also reported by Bagley and Farmer in their model). We further quantitatively characterize the interplay between ACS topology and rate constants including catalytic strengths, dissipation rate, etc., that results in ACS domination. We show that for ACS molecules to dominate above the background, the catalytic strength needs to be sufficiently large, growing exponentially with the size of the catalyst molecule.

We present a possible resolution to this problem via a hierarchy of nested ACSs. We show that if an ACS catalyzed by large molecules contains within it (or partially overlaps with) a smaller ACS catalyzed by smaller molecules (we refer to this as a `nested ACS' structure), the catalytic strength required for the large ACS to dominate comes down significantly. Effectively the small ACS reinforces the larger one. We construct a cascade of nested ACSs with each successive and larger ACS in the cascade containing (a part of) the smaller ACSs, thereby receiving reinforcement from them and in turn providing reinforcement to the next larger ACS in the cascade.  We show that when the network contains a cascade of nested ACSs with the catalytic strengths of molecules increasing gradually with their size (e.g., as a power law), a sparse subset of molecules including some very large molecules can come to dominate the system. We exhibit an example in which a large catalyst (with more than 400 monomers) arises with a significant population in the steady state starting from only monomers in the initial state.

Thus our work presents a possible resolution to the chicken and egg problem in chemistry posited by the existence of large molecules. It suggests an incremental bootstrap mechanism -- the existence of a hierarchy of nested ACSs in the chemistry -- through which a sparse subset of large molecules can naturally arise starting from plausible prebiotic initial conditions.
\end{abstract} 

\setcounter{secnumdepth}{3} \setcounter{tocdepth}{3}
\thispagestyle{plain}
\cleardoublepage
\tableofcontents

\newglossaryentry{FoodSet}{name=food set, description={A set of small molecular species that are presumed to be abundantly present in a prebiotic niche, denoted $\mathcal{F}$}}

\newglossaryentry{catalytic-strength}{name={catalytic strength}, description={The factor by which a catalyst enhances the rate of the reaction it catalyzes, denoted $\kappa$}, symbol={\ensuremath{\kappa}}}

\newglossaryentry{spontaneous-chemistry}{name={spontaneous chemistry}, description={The set of possible reactions, defined by non-zero rate constant $k^F_{XY}$ and $k^R_{XY}$}, plural={spontaneous chemistries}, sort={chemistry spontaneous}}

\newglossaryentry{connected-chemistry}{name={connected chemistry}, first={`Connected' chemistry}, description={A chemistry in which every molecule can be produced from the food set in some pathway consisting of a sequence of allowed reactions}, sort={chemistry connected}}

\newglossaryentry{fully-connected-chemistry}{name={fully connected chemistry}, first={`Fully connected' chemistry}, description={A chemistry in which all possible ligation and cleavage reactions are allowed: $k^F_{\mathrm{XY}} \neq 0$ and $k^R_{\mathrm{XY}} \neq 0$ for all $\mathrm{\mathbf X}$,$\mathrm{\mathbf Y}$}, sort={chemistry connected fully}}

\newglossaryentry{rev-chemistry}{name={reversible chemistry}, first={`Reversible' chemistry}, description={A chemistry for which each allowed reaction is reversible, \ie, for every reaction, $k^F \neq 0 \Leftrightarrow k^R \neq 0$}, sort={chemistry reversible}}

\newglossaryentry{homogeneous-chemistry}{name={homogeneous chemistry}, first={`Homogeneous' chemistry}, description={A chemistry in which all the nonzero rate constants, loss rates, and concentration of the food set molecules are independent of the species labels}, sort={chemistry homogeneous}}

\newglossaryentry{homogeneous-catalyzed-chemistry}{name={homogeneous catalyzed chemistry}, description={A catalyzed chemistry in which the catalytic strength of every catalyst for each reaction that it catalyzes is the same}, sort={chemistry homogeneous catalyzed}}

\newglossaryentry{sparse-chemistry}{name={sparse chemistry}, description={A chemistry that contains only a small fraction of all possible reactions}, sort={chemistry spontaneous}}

\newglossaryentry{catalyzed-chemistry}{name={catalyzed chemistry}, description={The set of catalyzed reactions together with the catalysts and their catalytic strengths}}

\newglossaryentry{finite-chemistry}{name={finite chemistry}, description={A chemistry in which all reactions that produce or consume any molecule of length larger than some positive integer $N$ are omitted}}

\newglossaryentry{standard-ic}{name={standard initial condition}, first={`standard' initial condition}, description={Initial condition in which all concentrations other than the food set are zero}}

\newglossaryentry{ACS}{name={ACS}, description={Autocatalytic set}}

\newglossaryentry{autocatalytic-set}{name={autocatalytic set}, description={A set of catalyzed one-way reactions such that (a) each catalyst of this chemistry gets produced by a reaction of the chemistry, and (b) every reactant in the chemistry is either a member of the food set ${\mathcal F}$ or gets produced in this chemistry}}

\newglossaryentry{background}{name={background}, description={The set of all molecules except the ACS product molecules and the food set}}

\newglossaryentry{ACS-length}{name={length of ACS}, description={The largest molecule produced in the ACS, denoted $L$}, see={ACS}}

\newglossaryentry{extremal-ACS}{name={extremal ACS}, description={An ACS of length $L$ is `extremal' if all reactions are catalyzed by the largest molecule (of size $L$) produced in the ACS}, see={ACS-length}}
\mainmatter
\pagestyle{fancy}

\thispagestyle{plain}
\cleardoublepage
\chapter{\label{Introduction}Introduction}

\begin{quote}
``{\it It is often said that all the conditions for the first production of a living organism are now present, which could have ever been present. But if (and oh! what a big if!) we could conceive in some warm pond, with all sorts of ammonia and phosphoric salts, light, heat, electricity, etc., present, that a protein compound was chemically formed ready to undergo still more complex changes, at the present day such matter would be instantly devoured or absorbed, which would not have been the case before living creatures were formed.}''
\begin{flushright}
-- Charles Darwin (1809-1882)
\end{flushright}
\end{quote}

\lettrine[lines=2, lhang=0, loversize=0.0, lraise=0.0]{E}{arth} is special in that it is the only known planet that supports life. Yet, life is ubiquitous on Earth. Within a rather short period after its creation, life emerged on Earth. Over the course of 3.5 -- 4 Gyr (gigayears) life has evolved to its current form. While biologists have made great progress in understanding how life `works', there has not been an equal progress on the question of life's origin. This thesis is an attempt to formulate and answer one of the many questions that relate to the problem of origin of life on Earth. More concretely we discuss, with the help of a mathematical model of primordial chemistries, the set of initial bootstrapping processes that could have led to the production of large molecules that were needed for the emergence of life. In this chapter, we discuss some of the facts known about life and the some of the questions that relate to its origin.

\section{The Earth's history}
\begin{figure}
    \begin{center}
        \includegraphics[width=5.75in]{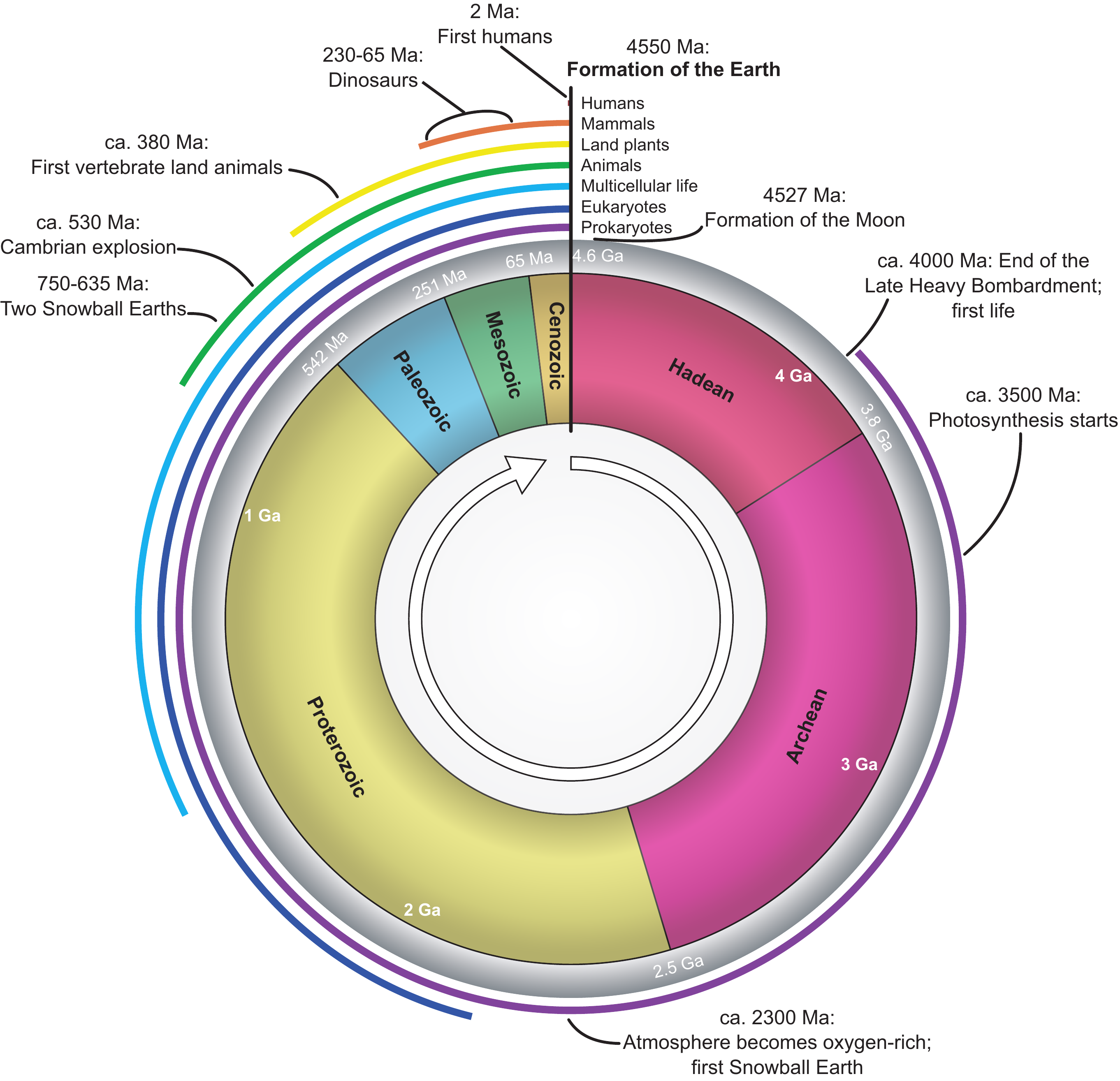}
    \end{center}
    \caption[Earth's geological time `clock'.]{{\bf Earth's geological time `clock'.} Earth began to form between 4.6 -- 4.5 Ga (gigayears ago) from a cloud of gas (mostly hydrogen and helium) and interstellar dust. It was formed under intense heat and pressure and was initially a molten planet. During the Hadean Period (up to nearly the first billion years of formation), the Earth was bombarded continuously by asteroids, meteors and comets. By 3.8 to 4.2 Ga, the Earth cooled and became a planet with an atmosphere and an ocean. Earth's atmosphere was formed mostly from water vapour, carbon dioxide, nitrogen, and small amounts of carbon monoxide, methane, ammonia, hydrochloric acid and sulfur produced by the constant volcanic eruptions. It had no free oxygen. It is believed that around this time first life forms appeared on Earth. Earth's initial life forms were bacteria, which could survive in the highly toxic atmosphere that existed during this time. Around 3.5 Ga, oxygen-forming photosynthesis began to occur. Once oxygen began to build up around 2.5 Ga, it made way for the emergence of more complex life as we know it today. About 530 Ma, `Cambrian Explosion' happened where life exploded developing almost all of the major groups of plants and animals in a relatively short time. The first modern human species emerged only about 2.2 Ma.
    Image source: \url{http://en.wikipedia.org/wiki/History_of_Earth}.}
    \label{Time-Line}
\end{figure}

\begin{figure}
    \begin{center}
        \includegraphics[width=5.75in]{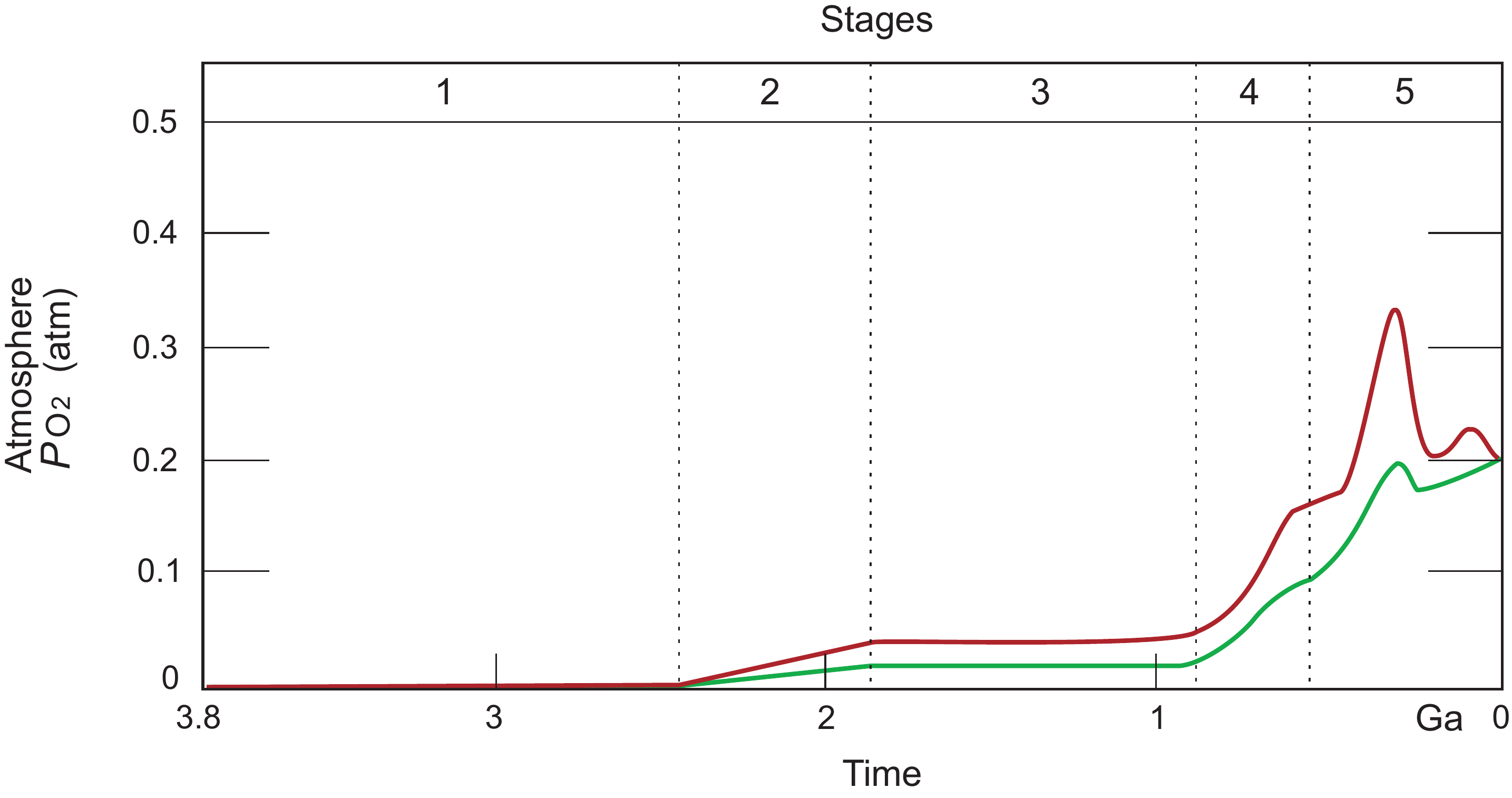}
    \end{center}
    \caption[Oxygenation of Earth's atmosphere.]{{\bf Oxygenation of Earth's atmosphere.} Red and green lines plot the range of the estimates of atmospheric oxygen, time is measured in billions of years ago (Ga). Initially atmosphere was devoid of any free oxygen (Stage 1: 3.85 -– 2.45 Ga). Though photosynthetic bacteria were producing oxygen, it was getting captured by organic matter and dissolved iron (and other reduced material) chemically. As these started to saturate, around 2.45 Ga, the `great oxygenation event' occurred. Oxygen was now released as free molecule and was absorbed in oceans and seabed rock (Stage 2: 2.45 -– 1.85 Ga). Later, oxygen started to gas out, but got absorbed by land surfaces and formation of the ozone layer (Stage 3: 1.85 -– 0.85 Ga). Around this period the first multi-cellular organisms started to appear. As the oxygen sinks got filled, the gas started to accumulate in the atmosphere (Stages 4 and 5: 0.85 Ga -– present). It is believe that high levels of oxygen led to the `Cambrian explosion' (which happened at the beginning of Stage 5).
    Image source: \url{http://en.wikipedia.org/wiki/Geological_history_of_oxygen}.}
    \label{Oxygenation}
\end{figure}

The age of the Earth is approximated to be 4.54 billion years ($\pm 1\%$) \cite{Manhes1980, Dalrymple1991, Dalrymple2001}. Fig. \ref{Time-Line} illustrates the major events in the history of Earth since its birth. Formed from a gas cloud made up of hydrogen and helium and interstellar dust under intense heat and pressure, at its birth Earth was a molten planet. Over the next few hundred million years, it cooled to form a solid crust \cite{Sleep2001, Valley2002, Harrison2005, O'Neil2008}. As it cooled further, clouds formed and rains created the ocean. It is also believed that part of the water that exists on Earth was delivered by impacting comets that contained ice. There are evidences to suggest that oceans formed on Earth between 4.4 --  4.2 Ga (gigayears ago) \cite{Wilde2001, Mojzsis2001, Sleep2001}. (It was earlier believed that it took much longer for Earth to cool enough to be able to support liquid water.) The Earth's atmosphere contained water vapour, carbon dioxide, nitrogen and smaller amounts of other gases \cite{Kasting2006}. It is estimated that in about 600 million years (around 3.8 Ga) first life forms appeared \cite{Mojzsis1996}. The earliest living organisms were microscopic bacteria, which show up in the fossil record as early as 3.5 Ga \cite{Schopf2002, Altermann2003, Furnes2004, Allwood2006, Wacey2011}. Some other studies suggest that life could have been present even earlier (as early as 4.25 Ga) \cite{Battistuzzi2004, Nemchin2008, Abramov2009}. The earliest life forms were chemoautotrophic, \ie, they relied on chemical sources for energy. Later, photosynthesis was adopted as a mechanism for energy production \cite{Kasting2002}. Photosynthetic  prokaryotic organisms emitted oxygen as a waste product \cite{Dutkiewicz2006}. This is believed to have caused the `great oxygenation event' at 2.45 Ga leading to release of oxygen into the atmosphere \cite{Bekker2004, Kopp2005, Holland2006, Buick2008, Rasmussen2008} (see Fig. \ref{Oxygenation}). There are studies that suggest alternate mechanisms for oxygenation of Earth's atmosphere \cite{Gaillard2011, Claire2006}. The presence of oxygen provided life with new opportunities. Organisms switched to aerobic metabolism, which is more efficient than anaerobic pathways. Multicellular life appeared soon after oxygen became available. As the oxygen level rose further life became more diversified and led to the `Cambrian explosion' around 530 Ma (million years ago). Humans emerged on Earth only 2.2 Ma.

\section{Life}
Life is one of the most complex phenomena known to human kind. Life involves several intricate processes, such as metabolism, genetic regulation, formation of physical structures, etc., that work in unison to allow organisms to harness energy, repair themselves, grow and replicate. Each of these processes is indispensable for life, and depends upon other processes to be able to successfully perform its task.

\subsection{\label{Life-Description}Life: An organized system}
Cells are the fundamental units of life. Cells respond to the environment, conduct internal processes, grow and replicate with variation. A cell contain several small and large molecules enclosed in a membrane. The membrane provides the cell a spatial form and separates the interior from the environment. The membrane contains several proteins which sense external signals, and allow a selective exchange of material between cell and the environment. The internal machinery of the cell is complex, involving metabolism, regulation and signaling.

\begin{figure}
    \begin{center}
        \subfloat[Transcription.]{\label{Transcription}\includegraphics[width=5in]{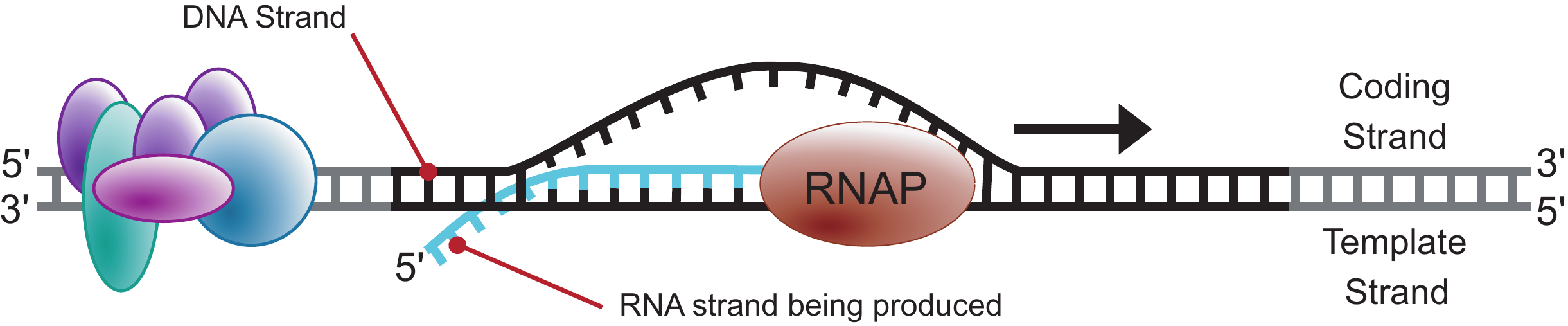}}\\\vspace{1cm}
        \subfloat[Translation.]{\label{Translation}\includegraphics[width=5.75in]{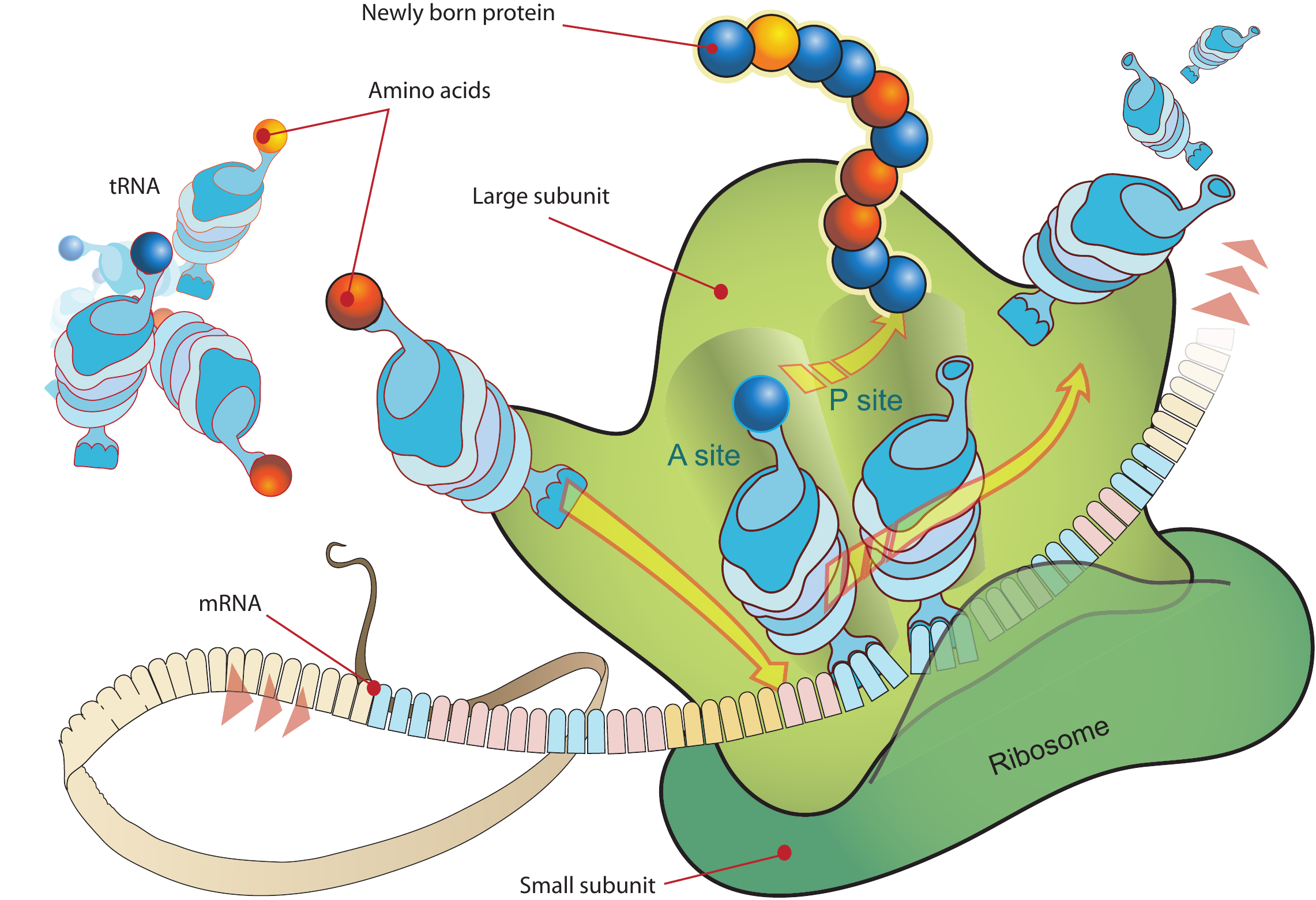}}
    \end{center}
    \caption[Gene expression and production of proteins.]{{\bf Gene expression and production of proteins.} {\bf (a)} The first step in gene expression is transcription. RNA polymerase (RNAP), an enzyme, binds to the DNA molecule and produces RNA using DNA genes as template. RNA polymerase first unzips the DNA double-strand by breaking the hydrogen bonds between complementary nucleotides and then adds complementary RNA nucleotides paired with DNA bases. RNA nucleotides join together by forming the sugar-phosphate backbone with the help from RNA polymerase. As the hydrogen bonds of the untwisted RNA+DNA helix break the newly synthesized RNA strand start to separate from DNA. {\bf (b)} Translation of mRNA and the synthesis of proteins by a ribosome. Ribosomes are made up of RNAs and proteins called ribonucleoproteins and are divided into two subunits. The smaller subunit binds to the mRNA, while the larger subunit binds to the tRNA and the amino acids. Each triplet of nucleotides on mRNA, called codon, corresponds to a binding site complementary to an anticodon triplet in tRNA. tRNAs with the same anticodon sequence always carry identical type of amino acid. The ribosome helps complementary tRNA to bind to mRNA and takes the amino acid from each tRNA, chaining them together to form protein. Images source: \url{http://en.wikipedia.org/wiki/Gene_expression}.}
\end{figure}

A cell converts simple food molecules absorbed from the environment (such as glucose, fructose, etc.) into other `building-block' molecules (such as amino acids, nucleotides, lipids, etc.) and produces energy, which is essential for the cell to grow and function. This process is called `metabolism'. Metabolism only involves the chemical conversion of small molecules called metabolites, but is dependent upon the availability of a set of `large' molecules called enzymes (a class of proteins that catalyze metabolic reactions). Enzymes govern which metabolic processes happen inside a cell. The instructions to produce these (and other) proteins are contained in the DNA of the cell. When needed this information is read and the corresponding proteins produced via a two step process called gene expression. In the first step, called `transcription', a machine called RNA polymerase reads a gene on a DNA molecule and assembles nucleotides in right linear order to produce the corresponding single-stranded messenger RNA (mRNA) molecule (Fig. \ref{Transcription}). Then, in the processe called `translation', another machine called ribosome reads the mRNA molecule and assembles amino acids in the right linear order to produce the corresponding protein molecule (Fig. \ref{Translation}). To be able to produce these macromolecules, RNA or the proteins, the cell needs the small building-block molecules that are the end product of metabolism -- the nucleotides and the amino acids (which in the first place require these macromolecules for their own production) \cite{Alberts2007}. It also needs the machines, RNA polymerase and ribosomes that are made up of a number of macromolecules.

The genetic code stored in DNA is ``interpreted'' during gene expression, and the properties of the expression (proteins, RNAs produced as result of gene expression) give rise to the organism's phenotype. This process also highlights the information flow inside the cell. In 1958, Francis Crick pointed out that the information flow can happen only from nucleic acids to proteins and such information cannot be transferred back from protein to either protein or nucleic acid \cite{CRICK1958, CRICK1970}. In other words a protein sequence can not be used to create a DNA in the cell. In the usual case the information flows from $\mathrm{DNA} \rightarrow \mathrm{RNA} \rightarrow \mathrm{proteins}$, and never from $\mathrm{protein} \rightarrow \mathrm{protein}$ or $\mathrm{protein} \rightarrow \mathrm{DNA}$ or RNA.

With the kind of elaborate machinery described above, life has to maintain a balance between various participating subcomponents. This requires regulation and synchronization of all the processes for life to be able to perform its tasks efficiently. This adds another level of complexity to the way life functions. Based on the environmental conditions (and some internal cues, such as cell division, etc.) the cells may express different sets of genes at different times. The ability to differentially express genes based on different environmental conditions gives cell the ability to adapt to different environments. The process of regulation also allows the same genome to express different phenotypes, as is the case for multicellular organisms where the same stem cell can differentiate into different adult cell types. Regulation involves both small and large molecules and it happens at all levels of gene expression.

A similarly large and complex biochemical machinery is needed to produce a copy of DNA at the time of cell division. The process, which is called DNA replication, involves several macromolecules, such as, DNA polymerase, DNA ligase, helicase, etc. In this process, DNA is used as a template on which these complex assemblies of proteins work to produce a copy of the DNA. Errors in copying can produce a non identical copy of DNA which results in two genetically different offsprings, producing the variation necessary for the evolution of life.

\subsection{Diversity and universality of life}
Life appears in various forms and sizes on Earth. The smallest cell measures a few hundred nanometers and the largest organism measures about tens of meters. Life can exist in the harshest of habitats ranging from extremely hot to extremely cold places. The diversity of life can be estimated from the fact that about $10^7 - 10^8$ different eukaryotic species inhabit Earth \cite{May1988, May1990}. A recent study put this number at $8.7 \pm 1.3$ million \cite{Mora2011}. Of this only about 10\% are known to us. In addition, it is estimated that there are about $4 - 6 \times 10^{30}$ prokaryotic cells that live on Earth \cite{Whitman1998}. One may imagine that such a diversity of organisms would require different mechanisms to operate. Though the specific pathways differ amongst different organisms, the biochemical machinery involved in encoding, transcription and translation is universal. It is believed that all life forms diverged from a common ancestor, called the Last Universal Common Ancestor (LUCA). Based on similarity of the genes, proteins, and metabolism, the tree of life has been constructed rooted on a common ancestor \cite{Woese1998, Doolittle1999, Glansdorff2008, Kim2011}.

\subsection{Minimum complexity of life}
{\it Mycoplasma genitalium}, a parasitic bacterium which lives in the primate genital and respiratory tracts, is the smallest known organism capable of independent growth and reproduction. The genome of {\it M. genitalium} consists of 521 genes of which 482 are protein encoding genes \cite{Fraser1995}. In comparison most bacterial genomes typically encode for more than 2000 different proteins. {\it E. coli}, the best studied bacterium, has about 4500 genes. The proteins produced from these genes are involved in metabolism, replication, as well as building and maintenance of structure. A theoretical study based on comparison of the genome of {\it M. genitalium} with that of {\it Haemophilus influenzae} suggested 256 genes that were close to the minimal set necessary and sufficient to sustain the existence of a modern-type cell \cite{Mushegian1996}. It has been estimated that a minimal bacterium will require about 206 protein coding genes based on an analysis of several free-living and endosymbiotic bacterial genomes \cite{Gil2004}. It is worth noting that according to this estimate even the simplest cells require over 200 proteins to function.

\section{Origin of Life: The chicken or the egg?}
\begin{quote}
``{\it If there has been a first man he must have been born without father or mother – which is repugnant to nature. For there could not have been a first egg to give a beginning to birds, or there should have been a first bird which gave a beginning to eggs; for a bird comes from an egg.}''
\begin{flushright}
-- Aristotle (384 -- 322 BC)
\end{flushright}
\end{quote}

\noindent The complexity of life has perplexed and enthralled us, more so the enigma of its origin \cite{Oparin1957, Schrodinger1992, Kauffman1993, Dyson1999, Cairns-Smith1990, Morowitz2004, Duve2005, Kaneko2006, Rasmussen2008a, Deamer2010}; \cite{Dyson1982, Orgel1998, Rode1999, Segre2000a, Bedau2000, Delaye2005, Brack2007, Shapiro2007, Stano2007, Orgel2008, Smith2008, Smith2008a, Smith2008b, Trefil2009, Noireaux2011}. It is difficult to imagine how an organization, as intricate as seen in life, could self-organize on a primordial Earth when none existed. We now review some experiments that have provided a perspective on what was possible on the prebiotic Earth.

\subsection{\label{Miller}Miller's experiment and other related works: The spontaneous production of certain building blocks of life}
\begin{figure}
    \begin{center}
        \includegraphics[width=5in]{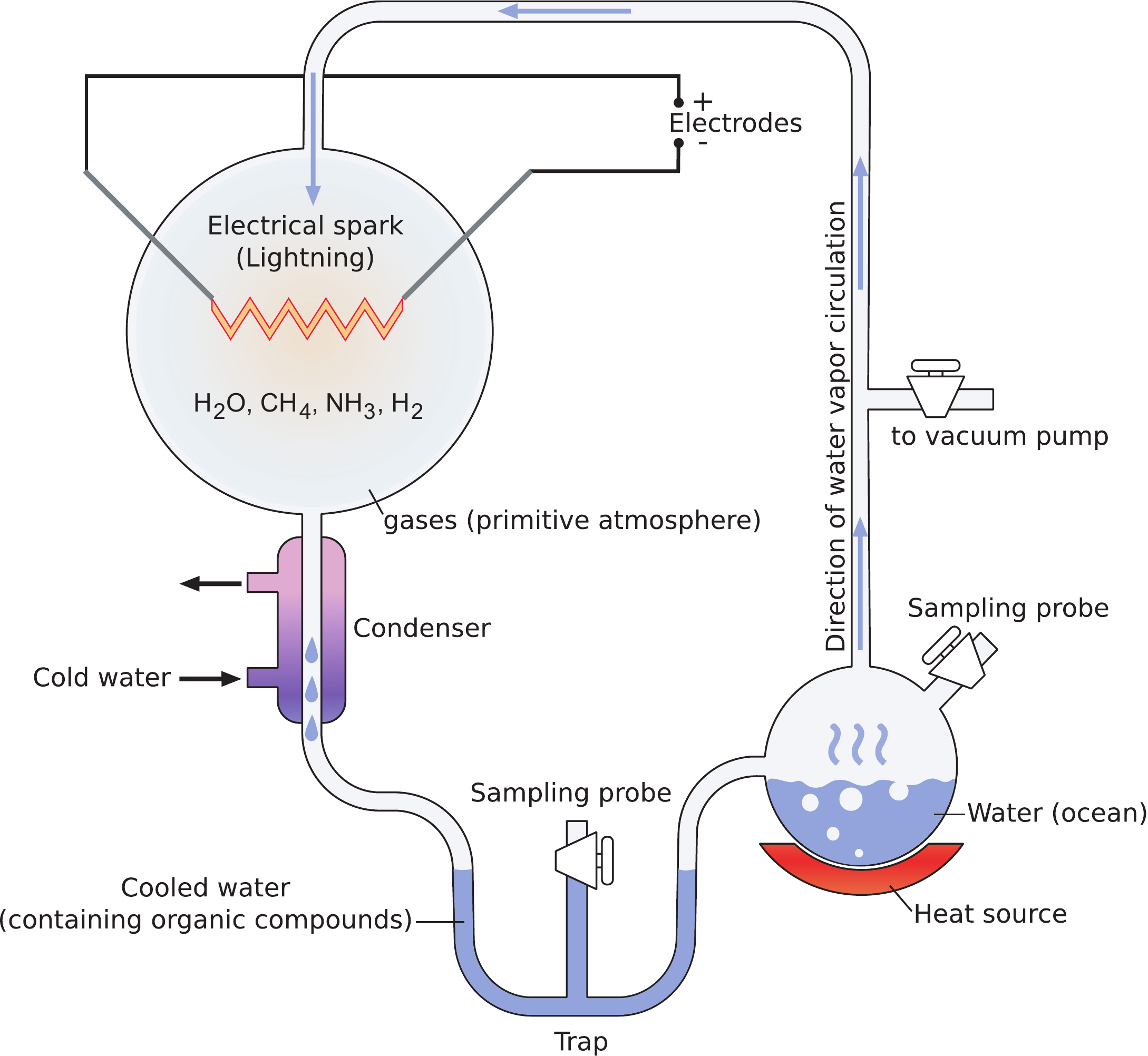}
    \end{center}
    \caption[Schematic of Miller's experiment.]{{\bf Schematic of Miller's experiment.} To simulate early Earth conditions, Miller used water (\ce{H2O}), methane (\ce{CH4}), ammonia (\ce{NH3}), and hydrogen (\ce{H2}) sealed inside a glass flask containing a pair of electrodes and connected in a loop to another flask containing liquid water. The liquid water was heated to induce evaporation and to add water vapour to the chemical mixture. The resulting gases were circulated around the apparatus, simulating the Earth's atmosphere. Sparks were fired between the electrodes to simulate lightning. The flask with heated water represents water on the Earth's surface and the recycled water vapour simulates water evaporating from lakes and seas, before going into the atmosphere and condensing into rain. The atmosphere was cooled again so that the water could condense and trickle back into the first flask in a continuous cycle. After one week of continuous operation, Miller observed a dark brown scum had collected in the lower flask. It was found to contain several types of amino acids, including glycine and alanine, together with sugars, tars, and various other unidentified organic chemicals. Image source: \url {http://en.wikipedia.org/wiki/Miller-Urey_experiment}.}
    \label{Miller-Expt}
\end{figure}

Miller's experiment was the first experiment to show that amino acids can be produced spontaneously in simulated primordial Earth conditions \cite{Miller1953} (see Fig. \ref{Miller-Expt} for the description of the experiment). Miller simulated an early Earth atmosphere that contained methane, ammonia, hydrogen gas, and water vapor. It has been successfully repeated by others who have also considered other possible primordial environmental conditions \cite{Miller1959, Lazcano1983, Miyakawa2002, Ricardo2004, Parker2011}. There have been other experiments that have demonstrated production of purines and pyrimidine ribonucleotides \cite{ORO1961, ORO1961a, ORO1962, Powner2009}.

These experiments are based on the studies that suggest a reducing or partially reducing atmosphere existed on the early Earth \cite{Ferris1975, Tian2005, Kasting2006, Zahnle2010}. Recent findings suggest that the primordial Earth's atmosphere may not be reducing as suggested earlier \cite{Trail2011}. A reducing environment may have existed locally, such as in the presence of iron or other metals in early oceans which could reduce nitrites to ammonia \cite{Henderson-Sellers1980, Summers1993}, or around deep sea hydrothermal vents or volcanic discharges \cite{Russell1997, Russell2004, Wachtershauser2006, Martin2007, Martin2008, Johnson2008} allowing a prebiotic chemistry to initiate.

Collectively these experiments indicate that several of the basic building-block molecules, such as amino acids, other molecules and possibly also nucleotides could be synthesized on the prebiotic Earth by natural processes.

\subsection{What came first?}
The above experiments as well as other work on prebiotic chemistry has led to several hypotheses about what were the first chemical steps towards life on Earth. Broadly these hypotheses have been classified into three categories: `metabolism first', `replication first', or `membranes first'. These three scenarios are not necessarily mutually exclusive. We briefly discuss the three scenarios below.

\begin{description}
  \item[Metabolism first:] In this hypothesis it is argued that a system of chemical reactions producing small organic molecules, metabolites, would have appeared first \cite{Wachtershauser1988, Wachtershauser1990, Dyson1999, Morowitz1999, Morowitz2000, Cody2000, Russell2004, Robinson2005, Shapiro2007, Guzman2009, Trefil2009, Morowitz2010}. One proposed candidate for the most primitive set of chemical reactions is the reductive tricarboxylic acid cycle (rTCA cycle). The TCA cycle is a universal structure found in the metabolic networks of most organisms and is the starting point in these networks for chemical pathways to synthesize all the building block molecules -- amino acids, nucleotides and lipids. The pathways leading to the synthesis of simple organic molecules might be promoted by natural catalysts such as minerals and porous surfaces in rocks, or as an autocatalytic system. The idea is that the origin of life was triggered by the accumulation of simple organic molecules in thermodynamically favorable conditions. The simple molecules formed as result of the above process can combine in various ways to form more complex molecules, the system as a whole constituting a primitive `metabolism'. The production of all the molecules involved in the primordial metabolism is envisaged as a self sustaining system, as it uses as input simple inorganic molecules readily available on the Earth and incrementally builds more complex molecules using catalysis. Eventually, simple peptides will be formed and would lead to better catalysts; as well as nucleic acids that would form the templates for information storage.
  \item[Replication first:] One of the fundamental properties of cells is to replicate. The replication of a whole cell is preceded by the replication of a single molecule, the DNA molecule, which carries information about all the genes. This information is central to how cell works and is transmitted from generation to generation through the above molecular replication. In the replication first scenario, it is believed that first the molecules capable of replication would have arisen. RNA is mooted as the candidate molecule for its ability to store, transmit, and duplicate information (like DNA) and act as catalyst for various chemical processes (ribozyme). It is believed that RNA was once capable of supporting independent life forms before giving rise to the DNA--RNA--protein world of today. This is called the `RNA world' hypothesis. DNA is thought to have taken over the role of data storage due to its increased stability at a later stage, while proteins, through a greater variety of amino acids, replaced RNA's role in specialized catalysis \cite{Gilbert1986, Orgel1986, Joyce1989, Bartel1999, Gesteland1999, Johnston2001, Joyce2002, Orgel2004, Noireaux2005, Lehmann2009, Cech2011}. The RNA world hypothesis suggests that RNA in modern cells is an evolutionary remnant of the RNA world that preceded it.
  \item[Membrane first:] In the membrane first scenario, it is hypothesized that physical enclosures were formed first. In water lipid molecules are known to spontaneously form micelles or vesicles. These vesicles allow selective interaction with the environment and can provide an isolated space for other systems to grow. The idea of lipid vesicles was first proposed by Oparin \cite{Oparin1957}. He proposed that coacervates, which are tiny spherical droplets of assorted organic molecules (specifically, lipid molecules) held together by hydrophobic forces from a surrounding liquid, would form first and followed by cell-like structure. A variant of this scenario is popularly called the `lipid world' \cite{Luisi1999, Segre2000, Segre2001, Monnard2002, Schrum2010}. These vesicles can grow by the addition of monomers and divide to produce two vesicles \cite{Shenhav2007, Zhu2009}. Additionally these vesicles can provide semi-permeable membranes for small molecules to pass through and enable chemical reactions to flourish inside the vesicle \cite{Furusawa2006, Carletti2008, Mansy2008}. As an alternate, proteinoids (protein-like molecules formed by inorganic polymerization of amino acids) have been proposed as the precursors to first living cells. When present in certain concentrations in aqueous solutions, proteinoids form small structures called microspheres or protocells  capable of growth and division \cite{Fox1974, Fox1977}.
\end{description}

None of these theories is self-sufficient and there are several issues that need to be resolved to have a comprehensive scenario for the origin of life \cite{Shapiro1999, Shapiro2007, Orgel2008, Vasas2010}. A common theme that these hypotheses are based upon is that a self sustaining system arises first, providing the right starting point for more complex organization to arise. It could be a metabolism, which gave birth to protein world and subsequently RNA/DNA world; or beginning with an RNA world, that later gave rise to protein and DNA life; or beginning with a lipid world and later giving rise to metabolism/replicating molecules. There are models that discuss co-evolution of more than one subsystem in prebiotic scenarios \cite{Rasmussen2003, Rasmussen2004, Furusawa2006, Szathmary2007, Piedrafita2010, Kamimura2010, Kondo2011}.

\section{Motivation}
Instead of asking specific historical questions about whether the TCA cycle, or RNA molecules, or lipid enclosures came first, one can posit some general questions at a higher level of abstraction. These questions are also motivated by empirical facts about living systems. Our work is motivated by three such observations about the structure and organization of the biochemical machinery of life.
\begin{enumerate}
  \item {\bf The existence of a chemical organization.} As described earlier a cell contains molecules with specific properties carrying out specific carefully orchestrated processes. Without these, it will not have the capability to uptake food sources from the environment, convert them to building blocks of life through chemical processes, grow, replicated and evolve. Life is an `organized' chemical system. Each of the process is well regulated and synchronized with other processes. The existence of such an organization is in itself interesting, but it also poses a question: how did such an organization arise in the first place?
  \item {\bf The question of a sparse biochemistry.} The set of molecules that participate in biochemical pathways is a very small subset of all organic molecules \cite{Morowitz2000, Srinivasan2009, Copley2010}. The Kyoto Encyclopedia of Genes and Genomes (KEGG) database lists 16,311 metabolites and small molecules in all characterized organisms (1,761 organisms) participating in 8,840 biochemical reactions \cite{KEGG, Kanehisa2010}. In a recent analysis of metabolism of {\it Aquifex aeolicus}, a reductive chemoautotroph and the smallest freely living bacteria, only 125 compounds were found in the core metabolism \cite{Srinivasan2009}. In comparison the number of small molecules weighing less than 300 Da has been estimated to be about 13.9 millon \cite{Fink2005} and all molecules of up to 11 atoms of C, N, O, and F possible under consideration of simple valency, chemical stability, and synthetic feasibility rules were found to be 26.4 million in number \cite{Fink2007a}. The fact that biochemistry uses a very sparse subset of molecules point to some underlying pruning principles. The primitive biochemistry need not have been as sparse as the modern biochemistry (selection has operated for long), but would have seeded the selection that exists now. We wish to understand at the level of dynamics, the selection rules that could have been operative in selecting such a sparse set of molecules. Could the chosen subset have some special properties that enable it to spontaneously arise in prebiotic conditions?
  \item {\bf Production of large molecules need large molecules: A chicken-and-egg problem in chemistry.} Large molecules, such as RNA and protein molecules, containing about a hundred to several thousand monomers, are essential in the functioning of life. As discussed in the Section \ref{Life-Description}, these molecules are responsible for the normal functioning of the cell. These molecules are produced inside a cell using machinery which is itself composed of such large molecules. The RNA polymerase reads the genes on DNA molecules and produces the corresponding messenger RNA molecules. The ribosome reads the messenger RNA molecule and produces the corresponding protein molecules. These machines, RNA polymerase and ribosome, are composed of macromolecules such as proteins and RNA molecules that are produced by the same process. When cells produce daughter cells, the latter are already endowed with these macromolecular assemblies at birth, from which they synthesize other molecules. Nowhere in the living world is there a natural process we know of that produces macromolecules and that does not itself use macromolecules. Hence the puzzle: how did the large molecules first arise? There has been much experimental work to explore mechanisms that could enhance the concentrations of monomers and synthesize long polymers \cite{Ferris1996, Unrau1998, Rode1999, Yokoyama2003, Ferris2004, Orgel2004, Budin2010}. While there is interesting progress, as yet there is no compelling scenario for the primordial origin of large molecules \cite{Benner2010, Marshall-Bowman2010}.
\end{enumerate}

One can ask if there are generic processes that might plausibly happen on the prebiotic Earth that could provide an answer to the above questions. Can one theoretically model such processes? Before we turn to models, we discuss one more empirically observed chemical property that seems to be fairly generic.

\subsection{Possible role of small molecule catalysis}
It has been observed that catalysis is a fairly ubiquitous property that arises in different kinds of molecules and even at small sizes. A catalyst is a molecule that speeds up the rate of a reaction without being consumed in the reaction. Enzymes are very powerful catalysts of metabolic reactions in a cell; they speed up the rate of the reactions they catalyze by factors ranging from $10^5$ to $10^{15}$ \cite{Berg2002}. They are also very specific catalysts, in that each enzyme catalyzes typically a single reaction only. Enzymes are very large protein molecules (containing from a few hundred to a few thousand amino acids) and were not available in the early stages of prebiotic chemical evolution.

However, small organic molecules which could have been produced by natural prebiotic processes are also known to have catalytic activity which is weaker and less specific. These include organocatalysts \cite{Barbas2008, MacMillan2008, Bertelsen2009}, peptides (small polymers of amino acids) \cite{Lee1996, Severin1997, Cordova2005, Cordova2006, Brack2007}, and RNA molecules \cite{Cech1986a, Symons1992, Chen2007, Lincoln2009, Turk2010}. Cofactors, which play an important role in catalyzing metabolic reactions (or their evolutionary predecessors) may also have had a role in prebiotic catalysis \cite{Srinivasan2009}.

\subsection{Autocatalytic sets}
The molecules produced in the spontaneous reactions from the prebiotic precursors (produced in Miller type experiements) may have included weak catalysts for other reactions involving the same set of chemicals as reactants, thereby producing in turn a new set of molecules. In this respect, autocatalysis has been mooted as an important self-organizing property. In chemistry, a reaction is said to be autocatalytic if one of its products catalyzes the reaction, e.g., \ce{A + B ->[\ce{C}] C}. An autocatalytic reaction has the consequence that the production of a small quantity of the product, C, results in a positive feedback that produces more of C.

A set of chemical reactions can be said to be ``collectively autocatalytic'' if the products of the set mutually catalyze all the reactions\footnote{A formal definition of autocatalytic sets relevant for our work is given later in Chapter \ref{ChapterACS}, Section \ref{Definition-ACS}.}. Autocatalytic sets (ACSs) were proposed by Eigen \cite{Eigen1971}, Kauffman \cite{Kauffman1971} and Rossler \cite{Rossler1971}. An ACS, like a single autocatalytic reaction, has the property that the production of a small amount of the products of the ACS can, in principle, lead to the collective production of an even greater amount of the same. Thus if an ACS is present in the chemistry this could lead to focusing of chemical resources preferentially into the products of the ACS. This could address the problem of why a sparse set of molecules has come to dominate biochemistry. Furthermore, an ACS is a cooperative organization, in that the products collectively reinforce each other's presence. It is conceivable that ACSs can therefore help us understand the cooperative nature of biological organization.

These properties have motivated many people to explore ACSs as a possible organization principle in the origin of life problem in variety of ways. ACSs have been used by many authors to study various aspects of self-organization, evolution and the origin of metabolism \cite{Farmer1986, Bagley1991, Bagley1991a, Stadler1993, Kauffman1993, Fontana1994, Jain1998, Jain2001, Hanel2005, Piedrafita2010, Filisetti2011}, the origin of replication \cite{Eigen1977, Szathmary2006, Ohtsuki2009}, and the origin and dynamics of  protocells \cite{Segre2000, Furusawa2006, Carletti2008, Sole2009, Kamimura2010}.

\subsection{Models of prebiotic autocatalytic organization}
We now discuss some models of autocatalytic organization that have motivated the present work. A model proposed by Farmer, Kauffman and Packard \cite{Farmer1986} discusses a mechanism based on ACSs via which a selected set of molecules can dominate the chemistry (see Section \ref{Bagley-Description}). Another model of network evolution by Jain and Krishna demonstrates how a complex autocatalytic organization can arise and grow when none exists to begin with \cite{Jain1998, Jain2001, Krishna2004a} (see Section \ref{Jain-Krishna-Description}).

We discuss these two models below and also state their limitations which we address in this work.

\subsubsection{\label{Bagley-Description}A model for autocatalytic chemical focussing}
The model was proposed by Farmer, Kauffman and Packard \cite{Farmer1986} and later discussed and analyzed in series of papers by Bagley et al \cite{Bagley1989}, Bagley \cite{Bagley1991b}, and Bagley and Farmer \cite{Bagley1991}. In the model they consider an artificial chemistry in which polymers could form by ligation of shorter polymers through spontaneous reactions as well as reactions catalyzed by other polymers in the chemistry. They considered a set of monomers $\{a, b, c, \ldots\}$ as the starting material, members of which could combine together to form polymers of different lengths. The polymers are represented as strings of the monomers ($a, b, c, \ldots$), such as, $accaaabd$, where the order in which the monomers combine differentiate a molecule. Furthermore they are considered to be oriented, so that $aab$ and $baa$ would represent different polynomials. The polymers undergo reactions of the form \ce{{\bf A} + {\bf B} <=>[\ce{{\bf E}}] {\bf C} + {\bf H}}, where {\bf A}, {\bf B}, {\bf C} and {\bf E} are the polymers produced in the chemistry and {\bf H} denotes water.

Bagley and Farmer analyzed the population dynamics of the molecular species in a homogeneous (well-stirred) environment and established some important properties of autocatalytic self-organization. When the food set (monomers) were supplied at a fixed input rate and the chemistry contained an ACS they showed that in a suitable range of parameters the concentrations of the ACS molecules dominated over the rest of the molecules (the background), thereby focusing the chemical resources of the system into a small set of molecules comprising the ACS. Bagley, Farmer and Fontana \cite{Bagley1991a} also considered a version of the model wherein new species could enter the system stochastically, resulting in an evolution of the autocatalytic organization.

The identity of polymers considered by Bagley and Farmer depends on the order in which monomers combine. Thus the number of molecules in their chemistry grows exponentially with the length of molecules ($\sim f^n$, $f$ is the number of distinct monomers, $n$ is the length of the polymer). This prevents one from studying large chemistries. In the model presented in this thesis, we ignore the order in which molecules combine allowing us to simulate much larger chemistries.

Though the Bagley-Farmer model was successful in demonstrating a mechanism of autocatalytic focusing, it only discusses this in cases where the largest polymers in the ACSs have about 15-20 monomers. Bagley and Farmer did not systematically investigate the problems that arise in generating much larger molecules in their chemistry. This problem was investigated by Ohtsuki and Nowak \cite{Ohtsuki2009} in a much simpler model in which they showed that for a large molecule to catalyze pathways of its own production starting from small food molecules, it is required to have a large catalytic strength that grows exponentially with the length of the catalyst. However, they did not present a solution to this problem.

\subsubsection{\label{Jain-Krishna-Description}A model of network evolution and emergence of organization}
Another motivation for our present model is the work of Jain and Krishna in which a simple model of catalytic network evolution was discussed \cite{Jain1998, Jain2001, Krishna2004a}.

In the model, they consider a prebiotic soup of chemical species of which some can catalyze the production of other species, forming a catalytic network. On a short time scale, they consider the population dynamics of the constituent chemical species. On a longer time scale evolutionary dynamics are considered in which one of the least populated species in the system is replaced with a new species. No qualitative change to the attractor happens by repeated replacement of species till a special structure, an autocatalytic set, arises in the network. An ACS is formed in the network when two or more species mutually catalyze the production of each other. Nodes downstream to an ACS also do well populationally, these nodes form the periphery of the ACS. This causes nodes outside the ACS to be selected for replacement. Soon, the ACS grows to include the entire network.

This model with its simplicity is able to demonstrate how a complex organization can build around an autocatalytic set over the course of network evolution. Importantly it also shows that this organization builds fairly quickly once an ACS appears in the system. One of the limitations of Jain-Krishna model is that it presupposes a buffered input for all the reactants required to produce the catalysts. One may envisage situations wherein a catalyst itself acts as reactant in some other reaction, causing its population to deplete. This in turn could effect the rates of the reaction in which this molecule act as a catalyst.

One of the motivations of the model discussed in the thesis was to study the Jain-Krishna model with the above mentioned resource limitation. In the model discussed in this thesis, we show that this affects the dynamics.

\section{Thesis organization}
The subsequent chapters in this thesis are organized as follows:
\begin{description}
  \item[Chapter 2] introduces our mathematical model describing the population dynamics of molecules in a stylized but prebiotically plausible artificial chemistry. Larger molecules are produced in this chemistry by the coalescing of smaller ones, with the smallest molecules, the monomers, assumed to be buffered. Some of the molecules produced in the chemistry can also act as the catalysts for the reactions of the chemistry. They do so by enhancing the rate of the reaction in proportion to their concentration by a factor that denotes their catalytic strength. Using mass action kinetics we write the dynamical equations describing populations of molecules in this chemistry. This chapter discusses in detail the examples of chemistries that contain one and two monomers in the food set.
  \item[Chapter 3] presents dynamical results from the model for the case of spontaneous chemistry, \ie, the chemistry in which none of the reactions is catalyzed. It shows that the steady state concentrations of the molecules fall exponentially with their length. It discusses the dependence of the steady state concentrations on the various parameters of the model. Analytical results in a region of phase space are also discussed.
  \item[Chapter 4] discusses the dynamics of the model when the chemistry contains some catalyzed reactions, and in particular autocatalytic sets. It defines autocatalytic sets in context of our model and discusses various dynamical features that arise in the system in the presence of an ACS in the chemistry. Using a specific example of an autocatalytic set we show that a sparse subset of molecules can dominate above the rest of the molecules in the chemistry (the background) under certain parameter regimes. We present an analytical understanding for this phenomena. We show how a large catalytic strength can counteract the effect of dissipation for the molecules produced in catalyzed reactions. We also show that ACSs can produce bistability in the population dynamics. This chapter also discusses the relation between the topology of ACSs, system parameters and the existence of bistability. The work described in this chapter extends the work of Bagley and Farmer on autocatalytic focusing.
  \item[Chapter 5] articulates the problem of producing large catalysts using ACSs. For a class of ACSs, referred to as the `extremal' ACSs, this chapter shows how the region of bistability changes with the length of the catalyst. It discusses that the catalytic strength required by a catalyst to dominate the chemistry starting from an initial condition relevant to the origin of life problem, grows exponentially with the size of the catalyst. This generalizes the results of Ohtsuki and Nowak to a larger class of chemistries and emphasizes the difficulty in producing large molecules via the ACS mechanism.
  \item[Chapter 6] introduces the mechanism of nested ACSs. It discusses in detail an example of two nested ACSs where the smaller ACS reinforces the large one. We discuss how the relative topologies of the two ACSs and the background chemistry affect the reinforcement received by the larger ACSs. This chapter discusses how the mechanism of nested ACSs can be used to overcome the problem of producing large molecules presented in Chapter 5.
  \item[Chapter 7] describes two algorithms, based on the discussion in Chapter 6, that produce a hierarchy of nested ACSs. It demonstrates that chemistries that contain a structure such as nested-ACSs can produce large molecules in prebiotically plausible conditions. We show an example where a molecule of size 441 (\ie, a molecule containing 441 units of monomer) dominates the chemistry starting from a initial condition where only monomers were present and at a catalytic strength that only grows as a power of length of the catalysts (as against exponentially needed otherwise).
  \item[Chapter 8] is a discussion of the results. It discusses some prebiotic scenarios  where our model could be applied. It also discusses limitations and caveats of the model and directions for further work.
  \item[Appendix A.] Though the model considers an infinite sized chemistry, it requires for numerical simulations to consider only a finite subset of the chemistry. This introduces an additional parameter $N$, the size of the largest molecule considered. This appendix presents the evidence for the $N$-independence of results for large $N$.
  \item[Appendix B.] The model discussed in this thesis considers a boundary condition in which the food set molecules are considered to be buffered, \ie, their concentrations remain a constant over the course of dynamics. We discuss in this appendix another boundary condition in which the monomers are supplied at a fixed input rate.
  \item[Appendix C] discusses the special case of our model under which it reduces the model of Ohtsuki and Nowak.
  \item[Appendix D] discusses examples of catalyzed chemistries that do not contain ACSs and the origin of multistability for an ACS.
  \item[Appendix E] presents a discussion on how the topology of an ACS, especially the assignment of the catalysts to its reactions, affects the dynamics.
  \item[Appendices F, G and H] list the reactions with their respective catalysts for various examples of catalyzed chemistry for $f=1$ and $2$ containing a cascade of nested ACSs generated using Algorithm 4 and 5 (described in Chapter \ref{ChapterHierarchyNestedACSs}).
  \item[Appendix I] describes various programs used to generate results reported in this thesis. The programs are available from the following link:\\\url{http://sites.google.com/site/varungiri/research/phd-thesis}
\end{description}

\thispagestyle{plain}
\cleardoublepage
\chapter{\label{ChapterModel}A model of pre-biotic chemistry}

\lettrine[lines=2, lhang=0, loversize=0.0, lraise=0.0]{I}{n} this chapter, we describe a model of pre-biotic artificial chemistry in which larger molecules are produced by coalescing of smaller ones via spontaneous and catalyzed reactions. In the following chapters, we use this model to study how large molecules, relevant for the origin of life, can arise in a pre-biotic scenario. We explore the model numerically and, in certain regimes, analytically.

\section{\label{SectionModel}The model}
The model is specified by describing the set of molecular species, their reactions, and the dynamical rate equations for their population dynamics. A special set of molecules, the `\gls{FoodSet}', denoted ${\mathcal F}$, consists of small molecular species, $f$ in number, that are presumed to be abundantly present in a prebiotic niche.
\begin{equation}
\label{foodset}
    \mathcal{F} = \{\mathrm{\mathbf A}_1, \mathrm{\mathbf A}_2, \ldots, \mathrm{\mathbf A}_f\}
\end{equation}
A general molecule, say $\mathrm{\mathbf A}$, is represented as an $f$-tuple of non-negative integers: $\mathrm{\mathbf A} = (a_1, a_2, \ldots, a_f)$, where $a_l$ is the number of monomers of type $\mathrm{\mathbf A}_l$ contained in $\mathrm{\mathbf A}$. The `size' or `length' $n$ of the molecule is defined as the total number of monomers of all types in it: $n = \sum_l a_l$. The \gls{FoodSet} molecules themselves are represented by the $f$-tuples $\mathrm{\mathbf A}_1 = (1,0,0,\ldots,0)$, $\mathrm{\mathbf A}_2 = (0,1,0,0,\ldots,0)$, $\ldots$, $\mathrm{\mathbf A}_f = (0,0,\ldots,0,1)$.

The identity of a molecule in the model is completely determined by the number of monomers of each type contained in the molecule; the order in which they appear is irrelevant. Thus the combinatorial diversity of distinct compounds containing a total of $n$ monomers (of all types) grows only as a power of $n$ ($\sim n^{f-1}$) instead of exponentially ($\sim f^n$ for strings) if the order had mattered. This simplification helps in picturizing the chemistry and significantly reducing the computational power needed to explore large values of $n$.

Starting from the food set, larger molecules in this chemistry are made through a series of ligation and cleavage reactions of the type $\mathrm{\mathbf B} + \mathrm{\mathbf C} \rightleftharpoons \mathrm{\mathbf A}$, where $\mathrm{\mathbf A}=(a_1, a_2, \ldots, a_f)$, $\mathrm{\mathbf B}=(b_1, b_2, \ldots, b_f)$ and $\mathrm{\mathbf C}=(c_1, c_2, \ldots, c_f)$, with $a_l = b_l+c_l$. If $x_{\mathrm A}$ denotes the concentration of the molecule $\mathrm{\mathbf A}$, using mass action kinetics, the rate equations are given by
\begin{subequations}
  \label{rateequation-generalmodel}
  \begin{align}
    \dot x_{\mathrm{A}} &= 0, & \quad \mathrm{if\ } \mathrm{\mathbf A} \in \mathcal{F}, \\
    \label {rateequation-generalmodel-nonF}
    &= \sum_{(\mathrm{B,C}) \in {\mathcal Q}_{\mathrm A}} v_{\mathrm{BC}} - \sum_{\mathrm B, B \ne A}  v_{\mathrm{AB}} - 2 v_{\mathrm{AA}} - \phi_{\mathrm{A}} x_{\mathrm{A}}, & \mathrm{otherwise}.
  \end{align}
\end{subequations}
It is assumed that in a well stirred prebiotic region the concentration of the members of the set $\mathcal{F}$ are constant (buffered)\footnote{In a prebiotic niche, such as a hydrothermal vent, even as the small molecules are used up in chemical reactions to form other products, natural geochemical processes replenish these molecules keeping their concentrations at a significant level. For modeling purposes we assume that their concentrations are constant in the region in question.}.
\begin{equation}
  \label{flux-v}
  v_{\mathrm{XY}}  = k^F_{\mathrm{XY}}x_{\mathrm X} x_{\mathrm Y} - k^R_{\mathrm{XY}}x_{\mathrm{Z}}
\end{equation}
is the net forward flux of the reaction pair $\mathrm{\mathbf X} + \mathrm{\mathbf Y} \rightleftharpoons \mathrm{\mathbf Z}$ with forward (ligation) rate constant $k^F_{\mathrm{XY}}$ and reverse (cleavage) rate constant $k^R_{\mathrm{XY}}$, $\phi_{\mathrm A}$ is the loss rate of $\mathrm{\mathbf A}$, and ${\mathcal Q}_{\mathrm A}$ represents the set of unordered pairs of molecules which can combine together to form $\mathrm{\mathbf A}$ (${\mathcal Q}_{\mathrm A} = \{(\mathrm{\mathbf B},\mathrm{\mathbf C}): b_l+c_l = a_l\ \forall\ l=1,2,\ldots,f\}$).

The two terms in the first sum, in Eq. (\ref{rateequation-generalmodel-nonF}), represent the formation (cleavage) of $\mathrm{\mathbf A}$ from (into) smaller molecules. The second sum and the following term represent the cleavage (formation) of larger molecules via reactions that produce (consume) $\mathrm{\mathbf A}$. The stoichiometric factor of 2 before $v_{\mathrm{AA}}$ arises because two molecules of $\mathrm{\mathbf A}$ are involved in the corresponding reaction pair.

The set of parameters $k^F_{\mathrm{XY}}$, $k^R_{\mathrm{XY}}$ that are non-zero define the set of possible reactions; collectively they define the `\gls{spontaneous-chemistry}' (`spontaneous' in the sense that the reactions are possible even in the absence of catalysts). A pair of ligation and cleavage reactions can be excluded from the chemistry by setting both $k^F_{\mathrm{XY}}$ and $k^R_{\mathrm{XY}}$ to zero. The scheme permits chemistries in which some reactions proceed in only one direction (ligation or cleavage) by setting only one of $k^F_{\mathrm{XY}}$ and $k^R_{\mathrm{XY}}$ to zero. However, we will primarily be interested in a chemistry in which each reaction is reversible. The existence of the cleavage reactions makes it more difficult for the long molecules to survive; thus it is more significant to demonstrate the appearance of long molecules in a model in which cleavage reactions are permitted than in one where only the forward (ligation) reactions are.

\subsection{\label{nomenclature}Nomenclature}
We classify \glspl{spontaneous-chemistry} based on their connectivity and heterogeneity of parameters.
\begin{description}
\item[\Gls{rev-chemistry}] A chemistry is called `reversible' if each allowed reaction is reversible, \ie, $k^F \neq 0 \Leftrightarrow k^R \neq 0$.
\item[\Gls{connected-chemistry}] A chemistry is `connected' if every molecule can be produced from the food set in some pathway consisting of a sequence of allowed reactions. \glsadd{fully-connected-chemistry}A chemistry is `fully connected' if all possible ligation and cleavage reactions are allowed: $k^F_{\mathrm{XY}} \neq 0$ and $k^R_{\mathrm{XY}} \neq 0$ for all $\mathrm{\mathbf X}$,$\mathrm{\mathbf Y}$. A fully connected chemistry is a \gls{rev-chemistry} by definition. \glsadd{sparse-chemistry}A chemistry is called `sparse' if it contains only a small fraction of all possible reactions. A sparse chemistry is not necessarily connected.
\item[\Gls{homogeneous-chemistry}] A chemistry is called `homogeneous' if all the non-zero rate constants, loss rates, and concentrations of the food set molecules are independent of the species labels: $k^F_{\mathrm{XY}} \neq 0 \implies k^F_{\mathrm{XY}} = k_f$, $k^R_{\mathrm{XY}} \neq 0 \implies k^R_{\mathrm{XY}} = k_r$ independent of ${\mathrm{\mathbf X}}$ and ${\mathrm{\mathbf Y}}$, $\phi_{\mathrm{X}} = \phi$ for all $\mathrm{\mathbf X}$, and, $x_{\mathrm {A}_1} = x_{\mathrm {A}_2} = \ldots = x_{\mathrm {A}_f} = A$.
\end{description}

\subsection{\label{subsection-Catalysis}Inclusion of catalysts}
We consider a simple scheme for inclusion of catalysts. We assume that a molecule enhances the rate of a reaction that it catalyzes in proportion to its own concentration. Thus, if $\mathrm{\mathbf C}$ is a catalyst of the reaction pair $\mathrm{\mathbf X} + \mathrm{\mathbf Y} \rightleftharpoons \mathrm{\mathbf Z}$, then the rate constants of this reaction pair, $k^F_{\mathrm{XY}}$ and $k^R_{\mathrm{XY}}$, are replaced
\begin{subequations}
  \label{catalyzed-rateconstants}
  \begin{align}
    k^F_{\mathrm{XY}} &\rightarrow k^F_{\mathrm{XY}}(1 + \kappa^{\mathrm{XY}}_\mathrm{C} x_\mathrm{C})\ \mathrm{and }\\
    k^R_{\mathrm{XY}} &\rightarrow k^R_{\mathrm{XY}}(1 + \kappa^{\mathrm{XY}}_\mathrm{C} x_\mathrm{C}),
  \end{align}
\end{subequations}
where $\kappa^{\mathrm{XY}}_\mathrm{C}$ is the `\gls{catalytic-strength}' of the catalyst for this reaction pair. The first term in the bracket, unity, represents the spontaneous reaction rate (which is present irrespective of whether the reaction is catalyzed or not), and the second term $\kappa^{\mathrm{XY}}_\mathrm{C} x_\mathrm{C}$ represents the enhancement of the reaction rate due to the catalyst. Note that in this scheme a catalyst enhances both the forward and reverse reaction rates by the same factor. If a reaction has multiple catalysts, $\kappa^{\mathrm{XY}}_\mathrm{C} x_\mathrm{C}$ is replaced by $\sum_\mathrm{C} \kappa^{\mathrm{XY}}_\mathrm{C} x_\mathrm{C}$, where the sum runs over all catalysts $\mathrm{\mathbf C}$ of the reaction in question. Typically, only a small subset of the spontaneous reactions will be catalyzed. The set of catalyzed reactions together with the catalysts and their catalytic strengths will be referred to as the `\gls{catalyzed-chemistry}'.

We refer to a \gls{catalyzed-chemistry} as \glsadd{homogeneous-catalyzed-chemistry}`homogeneous' if the catalytic strength of every catalyst for each reaction that it catalyzes is the same, \ie, if $\kappa^{\mathrm{XY}}_\mathrm{C} \neq 0$, then $\kappa^{\mathrm{XY}}_\mathrm{C} = \kappa$.

Eqs. (\ref{rateequation-generalmodel}) in conjunction with (\ref{flux-v}) and (\ref{catalyzed-rateconstants}) define a deterministic non-linear dynamical system on a complex network. The network structure is specified by the nonzero constants $k^F_{\mathrm{XY}}$, $k^R_{\mathrm{XY}}$, and $\kappa^{\mathrm{XY}}_\mathrm{C}$.

\subsection{\label{numerical-simulations}Numerical simulations}
While the chemistry under consideration is infinite, numerical simulations were done by choosing a finite number $N$ for the size of the largest molecule in the simulation. For a \gls{finite-chemistry} we exclude all the ligation reactions (and their reverse) that produce molecules of size larger than $N$. In principle this introduces another parameter, $N$, an artifact of the simulation. However, we find that most properties of physical interest become independent of $N$ when $N$ is sufficiently large (evidence for this is presented later in Appendix \ref{Appendix-N-independence}).

Our numerical work, which comprised of solving coupled ordinary differential equations, was mostly done using the CVODE solver library of the SUNDIALS (Suite of Nonlinear and Differential/Algebraic Equation Solvers) package v.2.3.0\footnote{\url{https://computation.llnl.gov/casc/sundials/main.html}} \cite{SUNDIALS}, and, for smaller $N$ values, using XPPAUT v.6.11\footnote{\url{http://www.math.pitt.edu/~bard/xpp/xpp.html}} \cite{XPPAUT}. Steady states obtained were verified using numerical root finders in Octave\footnote{\url{http://www.gnu.org/software/octave}} and Mathematica \cite{Mathematica7}.

\section{\label{subsectionModelF1}Chemistries with $f=1$}
The simplest version of the model ($f=1$) contains only a single monomer species. Here, instead of using a `1-tuple', we represent the molecule of size $n$ by the notation $\mathrm{\mathbf A}(n)$ for clarity. The concentration of the monomer,  $\mathrm{\mathbf A}(1)$, is represented by $x_1 (= A)$ and of the other molecules, $\mathrm{\mathbf A}(2), \mathrm{\mathbf A}(3), \ldots$ (dimers, trimers, etc.), is given by $x_2, x_3, \ldots$ respectively. Following Eqs. (\ref{rateequation-generalmodel}), the rate equations for the system are given by $\dot{x}_1 = 0$, and, for $n=2,3,\ldots$,
\begin{align}
    \label{rateequation-v-F1}
    \dot{x}_n =& \sum_{i\leq j,i+j=n}v_{ij} - \sum_{i=1, i \neq n}^\infty v_{in} - 2v_{nn} - \phi_n x_n \\
    \label{rateequationF1}
    \nonumber =& \sum_{i\leq j, i+j=n}(k^F_{ij}x_ix_j - k^R_{ij}x_n) + \sum_{i=1, i \neq n}^\infty(k^R_{in}x_{i+n} - k^F_{in}x_ix_n) + 2(k^R_{nn}x_{2n} - k^F_{nn}x_n^2) \\ & - \phi_n x_n,
\end{align}
where $v_{ij} = (k^F_{ij}x_ix_j - k^R_{ij}x_{i+j})$ is the net forward flux of the reaction pair $\mathrm{\mathbf A}(i) + \mathrm{\mathbf A}(j) \rightleftharpoons \mathrm{\mathbf A}(i+j)$, $k^F_{ij}$ and $k^R_{ij}$ being the forward and reverse rate constants, respectively.
\noindent Explicitly,
\begin{subequations}
  \begin{align}
    \nonumber
    \dot x_2 =& \left(k^F_{11}x_1x_1 - k^R_{11}x_{2}\right) - \left(k^F_{12}x_1x_2 - k^R_{12}x_{3}\right) - 2\left(k^F_{22}x_2^2 - k^R_{22}x_{4}\right) \\
              & - \left(k^F_{32}x_3x_2 - k^R_{32}x_{5}\right) - \ldots - \phi_2 x_2 \\
    \nonumber
    \dot x_3 =& \left(k^F_{12}x_1x_2 - k^R_{12}x_{3}\right) - \left(k^F_{13}x_1x_3 - k^R_{13}x_{4}\right) - \left(k^F_{23}x_2x_3 - k^R_{23}x_{5}\right) \\
              & - 2\left(k^F_{33}x_3x_3 - k^R_{33}x_{6}\right) - \left(k^F_{43}x_4x_3 - k^R_{43}x_7\right) - \ldots - \phi_3 x_3 \\
    \nonumber
    \dot x_4 =& \left(k^F_{13}x_1x_3 - k^R_{13}x_{4}\right) + \left(k^F_{22}x_2x_2 - k^R_{22}x_{4}\right) - \left(k^F_{14}x_1x_4 - k^R_{14}x_{5}\right) \\
    \nonumber
              & - \left(k^F_{24}x_2x_4 - k^R_{24}x_{6}\right) - \left(k^F_{34}x_3x_4 - k^R_{34}x_{7}\right) - 2\left(k^F_{44}x_4^2 - k^R_{44}x_{8}\right)\\
              & - \left(k^F_{54}x_5x_4 - k^R_{54}x_{9}\right) - \ldots - \phi_4 x_4 \\
    \nonumber
    \vdots
  \end{align}
\end{subequations}

For a \gls{finite-chemistry}, in simulating Eq. (\ref{rateequationF1}), all terms corresponding to reactions in which any molecule larger than $\mathrm{\mathbf A}(N)$ is produced or consumed are omitted. This results in $N$ molecules and $N-1$ coupled ordinary differential equations, given by:
\begin{align}
    \label{rateequation-N}\nonumber
    \dot x_n =& \sum_{i \leq j,i+j=n}\left(k^F_{ij}x_ix_j - k^R_{ij}x_n\right) - \sum_{i=1,i \neq n}^{(N-n)} \left(k^F_{in}x_ix_n - k^R_{in}x_{i+n}\right) \\
              &- 2\left(k^F_{nn}x_n^2 - k^R_{nn}x_{2n}\right) - \phi_n x_n
\end{align}
for $n=2,3,\ldots,N$ (the third bracket is absent for $n>N/2$). Note that $\infty$ is replaced by $(N-n)$ in the second summation so that the largest molecule produced is of length $N$. For the case of \gls{homogeneous-chemistry} Eq. (\ref{rateequation-N}) becomes
\begin{align}
    \label{rateequation-N-homogeneous}\nonumber
    \dot x_n =& \sum_{i \leq j,i+j=n}\left(k_fx_ix_j - k_rx_n\right) - \sum_{i=1,i \neq n}^{(N-n)} \left(k_fx_ix_n - k_fx_{i+n}\right) \\
              &- 2\left(k_fx_n^2 - k_rx_{2n}\right) - \phi x_n.
\end{align}

\subsection{\label{ExampleF1}An example with $f=1, N=6$}
The fully connected spontaneous chemistry has 9 reaction pairs of forward and reverse reactions:
\begin{center}
  R1: {\bf A}(1) + {\bf A}(1) \reactionrevarrow{\ensuremath{k^F_{11}}}{\ensuremath{k^R_{11}}} {\bf A}(2) \\
  R2: {\bf A}(1) + {\bf A}(2) \reactionrevarrow{\ensuremath{k^F_{12}}}{\ensuremath{k^R_{12}}} {\bf A}(3) \\
  R3: {\bf A}(1) + {\bf A}(3) \reactionrevarrow{\ensuremath{k^F_{13}}}{\ensuremath{k^R_{13}}} {\bf A}(4) \\
  R4: {\bf A}(2) + {\bf A}(2) \reactionrevarrow{\ensuremath{k^F_{22}}}{\ensuremath{k^R_{22}}} {\bf A}(4) \\
  R5: {\bf A}(1) + {\bf A}(4) \reactionrevarrow{\ensuremath{k^F_{14}}}{\ensuremath{k^R_{14}}} {\bf A}(5) \\
  R6: {\bf A}(2) + {\bf A}(3) \reactionrevarrow{\ensuremath{k^F_{23}}}{\ensuremath{k^R_{23}}} {\bf A}(5) \\
  R7: {\bf A}(1) + {\bf A}(5) \reactionrevarrow{\ensuremath{k^F_{15}}}{\ensuremath{k^R_{15}}} {\bf A}(6) \\
  R8: {\bf A}(2) + {\bf A}(4) \reactionrevarrow{\ensuremath{k^F_{24}}}{\ensuremath{k^R_{24}}} {\bf A}(6) \\
  R9: {\bf A}(3) + {\bf A}(3) \reactionrevarrow{\ensuremath{k^F_{33}}}{\ensuremath{k^R_{33}}} {\bf A}(6) \\
\end{center}
From Eq. (\ref{rateequation-N}), the 5 equations defining the dynamics are:
\begin{subequations}
 \begin{align}
    \nonumber
    \dot x_2 = & (k^F_{11} x_1^2 - k^R_{11} x_2)- (k^F_{12} x_1x_2 - k^R_{12}x_3) - 2(k^F_{22}x_2^2 - k^R_{22}x_4) - \\
               & (k^F_{23}x_2x_3 - k^R_{23}x_5) - (k^F_{24}x_2x_4 - k^R_{24}x_6) - \phi_2 x_2, \\
    \nonumber
    \dot x_3 = & (k^F_{12} x_1x_2 - k^R_{12} x_3) - (k^F_{13} x_1x_3 - k^R_{13}x_4) - (k^F_{23} x_2x_3 - k^R_{23}x_5) - \\
               & 2(k^F_{33}x_3^2 - k^R_{33}x_6) - \phi_3 x_3, \\
    \nonumber
    \dot x_4 = & (k^F_{13} x_1x_3 - k^R_{13} x_4) + (k^F_{22} x_2^2 - k^R_{22}x_4) - (k^F_{14} x_1x_4 - k^R_{14}x_5) - \\
               & (k^F_{24}x_2x_4 - k^R_{24}x_6) - \phi_4 x_4, \\
    \dot x_5 = & (k^F_{14} x_1x_4 - k^R_{14} x_5) + (k^F_{23} x_2x_3 - k^R_{23}x_5) - (k^F_{15} x_1x_5 - k^R_{15}x_6) - \phi_5 x_5, \\
    \dot x_6 = & (k^F_{15} x_1x_5 - k^R_{15} x_6) + (k^F_{24} x_2x_4 - k^R_{24}x_6) + (k^F_{33} x_3^2 - k^R_{33}x_6) - \phi_6 x_6.
 \end{align}
\end{subequations}
For a homogeneous chemistry, $k^F_{ij} = k_f$, $k^R_{ij} = k_r$ and $\phi_i = \phi$, for all $i,j$, the equations become
\begin{subequations}
 \begin{align}
    \dot x_2 =& k_f x_1^2 - k_r x_2- k_fx_2(x_1 + 2x_2 + x_3 + x_4) + k_r(x_3 + 2x_4 + x_5 + x_6) - \phi x_2 \\
    \dot x_3 =& k_f x_1x_2 - k_r x_3 - k_fx_3(x_1 + x_2 + 2x_3) + k_r(x_4 + x_5 + 2x_6) - \phi x_3 \\
    \dot x_4 =& k_f(x_1x_3 + x_2^2) - 2k_rx_4 - k_fx_4(x_1 + x_2) + k_r(x_5 + x_6) - \phi x_4 \\
    \dot x_5 =& k_f(x_1x_4 + x_2x_3) - 2k_rx_5 - k_fx_5(x_1) + k_r(x_6) - \phi x_5 \\
    \dot x_6 =& k_f(x_1x_5 + x_2x_4 + x_3^2) - 3k_rx_6 - \phi x_6
 \end{align}
\end{subequations}

\subsubsection{\label{f1-catalyst}An example of inclusion of catalyst}
Say reaction pairs R1 and R4 of the previous example are catalyzed by ${\mathrm A}(4)$ with catalytic strengths $\kappa^{11}_4$ and $\kappa^{22}_4$ respectively. Then the equations for $x_2$ and $x_4$ get modified to
\begin{subequations}
 \begin{align}
    \nonumber
    \dot x_2 =& (1 + \kappa^{11}_4x_4)(k^F_{11} x_1^2 - k^R_{11} x_2) - (k^F_{12} x_1x_2 - k^R_{12}x_3)
    \\ \nonumber & - 2(1 + \kappa^{22}_4x_4)(k^F_{22}x_2^2 - k^R_{22}x_4) - (k^F_{23}x_2x_3 - k^R_{23}x_5)
    \\ & - (k^F_{24}x_2x_4 - k^R_{24}x_6) - \phi_2 x_2, \\
    \nonumber
    \dot x_4 =& (k^F_{13} x_1x_3 - k^R_{13} x_4) + (1 + \kappa^{22}_4x_4)(k^F_{22} x_2^2 - k^R_{22}x_4) - (k^F_{14} x_1x_4 - k^R_{14}x_5) \\ & - (k^F_{24}x_2x_4 - k^R_{24}x_6) - \phi_4 x_4.
 \end{align}
\end{subequations}

The equations for $\dot x_3$, $\dot x_5$ and $\dot x_6$ remain as before. If the catalyzed chemistry is homogeneous, then $\kappa^{11}_4 = \kappa^{22}_4 = \kappa$.

\section{\label{subsectionModelF2}Chemistries with $f=2$}
For a $f=2$ chemistry, ${\mathcal F} = \{\mathrm{\mathbf A}_1, \mathrm{\mathbf A}_2\}$, a general molecule is given by $\mathrm{\mathbf A} = (a_1, a_2)$, and the monomers are given by the 2-tuples (1,0) and (0,1). There are $(n+1)$ molecules of length $n$, given by the 2-tuples: $(n,0),(n-1,1),\ldots,(1,n-1),(0,n)$. A \gls{finite-chemistry}, has a total of $\frac{(N+1)(N+2)}{2}-1$ molecules (including monomers), and hence $\frac{(N+1)(N+2)}{2}-3$ rate equations.

Since the number of species goes as $N^2/2$ and the number of reactions as $\sim N^4$ for a chemistry with $f=2$, computational limitations require us to work with a smaller $N$ than for $f=1$. Qualitative conclusions nevertheless appear to be $N$ independent.

\subsection{\label{ExampleF2}An example with $f=2, N=3$}
The fully connected spontaneous chemistry has 9 reaction pairs which we label as $\mathrm{R1}, \mathrm{R2},\ldots, \mathrm{R9}$.
\begin{center}
  R1: (0,1) + (0,1) \reactionrevarrow{}{} (0,2) \\
  R2: (0,1) + (1,0) \reactionrevarrow{}{} (1,1) \\
  R3: (1,0) + (1,0) \reactionrevarrow{}{} (2,0) \\
  R4: (0,1) + (0,2) \reactionrevarrow{}{} (0,3) \\
  R5: (0,1) + (1,1) \reactionrevarrow{}{} (1,2) \\
  R6: (1,0) + (0,2) \reactionrevarrow{}{} (1,2) \\
  R7: (0,1) + (2,0) \reactionrevarrow{}{} (2,1) \\
  R8: (1,0) + (1,1) \reactionrevarrow{}{} (2,1) \\
  R9: (1,0) + (2,0) \reactionrevarrow{}{} (3,0) \\
\end{center}
Using Eqs. (\ref{rateequation-generalmodel}) suitably modified for a finite chemistry, we can write the following rate equations:
\begin{subequations}
\label{f2-fluxequations}
\begin{align}
    \dot x_{(0,2)} &= v_{\mathrm R1} - v_{\mathrm R4} - v_{\mathrm R6} - \phi_{(0,2)} x_{(0,2)} \\
    \label{f2-fluxequations-11}\dot x_{(1,1)} &= v_{\mathrm R2} - v_{\mathrm R5} - v_{\mathrm R8} - \phi_{(1,1)} x_{(1,1)} \\
    \dot x_{(2,0)} &= v_{\mathrm R3} - v_{\mathrm R7} - v_{\mathrm R9} - \phi_{(2,0)} x_{(2,0)} \\
    \dot x_{(0,3)} &= v_{\mathrm R4} - \phi_{(0,3)} x_{(0,3)} \\
    \dot x_{(1,2)} &= v_{\mathrm R5} + v_{\mathrm R6} - \phi_{(1,2)} x_{(1,2)} \\
    \dot x_{(2,1)} &= v_{\mathrm R7} + v_{\mathrm R8} - \phi_{(2,1)} x_{(2,1)} \\
    \dot x_{(3,0)} &= v_{\mathrm R9} - \phi_{(3,0)} x_{(3,0)}.
\end{align}
\end{subequations}
$v_{\mathrm Ri}$ is the net forward flux of the reaction pair R$i$. Eqs. (\ref{f2-fluxequations}) can be expanded to get the equations in terms of the $x$'s. For example, Eq. (\ref{f2-fluxequations-11}) becomes
\begin{align}
    \nonumber
    \dot x_{(1,1)} =& \left(k^F_{\mathrm R2}x_{(0,1)}x_{(1,0)} - k^R_{\mathrm R2}x_{(1,1)}\right) - \left(k^F_{\mathrm R5}x_{(0,1)}x_{(1,1)} - k^R_{\mathrm R5}x_{(1,2)}\right) \\
       &  - \left(k^F_{\mathrm R8}x_{(1,0)}x_{(1,1)} - k^R_{\mathrm R8}x_{(2,1)}\right) - \phi_{(1,1)} x_{(1,1)}
\end{align}
where, $k^F_{{\mathrm R}i}$ and $k^R_{{\mathrm R}i}$ are the forward and reverse rate constant of reaction ${\mathrm R}i$, respectively.

When the catalysts are included, the reactions that involve catalysts are changed in the same way as in $f=1$ case (see Section \ref{f1-catalyst}).

\section{\label{rateeq-in-dimless-var}Rate equations in terms of dimensionless variables}
The rate equations (Eqs. (\ref{rateequation-generalmodel})) can be cast in dimensionless form by introducing a concentration scale $\omega$ and a time scale $\tau$. We discuss below the case of a homogeneous\glsadd{homogeneous-chemistry} \gls{spontaneous-chemistry} with $f=1$.

We define dimensionless quantities $u_n \equiv x_n/\omega$, $t' \equiv t/\tau$, $k_f' \equiv k_f \omega\tau$, $k_r' \equiv k_r\tau$ and $\phi' \equiv \phi\tau$. Then
\begin{equation}
    \dot x_n = \frac{dx_n}{dt} = \frac{d(u_n\omega)}{d(t'\tau)} = \frac{\omega}{\tau}\frac{du_n}{dt'},
\end{equation}
and Eq. (\ref{rateequationF1}) becomes
\begin{align}\label{rateeqndimless}
    \nonumber
    \frac{\omega}{\tau}\frac{du_n}{dt'} =& \sum_{i \leq j,i+j=n}\left(\frac{k_f'}{\omega\tau}(\omega u_i)(\omega u_j) - \frac{k_r'}{\tau}(\omega u_n)\right) \\ \nonumber
    & - \sum_{i=1,i \neq n}^{\infty}\left(\frac{k_f'}{\omega\tau} (\omega u_i)(\omega u_n) - \frac{k_r'}{\tau}(\omega u_{i+n})\right) \\
    & - 2\left(\frac{k_f'}{\omega\tau}(\omega u_n)^2 - \frac{k_r'}{\tau}(\omega u_{2n})\right) - \frac{\phi'}{\tau} (\omega u_n) \\
    \nonumber
    \frac{du_n}{dt'} =& \sum_{i \leq j,i+j=n}\left(k_f' u_i u_j - k_r'u_n\right) - \sum_{i=1,i \neq n}^{\infty}\left(k_f'u_i u_n - k_r'u_{i+n}\right)\\
    & - 2\left(k_f'u_n^2 - k_r'u_{2n}\right) - \phi' u_n
\end{align}

Without loss of generality one can choose $\omega = x_1 = A$ and (whenever $k_r \neq 0$) $\tau = 1/k_r$. Then the dimensionless concentration variables satisfy the same equations as before, but with $A = k_r =1$. There are now only
two independent dimensionless parameters, $k_f'$ and $\phi'$. The dependence on all 4 parameters can be recovered at the end by replacing $u_n$ by $x_n/A$, $t'$ by $tk_r$, $k_f'$ by $k_f A/k_r$ and $\phi'$ by $\phi/k_r$. One may also choose $\tau = 1/\phi$ (whenever $\phi \neq 0$). In that case the two independent dimensionless parameters will be $k_f'$ and $k_r'$.

\section{Comparison with existing models}
The model presented here is closely related to two existing models, one discussed by Bagley and Farmer \cite{Bagley1989, Bagley1991} and the other by Ohtsuki and Nowak \cite{Ohtsuki2009}. The differences between the present model and the existing ones is discussed below.

\subsubsection*{\label{Comparison-Bagley}\bf Comparison with the model of Bagley and Farmer}
The main differences between the present model and that studied by Bagley and Farmer \cite{Bagley1989, Bagley1991} are:
\begin{itemize}
  \item A simpler representation of molecules (we do not consider molecules as strings): Bagley and Farmer consider a polymeric chemistry in their model. The molecules produced in this chemistry are differentiated not only based upon their length but also the order in which the monomers appear in them, \ie, $aab$, $aba$, $baa$ are considered to be three different molecules. We have only differentiated molecules based on their length and the number of monomers of each type they contain. The order in which the monomers appear is not considered. For a given maximum size of molecule $N$, this results in a chemistry that contains much fewer molecules and reactions among them, making it possible to numerically investigate chemistries with much larger sized molecules.
  \item A simpler treatment of catalysis (we do not consider intermediate complexes): They consider a detailed mechanism of catalysis including the production of intermediate complexes. Their dynamical equations included equations for these intermediate complexes. We consider an approximate and much simpler treatment for inclusion of catalysis.
  \item We ignore the effects coming from small populations containing a discrete number of molecules: Bagley and Farmer discuss the effects that arise because of the stochasticity in molecular populations. We consider only continuous dynamics and ignore the effects arising out of finite number of molecules.
  \item Their results are at fixed input rate of food set molecules, whereas ours are primarily at buffered concentration of food set molecules (we also show results for a fixed input rate in Appendix \ref{Appendix-Bagley}).
\end{itemize}
We reproduce the main phenomenon of ACS dominance that Bagley and Farmer observed, but the relative simplicity of the present model allows us to explore other phenomena that they do not report about (this includes a multistability in the dynamics and the possibility of building large molecules through nested ACSs).

\subsubsection*{\bf Comparison with the model of Ohtsuki and Nowak}
The main differences with the model of Ohtsuki and Nowak \cite{Ohtsuki2009} are
\begin{itemize}
  \item A much richer spontaneous chemistry of ligation reactions and the inclusion of reverse reactions: Ohtsuki and Nowak considered chemistries that only contained forward reactions. Also, larger molecules in their chemistry get produced only by successive addition of monomers. We have considered chemistries that include reverse reactions and have multiple ways for production of molecules. This makes an analytical treatment of the system more difficult.
  \item A much more general class of catalyzed chemistries, instead of a single catalyst (which allows us to talk of nested ACSs, in particular): They have only considered the cases wherein one molecule catalyses all the reactions in its production pathway; they do not discuss cases with more than one catalyst. We consider various topologies of catalyzed chemistries and discuss how the choice of topology affect dynamics.
\end{itemize}
With a specific choice of parameters our $f=1$ model reduces exactly to their `symmetric prelife' model with a catalyst. This is discussed in Appendix \ref{Appendix-Nowak}. In spite of greater complexity we are able to numerically reproduce their main results in a much more general setting, and also provide an analytical understanding of the results based on certain approximations.

\thispagestyle{plain}
\cleardoublepage
\chapter{\label{ChapterSpontaneousChemistry}Steady state properties of the spontaneous chemistry}

\lettrine[lines=2, lhang=0, loversize=0.0, lraise=0.0]{T}{his} chapter discusses the behaviour of the model described in Chapter \ref{ChapterModel} when the chemistry only contains spontaneous reactions. We first discuss a \gls{spontaneous-chemistry} that is fully connected\glsadd{fully-connected-chemistry} and homogeneous\glsadd{homogeneous-chemistry}. This chemistry has 4 parameters, $k_f$, $k_r$, $\phi$, and $A$, the concentration of the monomers. The results for $\phi=0$ are discussed first, for which we also present some analytical results. We then describe results for a general $\phi$ and show that the concentration of large molecules fall exponentially with their length. We have checked that introducing irreversible reactions and bringing in a small amount of heterogeneity does not change the conclusions. Some results for sparse\glsadd{sparse-chemistry} and heterogeneous chemistries are also presented.

In this chapter, and the remainder of this thesis, we discuss in detail results for the chemistries with $f = 1$. We also give some results for $f=2$. The qualitative features of the model can be generalized for any values of $f$.

\section{Behaviour of the system in the limiting case of $\phi=0$}
\subsection{\label{phi0-analytical-solution}Analytical solution}
For $\phi = 0$, the following exact analytical solution for the steady state concentrations exists for homogeneous and connected uncatalyzed chemistries with $f=1$:
\begin{equation}\label{sszerophi}
    x_n = A \left(\frac{k_f A}{k_r}\right)^{n-1}.
\end{equation}
To see that this is a fixed point, note that when Eq. (\ref{sszerophi}) holds, then $v_{ij} = k_fx_ix_j - k_r x_{i+j} = 0$ for all $i,j = 1,2,\ldots$; hence the r.h.s. of Eq. (\ref{rateequation-v-F1}) vanishes (at $\phi = 0$). The same argument applies for a general $f$; for $f=2$ the expression becomes,
\begin{equation}\label{sszerophi-f2}
    x_{(n_1,n_2)}= A^{n_1}B^{n_2}\left(\frac{k_f}{k_r}\right)^{(n_1+n_2)-1},
\end{equation}
where $A=x_{(1,0)}$ and $B=x_{(0,1)}$ are the concentrations of the two monomers. In the case when the two monomers have same buffered concentrations (case of homogeneous chemistries), \ie, $A=B$, Eq. (\ref{sszerophi-f2}) reduces to Eq. (\ref{sszerophi}) with $n=n_1+n_2$.

\subsection{Numerical results}
\begin{figure}
    \begin{center}
        \subfloat[]{\label{noacs-phi0-a}\includegraphics[height=4.75in,angle=-90,trim=0.5cm 0.5cm 0.0cm 0.3cm,clip=true]{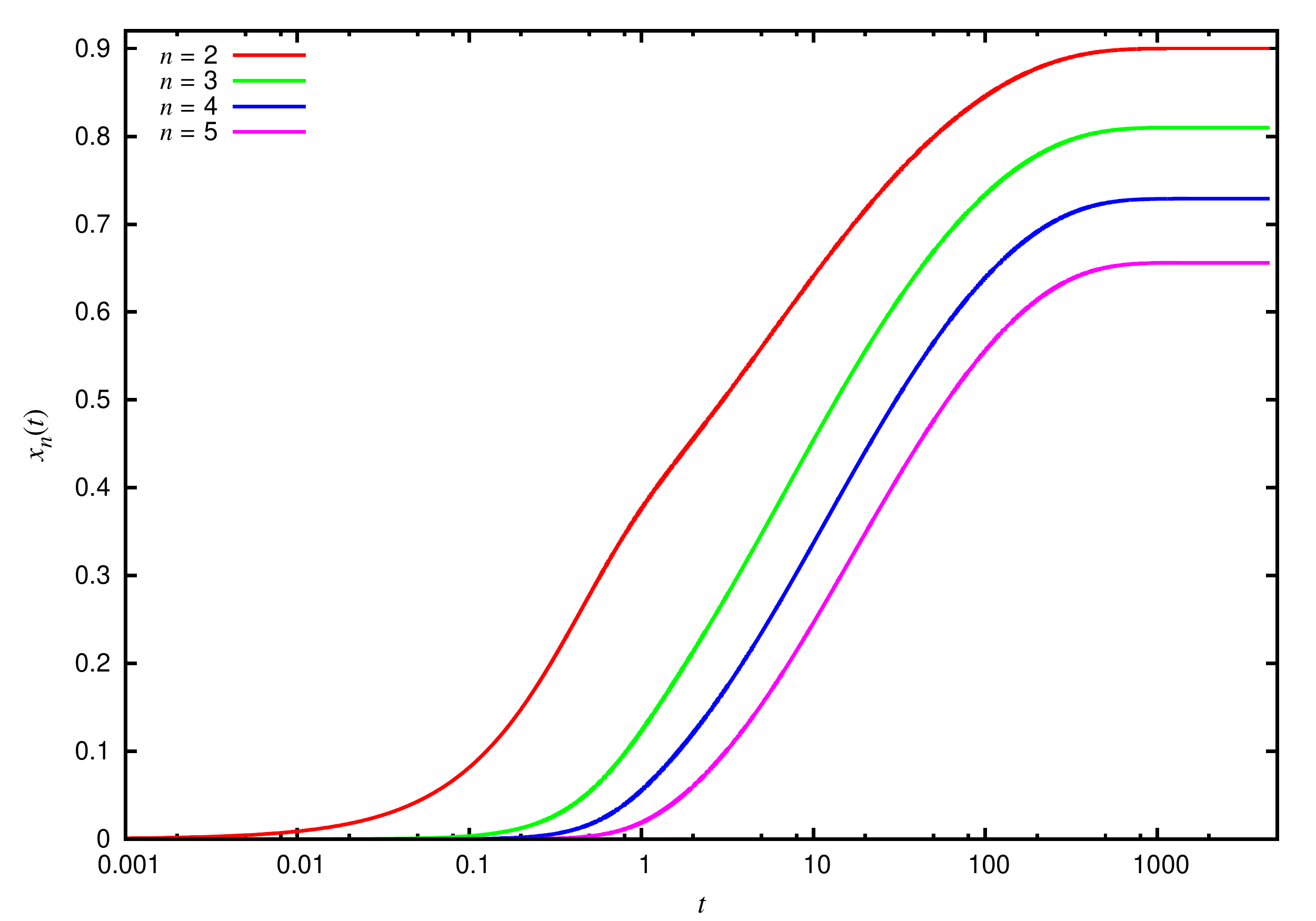}} \\
        \subfloat[]{\label{noacs-phi0-b}\includegraphics[height=4.75in,angle=-90,trim=0.5cm 0.5cm 0.0cm 0.3cm,clip=true]{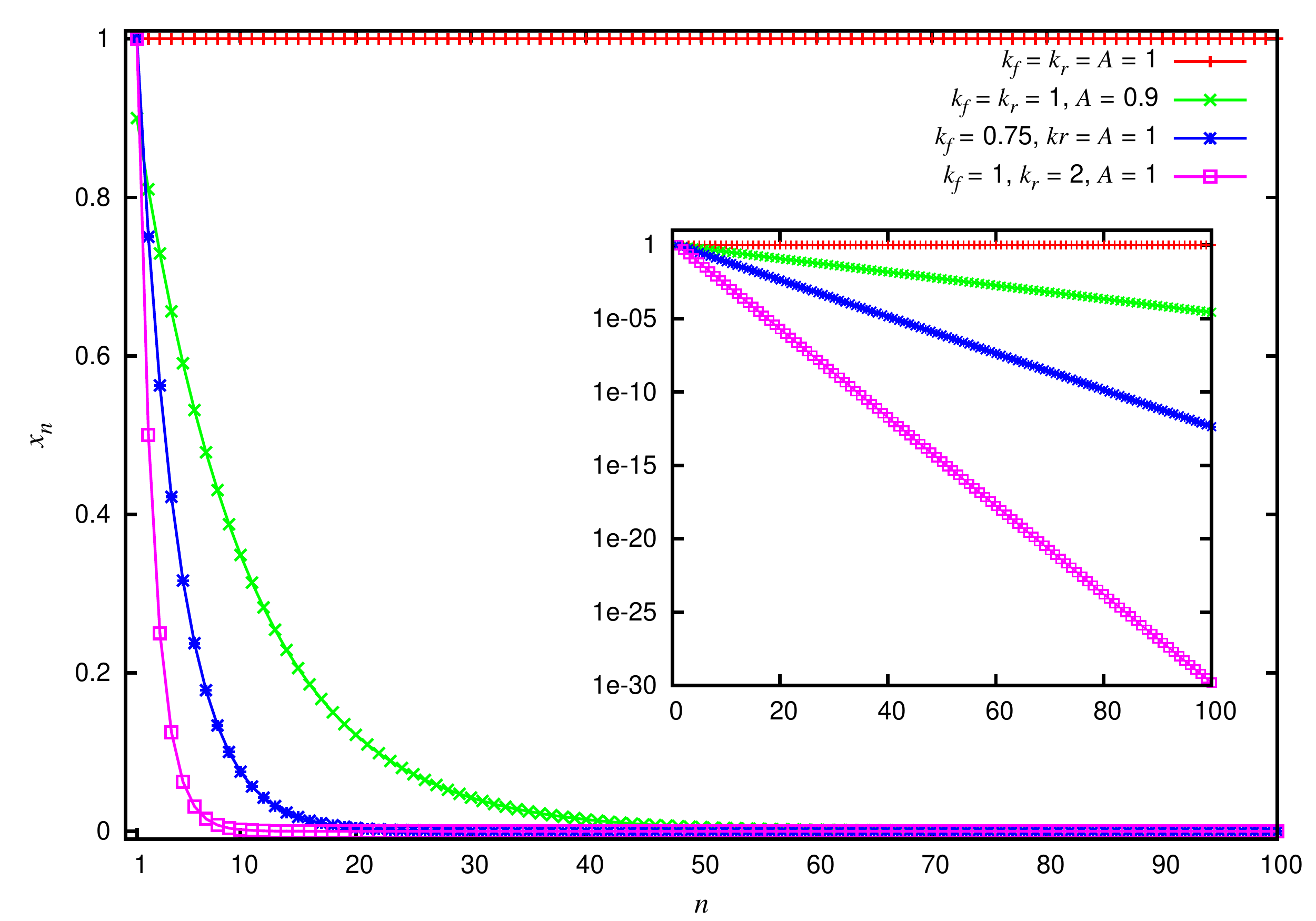}}
    \end{center}
    \caption[Concentrations in uncatalyzed chemistries with a single food source for $\phi=0$.]{{\bf Concentrations in uncatalyzed chemistries with a single food source for $\phi=0$.} {\bf (a)} Evolution of concentrations with time for a chemistry with $k_f = 0.9, k_r = A = 1$. {\bf (b)} Steady state concentration $x_n$ as a function of molecule size $n$. The concentrations fall exponentially with length. Inset shows the same on a semi-log plot; the straight lines are evidence of exponential damping of $x_n$ (Eq. (\ref{ss1})), with $\Lambda = 1, 0.9, 0.75, 0.5$ for the four cases, respectively. $\Lambda$ is computed from the slope of a straight line fit. The value of parameters are indicated in the key. For simulation purposes, the size of the largest molecule was taken to be $N=100$.}
    \label{noacs-phi0}
\end{figure}

Starting from the initial condition in which all concentrations other than that of the food set molecules are zero, \ie,
\begin{equation}
    \label{standard-ic-eq}
    x_n = 0, n = 2, 3, 4, \ldots
\end{equation}
(we refer to this as the \gls{standard-ic}), in a fully connected and homogeneous spontaneous chemistry with $f=1$ and $\phi=0$, the concentrations were found to increase monotonically and reach a steady state (Fig. \ref{noacs-phi0-a}). The plot of steady state concentration $x_n$ versus $n$ is shown in Fig. \ref{noacs-phi0-b}. The graph of $x_n$ versus $n$ on a semi-log plot (inset of Fig \ref{noacs-phi0-b}) was found to be a straight line, given by the expression
\begin{equation}\label{ss1}
x_n = c\Lambda^{n} = ce^{-\gamma (n)},
\end{equation}
where, $c$ and $\Lambda=e^{-\gamma}$ are constants. $\Lambda(\phi =0)$, determined by numerically fitting the slope, was found to be equal to $k_fA/k_r$, consistent with the Eq. (\ref{sszerophi}). Hence, whenever $k_f A < k_r$, the steady state concentrations are exponentially damped. When $k_f A > k_r$, the steady state concentrations increase exponentially (for $k_f A > k_r$, the numerical integration does not converge as the steady state solution is numerically very large for large $n$).

\begin{figure}
    \begin{center}
        \subfloat[]{\label{noacs-2d-phi0-a}\includegraphics[height=5in,angle=-90,trim=1.75cm 1.0cm 1.75cm 0.1cm,clip=true]{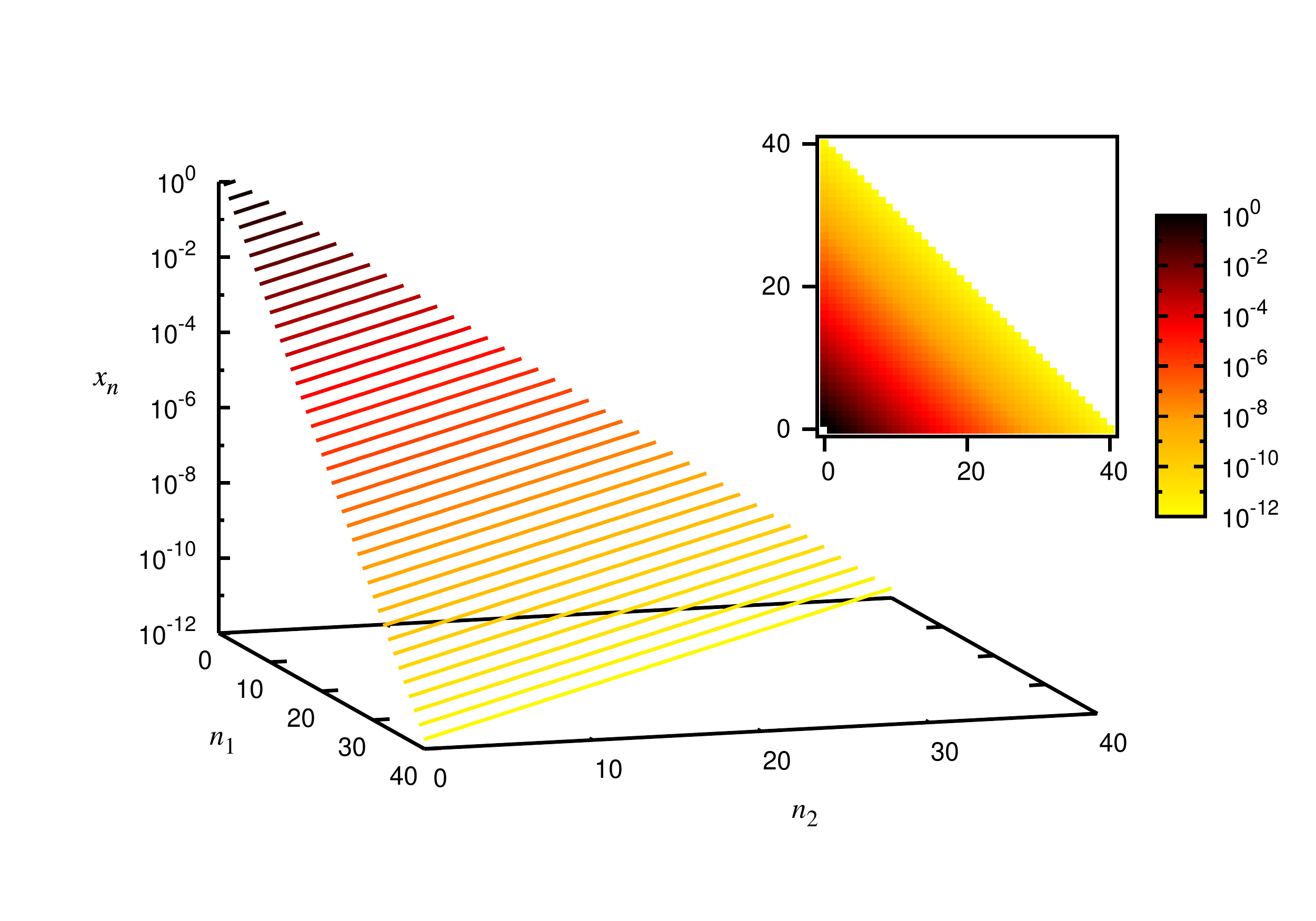}} \\
        \subfloat[]{\label{noacs-2d-phi0-b}\includegraphics[height=5in,angle=-90,trim=1.75cm 1.0cm 1.75cm 0.1cm,clip=true]{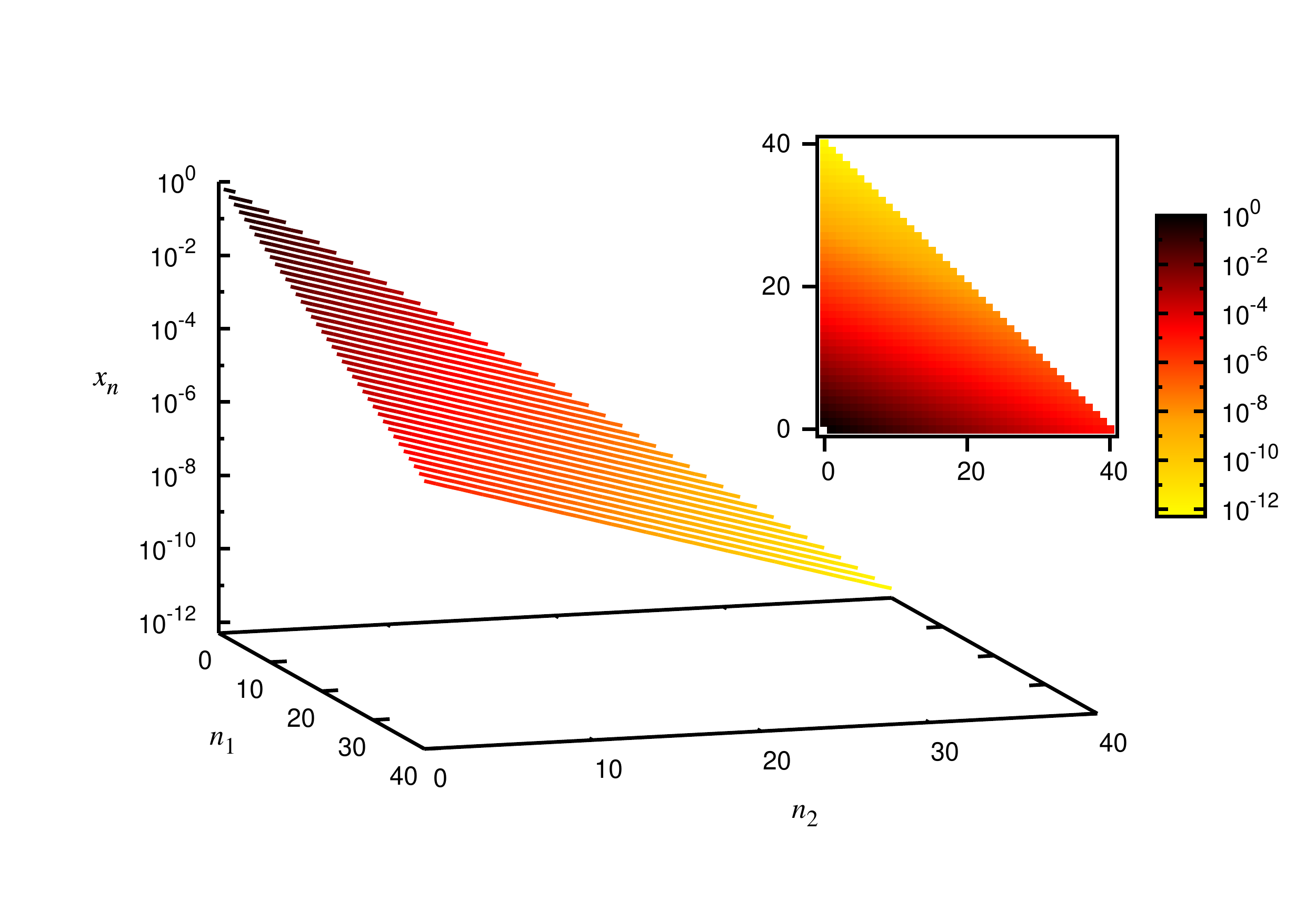}}
    \end{center}
    \caption[Concentrations in uncatalyzed chemistries with two food sources for $\phi=0$.]{{\bf Concentrations in uncatalyzed chemistries with two food sources for $\phi=0$.} The 3D plots show the steady state concentration $x_n$ of the molecules $n=(n_1,n_2)$ as a function of $n_1$ and $n_2$. The insets show a `top view' of the $(n_1,n_2)$ plane with $x_n$ indicated in a colour map on a logarithmic scale. {\bf(a)} The two monomers have same buffered concentrations, $x_{(1,0)} = x_{(0,1)} = 1$ (homogeneous chemistry). $k_f = 1, k_r = 2, N = 40$. {\bf(b)} Two monomers have different concentrations. $k_f = k_r = 1, x_{(1,0)} = 0.75, x_{(0,1)} = 0.5, N = 40$. The results in two cases are found to be consistent with the Eq. (\ref{sszerophi-f2}).}
    \label{noacs-2d-phi0}
\end{figure}

Similar results hold when more than one monomer is present in the system. For $f=2$, the steady state concentration profile is shown in Fig. \ref{noacs-2d-phi0-a}, when the concentration of the two monomers is the same and in Fig. \ref{noacs-2d-phi0-b}, when they have have different buffered concentrations. The results are found to be consistent with the Eq. (\ref{sszerophi-f2}).

Note that these results are independent of the size of the largest molecules in the chemistry, $N$; $x_n$'s are completely determined by the values of parameters $k_f, k_r$ and $A$.

\section{Behaviour of the system for $\phi>0$}
\begin{figure}
    \begin{center}
        \subfloat[]{\label{noacs-a}\includegraphics[height=4.75in,angle=-90,trim=0.3cm 0.5cm 0cm 0.3cm,clip=true]{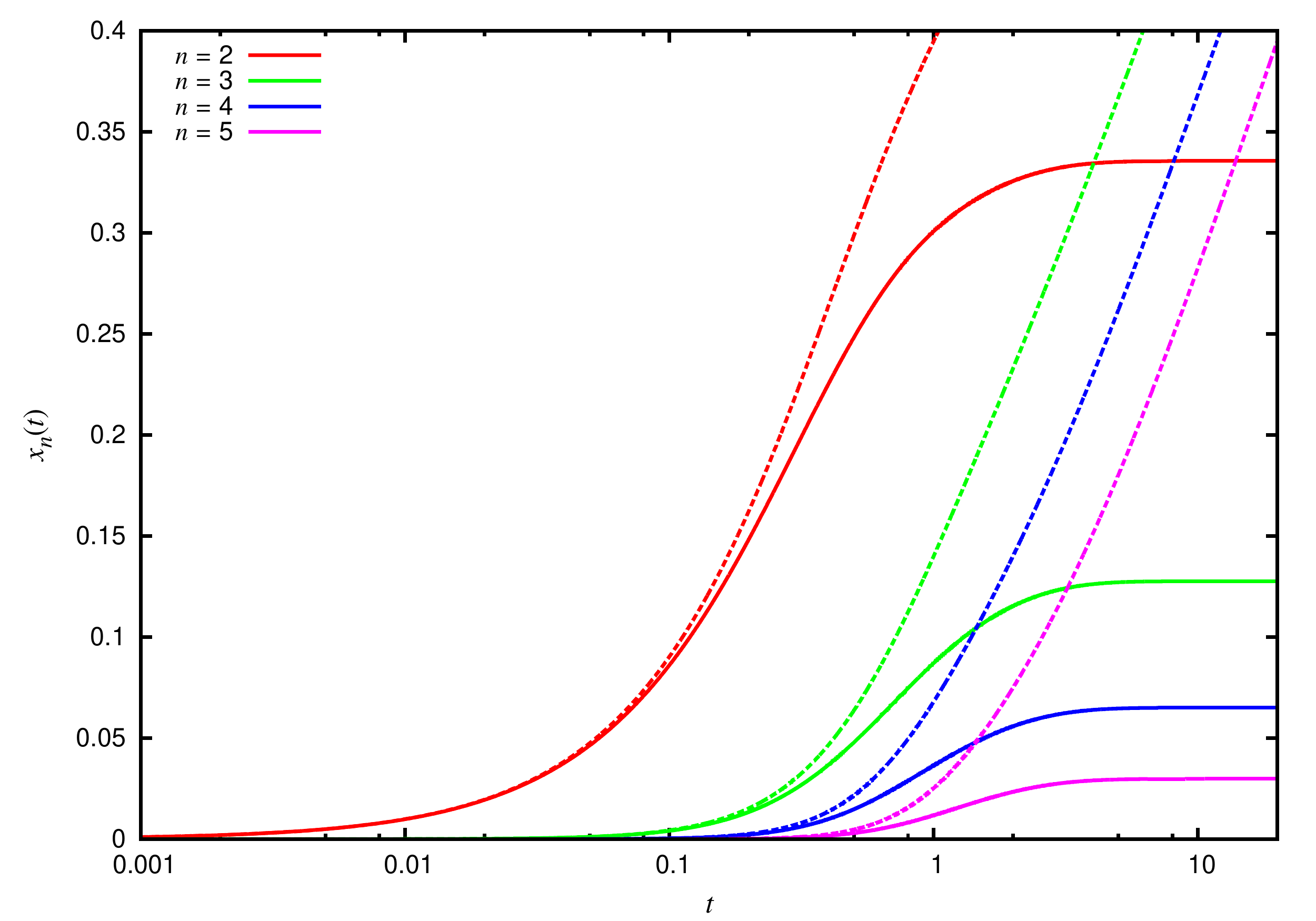}} \\
        \subfloat[]{\label{noacs-b}\includegraphics[height=4.75in,angle=-90,trim=0.5cm 0.5cm 0cm 0.3cm,clip=true]{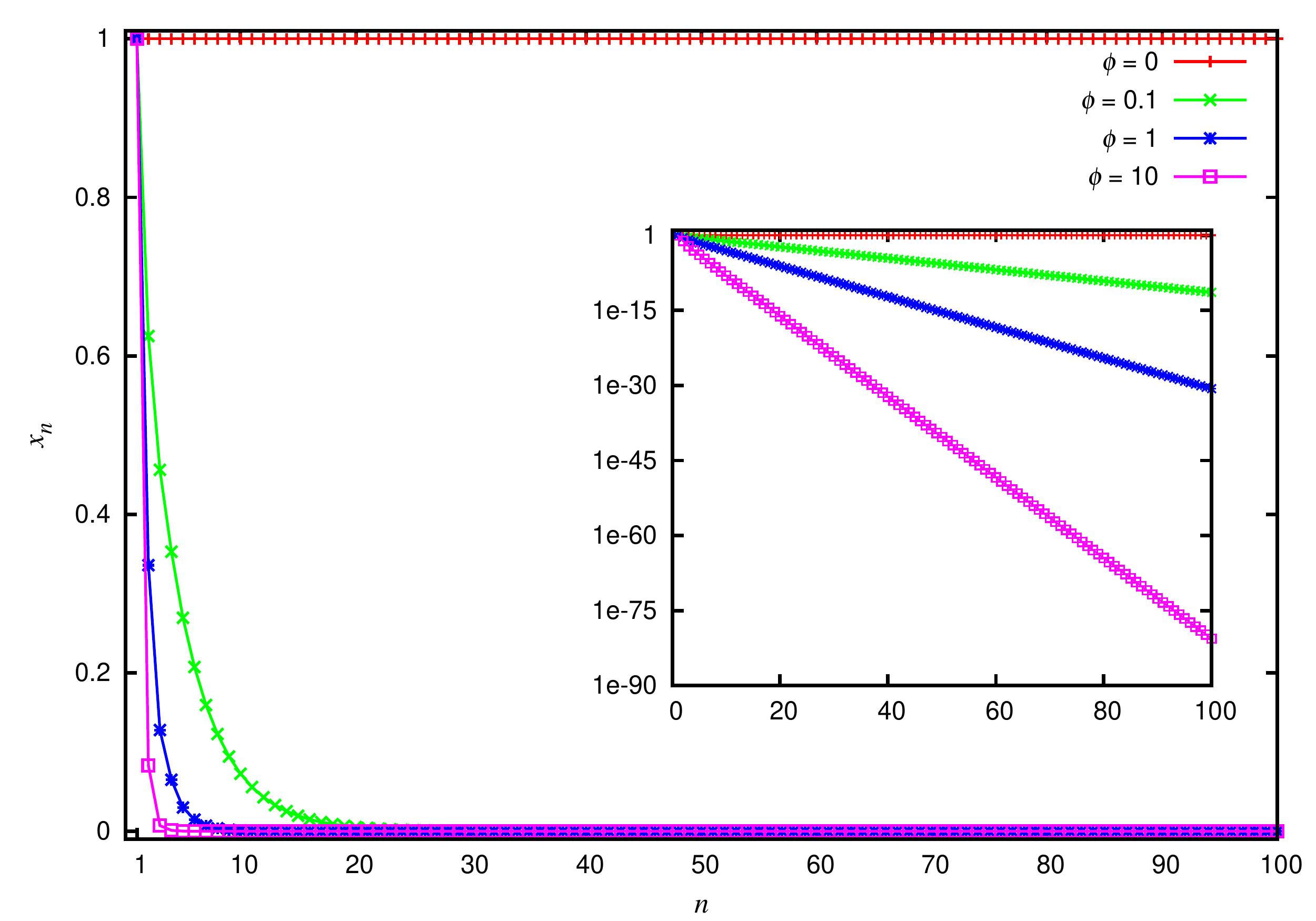}}
    \end{center}
    \caption[Concentrations in uncatalyzed chemistries with a single food source.]{{\bf Concentrations in uncatalyzed chemistries with a single food source.} {\bf(a)} Evolution of concentrations with time for a chemistry with $k_f = k_r = A = \phi = 1, N=100$. Dashed curves show the same when $\phi=0$ (other parameters remaining the same). For small $t$ the $\phi = 1$ curves are similar to $\phi=0$; as the concentrations become large the effect of finite $\phi$ starts to show. {\bf(b)} Steady state concentration as a function of molecule size. Parameters take the same values as in (a) except that four values of $\phi$ are shown, $\phi = 0, 0.1, 1, 10$. Inset shows the same on a semi-log plot; $\Lambda = 1, 0.77, 0.49, 0.16$ for the four cases, respectively. $\Lambda$ is computed from the slope of a straight line fit after ignoring the smaller molecules (up to $n=4$ in this case).}
    \label{noacs}
\end{figure}

When $\phi > 0$ we do not have an exact analytical solution. As in the case for $\phi=0$, for a fully connected and homogeneous spontaneous chemistry ($f=1$), starting from the \gls{standard-ic} (Eq. (\ref{standard-ic-eq})), the concentrations were found to increase monotonically and reach a steady state (Fig. \ref{noacs-a}). The steady state values, which now depend upon the value of $\phi$, were comparatively smaller. The plot of steady state concentration $x_n$ versus $n$ for different values of $\phi$ is shown in Fig. \ref{noacs-b}. The semi-log plot of $x_n$ versus $n$ (inset of Fig \ref{noacs-b}) was now found to be approximately a straight line for large $n$. The small molecules deviate from the straight line. $\Lambda$, calculated in this case by fitting the slope ignoring small molecules, decreases monotonically as $\phi$ increases (Fig. \ref{noacs-b}).

\begin{figure}
    \begin{center}
        \subfloat[]{\label{noacs-lambda-a}\includegraphics[height=4.75in,angle=-90,trim=0.3cm 0.5cm 0cm 0.3cm,clip=true]{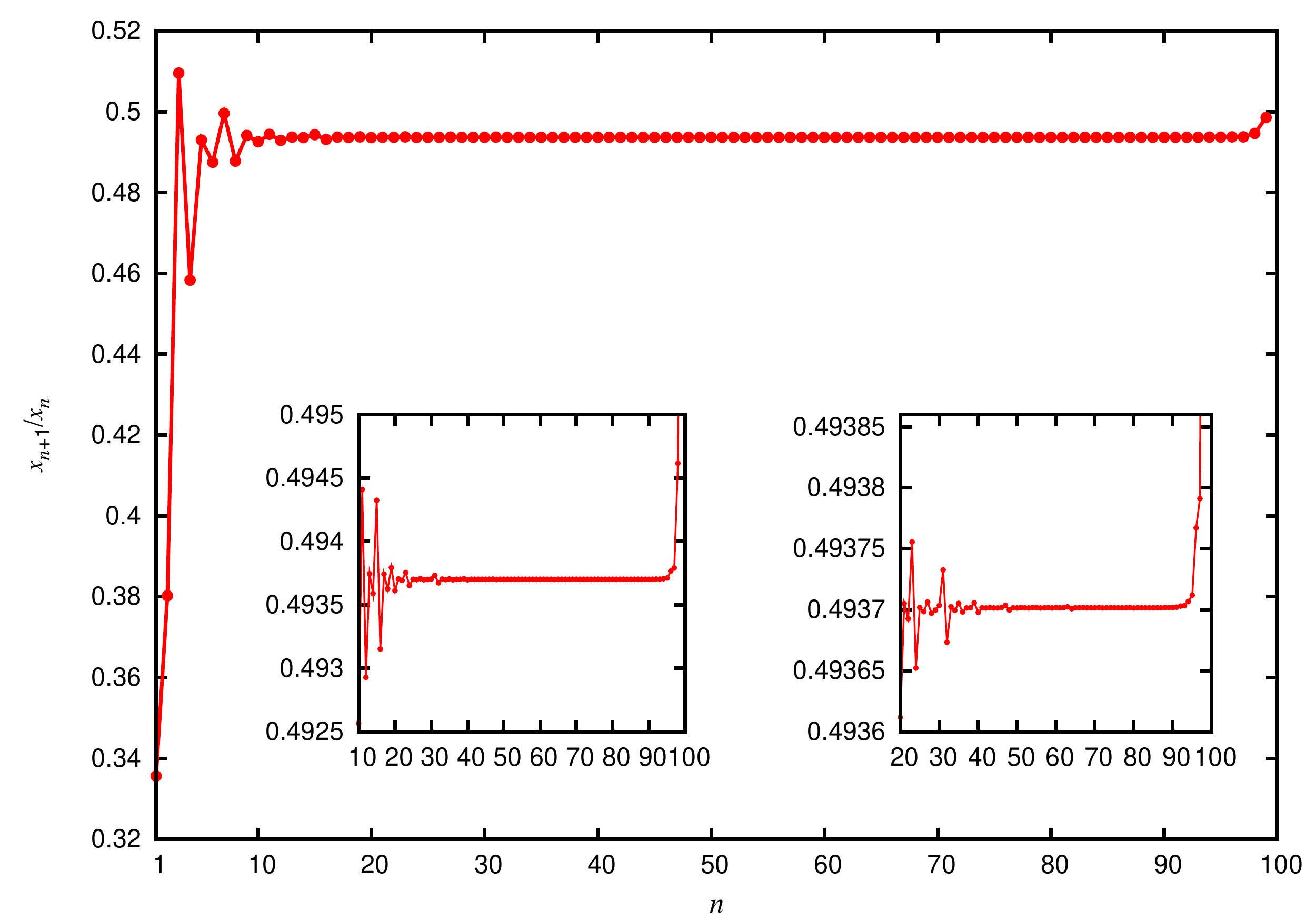}} \\
        \subfloat[]{\label{noacs-lambda-b}\includegraphics[height=4.75in,angle=-90,trim=0.3cm 0.5cm 0cm 0.3cm,clip=true]{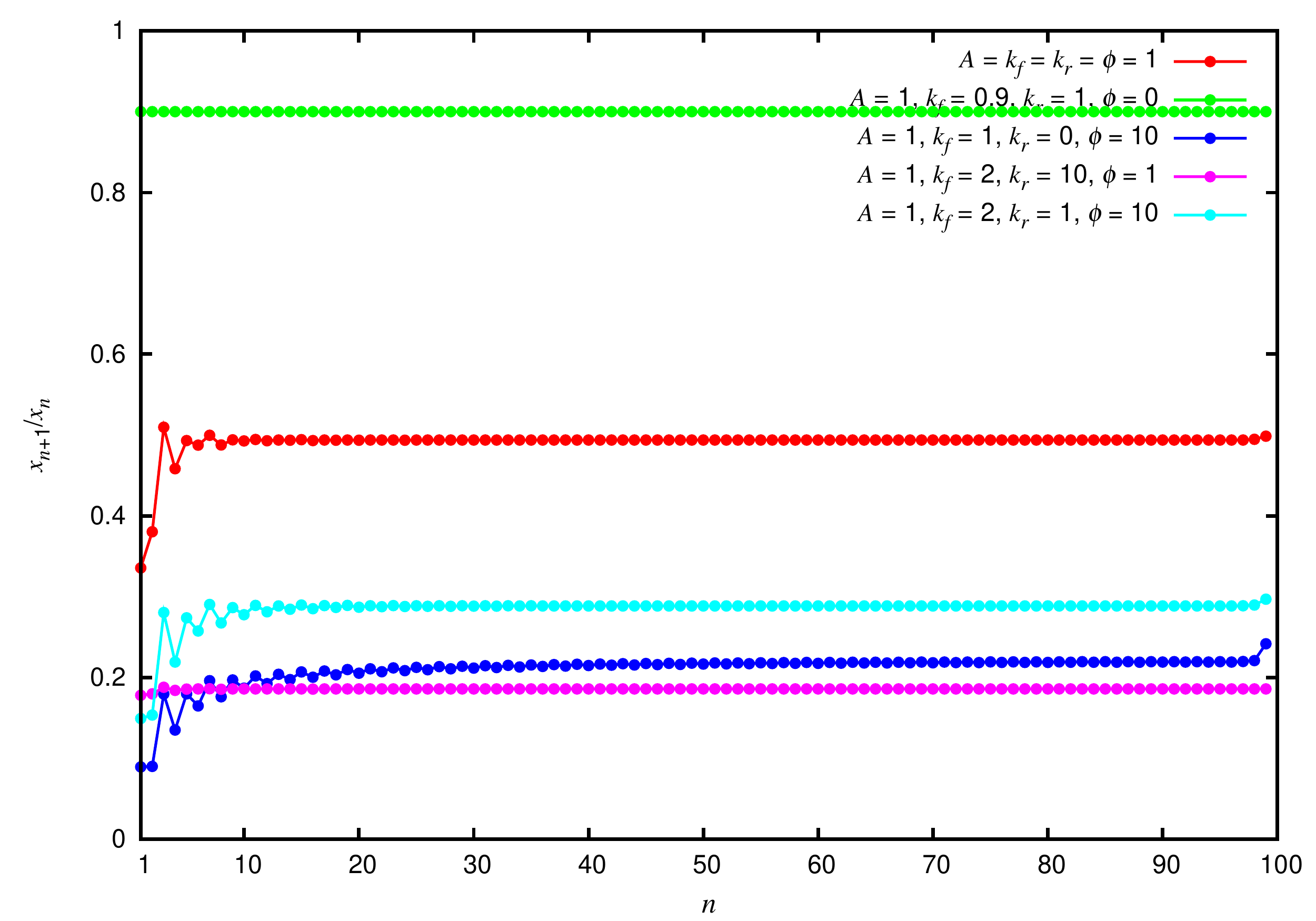}}
    \end{center}
    \caption[The ratio of steady state concentration of successive molecules, $x_{n+1}/x_n$, versus $n$.]{{\bf The ratio of steady state concentration of successive molecules, $x_{n+1}/x_n$, versus $n$.} {\bf(a)} The ratio $x_{n+1}/x_n$ is not strictly a constant but depends upon $n$. It oscillates about a mean, becoming larger for even values of $n+1$ (powers of 2 show even larger variation). For larger values of $n$, the variation in ratio becomes small (see insets) and thus the ratio can be approximated by a constant. The effects arising because of the finiteness of chemistry are also visible when $n \sim N$. Parameters take the same values as in Fig. \ref{noacs-a}. {\bf(b)} The ratio $x_{n+1}/x_n$ versus $n$ is plotted for 5 different set of parameters (values are indicated in the key).}
    \label{noacs-lambda}
\end{figure}

The ratio of steady state concentration of successive molecules, $x_{n+1}/x_{n}$ ($=\Lambda$ as given by Eq. (\ref{ss1})), is not strictly a constant but depends upon $n$ (Fig. \ref{noacs-lambda}). The ratio oscillates about a mean, and is larger when $n+1$ is even (more so when it is a power of 2) than when $n+1$ is odd; the molecules that are made up of two smaller molecules of the same size tend to have higher populations. The variation becomes quite small as $n$ increases (insets of Fig. \ref{noacs-lambda-a}). Thus Eq. (\ref{ss1}) can still be used to approximate the solution that characterizes the steady state concentrations. Numerically, we find that $\Lambda (\phi>0)$ drops to below 1 even when $k_fA/k_r > 1$.

\begin{figure}
    \begin{center}
        \subfloat[]{\label{noacs-2d-a}\includegraphics[height=5in,angle=-90,trim=1.75cm 1.0cm 1.75cm 0.1cm,clip=true]{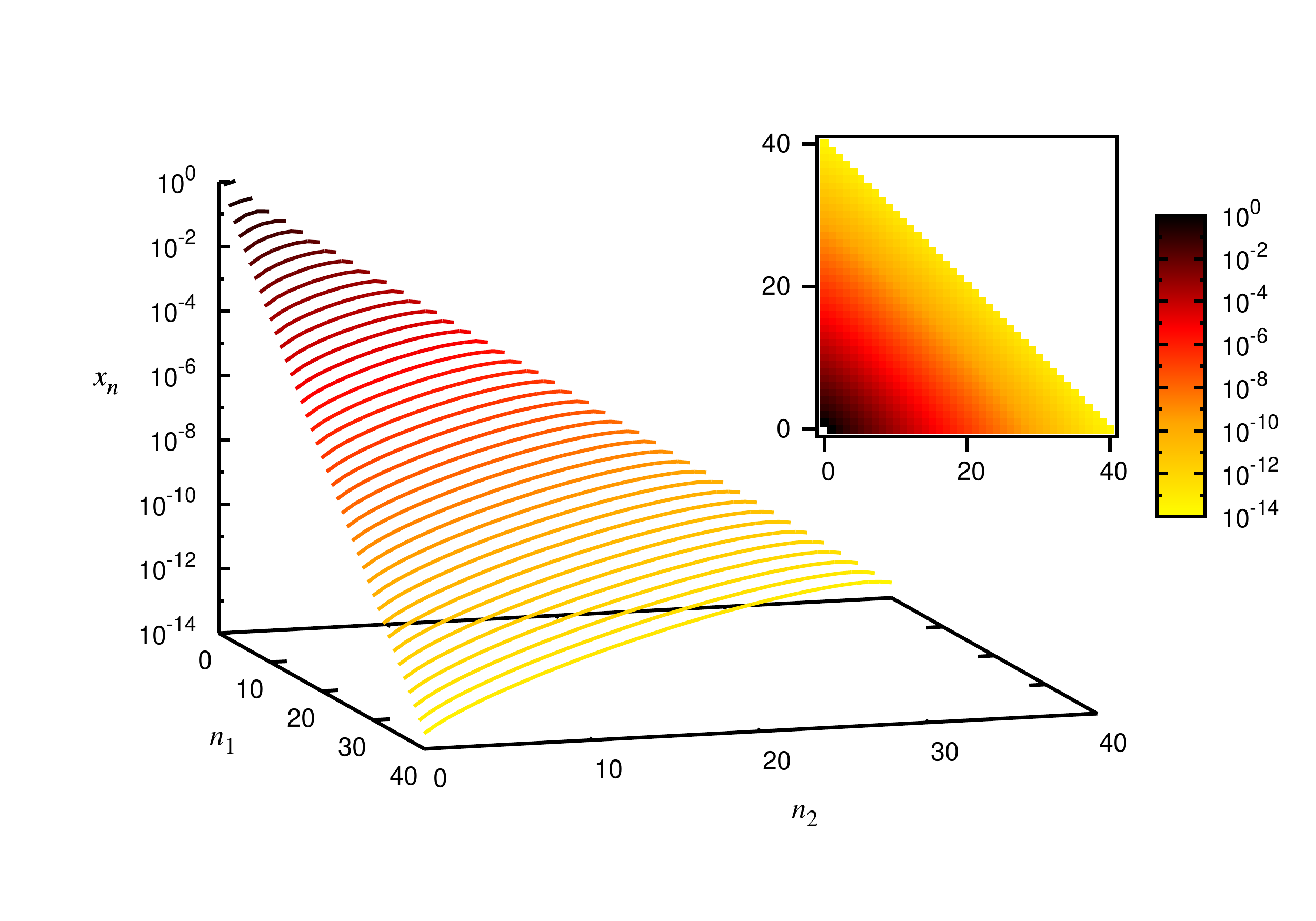}} \\
        \subfloat[]{\label{noacs-2d-b}\includegraphics[height=5in,angle=-90,trim=1.75cm 1.0cm 1.75cm 0.1cm,clip=true]{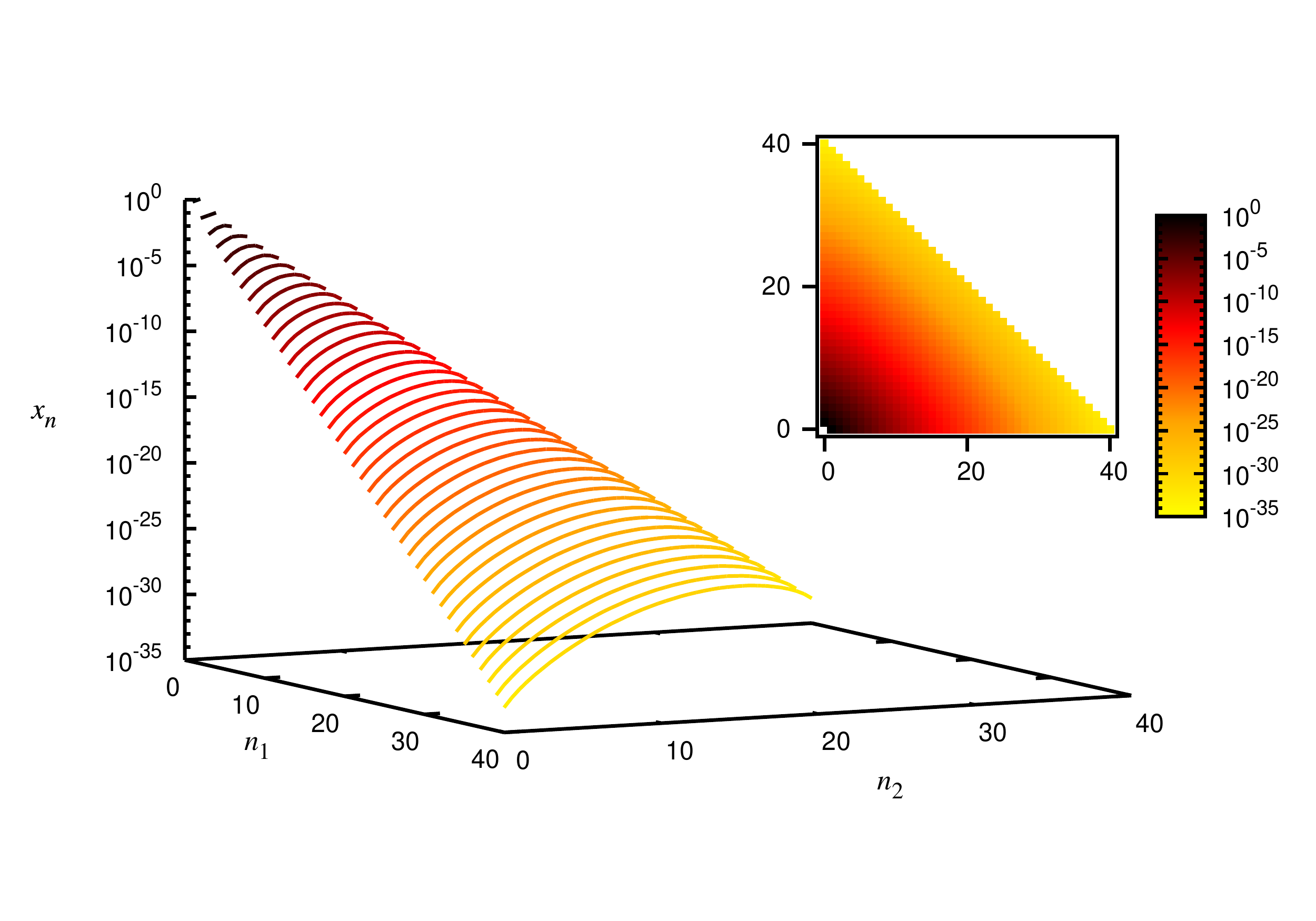}}
    \end{center}
    \caption[Steady state concentration profile in an uncatalyzed chemistry with $f=2$.]{{\bf Steady state concentration profile in an uncatalyzed chemistry with $f=2$.} The 3D plots show the concentration $x_n$ of the molecule $n=(n_1,n_2)$ as a function of $n_1$ and $n_2$ in the steady state, for an uncatalyzed chemistry with $k_f = k_r = x_{(1,0)} = x_{(0,1)}=1, N=40$ for two values of $\phi=$ {\bf (a)} 1 and {\bf (b)} 10. The inset shows a `top view' of the ($n_1$,$n_2$) plane with $x_n$ indicated in a colour map on a logarithmic scale.}
    \label{noacs-2d}
\end{figure}

Similar results hold for $f=2$; steady state concentration profiles for two different values of $\phi$ are shown in Fig. \ref{noacs-2d}. `Diagonal entries' ($n_1 = n_2$) have higher concentrations in homogeneous chemistries because there are more reaction pathways to build molecules with equal numbers of both monomers than unequal, and the larger number of production pathways for the diagonal molecules are able to counteract the effect of dissipation more effectively.

The values of $\Lambda$ extracted from the slope of the straight line fit are found to be $N$ independent for sufficiently large $N$. A detailed discussion for $N$-independence of this and other results is provided in the Appendix \ref{Appendix-N-independence}.

\subsection{Dependence on parameters}
\begin{figure}
  \begin{center}
    \includegraphics[height=5in,angle=-90,trim=1.75cm 1.5cm 2cm 1cm,clip=true]{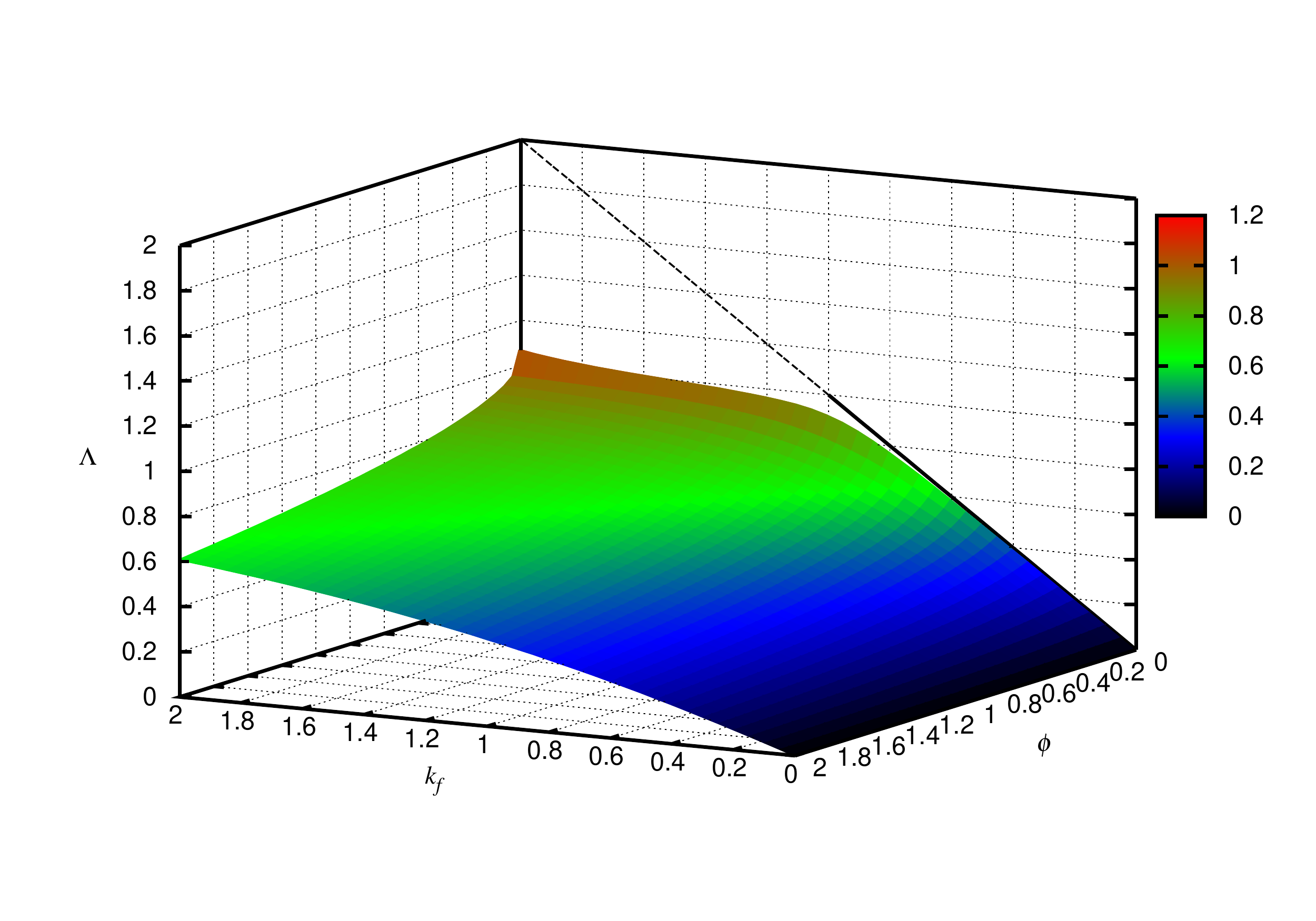}
  \end{center}
  \caption[Behaviour of $\Lambda$ as a function of $k_f$ and $\phi$.]{{\bf Behaviour of $\Lambda$ as a function of $k_f$ and $\phi$.} The figure shows the dependence of $\Lambda$ on $k_f$ and $\phi$ for an uncatalyzed chemistry, keeping $A = k_r = 1$, $N=100$. The curved surface was made with parameter values in the range $0 \leq k_f \leq 2$, $0.01 \leq \phi \leq 2$. $\Lambda$ is found to be a monotonically increasing function of $k_f$ and a monotonically decreasing function of $\phi$. For $\phi=0$, there is an analytical solution $\Lambda = k_f$ (see Eq. (\ref{sszerophi})). This was verified numerically in the range $0 \leq k_f \leq 1$ (see solid line at $\phi=0$). In the region $k_f > 1$ the numerical integration does not converge at $\phi=0$ as the steady state solution (Eq. (\ref{sszerophi})) $x_n = A\Lambda^{n-1} = Ak_f^{n-1}$ is numerically very large for large $n$. The dotted extension of the line ($1 < k_f \le 2$) is simply the analytical result. Note that for most of the phase-space $\Lambda < 1$, except for very small values of $\phi$.}
  \label{Kf-Phi-Phasespace}
\end{figure}

\begin{figure}
  \begin{center}
    \includegraphics[height=5in,angle=-90,trim=1.75cm 1.5cm 2cm 1cm,clip=true]{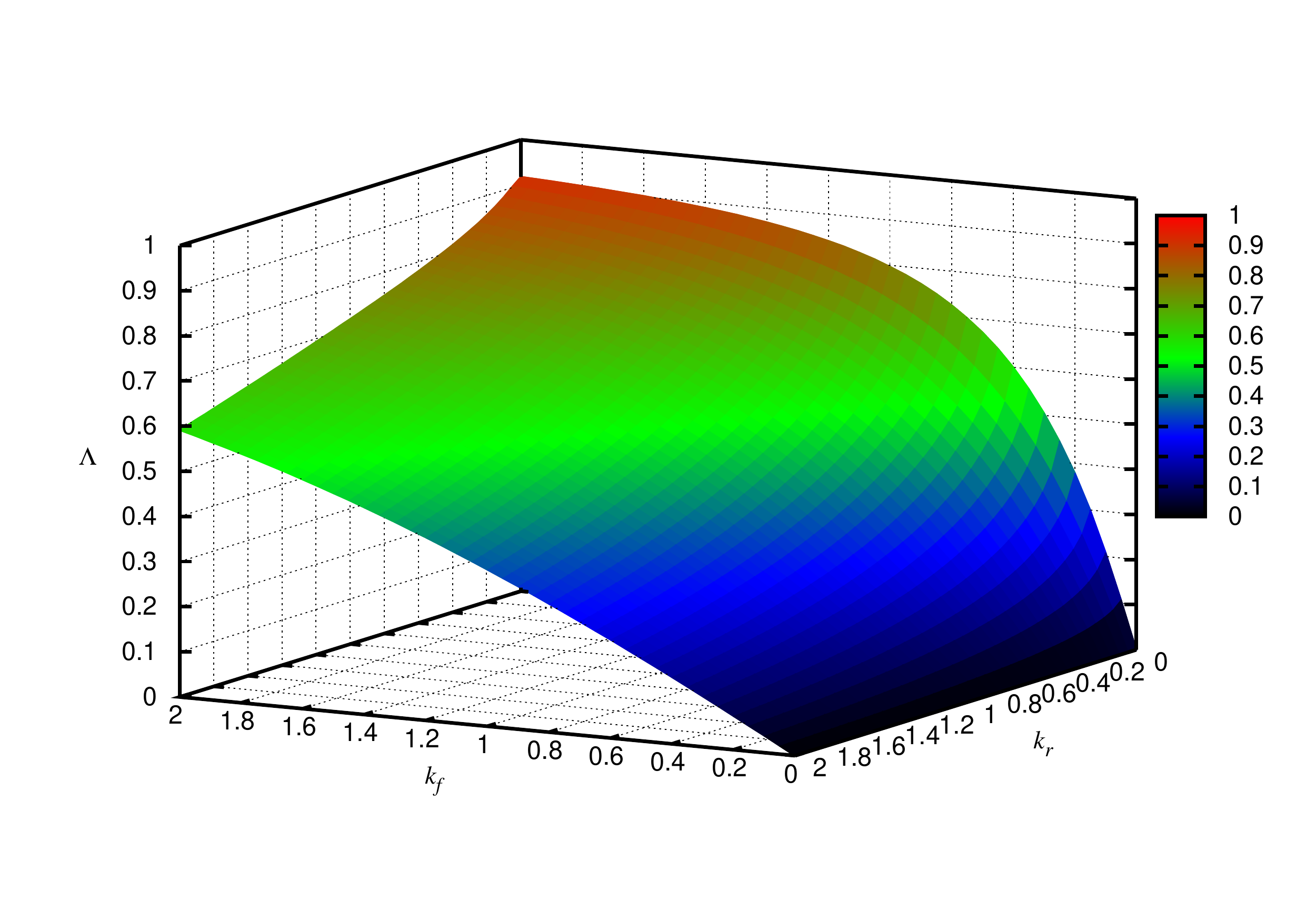}
  \end{center}
  \caption[Behaviour of $\Lambda$ as a function of $k_f$ and $k_r$.]{{\bf Behaviour of $\Lambda$ as a function of $k_f$ and $k_r$.} The figure shows the dependence of $\Lambda$ on $k_f$ and $k_r$ for an uncatalyzed chemistry, keeping $A = \phi = 1$, $N=100$. $\Lambda$ is found to be a monotonically increasing function of $k_f$ and a monotonically decreasing function of $k_r$.}
  \label{Kf-Kr-Phasespace}
\end{figure}

$\Lambda$ is found to be a monotonically increasing function of $k_f$ and $A$, and a monotonically decreasing function of $k_r$ and $\phi$. This corresponds to the intuition that an increased ligation rate favours large molecules and an increased cleavage or dissipation rate disfavours them. Recall the discussion on casting the rate equations in terms of dimensionless variables presented in Section \ref{rateeq-in-dimless-var}. For homogeneous chemistries, whenever $k_r \neq 0$, the system can be described by two dimensionless parameters, $k_f'$ and $\phi'$; or whenever $\phi \neq 0$, by $k_f'$ and $k_r'$. The behaviour of $\Lambda$ as a function of $k_f$ and $\phi$ (keeping $A = k_r = 1$) is shown in Fig. \ref{Kf-Phi-Phasespace}, and as a function of $k_f$ and $k_r$ (keeping $A = \phi = 1$) is shown in Fig. \ref{Kf-Kr-Phasespace}.

When $k_r=0$ and $\phi=0$ both the above choices for scaling fail. In this case the system has no steady state for any finite $N$ as $\dot x_N$ is always positive (in Eq. (\ref{rateequation-N}) for $n=N$, the longest molecule has only a positive production term, thus system has no steady state). In all other cases the uncatalyzed chemistry seems to have a global fixed point attractor (all initial conditions tested lead to the same steady state).

\section{\label{section-saprse-spont-chem}Results for a sparse chemistry}
As defined earlier (in Section \ref{nomenclature}) a \gls{sparse-chemistry} is one that contains only a small subset of all possible reactions. There are several ways of constructing a sparse chemistry; we use the following algorithm:\\
{\bf Algorithm to generate sparse chemistries:} We prune the reaction set to only $k$ ligation reactions per molecule, randomly chosen from all possible ligation reaction producing the molecule. For molecules too small to have $k$ ligation reactions, all ligation reactions are retained. (Note that while we refer to only the ligation reactions and not cleavage reactions for the purpose of defining the degree of a molecule, in our simulations all reactions are treated as reverse reactions. That is, for every ligation reaction included in the chemistry the reverse (cleavage) reaction is also present.)
Thus we define a chemistry with degree $k$ as one in which the number of ligation reactions producing a molecule is $k$ obtained as a result of above pruning algorithm.

\begin{figure}
    \begin{center}
        \subfloat[]{\includegraphics[height=4.75in,angle=-90,trim=0.5cm 0.5cm 0.0cm 0.3cm,clip=true]{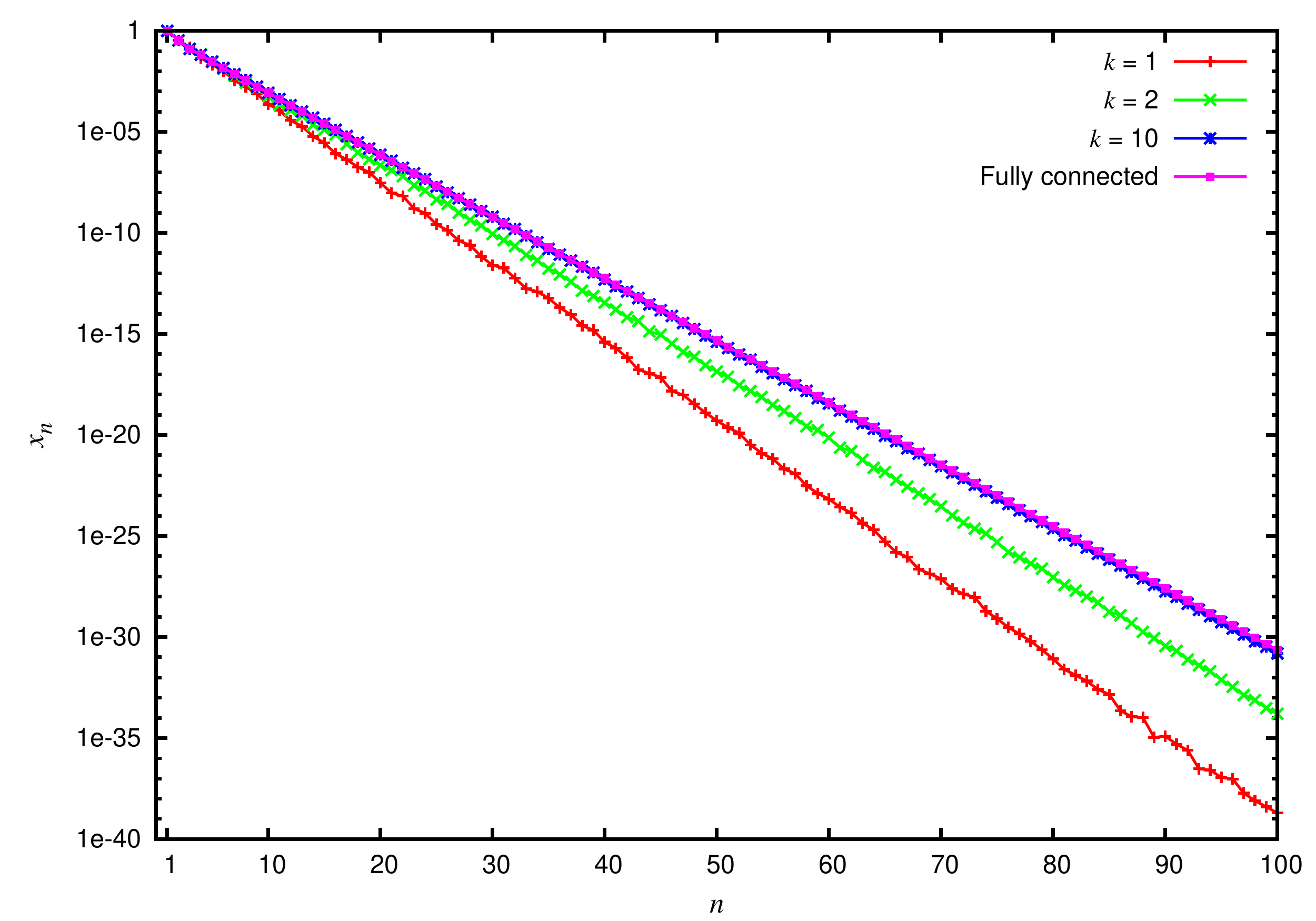}}\\
        \subfloat[]{\includegraphics[height=4.75in,angle=-90,trim=1.75cm 1.0cm 1.75cm 3.0cm,clip=true]{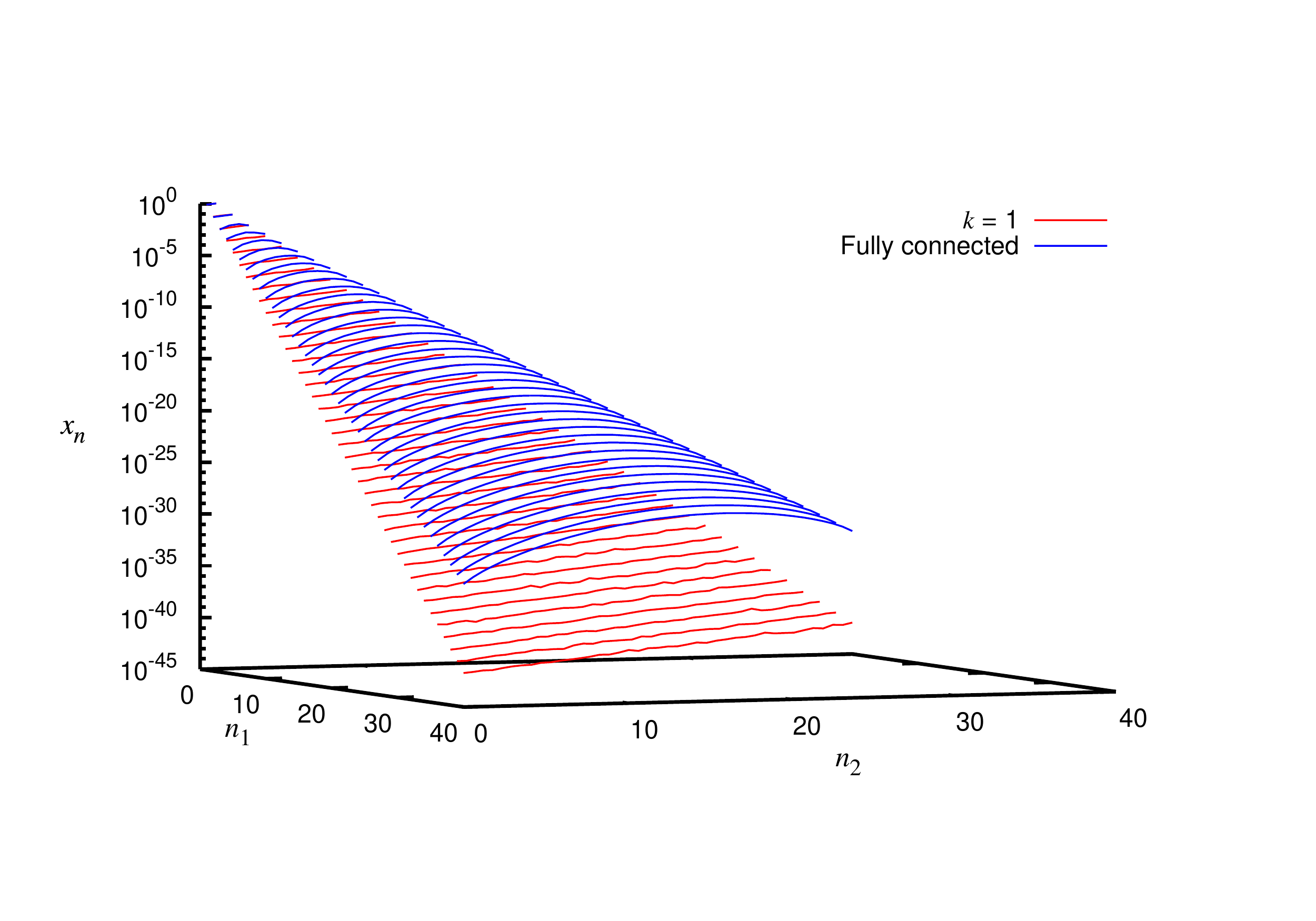}}
    \end{center}
    \caption[Concentrations in sparse spontaneous chemistries.]{{\bf Concentrations in sparse spontaneous chemistries.} {\bf(a)} Steady state concentration $x_n$ as a function of $n$ is shown on a semi-log plot for sparse chemistries with $f=1$ and degrees, $k=1, 2$ and $10$. Also shown for comparison are the steady state profiles for a fully connected chemistry. $A = k_f = k_r = \phi = 1$, $N=100$. {\bf(b)} Steady state concentration profile for sparse chemistry with $f=2$ and degree, $k=1$. $k_f = k_r = x_{(1,0)} = x_{(0,1)}=1$, $\phi = 10$, $N = 40$. In comparison to a fully connected chemistry, the steady state concentrations are smaller for a sparse chemistry.}
    \label{noacs-sparse}
\end{figure}

For a sparse spontaneous chemistry, the plot of steady state concentration $x_n$ versus $n$ is shown in Fig. \ref{noacs-sparse}. The steady state concentration of molecules are smaller in a chemistry with smaller average degree as compared to fully connected chemistry as the number of ways is which a molecule can be produced are fewer.

\section{\label{section-spont-hetero}Results for a heterogeneous chemistry}
In the earlier sections we have discussed chemistries in which all forward and reverse rate constants take same values, respectively. In real chemistries, the rate constants depend upon various physical parameters such as the energy of the reactants and products, the activation energy of the reaction, etc. The difference in the energies of the reactants and the products determine the ratio of forward and reverse rate constants. Whereas, their actual value is determined by the activation energy of the reaction, and is given by `Arrhenius' equation,
\begin{equation}
    \label{arrhenius-eq}
        k = Ze^{-E_a},
\end{equation}
where, $Z$ is the pre-exponential factor and $E_a$ is the activation energy of the reaction in the units of $k_BT$.

\begin{figure}
    \begin{center}
        \subfloat[]{\label{activation-energy-a}\includegraphics{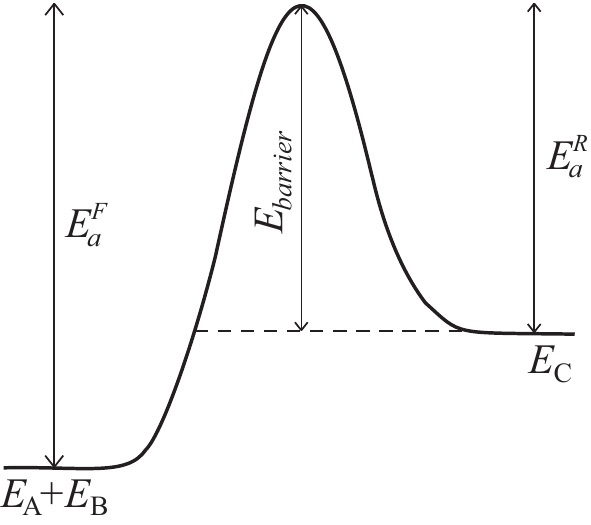}} \qquad\qquad
        \subfloat[]{\label{activation-energy-b}\includegraphics{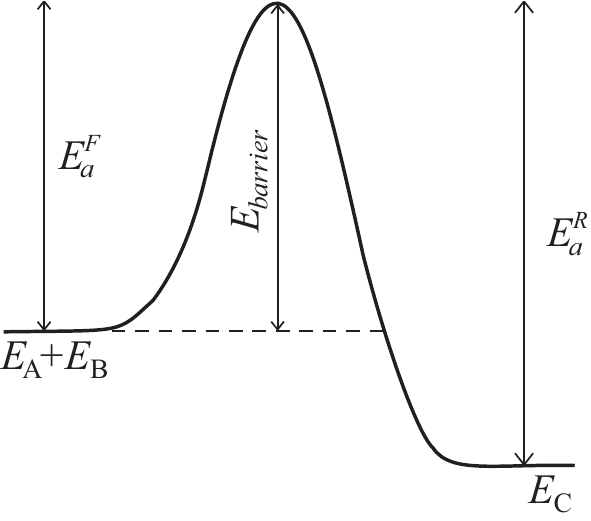}}
    \end{center}
    \caption[Activation energy for a reversible reaction pair.]{{\bf Activation energy for a reversible reaction pair.} For a reversible reaction pair, $\mathrm{\mathbf A} + \mathrm{\mathbf B} \rightleftharpoons \mathrm{\mathbf C}$, the activation energy for forward reaction is defined as $E^F_a$ and for reverse reaction $E^R_a$. $E_\mathrm{A} + E_\mathrm{B}$ represents the energy of the reactants and $E_\mathrm{C}$, energy of product. The barrier height is defined as the lower of two activation energies. {\bf(a)} $E_{barrier} = E^R_a$. {\bf(b)} $E_{barrier} = E^F_a$.}
    \label{activation-energy}
\end{figure}

For a reversible reaction, $\mathrm{\mathbf A} + \mathrm{\mathbf B} \rightleftharpoons \mathrm{\mathbf C}$, the two rate constants are given as,
\begin{equation}\label{hetro-rates}
        k^F_\mathrm{AB} = Z_1e^{-E^F_a},
        k^R_\mathrm{AB} = Z_2e^{-E^R_a},
\end{equation}
where, $E^F_a$ ($E^R_a$) is the activation energy for the forward (respectively, reverse) reaction. The difference between $E^F_a$ and $E^R_a$ is equal to the difference between the energy levels of the reactants and the products (as defined for the forward reaction), \ie, $E^F_a - E^R_a = E_\mathrm{C} - (E_\mathrm{A} + E_\mathrm{B})$. $E_\mathrm{X}$ is the energy of molecule $\mathbf X$. For every reaction pair we define barrier height as the smaller of the two activation energies (Fig. \ref{activation-energy}).

We employ following scheme to assign reaction rates. We first assign each molecule an energy. For a molecule of length $n = 2, 3, 4, \ldots$, the energy, $E_n$, is given as,
\begin{equation}\label{energy}
    E_n = n\epsilon_1 \pm \eta_1\epsilon_2,
\end{equation}
where, $\epsilon_1$ is the energy of the monomer, $E_1 = \epsilon_1$, and $\epsilon_2$ is the scale of variation in energies. All energies are expressed in units of $k_BT$. $\eta_1$ is a random number in $[0,1]$. We then assign barrier height to each reaction $\mathrm{\mathbf A} + \mathrm{\mathbf B} \rightleftharpoons \mathrm{\mathbf C}$ as
\begin{equation}
    E_{barrier} = \eta_2\epsilon_3,
\end{equation}
where, $\epsilon_3$ is the scale of the barrier heights and $\eta_2$ is a random number in $[0,1]$. $E_{barrier}$ is used to determine the forward and the reverse activation energies using the energies of the reactants and the products as,
\\if $E_\mathrm{A} + E_\mathrm{B} < E_\mathrm{C}$, $E^F_a = (E_\mathrm{A} + E_\mathrm{B}) + E_{barrier}, E^R_a = E_{barrier}$ (see Fig \ref{activation-energy-a}), and
\\if $E_\mathrm{A} + E_\mathrm{B} > E_\mathrm{C}$, $E^F_a = E_{barrier}, E^R_a = (E_\mathrm{A} + E_\mathrm{B}) + E_{barrier}$ (see Fig \ref{activation-energy-b}).\\
The reactions rates are calculated using Eq. (\ref{hetro-rates}). Thus the reaction rates are determined by the parameters $\epsilon_1, \epsilon_2, \epsilon_3, Z_1, Z_2$, and the random numbers $\eta_1$ (one for each molecule) and $\eta_2$ (one for each reaction).

The Steady state concentration profile for a heterogeneous spontaneous chemistry with $f=1$ is shown in Fig. \ref{noacs-hetero-ssc}. It shows that even in the presence of a small amount of heterogeneity in the system, while there are ups and downs in the steady state concentration profile, the steady state concentrations of molecules still generally fall exponentially with their lengths. We still find a single global fixed point attractor. The energies assigned to the molecules are shown in Fig. \ref{noacs-hetero-energy} and the forward and reverse reaction rates in Fig. \ref{noacs-hetero-kf} and \ref{noacs-hetero-kr} respectively.

\begin{figure}
    \begin{center}
        \subfloat[]{\label{noacs-hetero-ssc}\includegraphics[height=4.75in,angle=-90,trim=0.25cm 0.25cm 0.0cm 0.3cm,clip=true]{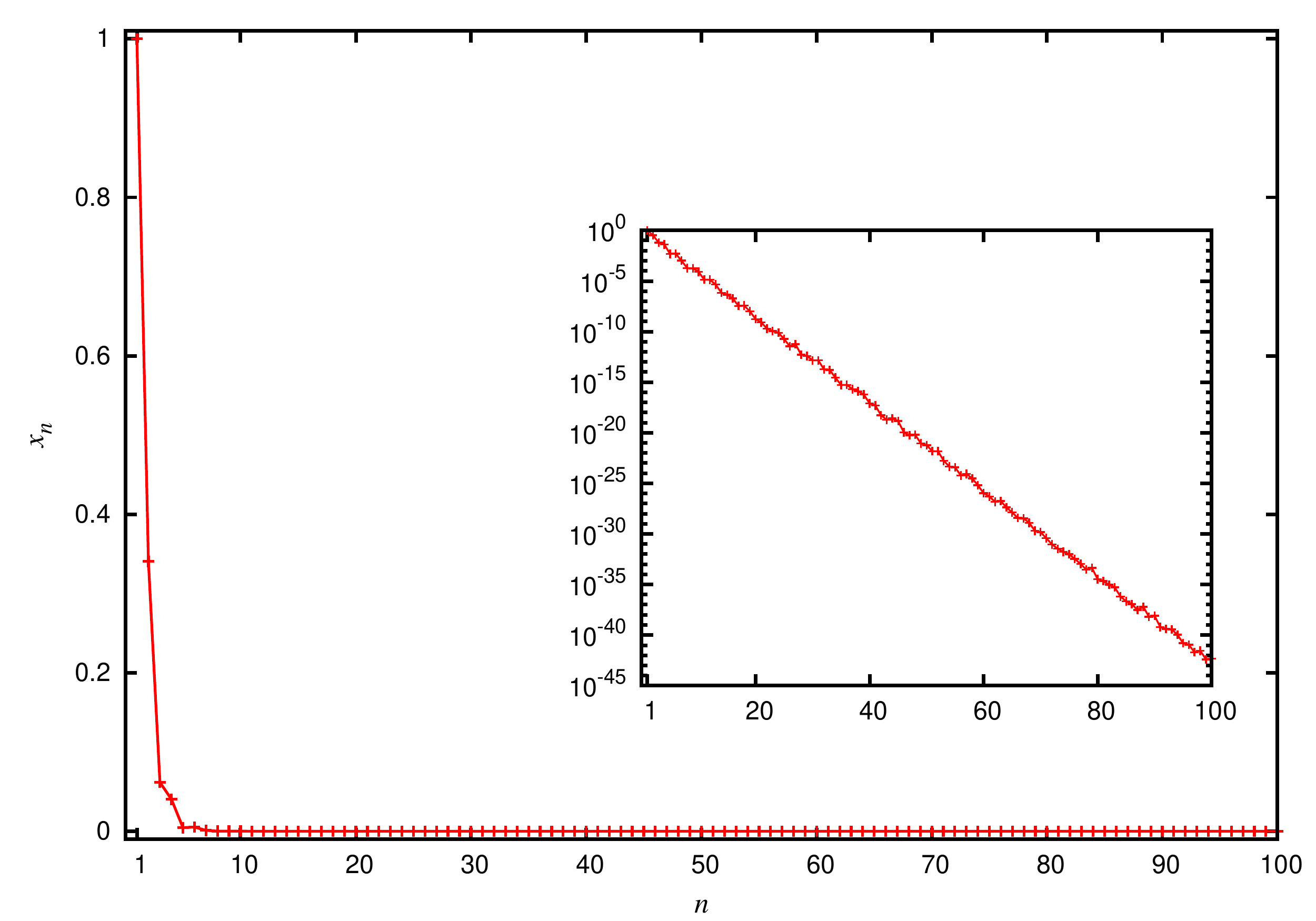}} \\
        \subfloat[]{\label{noacs-hetero-energy}\includegraphics[height=4.75in,angle=-90,trim=0.25cm 0.25cm 0.0cm 0.3cm,clip=true]{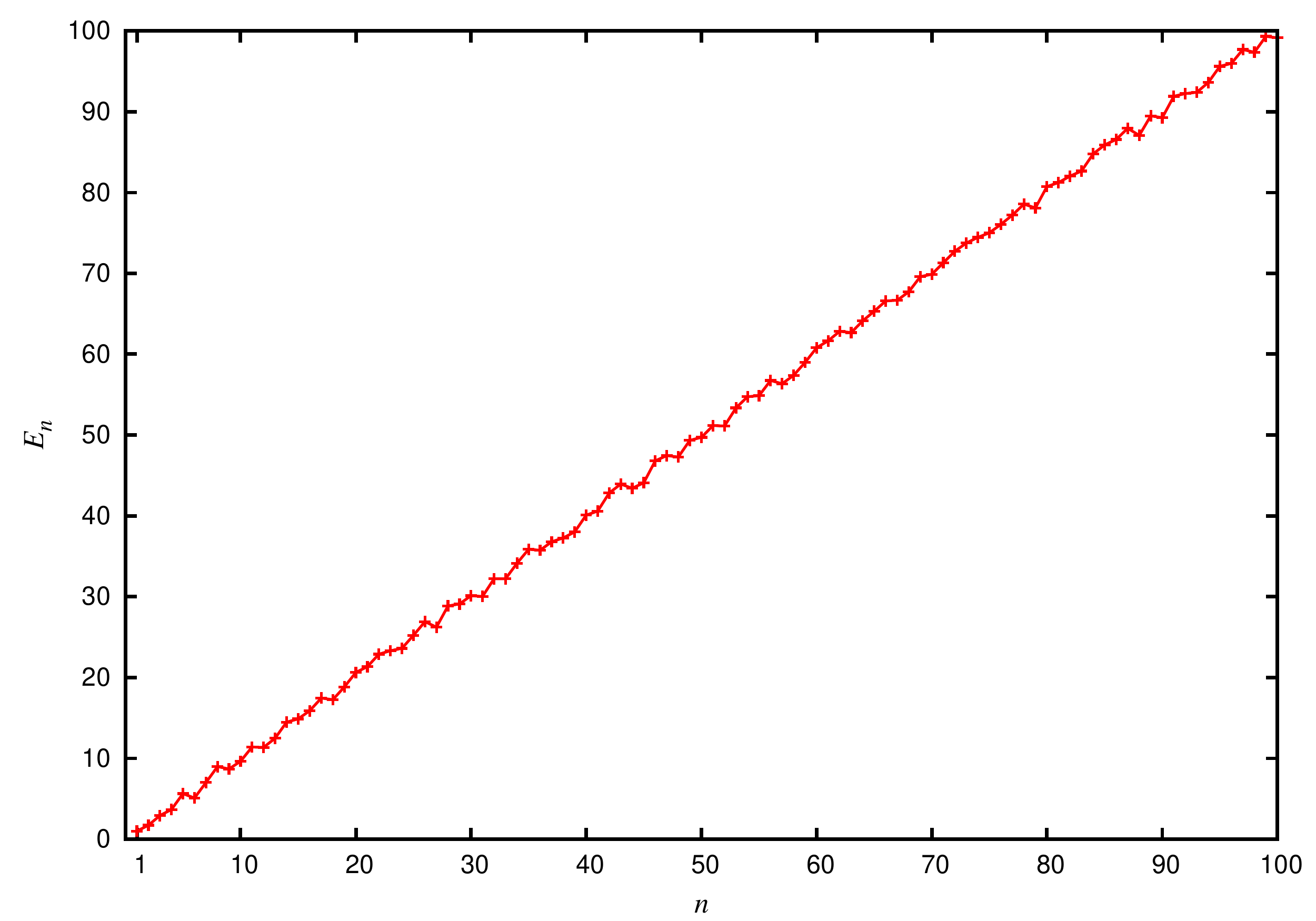}}
    \end{center}
    \caption[Heterogeneous spontaneous chemistries with single food source.]{{\bf Heterogeneous spontaneous chemistries with single food source.} {\it Continued.}}
\end{figure}

\captionsetup{list=no}
\begin{figure}
    \ContinuedFloat
    \begin{center}
        \subfloat[]{\label{noacs-hetero-kf}\includegraphics[height=3.5in,angle=-90,trim=1cm 4cm 1cm 4cm,clip=true]{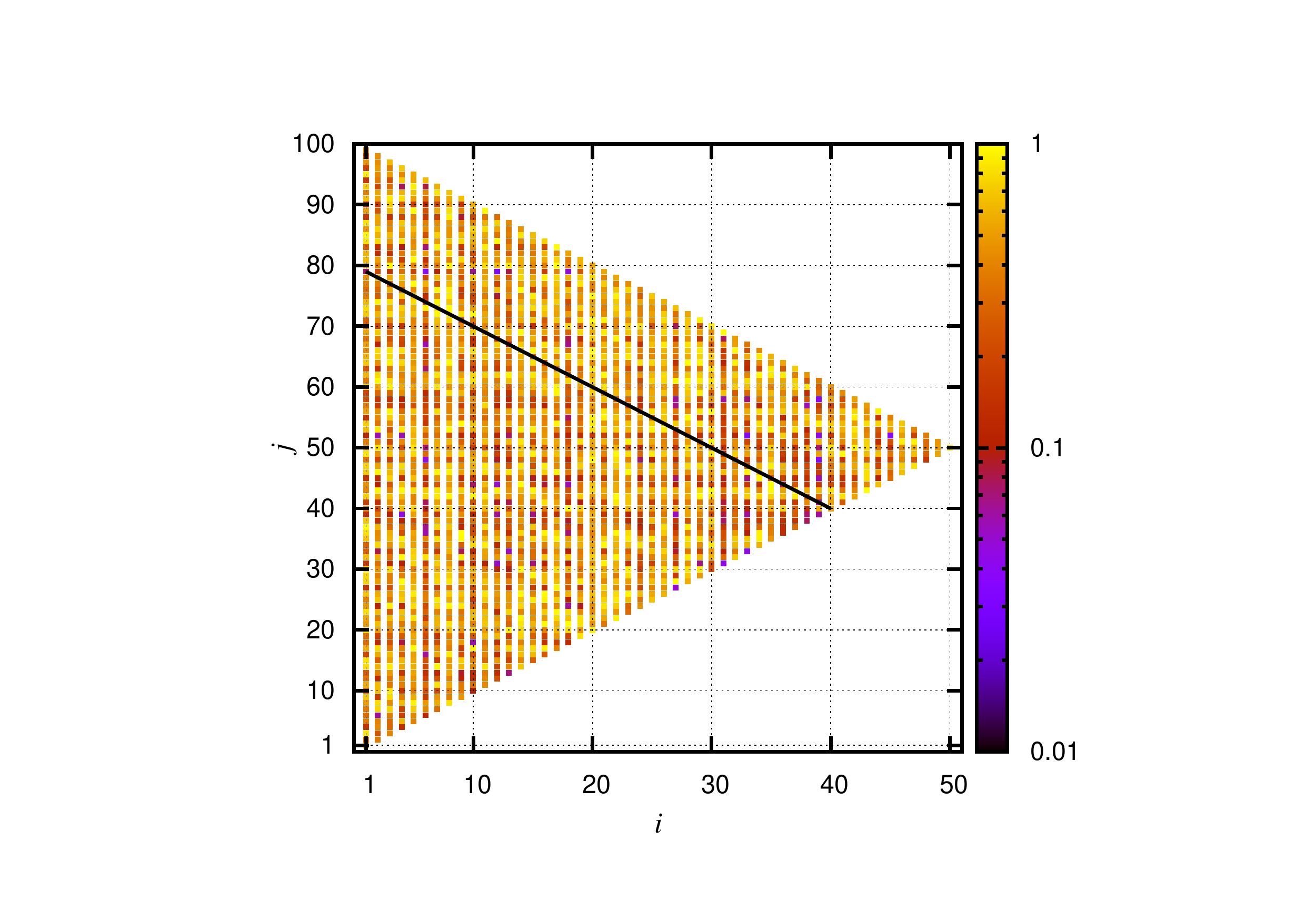}}\\
        \subfloat[]{\label{noacs-hetero-kr}\includegraphics[height=3.5in,angle=-90,trim=1cm 4cm 1cm 4cm,clip=true]{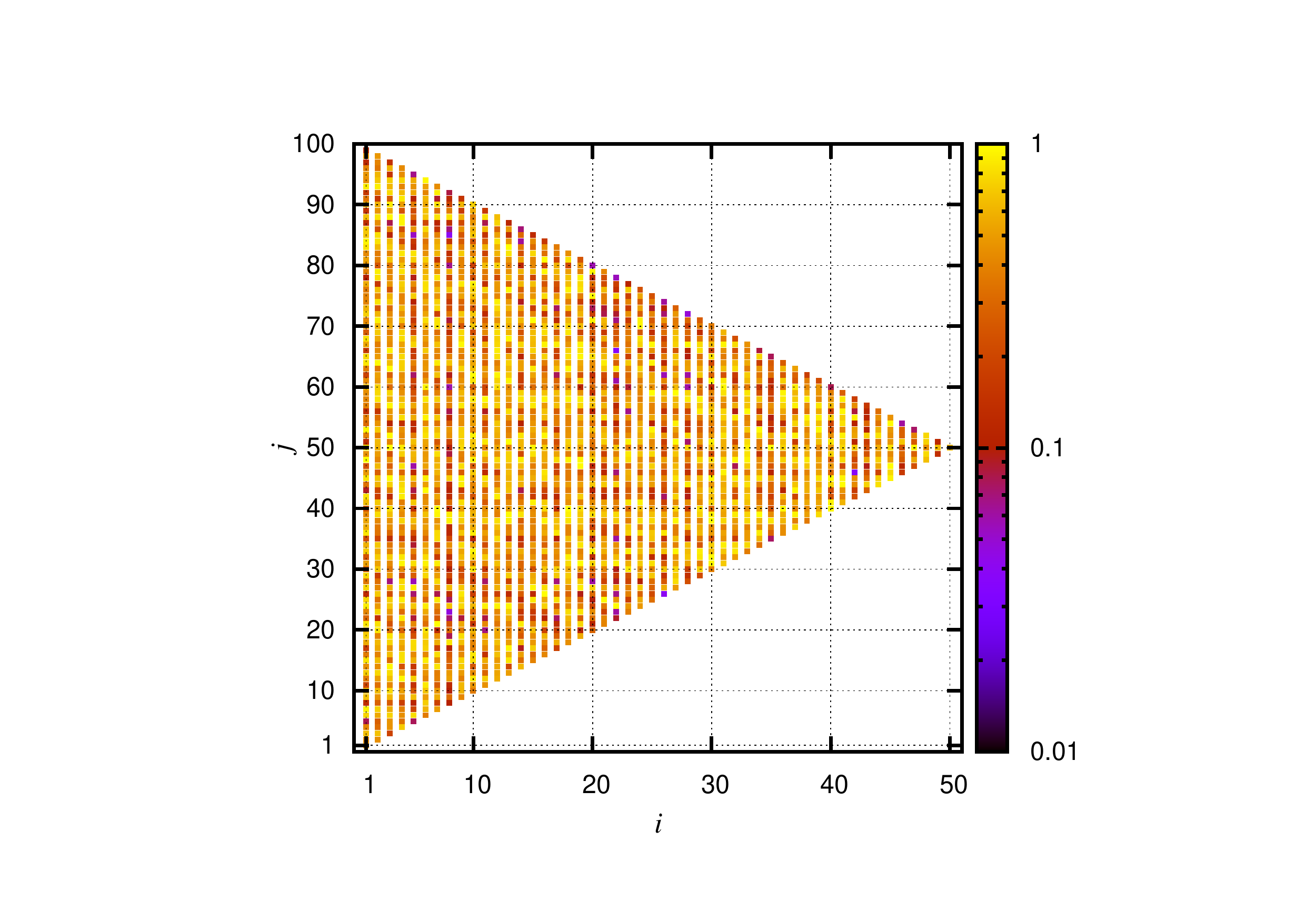}}
    \end{center}
    \caption[]{{\bf Heterogeneous spontaneous chemistries with single food source.} The forward ($k^F_{ij}$) and reverse ($k^R_{ij}$) reaction rates for every reaction ($\mathbf A(i) + \mathbf A(j) \rightleftharpoons \mathbf A(i+j)$) were assigned using parameter values $\epsilon_1 = \epsilon_2 = \epsilon_3 = Z_1 = Z_2 = 1$. {\bf(a)} Steady state concentration $x_n$ as a function of $n$. Inset shows the same on a semi-log plot. $A = \phi = 1$, $N=100$. {\bf(b)} Energy, $E_n$ of the molecule of length $n$ assigned using Eq. (\ref{energy}). The value of {\bf(c)} $k^F_{ij}$ and {\bf(d)} $k^R_{ij}$ indicated in a color map on a logarithmic scale. The solid black line in (c) indicates all the forward rate constants for the reactions which produce $\mathbf A(80)$ ($i + j = 80$).}
    \label{noacs-hetero}
\end{figure}
\captionsetup{list=yes}

We remark that instead of the above scheme for choosing $k^F_\mathrm{AB}$ and $k^R_\mathrm{AB}$ if we simply choose $k^F_\mathrm{AB} = k_f + \eta^f_\mathrm{AB}, k^R_\mathrm{AB} = k_f + \eta^r_\mathrm{AB}$ where $\eta^f_\mathrm{AB}$ and $\eta^r_\mathrm{AB}$ are random numbers drawn uniformly from the range $[-\eta,\eta]$ for some small $\eta$, then a behaviour similar to Fig. \ref{noacs-hetero-ssc} is obtained.

\section{Summary}
We show that for the model presented in Chapter \ref{ChapterModel}, in presence of buffered food set molecules, the concentrations of large molecules produced via a spontaneous chemistry decline exponentially with the size of the molecule. This system is found to have a global fixed point attractor. We also discussed chemistries that are sparse and heterogeneous. We find this qualitative feature is robust to small heterogeneity present in the chemistry and also holds for sparse chemistries. It may be interesting to relate this model to work on monotone dynamical systems \cite{Angeli2009, Craciun2011}. In subsequent chapters we discuss the role of catalysis in production of large molecules.

\thispagestyle{plain}
\cleardoublepage
\chapter{\label{ChapterACS}Catalyzed chemistries and autocatalytic sets}
\lettrine[lines=2, lhang=0, loversize=0.0, lraise=0.0]{T}{his} chapter discusses the model when the chemistries contain some catalyzed reactions in addition to the spontaneous reactions. We study the dynamics in presence of `autocatalytic sets' (ACSs). We begin by defining ACSs and then discuss various dynamical properties exhibited by the chemistries containing them including the existence of multistability. We show how a small subset of molecules can be produced in large concentration (compared to other molecules in the chemistry) in presence of an ACS.

\section{\label{Definition-ACS}Autocatalytic Sets (ACSs)}
Consider a set $S$ of catalyzed one-way reactions. `One-way' means that each reaction in $S$ is either a ligation or cleavage reaction. Thus the set of reactants and the set of products are unambiguously defined for each reaction and the two sets are distinguished. The presence of a given ligation or cleavage reaction in $S$ does not mean that its reverse is also necessarily a member of $S$. Let $P(S)$ be the union of sets of products of all reactions in $S$, and $R(S)$ the union of sets of reactants of all reactions in $S$. We exclude the food set molecules from both $P(S)$ and $R(S)$. We will refer to the set $S$ of catalyzed reactions as an \gls{autocatalytic-set} (\gls{ACS}) if (a) $P(S)$ includes a catalyst for every reaction in $S$, and (b) $R(S) \subset P(S)$. The latter condition implies that all members of $R(S)$ can be produced from the food set by (recursively) applying reactions from within $S$. An ACS thus ensures the existence of a catalyzed pathway, starting from the food set, for the production of each of its products \cite{Eigen1971,Kauffman1971,Rossler1971}. Alternative valid definitions of an ACS can be given (see \cite{Wachtershauser1990, Morowitz2000} for one such); the above definition suffices for our present purposes.

Note that if $S$ is an ACS, then its extension, $S'$, that additionally includes the reverse of some reactions in $S$, is also trivially an ACS, as in our scheme a catalyst works for both forward and reverse reactions if both exist in the chemistry.

\section{An example of chemistries with autocatalytic sets}
\begin{figure}
    \begin{center}
        \includegraphics[height=6.25in]{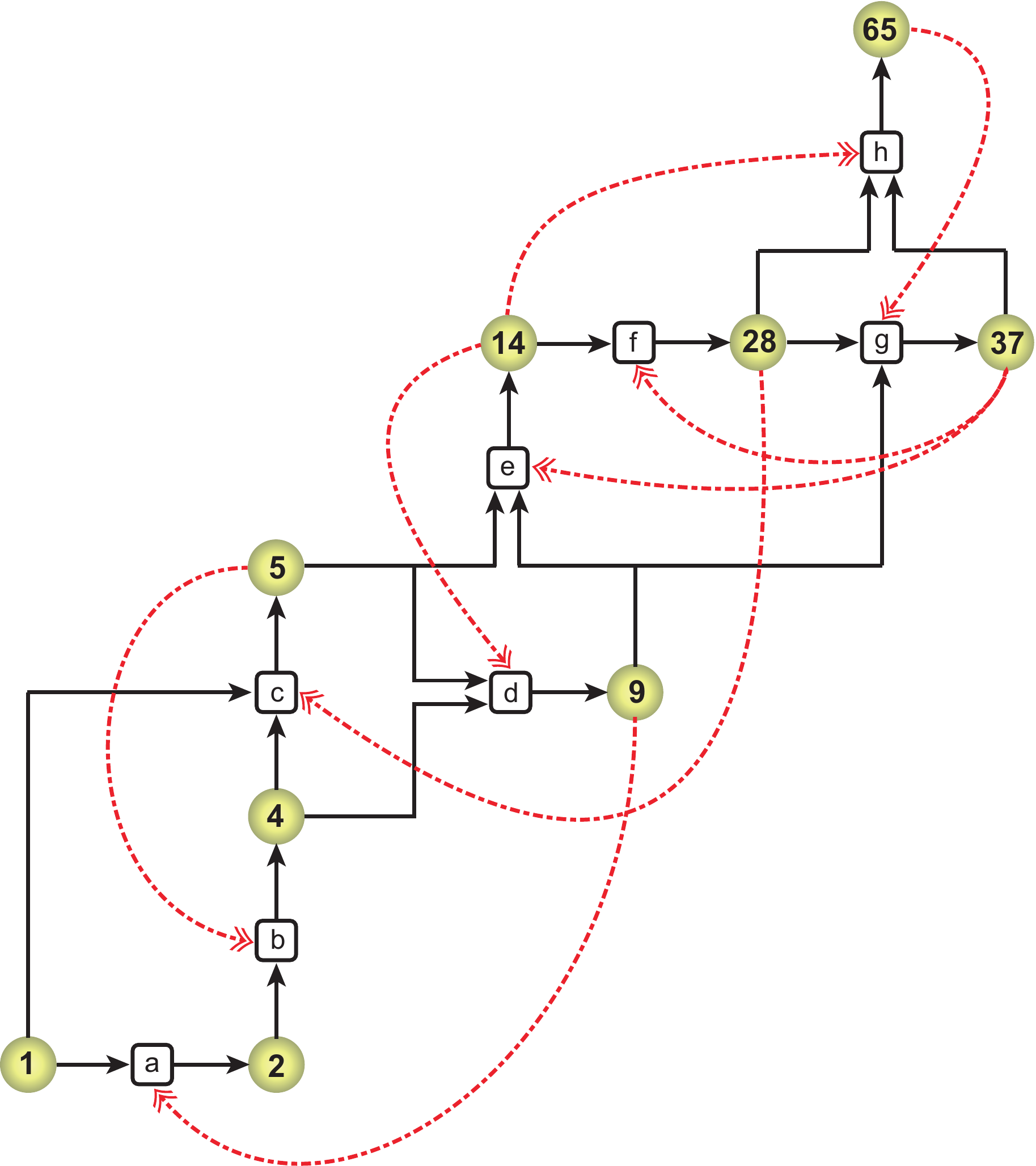}
    \end{center}
    \caption[Pictorial representation of the catalyzed chemistry in Eqs. (\ref{acs65-definition}), referred to as ACS65.]{{\bf Pictorial representation of the catalyzed chemistry in Eqs. (\ref{acs65-definition}), referred to as ACS65.} This is a directed bipartite graph with two types of links. Circular nodes represent molecules and rectangular nodes represent reactions. The label inside the nodes identify the nodes (molecule size $n$ for circular nodes and reaction sub-equation label of Eq. (\ref{acs65-definition}) for rectangular nodes). A black solid arrow from a molecule to a reaction node indicates that the former is a reactant in the latter, and one from a reaction to a molecule node that the latter is a product of the former. A red dashed arrow from a molecule to a reaction node indicates that the former is a catalyst for the latter. To avoid visual clutter some black arrows starting from molecule nodes are shown to branch out into more than one arrow. (For example, the arrow from molecule node 5 branches into reaction nodes d and e; this means that molecule 5 is a reactant in both reactions. This structure should not be construed as a bi-directional link between reaction nodes d and e.) The figure only represents the ligation reactions in the catalyzed chemistry; the reverse (cleavage) reactions are not shown.}
    \label{acs-eg-network}
\end{figure}

As a specific example to display certain generic properties, we consider the following catalyzed chemistry (for a diagrammatic representation of the network topology see Fig. \ref{acs-eg-network}):
\begin{subequations}
  \label{acs65-definition}
  \begin{eqnarray}
    \label {acs65-rct1} \mathrm{\mathbf A}(1) + \mathrm{\mathbf A}(1) & \reactionrevarrow{\ensuremath{\mathrm{\mathbf A}(9)}}{} & \mathrm{\mathbf A}(2) \\
    \mathrm{\mathbf A}(2) + \mathrm{\mathbf A}(2) & \reactionrevarrow{\ensuremath{\mathrm{\mathbf A}(5)}}{} & \mathrm{\mathbf A}(4) \\
    \label {acs65-rct3} \mathrm{\mathbf A}(1) + \mathrm{\mathbf A}(4) & \reactionrevarrow{\ensuremath{\mathrm{\mathbf A}(28)}}{} & \mathrm{\mathbf A}(5) \\
    \mathrm{\mathbf A}(4) + \mathrm{\mathbf A}(5) & \reactionrevarrow{\ensuremath{\mathrm{\mathbf A}(14)}}{} & \mathrm{\mathbf A}(9) \\
    \mathrm{\mathbf A}(5) + \mathrm{\mathbf A}(9) & \reactionrevarrow{\ensuremath{\mathrm{\mathbf A}(37)}}{} & \mathrm{\mathbf A}(14) \\
    \mathrm{\mathbf A}(14) + \mathrm{\mathbf A}(14) & \reactionrevarrow{\ensuremath{\mathrm{\mathbf A}(37)}}{} & \mathrm{\mathbf A}(28) \\
    \mathrm{\mathbf A}(9) + \mathrm{\mathbf A}(28) & \reactionrevarrow{\ensuremath{\mathrm{\mathbf A}(65)}}{} & \mathrm{\mathbf A}(37) \\
    \mathrm{\mathbf A}(28) + \mathrm{\mathbf A}(37) & \reactionrevarrow{\ensuremath{\mathrm{\mathbf A}(14)}}{} & \mathrm{\mathbf A}(65).
  \end{eqnarray}
\end{subequations}

\begin{figure}
    \begin{center}
        \subfloat[]{\label{acs-eg-ssc}\includegraphics[height=4.75in,angle=-90,trim=0.5cm 0.5cm 0cm 0.3cm,clip=true]{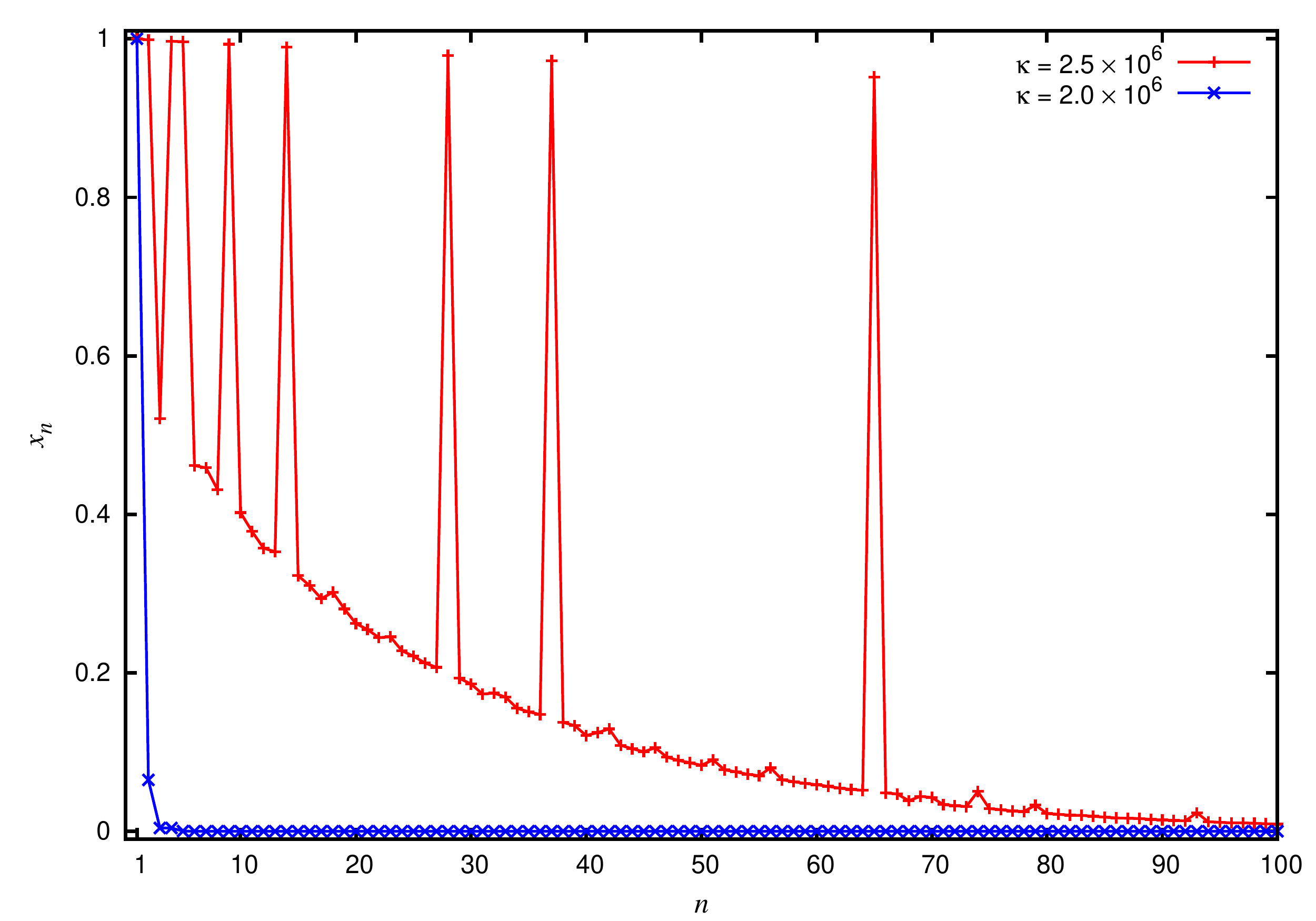}} \\
        \subfloat[]{\label{acs-eg-time}\includegraphics[height=4.75in,angle=-90,trim=0.5cm 0.5cm 0cm 0.3cm,clip=true]{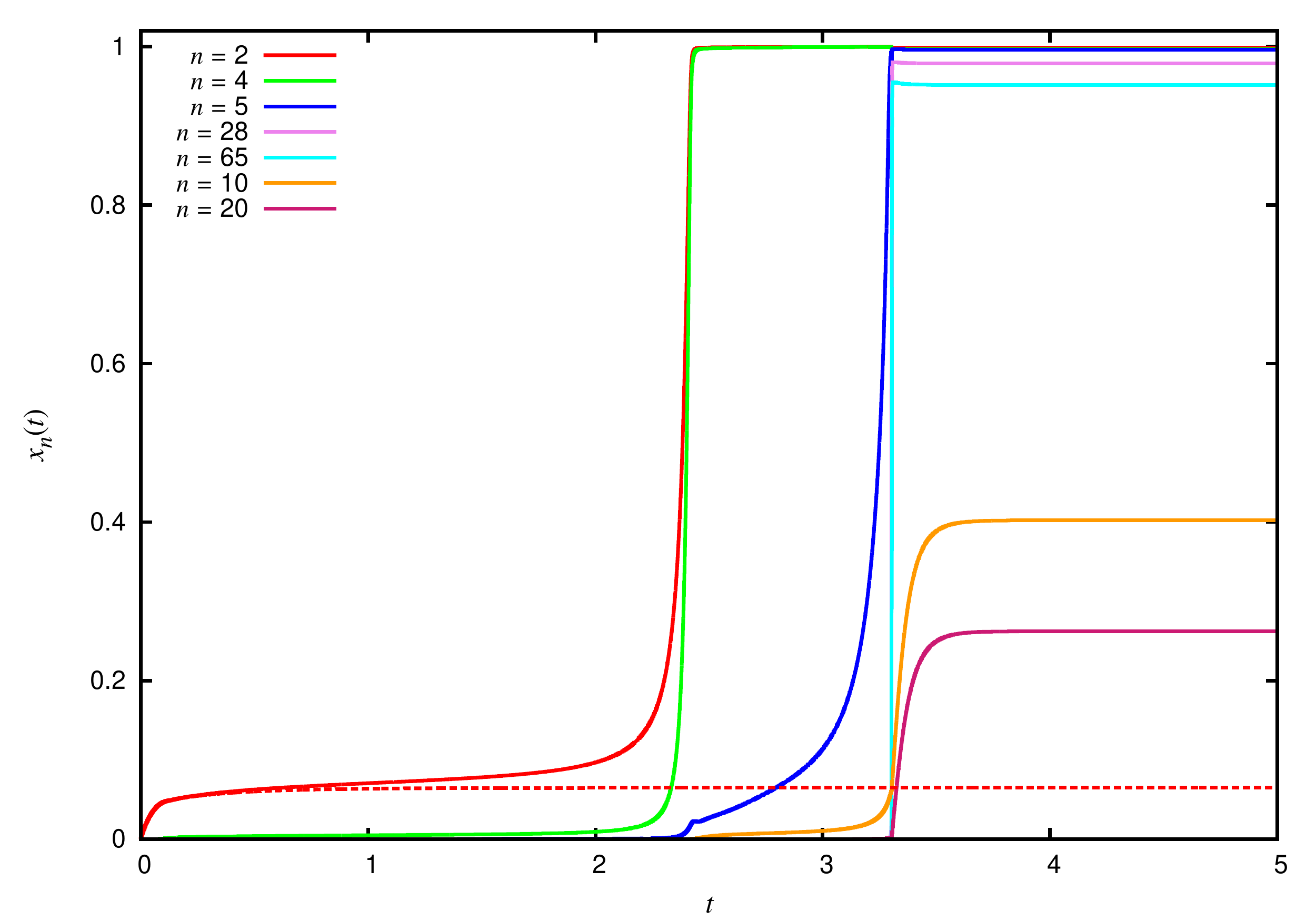}}
    \end{center}
    \caption[Steady state concentration profile for ACS65 (Eqs. (\ref{acs65-definition})).]{{\bf Steady state concentration profile for ACS65 (Eqs. (\ref{acs65-definition})).} $k_f = k_r = A = 1, \phi=15, N=100.$ {\bf (a)} The steady state concentration profile for two values of $\kappa$. There exists a threshold value of $\kappa$ above which ACS dominates the background. For $\kappa = 2.5 \times 10^6$ ACS molecules have significantly higher concentrations as compared to their neighbours. For a lower value of $\kappa$ $(=2 \times 10^6)$ this phenomenon disappears and the system behaves as if it only contains spontaneous reactions. {\bf (b)} Evolution of concentration with time for some ACS molecules ($n=2,4,5,28,65$) and background molecules ($n=10,20$) is shown for $\kappa = 2.5 \times 10^6$. The concentration of ACS molecules sharply rise to higher values as and when right catalysts and/or reactants get produced in sufficient concentrations. The dotted line shows the time evolution of $\mathbf A(2)$ for $\kappa = 2 \times 10^6$.}
    \label{acs-eg}
\end{figure}

Note that this set of reactions constitutes an ACS (which we will refer to as ACS65). If any one reaction pair is deleted from the set, it is no longer an ACS. For the moment, for simplicity, we consider the case where the catalytic strengths of all the catalyzed reactions are equal (`homogeneous' catalytic strengths): $\kappa^{1,1}_9 = \kappa^{2,2}_{5} =
\kappa^{1,4}_{28} =\kappa^{4,5}_{14} =\kappa^{5,9}_{37} =\kappa^{14,14}_{37} =\kappa^{9,28}_{65} =\kappa^{28,37}_{14} =\kappa$, and all other $\kappa^{ij}_m = 0$. (For clarity, in view of double digit indices, we have introduced a comma between the pair of indices in the superscript.) Fig. \ref{acs-eg-ssc} describes the steady state concentrations, starting from the standard initial condition (Eq. (\ref{standard-ic-eq})), for the chemistry that contains these eight catalyzed reactions in addition to all the reactions of the fully connected spontaneous chemistry. At $\kappa=2.5 \times 10^6$ the ACS product molecules `dominate' over the \gls{background} (the `background' being defined as the set of all molecules except the ACS product molecules and the food set), in the sense that the ACS molecules have significantly larger populations than their `neighbours', namely, the background molecules of similar size \cite{Bagley1991}. There is a fairly sharp threshold value of $\kappa$ above which ACS domination appears, as evident from the comparison with the lower curve in Fig. \ref{acs-eg-ssc} drawn for $\kappa=2.0 \times 10^6$. Time evolution of concentrations for a few ACS and background molecules is shown in Fig. \ref{acs-eg-time}. As the right catalysts and/or reactants become available the concentrations of ACS molecules jump sharply to higher values.

\begin{figure}
    \begin{center}
        \includegraphics[height=4.75in,angle=-90,trim=0.5cm 0.5cm 0cm 0.3cm,clip=true]{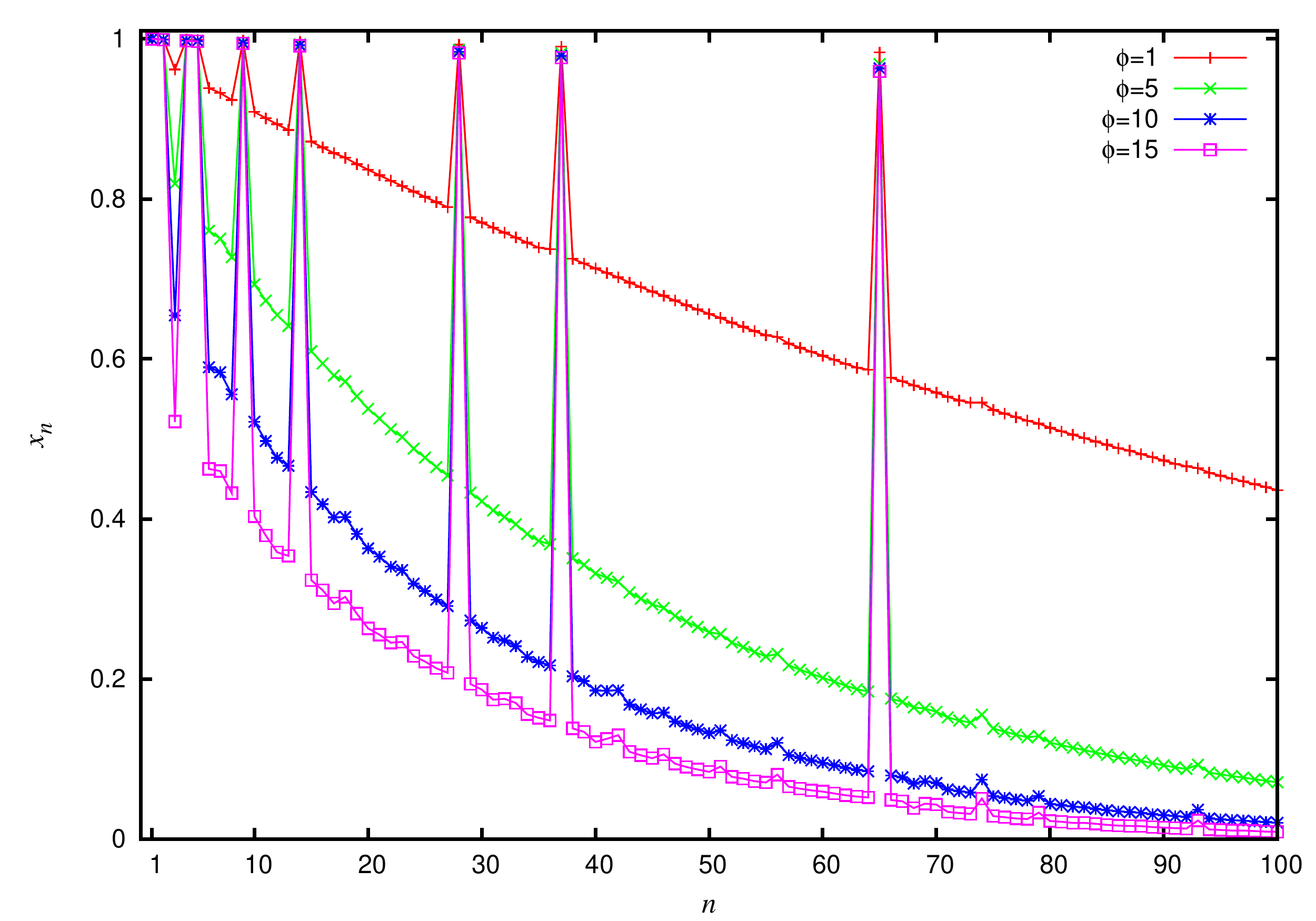}
    \end{center}
    \caption[Steady state concentration profile for ACS65 for different values of $\phi$.]{{\bf Steady state concentration profile for ACS65 for different values of $\phi$.} The concentration profile for four values of $\phi=1,5,10,15$ is shown for $\kappa = 3.0 \times 10^6$. As $\phi$ increases the concentration of molecules in background decreases, though the concentrations of ACS molecules remains largely unaffected.}
    \label{acs-eg-phi}
\end{figure}

\begin{figure}
    \begin{center}
        \includegraphics[height=4.75in,angle=-90,trim=0.5cm 0.5cm 0cm 0.3cm,clip=true]{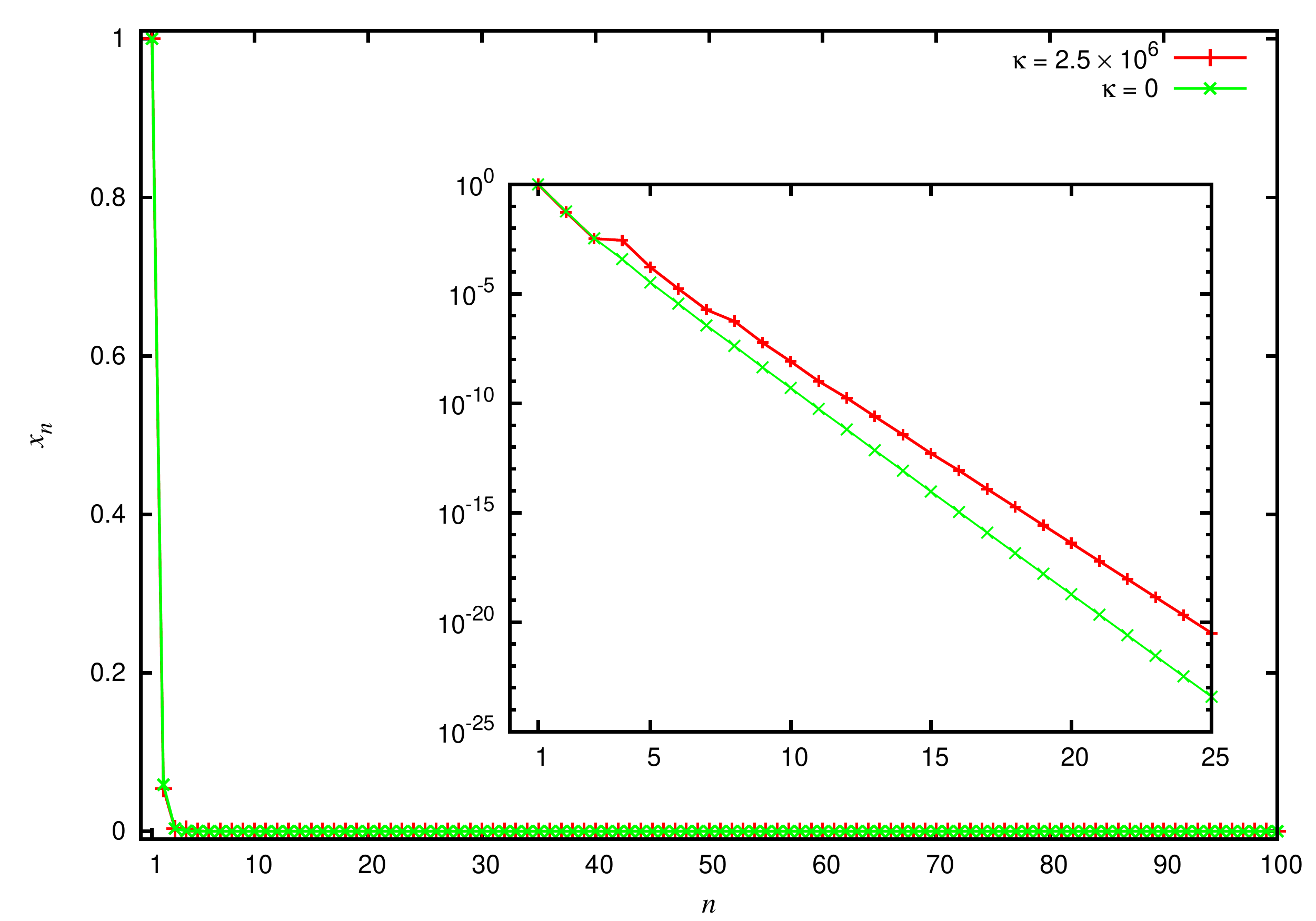}
    \end{center}
    \caption[Steady state concentration for ACS65 with reaction (\ref{acs65-rct1}) deleted.]{{\bf Steady state concentration for ACS65 with reaction (\ref{acs65-rct1}) deleted.} $k_f = k_r = A = 1, \phi=15, N=100.$ The concentration profile for $\kappa=2.5 \times 10^6$ but with reaction (\ref{acs65-rct1}) removed from the ACS65 (red curve) compared with the profile for the spontaneous chemistry, $\kappa=0$ (green curve). The inset shows the same with $x_n$ on a logarithmic scale. On the linear scale the two curves are indistinguishable.}
    \label{acs-eg-rct-del}
\end{figure}

Fig. \ref{acs-eg-phi} shows that the steady state background concentrations decline as $\phi$ increases, while the ACS concentrations are relatively unaffected in this regime (thus ACS domination increases).  If catalyzed production pathways from the food set to other molecules are broken somewhere, the concentration of the latter molecules declines significantly. This is evident from Fig. \ref{acs-eg-rct-del} for which only one reaction pair (\ref{acs65-rct1}) is deleted from the catalyzed chemistry (which now contains no ACS) while others are catalyzed at the same strength as before.

\begin{figure}
    \begin{center}
        \includegraphics[height=5in,angle=-90,trim=2.5cm 0.5cm 2.5cm 0.1cm,clip=true]{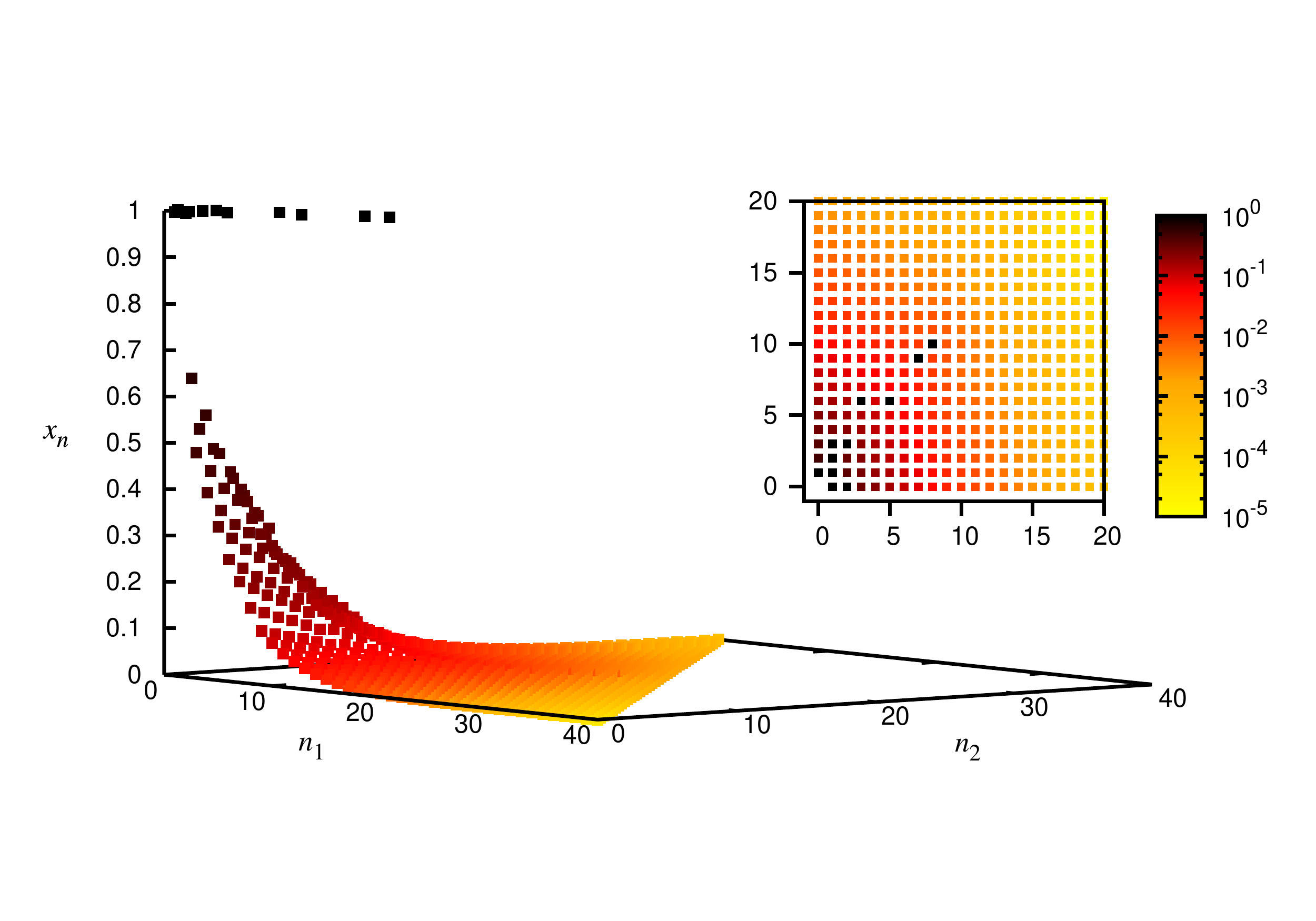}
    \end{center}
    \caption[Steady state concentration profile for ACS(8,10) in a chemistry with $f=2$.]{{\bf Steady state concentration profile for ACS(8,10) in a chemistry with $f=2$.} The 3D plots show the steady state concentration $x_n$ of the molecules $n=(n_1,n_2)$ as a function of $n_1$ and $n_2$. The inset shows a `top view' of the $(n_1,n_2)$ plane with $x_n$ indicated in a colour map on a logarithmic scale. $k_f = k_r = x_{(1,0)} = x_{(0,1)} = 1, \phi = 10, \kappa=10^6, N=40.$}
    \label{acs-eg-2d}
\end{figure}

ACS domination at a sufficiently high catalytic strength also occurs when there is more than one monomer. An example with $f=2$, referred to as ACS(8,10), is given below (Eqs. \ref{acs8-10-definition}) and the steady state concentrations shown in Fig. \ref{acs-eg-2d}.
\begin{subequations}
  \label{acs8-10-definition}
  \begin{eqnarray}
   (0,1) + (1,0) & \reactionrevarrow{(1,3)}{} & (1,1) \\
   (1,0) + (1,0) & \reactionrevarrow{(3,6)}{} & (2,0) \\
   (0,1) + (1,1) & \reactionrevarrow{(5,6)}{} & (1,2) \\
   (0,1) + (1,2) & \reactionrevarrow{(8,10)}{} & (1,3) \\
   (1,1) + (1,2) & \reactionrevarrow{(1,3)}{} & (2,3) \\
   (1,3) + (2,3) & \reactionrevarrow{(1,2)}{} & (3,6) \\
   (2,0) + (3,6) & \reactionrevarrow{(2,3)}{} & (5,6) \\
   (2,3) + (5,6) & \reactionrevarrow{(3,6)}{} & (7,9)  \\
   (1,1) + (7,9) & \reactionrevarrow{(2,3)}{} & (8,10).
  \end{eqnarray}
\end{subequations}

\subsection{\label{kappa-infty}Understanding why ACS concentrations are large (the $\kappa \rightarrow \infty$ limit)}
The above features are generic for a large class of ACSs. It is instructive to consider the $\kappa \rightarrow \infty$ limit which we discuss analytically. We discuss the case when $f=1$ but the argument can be extended to a general $f$. When $\kappa$ is nonzero, the terms in Eq. (\ref{rateequation-v-F1}) corresponding to catalyzed reactions get modified. The net flux $v$ of such reaction pairs on the r.h.s. (for brevity we are omitting the subscript $ij$ in $v_{ij}$) is replaced by $(1 + \kappa \sum_m x_m)v$, where the sum over $m$ is a sum over all catalysts of the reaction pair. Now let the set $S$ of catalyzed reactions be an ACS. Then, if $\mathrm{\mathbf A}(n) \in P(S)$ the r.h.s. of $\dot{x}_n$ contains at least one such catalyzed term, while if $\mathrm{\mathbf A}(n) \notin P(S)$ $\dot{x}_n$ contains no such term. For example, for the above mentioned ACS, ACS65, we have
\begin{subequations}
  \label{kappainftyrateeqns}
  \begin{eqnarray}
  \label{kappainftyrateeqns-x2dot}
  \dot{x}_2 & \simeq & \kappa (x_9 v_{1,1} - 2 x_5 v_{2,2}) + (\mathrm{terms\ independent\ of \ }\kappa)\\
  \label{kappainftyrateeqns-x4dot}
  \dot{x}_4 & \simeq & \kappa (x_5 v_{2,2} - x_{28} v_{1,4} - x_{14} v_{4,5}) + (\kappa^0 \mathrm {\ terms})\\
  \label{kappainftyrateeqns-x5dot}
  \dot{x}_5 & \simeq & \kappa (x_{28} v_{1,4} - x_{14} v_{4,5} - x_{37} v_{5,9}) + (\kappa^0 \mathrm{\ terms})\\
  \label{kappainftyrateeqns-x9dot}
  \dot{x}_9 & \simeq & \kappa (x_{14} v_{4,5} - x_{37} v_{5,9} - x_{65} v_{9,28}) + (\kappa^0 \mathrm {\ terms})\\
  \label{kappainftyrateeqns-x14dot}
  \dot{x}_{14} & \simeq & \kappa (x_{37} v_{5,9} - 2 x_{37} v_{14,14}) + (\kappa^0 \mathrm{\ terms})\\
  \label{kappainftyrateeqns-x28dot}
  \dot{x}_{28} & \simeq & \kappa (x_{37} v_{14,14} - x_{65} v_{9,28} - x_{14} v_{28,37}) + (\kappa^0 \mathrm{\ terms})\\
  \label{kappainftyrateeqns-x37dot}
  \dot{x}_{37} & \simeq & \kappa (x_{65} v_{9,28} - x_{14} v_{28,37}) + (\kappa^0 \mathrm{\ terms})\\
  \label{kappainftyrateeqns-x65dot}
  \dot{x}_{65} & \simeq & \kappa (x_{14} v_{28,37}) + (\kappa^0 \mathrm{\ terms}),
 \end{eqnarray}
\end{subequations}
while the rate equations for all other (non ACS) molecules ($\dot{x}_3$, $\dot{x}_6$, etc.) have no terms proportional to $\kappa$. In a steady state solution the r.h.s. of Eqs. (\ref{kappainftyrateeqns}) is zero, and to leading order in the $\kappa \rightarrow \infty$ limit we must set the coefficients of $\kappa$ to zero. The coefficients involve only the ACS fluxes $v_{ij}$ and catalyst concentrations. Each coefficient is a sum of terms, and each term is proportional to an ACS flux $v_{ij}$. Thus $v_{ij} = 0$ for the ACS fluxes provides a steady state solution in the $\kappa \rightarrow \infty$ limit. Numerically we find that when $\kappa$ is sufficiently high the rate equations converge to this solution starting from the standard initial condition. Now $v_{ij} = k_f x_i x_j - k_r x_{i+j}$, therefore $v_{ij} = 0$ implies $x_{i+j} = k_f x_i x_j / k_r$ for the members of $P(S)$. Since by definition there is a catalyzed pathway from the food set to every ACS product, we can recursively express the steady state concentration of every ACS molecule in terms of $x_1 = A$: $x_n = A (k_f A/k_r)^{n-1}$.

It is evident that this argument applies whenever the set $S$ of catalyzed reactions is an ACS; thus for every member of $P(S)$, $x_n \simeq A (k_f A/k_r)^{n-1}$ is a steady state solution of the rate equations in the limit $\kappa \rightarrow \infty$. This is corroborated numerically: in Fig. \ref{acs-eg-phi} since $A = k_f = k_r = 1$, all the eight ACS products should have $x_n = 1$ in this limit; the numerical result at $\kappa = 3 \times 10^6$ is not too far from this limiting analytical value.

\subsubsection*{A strong ACS counteracts dissipation}
Recall from Eq. (\ref{sszerophi}) and the discussion in Section \ref{phi0-analytical-solution} that every molecule in a homogeneous connected uncatalyzed chemistry has the steady state concentration $x_n = A (k_f A/k_r)^{n-1}$ when there is no dilution flux or dissipation ($\phi = 0$), and a smaller concentration when there is dissipation ($\phi > 0$). We have observed above that an ACS with a sufficiently large $\kappa$ can boost the steady state concentrations of its members, even when $\phi > 0$, to the same level. The expression $x_n = A (k_f A/k_r)^{n-1}$ seems to represent an upper limit on the steady state concentration of $\mathrm{\mathbf A}(n)$, which can be approached either when dissipation goes to zero, or, when there is dissipation, by membership of an ACS whose catalytic strength becomes very large.

When the reaction pair $\mathrm{\mathbf A}(1) + \mathrm{\mathbf A}(1) \rightleftharpoons \mathrm{\mathbf A}(2)$, in ACS65, is
not catalyzed the production of $\mathrm{\mathbf A}(2)$ takes place at a much smaller rate, the spontaneous rate. Therefore its concentration is much smaller, and hence so are the concentrations of the larger molecules.

When $\mathrm{\mathbf A}(n)$ belongs to the background the r.h.s. of $\dot{x}_n$ contains no term proportional to $\kappa$, and all the $\kappa$-independent terms have to be kept, including the $\phi x_n$ term. Thus its steady state concentration depends upon $\phi$, and as in the case of the uncatalyzed chemistry, declines more rapidly with $n$ when $\phi$ increases.

\subsection{Multistability in the ACS dynamics and ACS domination}
\begin{figure}
    \begin{center}
        \includegraphics[height=4.75in,angle=-90,trim=0.5cm 0.5cm 0cm 0.3cm,clip=true]{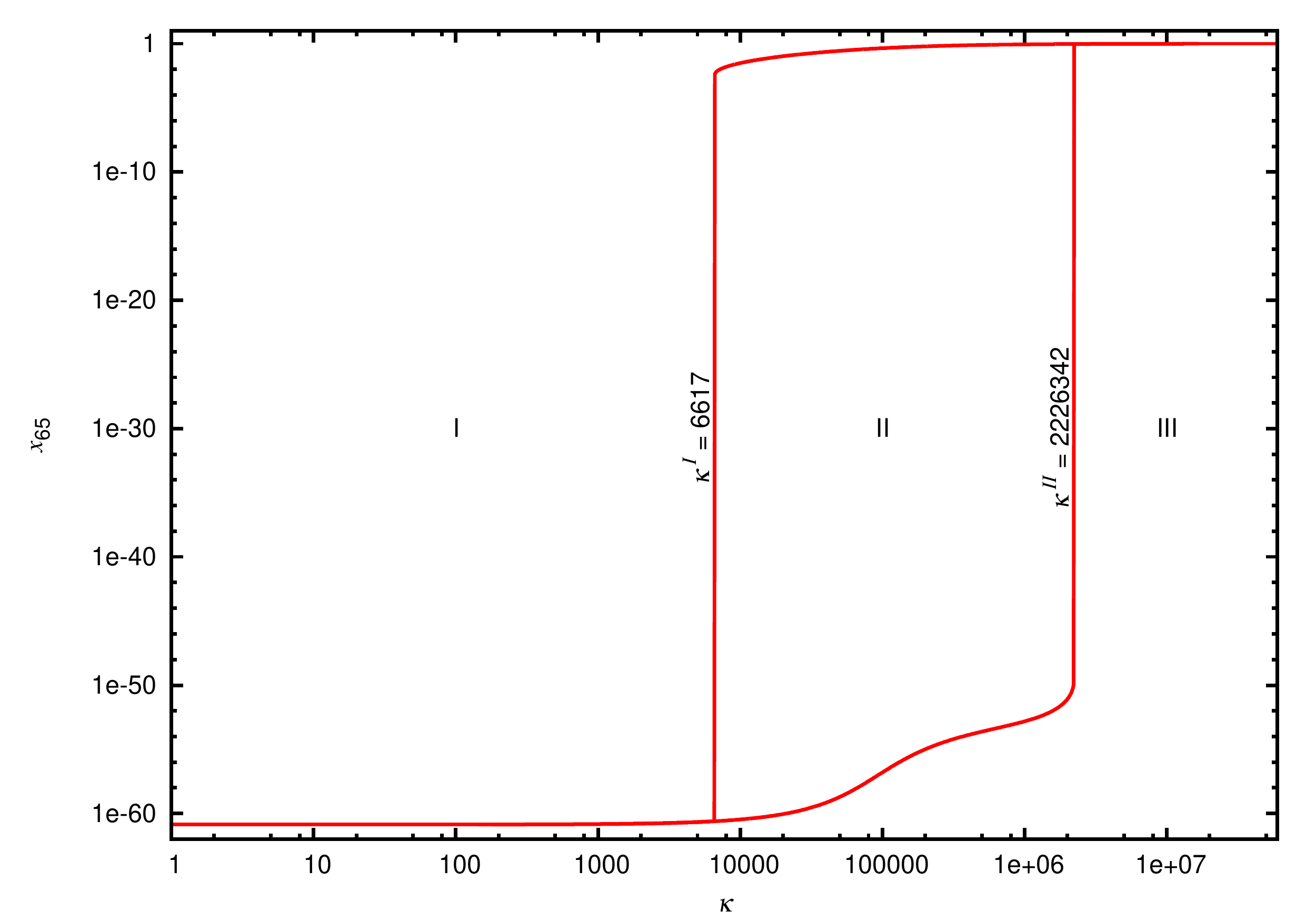}
    \end{center}
    \caption[Bistability in the dynamics of ACS65.]{{\bf Bistability in the dynamics of ACS65.} `Hysteresis curve' of the steady state concentration of $\mathbf A(65)$ versus $\kappa$ for $k_f = k_r = A = 1, \phi=15, N=100$. The curve is obtained by using two different initial conditions (i) the standard initial condition $x_n=0$ for all $n\geq2$, and (ii) a `high' initial condition $x_n=1$ for all $n\geq2$. In region I ($\kappa < \kappa^{I}=6617$) both initial conditions lead to a single fixed point in which $x_{65}$ is very low, $10^{-60}$. In region III ($\kappa > \kappa^{II} = 2226342$) both initial conditions again lead to a single fixed point but in this fixed point $x_{65}$ is high, close to unity. In region II ($\kappa^{I} \leq \kappa \leq \kappa^{II}$) the initial condition (i) leads to the lower fixed point and (ii) leads to the upper one. The transitions are very sharp, e.g., at $\kappa=2226341$ the system is numerically clearly seen in region II and at 2226343 in region III.}
    \label{acs-eg-bis}
\end{figure}

The reason for the sudden change in the qualitative character of the steady state profile as $\kappa$ is increased (as shown in Fig. \ref{acs-eg-ssc}) is a bistability in the chemical dynamics due to the presence of the ACS. Fig. \ref{acs-eg-bis} shows three regions in the phase diagram of the chemistry containing ACS65, separated by values $\kappa^{I}$ and $\kappa^{II}$ of $\kappa$. For $0 \leq \kappa < \kappa^{I}$ (region I), the dynamics starting from both the initial conditions mentioned in the figure caption converged to the same attractor configuration, which is a fixed point in which the large ACS molecules have a very small concentration which are comparable to the background concentrations (the concentration declines exponentially with $n$). For $\kappa^{II} < \kappa$ (region III), again they converge to a single attractor, a fixed point in which the ACS molecules have a significant concentration which approaches $x_n = A (k_f A/k_r)^{n-1}$ as $\kappa \rightarrow \infty$ while the background molecules have significantly lower concentration. In the range $\kappa^{I} \leq \kappa \leq \kappa^{II}$ (region II), they converge to two different stable attractors, both fixed points for the ACS under discussion.

This phase structure implies that if we start from the standard initial condition and consider the steady state profile to which the system converges for different values of $\kappa$, we will see a sharp change in the steady state profile as $\kappa$ is increased from a value slightly below $\kappa^{II}$ to a value slightly above $\kappa^{II}$. Below $\kappa^{II}$ the large ACS molecules will be essentially absent in the steady state, and above $\kappa^{II}$ they will be present in large numbers and will dominate over the background.

Therefore, following the nomenclature of Ohtsuki and Nowak \cite{Ohtsuki2009}, who observed a similar bistability in their model with a single catalyst, we refer to $\kappa^{II}$ as the `initiation threshold' of the ACS. Similarly $\kappa^{I}$ will be referred to as the `maintenance threshold' of the ACS, because once the ACS has been initiated, $\kappa$ can come down to as low a value as $\kappa^{I}$, and the ACS will continue to dominate.

\begin{figure}
    \begin{center}
        \subfloat[]{\label{acs-eg-multi-a}\includegraphics[height=4.75in,angle=-90,trim=0.5cm 0.5cm 0cm 0.3cm,clip=true]{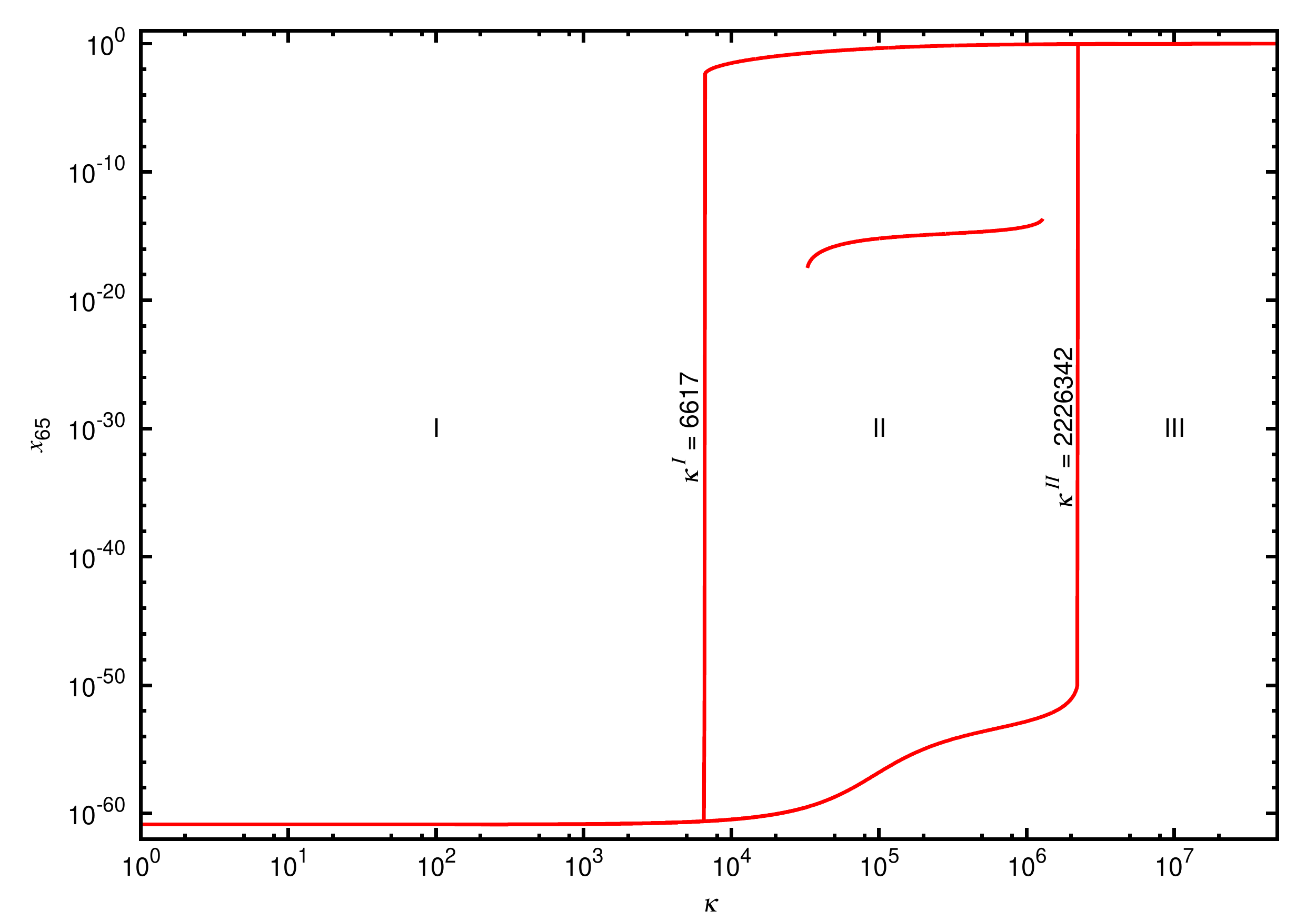}} \\
        \subfloat[]{\label{acs-eg-multi-b}\includegraphics[height=4.75in,angle=-90,trim=0.5cm 0.5cm 0cm 0.3cm,clip=true]{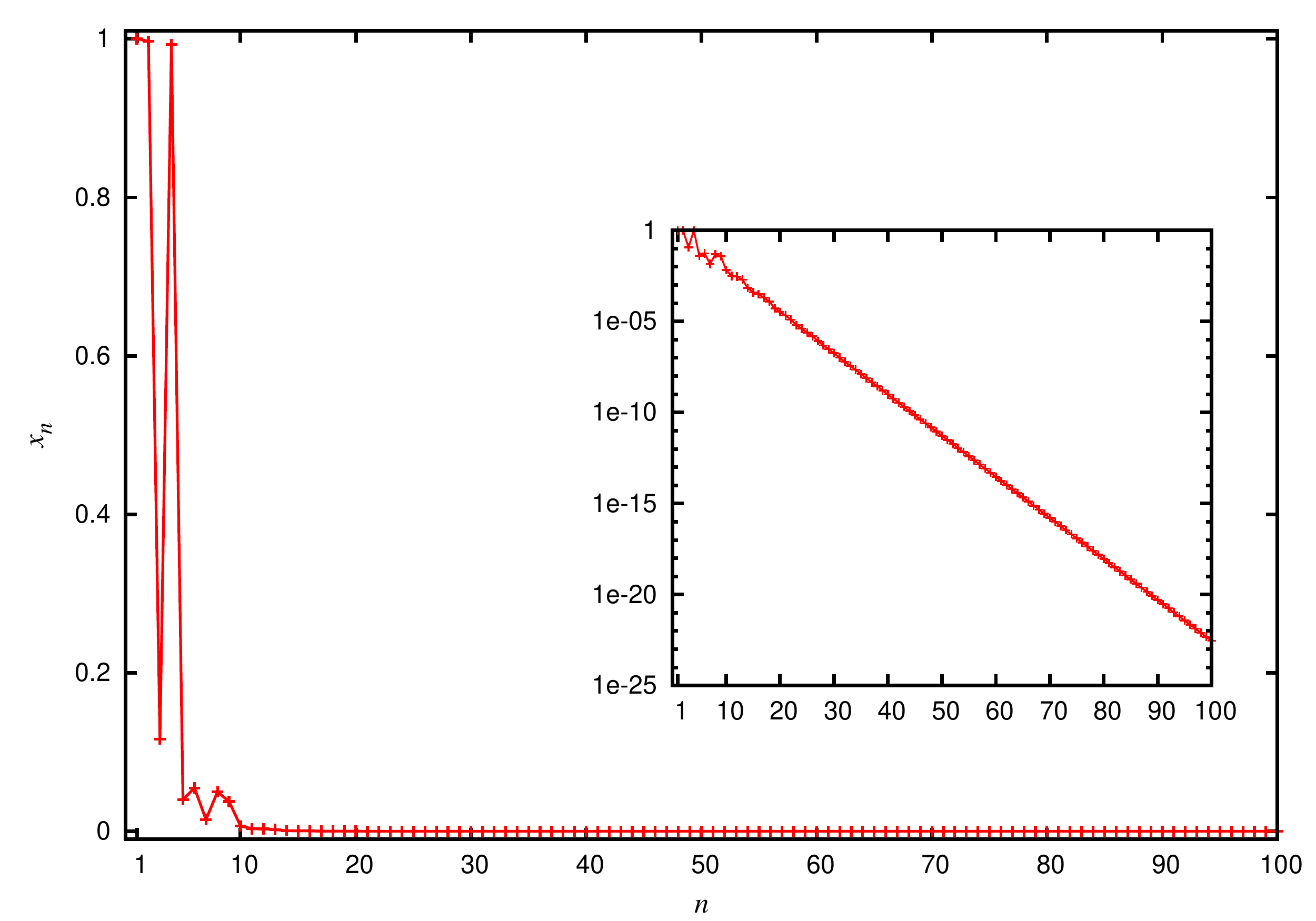}}
    \end{center}
    \caption[Multistability in the dynamics of ACS65.]{{\bf Multistability in the dynamics of ACS65.} {\bf(a)} Same as Fig. \ref{acs-eg-bis} but with the additional stable state in Region II shown. This steady state is achieved by initial conditions that are intermediate to the ones discussed in Fig. \ref{acs-eg-bis}. {\bf(b)} The steady state concentration profile in the intermediate steady state. $k_f = k_r = A = 1, \phi = 15, \kappa = 5 \times 10^5, N=100$. The initial condition used is $x_i = 10^{-5}$ for $i=2,3,\ldots,N$.}
    \label{acs-eg-multi}
\end{figure}

We remark that using other initial conditions we have found at least one more stable fixed point in a part of region II which has intermediate values of $x_{65}$ (Fig. \ref{acs-eg-multi-a}, indicating that this system has multistability. The steady state concentration profile for a system in the intermediate attractor is shown in Fig. \ref{acs-eg-multi-b}. An intuitive reason for appearance of this attractor is discussed in Appendix \ref{Appendix-Catalyzed-Chemistries-wo-ACS}.

\section{\label{acs-bistability}Bistability in simple ACSs}
In general $\kappa^{I}$ and $\kappa^{II}$ depend upon the other parameters, as well as the topology of the catalyzed and spontaneous chemistries. The phase structure is exhibited in more detail for a simpler example in Fig. \ref{acs4}, where the catalyzed chemistry consists of only two reaction pairs:
\begin{subequations}
 \label{acs4-definition}
 \begin{eqnarray}
    \label {acs4-rct1} \mathrm{\mathbf A}(1) + \mathrm{\mathbf A}(1) & \reactionrevarrow{\ensuremath{\mathrm{\mathbf A}(4)}}{} & \mathrm{\mathbf A}(2) \\
    \label {acs4-rct2} \mathrm{\mathbf A}(2) + \mathrm{\mathbf A}(2) & \reactionrevarrow{\ensuremath{\mathrm{\mathbf A}(4)}}{} & \mathrm{\mathbf A}(4),
\end{eqnarray}
\end{subequations}
which constitute an ACS (called ACS4). This system, investigated numerically using XPPAUT, shows bistability. For a fixed $\phi$ the bistability diagram is shown in Fig. \ref{acs4-a}. The dependence of $\kappa^{I},\kappa^{II}$ on $\phi$ is exhibited in Fig. \ref{acs4-b}. For a given $k_f$, there is critical value of $\phi(=\bar\phi)$ at which the $\kappa^{I}$ and $\kappa^{II}$ curves meet, below which there is no bistability. The locations of the phase boundaries, the $\kappa^{I}$ and $\kappa^{II}$ curves, depend upon the specific underlying chemistries (catalyzed and spontaneous) as well as the ACS topology. The steady state profiles are shown at sample points in the phase space in Fig. \ref{acs4-c}. For $\phi>\bar\phi$ it can be seen, that as in the case of the larger ACS discussed earlier, if we start from the standard initial condition, the largest molecule of the catalyzed chemistry, here $\mathrm{\mathbf A}(4)$, dominates over the background in the steady state only for $\kappa > \kappa^{II}$ (e.g., the panel marked 3 in Fig. \ref{acs4-c}). In the range $\kappa^{I} \leq \kappa \leq \kappa^{II}$, it dominates only if we start from initial conditions where it has a large enough value to begin with (panel 2b), but not if we start from the standard initial condition (panel 2a). It does not dominate for any initial condition if $\kappa < \kappa^{I}$ (panel 1). If $\phi$ is below $\bar\phi$, there is a single attractor with no significant ACS dominance if $\kappa$ is small (panel 4), or if $\kappa$ is large (panel 5), ACS dominance exists but is not very pronounced as the background concentrations are also substantial. The dependence of $\kappa^{I}, \kappa^{II}$ on both $\phi$ and $k_f$ is shown Fig. \ref{acs4-d}. For a smaller $k_f$, bistability appears at smaller values of $\phi$.\footnote{A technical remark on the definition and calculation of $\kappa^{I}$ and $\kappa^{II}$: We have defined $\kappa^{I}$ and $\kappa^{II}$ as the values of $\kappa$ that mark the two extremes of the bistable region. To calculate the limits of bistability, for each value of $\kappa$, we start from two initial conditions, (i) the standard initial condition, $x_n=0\ \forall\ n\ge2$, and (ii) a `high' initial condition, $x_n=1\ \forall\ n\ge2$, and integrate the system to identify the respective attractors these initial conditions lead to. $\kappa^{II}$ is the value of $\kappa$ upto which the system remains in the lower attractor starting from the initial condition (i), and, $\kappa^{I}$ is the value of $\kappa$ at which the system first moves to the higher attractor starting from the initial condition (ii). Alternatively, we can define $\kappa^{I}$ and $\kappa^{II}$ by using an initial condition which tracks the attractor at a neighbouring value of $\kappa$, \ie, the steady state concentrations calculated at a $\kappa$ value are used as the initial condition to calculate the attractor for $\kappa \pm \delta$ (for small $\delta$). Starting from the lower attractor and using initial conditions that track the attractor as $\kappa$ is increased, one can define $\kappa^{II}$ as the largest value of $\kappa$ for which the lower attractor exists. Similarly, starting from the higher attractor and using initial conditions that track this attractor as $\kappa$ is decreased, one can define $\kappa^{I}$ as the lowest value of $\kappa$ for which the higher attractor exists. XPPAUT uses the latter definition of $\kappa^{I}$ and $\kappa^{II}$. In principle the values of $\kappa^{I}$ and $\kappa^{II}$ obtained using the two approaches can be different. For the case of ACS4, these two values happen to be same at least upto 4 significant figures. The definition of $\kappa^{II}$ that we have followed based on the standard initial condition is more relevant for the origin of life problem.}

\begin{figure}
    \begin{center}
        \subfloat[]{\label{acs4-a}\includegraphics[height=4.75in,angle=-90,trim=0.5cm 0.5cm 0cm 0.3cm,clip=true]{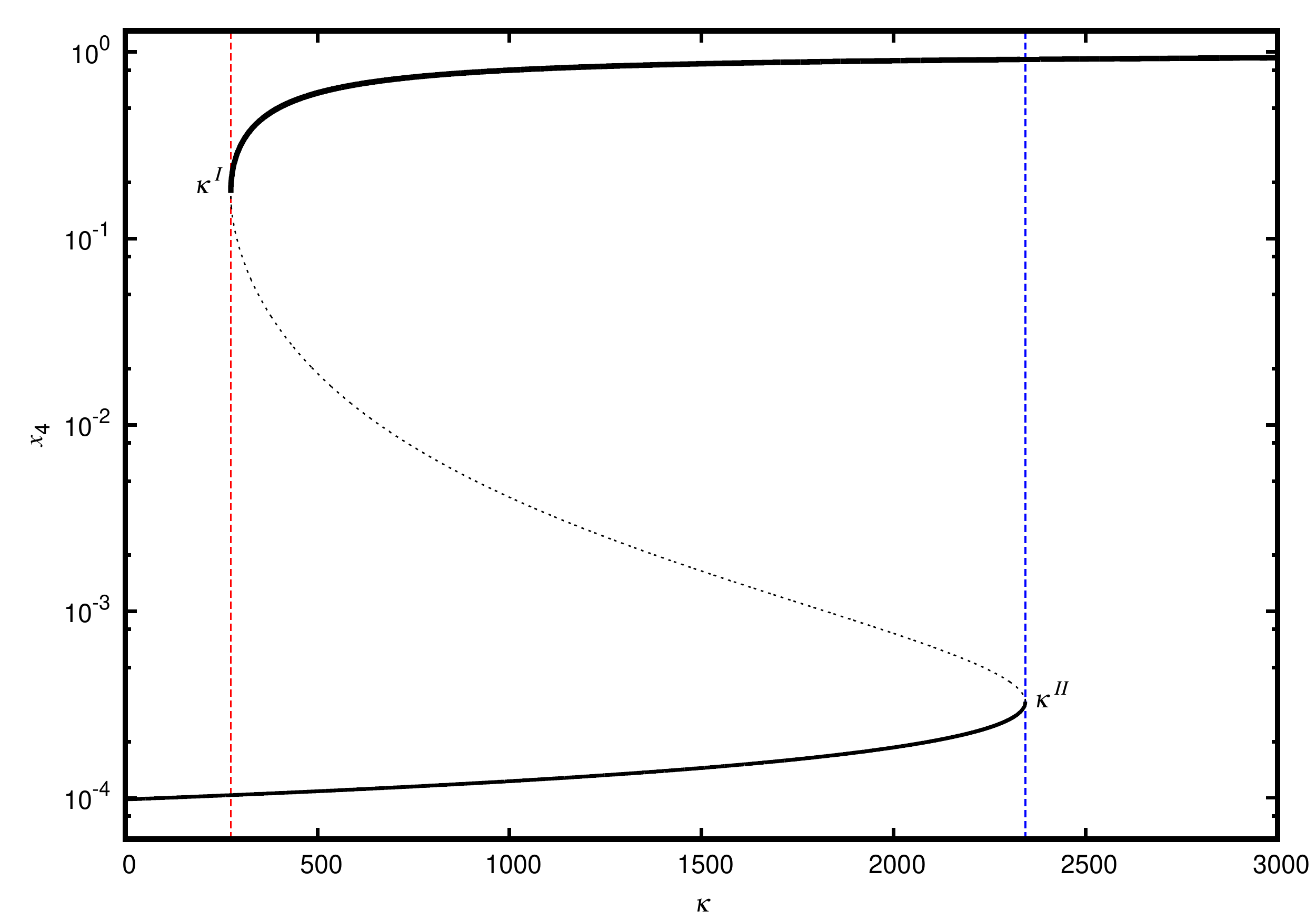}}\\
        \subfloat[]{\label{acs4-b}\includegraphics[height=4.75in,angle=-90,trim=0.5cm 0.5cm 0cm 0.3cm,clip=true]{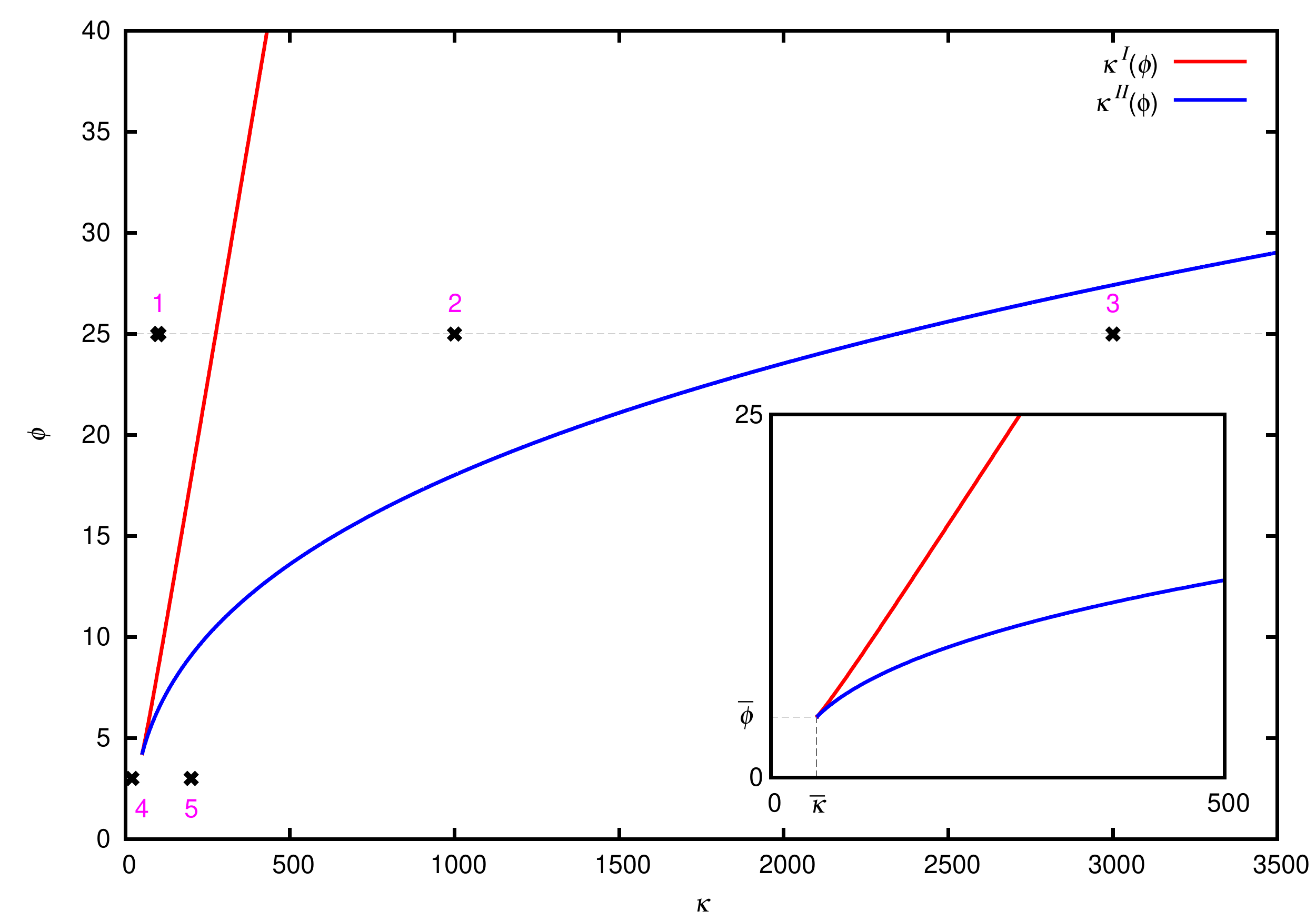}}
    \end{center}
    \caption[Phase diagram and concentration profiles for ACS4.]{{\bf Phase diagram and concentration profiles for ACS4.} {\it Continued.}}
\end{figure}

\captionsetup{list=no}
\begin{figure}
    \ContinuedFloat
    \begin{center}
        \subfloat[]{\label{acs4-c}\includegraphics[width=4.75in,trim=0.25cm 0.25cm 0cm 0.3cm,clip=true]{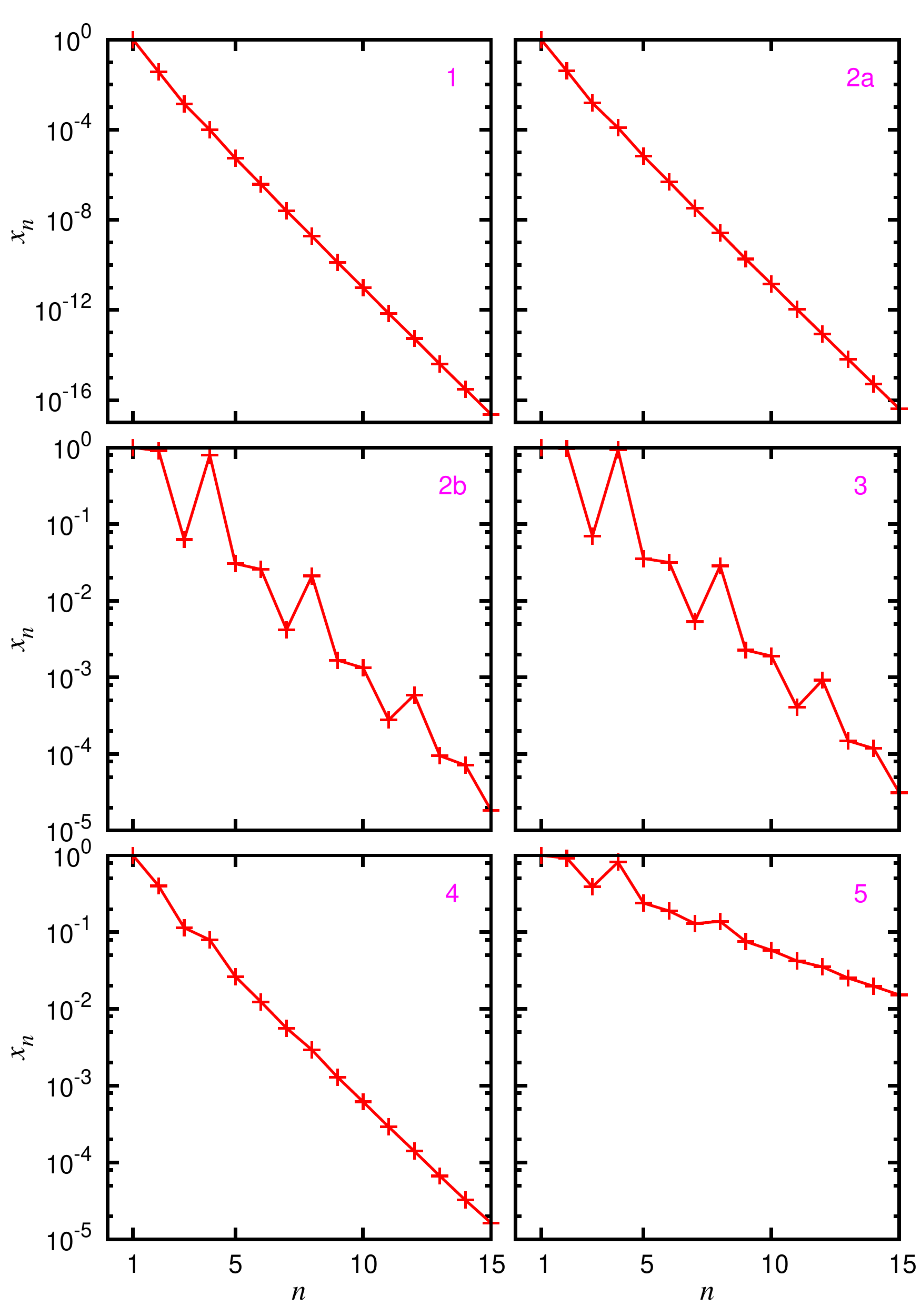}}
    \end{center}
    \caption[]{{\bf Phase diagram and concentration profiles for ACS4.} {\it Continued.}}
\end{figure}

\begin{figure}
    \ContinuedFloat
    \begin{center}
        \subfloat[]{\label{acs4-d}\includegraphics[height=4.75in,angle=-90,trim=0.5cm 0.5cm 0cm 0.3cm,clip=true]{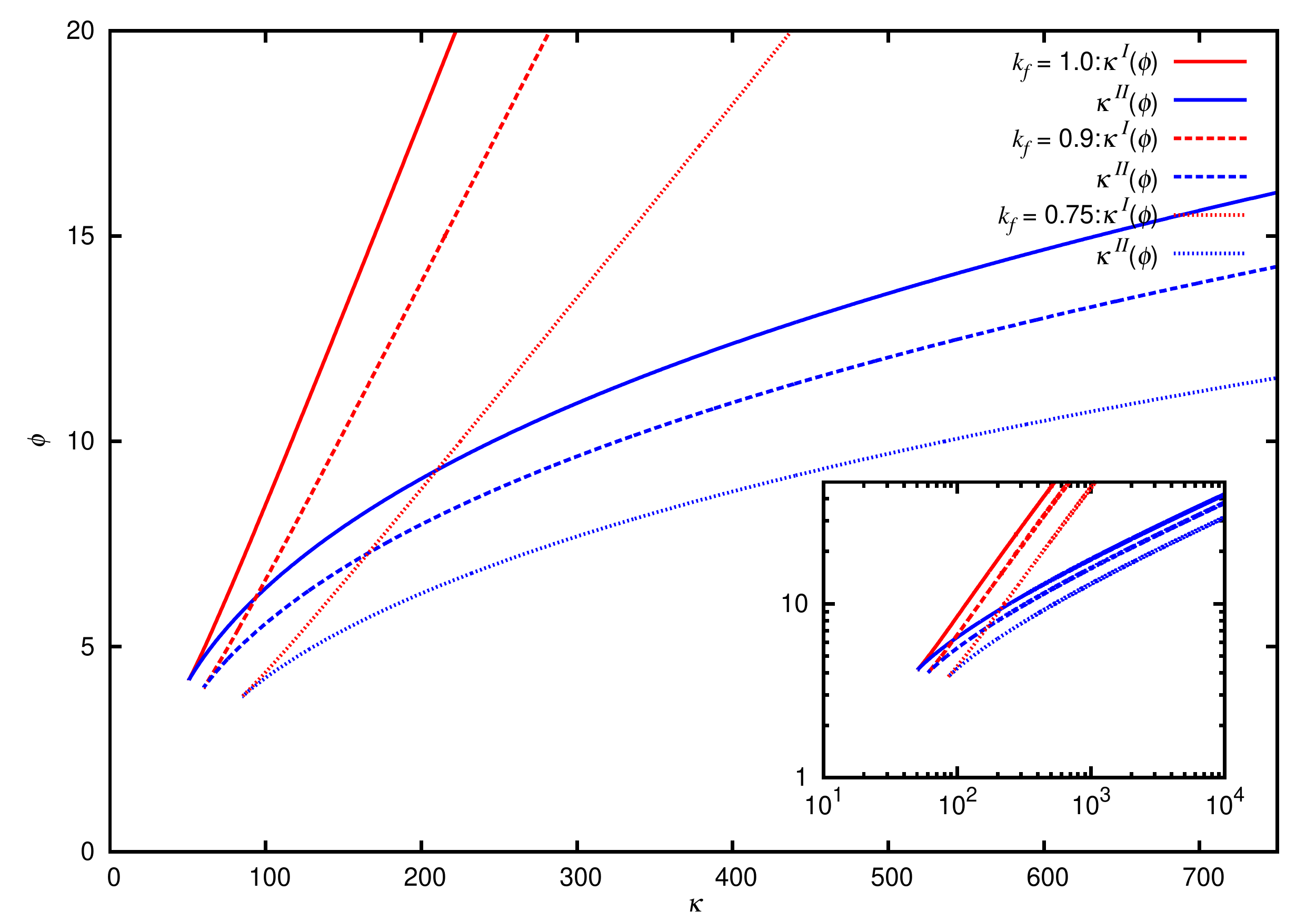}}
    \end{center}
    \caption[]{{\bf Phase diagram and concentration profiles for ACS4.} {\bf (a)} The steady state concentration $x_4$ versus $\kappa$ for $k_f = k_r = A = 1, \phi = 25, N=15$. The bistable region exists for the range $\kappa^{I} \leq \kappa \leq \kappa^{II}$ in which different initial conditions lead to two distinct steady state values of $x_4$. The solid black curves correspond to the two stable fixed points, and the dotted black curve to the unstable fixed point. {\bf (b)} The dependence of $\kappa^{I}$ (red curve) and $\kappa^{II}$ (blue curve) on $\phi$ for $k_f = k_r = A = 1, N=15$. The bistable region lies between the two curves; in the rest of the phase space the system has a single fixed point. The inset shows the location of the critical point $(\bar{\kappa},\bar{\phi})$; there is no bistability for $\phi < \bar{\phi}$. {\bf (c)} The steady state concentration profile of molecules shown at five representative points in the phase space (numbered 1 through 5 and marked in (b)). Note that at the phase point 2 that lies between the $\kappa^{I}$ and $\kappa^{II}$ curves there are two steady state profiles corresponding to the two stable fixed points of the system. The figure marked 2a shows the profile starting from the standard initial condition, and 2b from the initial condition where $x_n = 1$ for all $n$. {\bf (d)} Dependence of the phase boundaries on $k_f$ for $k_r = A = 1, N = 15$, with the inset showing the behaviour on a log-log plot.}
    \label{acs4}
\end{figure}
\captionsetup{list=yes}

We remark that while bistability seems to be quite generic in homogeneous chemistries containing ACSs, the existence of an ACS does not guarantee that bistability exists somewhere in phase space. For example consider the simplest possible chemistry ($N=2$) containing only the monomer (which is buffered) and the dimer. If we assume that the sole reaction pair $\mathrm{\mathbf A}(1) + \mathrm{\mathbf A}(1) \rightleftharpoons \mathrm{\mathbf A}(2)$ is catalyzed by $\mathrm{\mathbf A}(2)$, the catalyzed chemistry is trivially an ACS and the only rate equation is $\dot{x}_2 = k_f A^2 (1 + \kappa x_2) - k_r x_2 (1 + \kappa x_2) - \phi x_2$. The system can be solved exactly and always goes to a global fixed point attractor starting from any initial condition $x_2(0) \geq 0$. However, the $N=3$ chemistry defined by the two catalyzed reactions
\begin{subequations}
 \label{acs3-definition-eg}
 \begin{eqnarray}
    \label {acs3-eg-rct1} \mathrm{\mathbf A}(1) + \mathrm{\mathbf A}(1) & \reactionrevarrow{\ensuremath{\mathrm{\mathbf A}(3)}}{} & \mathrm{\mathbf A}(2) \\
    \label {acs3-eg-rct2} \mathrm{\mathbf A}(1) + \mathrm{\mathbf A}(2) & \reactionrevarrow{\ensuremath{\mathrm{\mathbf A}(3)}}{} & \mathrm{\mathbf A}(3),
 \end{eqnarray}
\end{subequations}
does exhibit bistability at a sufficiently large $\phi$. Ohtsuki and Nowak \cite{Ohtsuki2009} had also found a lower limit on catalyst size for bistability to exist in their model. Similar results hold for the $f=2$ case. From our simulations a general observation seems to be that bistability is ubiquitous at sufficiently large values of $\phi$ in homogeneous chemistries whenever the smallest catalyst is large enough compared to the food set. When it does exist it seems to provide a crisp criterion for `ACS domination', including `initiation' ($\kappa > \kappa^{II}$) and `maintenance' ($\kappa \geq \kappa^{I}$).

ACSs with the same set of reactions (and hence the same product set $P(S)$) may differ from each other in the assignment of catalysts to the reaction. We remark that the assignment of catalysts to the reactions of an ACS affects the dynamics, specifically the existence and extent of bistability. A detailed discussion of an example of how the assignment of catalysts effects the dynamics is given in Appendix \ref{Appendix-ACS-topology-and-Bistability}.

There exists a substantial literature including results on the conditions for bistability in chemical systems\cite{Craciun2006, Wilhelm2007, Ramakrishnan2008}. It would be interesting to apply some of those results to models of the kind being studied here, which involve a large number of molecular species.

\section{\label{ACS-Sparse-chem}ACSs and sparse chemistries}
We now discuss the dynamics of ACSs when the underlying chemistry is not the fully connected chemistry discussed as far, but is a sparse chemistry. We generate the sparse spontaneous chemistries with the following modification to the algorithm described in Section \ref{section-saprse-spont-chem}. For every reaction in the catalyzed chemistry we include the same reaction (without the catalyst) as part of the spontaneous chemistry. Thus if a molecule is produced in $m$ catalyzed ligation reactions, it is automatically also produced in $m$ spontaneous ligation reactions. If $m$ is less than the degree of chemistry, $k$, another $k-m$ spontaneous reactions are chosen to be added randomly from the remaining ligation reactions producing this molecule.

\begin{figure}
    \begin{center}
        \subfloat[]{\includegraphics[height=4.75in,angle=-90,trim=0.5cm 0.5cm 0cm 0.3cm,clip=true]{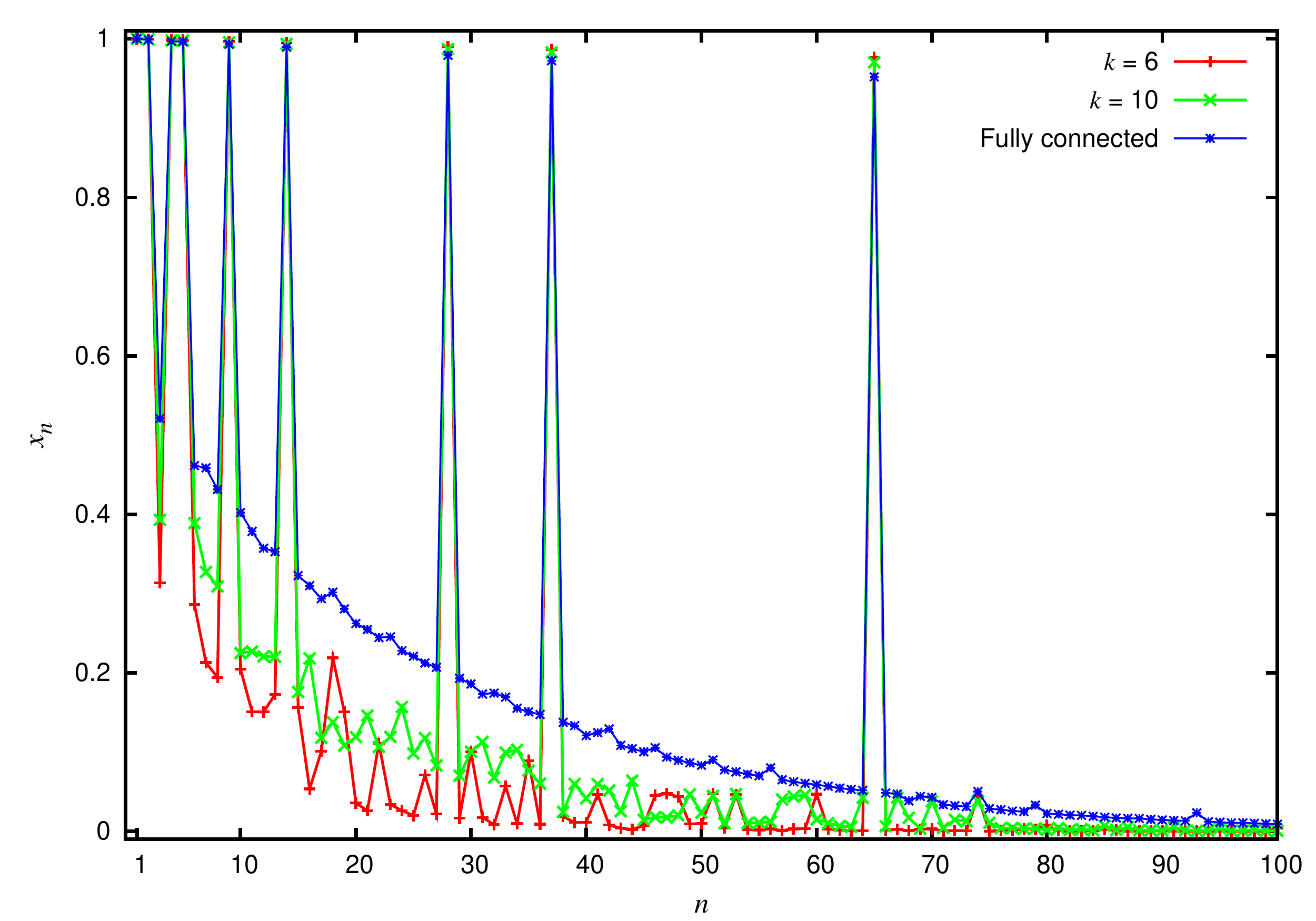}} \\
        \subfloat[]{\includegraphics[height=4.75in,angle=-90,trim=0.5cm 0.5cm 0cm 0.3cm,clip=true]{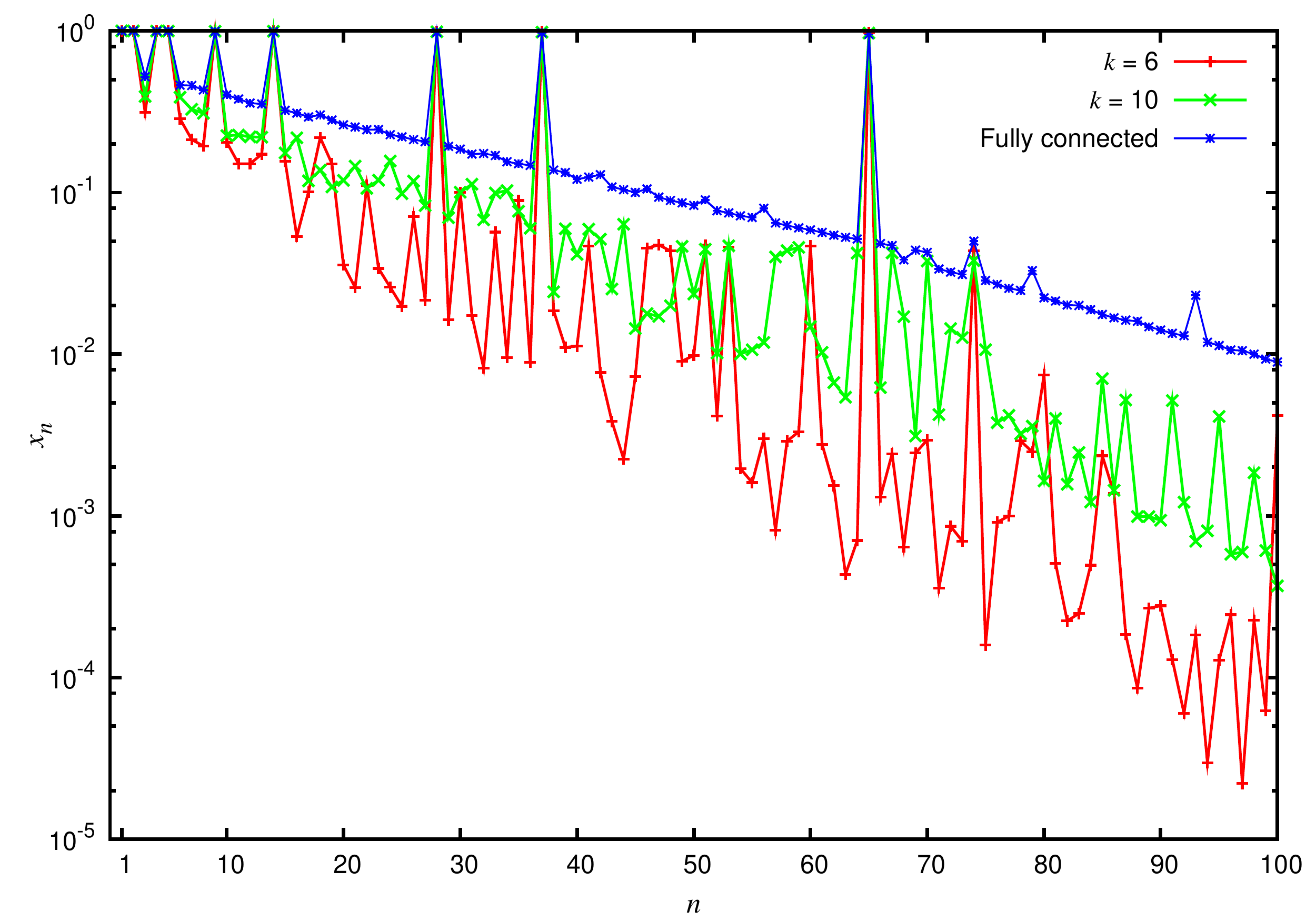}}
    \end{center}
    \caption[Steady state concentration profile of ACS65 with sparse spontaneous chemistry.]{{\bf Steady state concentration profile of ACS65 with sparse spontaneous chemistry.} {\bf (a)} The steady state concentrations, $x_n$ versus $n$ for ACS65 is shown for sparse spontaneous chemistries with $k = 6$ and $10$. For comparison the steady state profile for a fully connected spontaneous chemistry is also shown. $A = k_f = k_r = 1, \phi = 15, \kappa = 2.5 \times 10^6$. {\bf (b)} Same as (a) with concentrations on logarithmic scale.}
    \label{acs-eg-sparse}
\end{figure}

The steady state concentration profiles for ACS65 with spontaneous chemistries with degree $k=6$ and $k=10$ is shown in Fig. \ref{acs-eg-sparse}. As sparseness increases ($k$ decreases) the background concentrations decline. This happens as there are fewer pathways to produce the background molecules. There is also a larger variation in their concentrations because their production reactions are chosen randomly, and background molecules produced in reactions that happen to involve the ACS molecules as reactants do better than others. The ACS molecules are seen to dominate more strongly over the background in sparser chemistries; this is because there are a fewer production pathways to the background that drain their concentrations. Similar results are also seen in chemistries with $f=2$.

Sparseness can also effect the value of $\kappa^{II}$. As the chemistry becomes more sparse, the number of ways in which a molecule is produced and hence its concentration in the steady state accessible from the standard initial condition decreases. If a molecule that acts as a catalyst is not produced in sufficient concentration that can cause the initiation of the ACS domination to happen at a larger value of $\kappa$.

\section{ACSs and heterogeneous chemistries}
\begin{figure}
    \begin{center}
        \includegraphics[height=4.75in,angle=-90,trim=0.25cm 0.5cm 0cm 0.3cm,clip=true]{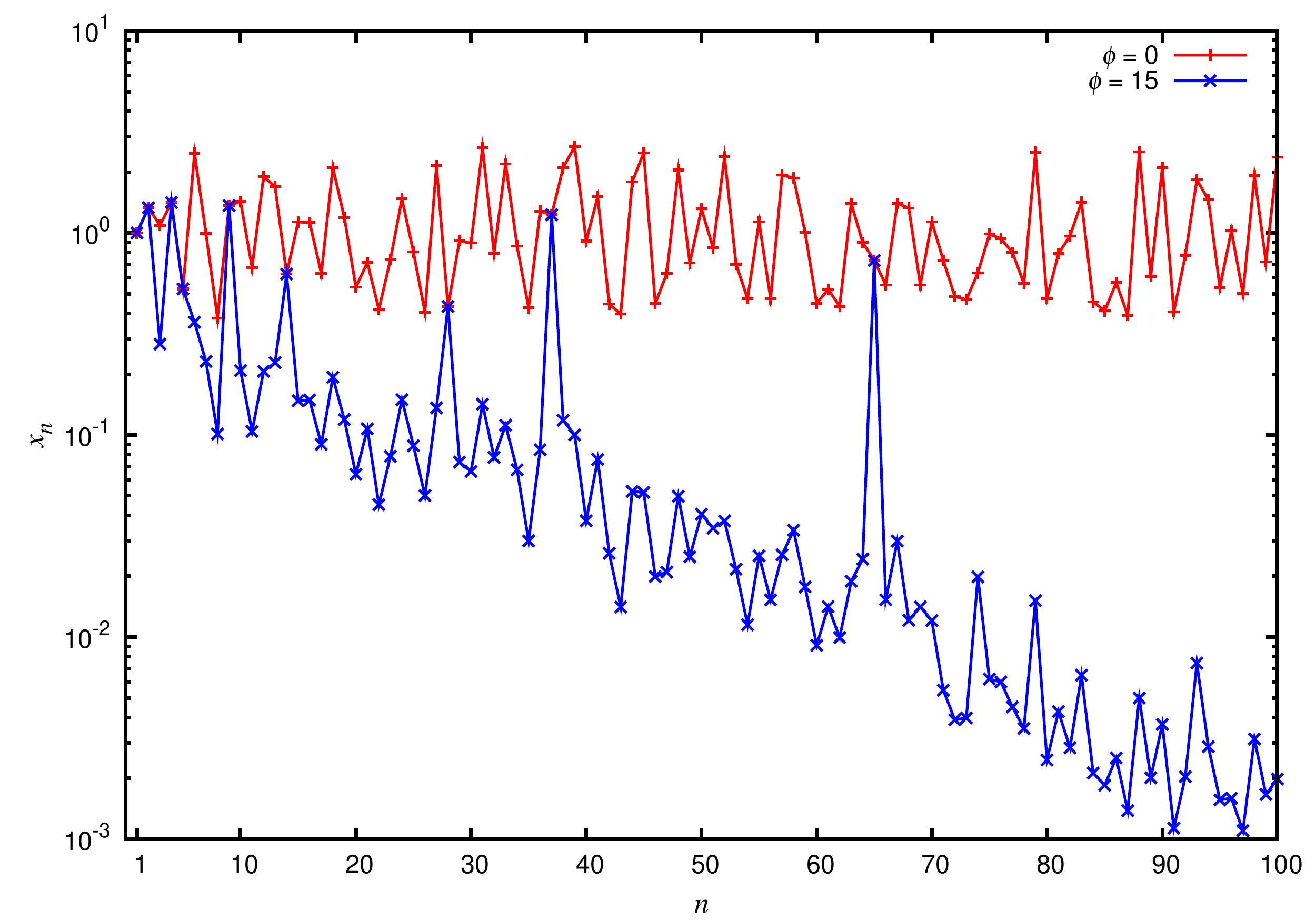}
    \end{center}
    \caption[Steady state concentration profile for ACS65 in a heterogeneous chemistry.]{{\bf Steady state concentration profile for ACS65 in a heterogeneous chemistry.} The rate constants for the chemistry were generated using $\epsilon_1 = \epsilon_2 = \epsilon_3 = Z_1 = Z_2 = 1$ (the parameters and the random seed used to generate the chemistry are same as in Fig. \ref{noacs-hetero}). The red curve plots the steady state concentrations for the heterogeneous chemistry at $\phi=0$. (The concentrations at $\phi=0$ are independent of presence of an ACS). The blue curve shows the steady state concentrations of the molecules when ACS65 is present in the chemistry, here $\phi = 15$ and $\kappa=3.5 \times 10^8$. The maximum concentration ACS members are able to achieve is still govern by their steady state concentration value at $\phi=0$. In both cases $A = 1$.}
    \label{acs-eg-hetro}
\end{figure}

In Section \ref{section-spont-hetero}, we discussed the effect of heterogeneity in the rate constants on the steady state concentrations in a spontaneous chemistry. For the same chemistry (rate constants) we study the effect of inclusion of catalysts. The steady state concentration profile for ACS65 with heterogeneous rate constants is shown in Fig. \ref{acs-eg-hetro}. Again, the molecules produced in the ACS have larger concentrations. The maximum concentration that any ACS member achieves is limited by its concentration in the equilibrium case of $\phi=0$. When $\phi>0$, members of the ACS have concentrations very close to their limiting values, whereas the background molecules have much lower values in comparison.

\section{Summary}
We have shown in this chapter that in the presence of an ACS at sufficiently large values of $\kappa$, a small subset of molecules that constitute the ACS can dominate the chemistry, \ie, have much larger concentrations as compared to the background (which consists of other molecules in the system produced via spontaneous reactions). We present an analytical reasoning, in the $\kappa \rightarrow \infty$ limit, for the domination of ACS molecules. The chemistries which contain ACSs usually exhibit a rich dynamical behaviour such as the existence of multistability. We find that the qualitative features of these results are robust to a small heterogeneity present in the system and also hold for sparse chemistries. This provides a mechanism of `chemical selection' that allows a subset of molecules in a large set of molecules (comprising the whole chemistry) to dominate over the rest. We now turn to the question: can we build a sparse subset containing large molecules using ACSs? In the following chapter we discuss the problem one encounters in production of large molecules using ACSs and provide a solution to the problem in later chapters.

\thispagestyle{plain}
\cleardoublepage
\chapter{\label{ChapterProblemLargeMolecules}Using ACSs to produce large molecules: The problem}

\lettrine[lines=2, lhang=0, loversize=0.0, lraise=0.0]{W}{e} have seen that in uncatalyzed chemistries, the concentrations of the large molecules remain exponentially small ($x_n \sim e^{-\gamma n}, \gamma > 0$). In catalyzed chemistries, especially in the presence of an ACS, a few specific large molecules produced by the ACS can acquire a high population. However, this seems to require a large catalytic strength for the catalysts. For example, for ACS65 this happens at $\kappa > \kappa^{II} = 2226342$ (see Fig. \ref{acs-eg-bis}), starting from the standard initial condition. The fact that such a large catalytic strength is needed to produce appreciable concentrations of molecules of even moderate length like $n=65$ could be a problem for the ACS mechanism to produce large molecules in the kind of prebiotic scenario we are considering. In this chapter we characterize the problem somewhat more quantitatively by determining how $\kappa^{II}$ depends upon the size $n$ of the catalysts in the ACS.

\section{Scaling of $\kappa^I$ and $\kappa^{II}$ with the size of the catalyst}
The values of $\kappa^{I}$, $\kappa^{II}$ for an ACS (for a fixed set of parameters) depend upon its topology. The topology of the ACS includes the set of catalyzed reactions and the assignment of catalysts to each of the catalyzed reactions. We define the `length' $L$ of an ACS as the size of (\ie, the total number of monomers of all types in) the largest molecule produced in the ACS\glsadd{ACS-length}. An `extremal' ACS\glsadd{extremal-ACS} of length $L$ will be referred to as one in which all reactions belonging to the ACS are catalyzed by the same molecule which is the largest molecule (of size $L$) in the ACS. For concreteness, since we are interested in the dependence of $\kappa^{II}$ on the catalyst size, we consider only extremal ACSs of length $L$. We assume that the catalyst has the same catalytic strength $\kappa$ for all the reactions in the ACS. We wish to determine the bistable region for such ACSs and in particular how the values of $\kappa^{I}$ and $\kappa^{II}$ depend upon $L$. These values depend upon the precise set of catalyzed reactions constituting the ACS. For illustrative purposes we consider three different ways of generating the ACS described below as Algorithm 1, 2 and 3, which generate ACSs with different characteristic structure.

\subsection{\label{Algos-1-2-3}Generating extremal ACSs of length $L$}
We describe three different algorithms used for generating a set of reactions that provide a pathway to produce a molecule of a given length $L$ from the food set.
\begin{description}
  \item[Algorithm 1:] {\bf Incremental, smallest-stepsize, deterministic construction.} Each molecule of size $n$ ($n=2,3,\ldots,L$) is produced in the reaction $\mathbf{A}(1) + \mathbf{A}(n-1) \rightarrow \mathbf{A}(n)$. All such reactions for $n=2,3,\ldots,L$ are included in the ACS. \\
      {\it Example:} For $L=5$ following reactions are included
      \begin{align*}
        \mathbf{A}(1) + \mathbf{A}(1) \rightarrow \mathbf{A}(2) \\
        \mathbf{A}(1) + \mathbf{A}(2) \rightarrow \mathbf{A}(3) \\
        \mathbf{A}(1) + \mathbf{A}(3) \rightarrow \mathbf{A}(4) \\
        \mathbf{A}(1) + \mathbf{A}(4) \rightarrow \mathbf{A}(5)
      \end{align*}
  \item[Algorithm 2:] {\bf Shortest path, top-down, deterministic construction.} Start with $\mathbf{A}(L)$. If $L$ is even, it is produced in the reaction $\mathbf{A}(L/2) + \mathbf{A}(L/2) \rightarrow \mathbf{A}(L)$. If $L$ is odd, say $L=2m+1$, then $\mathbf{A}(L)$ is produced in the reaction $\mathbf{A}(m) + \mathbf{A}(m+1) \rightarrow \mathbf{A}(L)$. The same algorithm is used to select a reaction for the production of the precursor(s) of $\mathbf{A}(L)$ (namely, for $\mathbf{A}(L/2)$ if $L$ is even, and for each of $\mathbf{A}(m)$ and $\mathbf{A}(m+1)$ if $L$ is odd ($=2m+1$)), and is iterated for their precursors, etc., until the only reactant appearing in the reactions is the food set molecule, A(1). All the production reactions mentioned above are included in the ACS.\\
      {\it Example:} For $L=10$ following reactions are included
      \begin{align*}
        \mathbf{A}(5) + \mathbf{A}(5) & \rightarrow \mathbf{A}(10) \\
        \mathbf{A}(2) + \mathbf{A}(3) & \rightarrow \mathbf{A}(5) \\
        \mathbf{A}(1) + \mathbf{A}(2) & \rightarrow \mathbf{A}(3) \\
        \mathbf{A}(1) + \mathbf{A}(1) & \rightarrow \mathbf{A}(2)
      \end{align*}
  \item[Algorithm 3:] {\bf Incremental, random construction.} In this method, starting from the food set $\mathcal{F}$ the set of reactants $R_k$ is sequentially enlarged step by step ($\mathcal{F}=R_0 \subset R_1 \subset R_2 \subset \ldots $) to include larger molecules and a reaction chosen randomly until a product of length $L$ is obtained. $R_k$ denotes the set of reactants at step $k$ in the algorithm, and $L_k$ the size of the largest molecule in $R_k$. At step $k$, pick a pair of molecules (say $\mathbf{X}$ and $\mathbf{Y}$) randomly from $R_k$, and determine the product $\mathbf{Z}$ formed if they were to be ligated ($\mathbf{X}$ being the same as $\mathbf{Y}$ is allowed). If the size of this product, $L(\mathbf{Z})$, is $\leq L_k$ or $> L$, discard the pair and choose another pair. If $L_k < L(\mathbf{Z}) < L$, add the ligation reaction $\mathbf X + \mathbf Y \rightarrow \mathbf Z$ to the ACS, define $R_{k+1} = R_k \cup \{ \mathbf{Z} \} $, and iterate the procedure. If $L(\mathbf{Z}) = L$, add the ligation reaction $\mathbf X + \mathbf Y \rightarrow \mathbf Z$ to the ACS, and stop. Initially ($k=0$) the reactant set is just the food set, $R_0 = \mathcal{F}$.
\end{description}

To complete the extremal ACS, we assign $\mathbf{A}(L)$ as the catalyst of all the reactions generated using any of the above algorithms. For the simulations reported in the paper, the reverse of each reaction is also included as a reaction catalyzed by the same catalyst. For concreteness we have described the Algorithms 1 and 2 for the single monomer case; their generalizations to $f=2$ have also been considered by us. Algorithm 3, as described above, can be used for any $f$.

\subsection{\label{Exponential-Kappa}The requirement of exponentially large catalytic strengths}
\begin{figure}
    \begin{center}
        \subfloat[]{\includegraphics[height=4.75in,angle=-90,trim=0.5cm 0.5cm 0cm 0.3cm,clip=true]{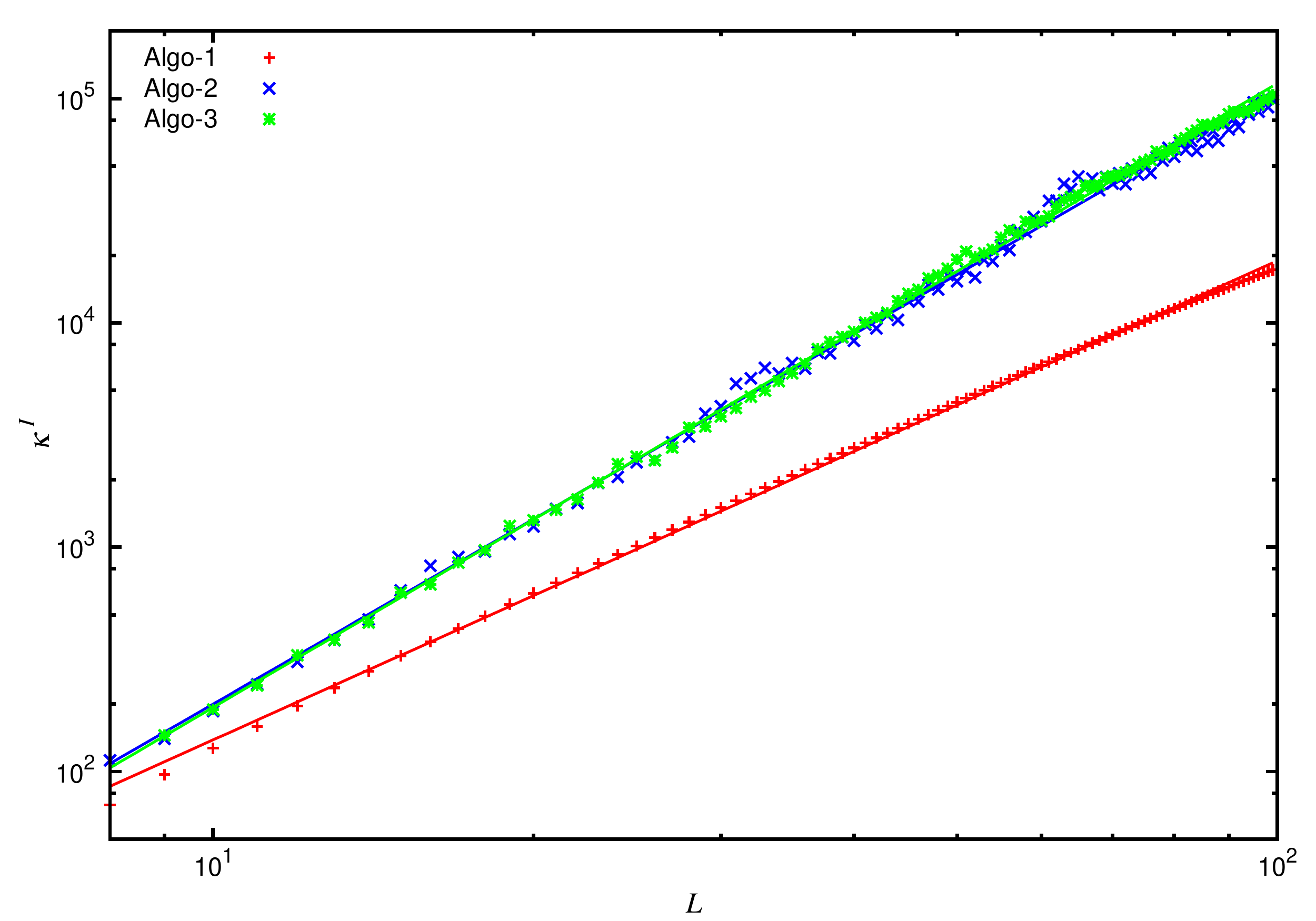}}\\
        \subfloat[]{\includegraphics[height=4.75in,angle=-90,trim=0.5cm 0.5cm 0cm 0.3cm,clip=true]{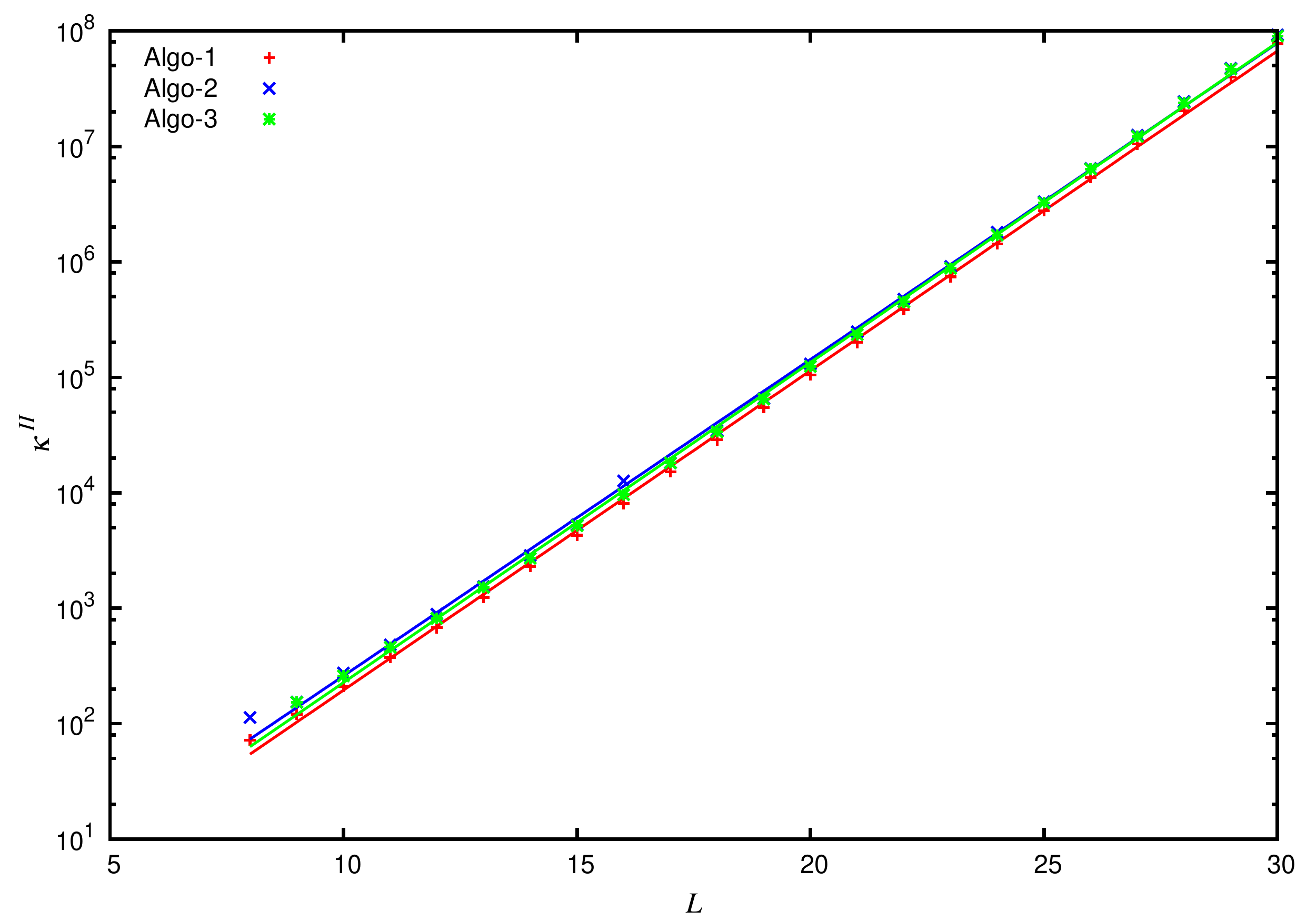}}
    \end{center}
    \caption[The dependence of the bistable region on catalyst length $L$.]{{\bf The dependence of the bistable region on catalyst length $L$.} {\bf (a)} The dependence of $\kappa^{I}$ on $L$. {\bf (b)} The dependence of $\kappa^{II}$ on $L$. Simulations were done for extremal ACSs of length $L$ generated by three algorithms discussed above, represented in the figure by different colours. For each $L$ the ACS in question has the property that the largest molecule produced in the ACS has $L$ monomers and catalyzes all the reactions in the ACS. All simulations were done for $k_f = k_r = A = \phi = 1$. $N=100$ in all cases except the $\kappa^{I}$ curve for Algorithm 1, where $N=200$, because in this case `finite-$N$' effects were quite significant at $N=100$. The figures suggest an approximate power law growth of $\kappa^{I}$ and exponential growth of $\kappa^{II}$ with $L$.}
    \label{kappa1-2-vs-L}
\end{figure}

We constructed ACSs of different values of $L$ using the above three algorithms and determined the $\kappa^{I}$ and $\kappa^{II}$ values numerically. These are plotted in Fig. \ref{kappa1-2-vs-L}. It is evident that $\kappa^{I}$ increases with $L$ according to a power law
\begin{equation}
    \label{kappa1-powerlaw-L}
    \kappa^{I} \sim L^{\alpha}
\end{equation}
with $\alpha$ ranging from 2.1 to 2.8 for the three algorithms, while $\kappa^{II}$ increases exponentially,
\begin{equation}
    \label{kappa2-exponential-L}
    \kappa^{II} \sim e^{\rho L},
\end{equation}
with $\rho \approx 0.64$ for all the algorithms. $\alpha$ and $\rho$ depend upon the other parameters. In particular we find that $\rho$ increases with $\phi$, \ie, the catalytic strength needed for large molecules to arise increases faster with the size of catalyst at larger values of dissipation. This generalizes, to a much larger class of models, the results of Ohtsuki and Nowak\cite{Ohtsuki2009}, who found a linear dependence of $\kappa^{I}$ on $L$ and an exponential dependence of $\kappa^{II}$.

\subsection{\label{exponential-initaition-threshold}Why the initiation threshold is exponentially large}
\begin{figure}
    \begin{center}
        \subfloat[]{\label{kappa2-xL-relationship-a}\includegraphics[height=4.75in,angle=-90,trim=0.5cm 0.5cm 0cm 0.3cm,clip=true]{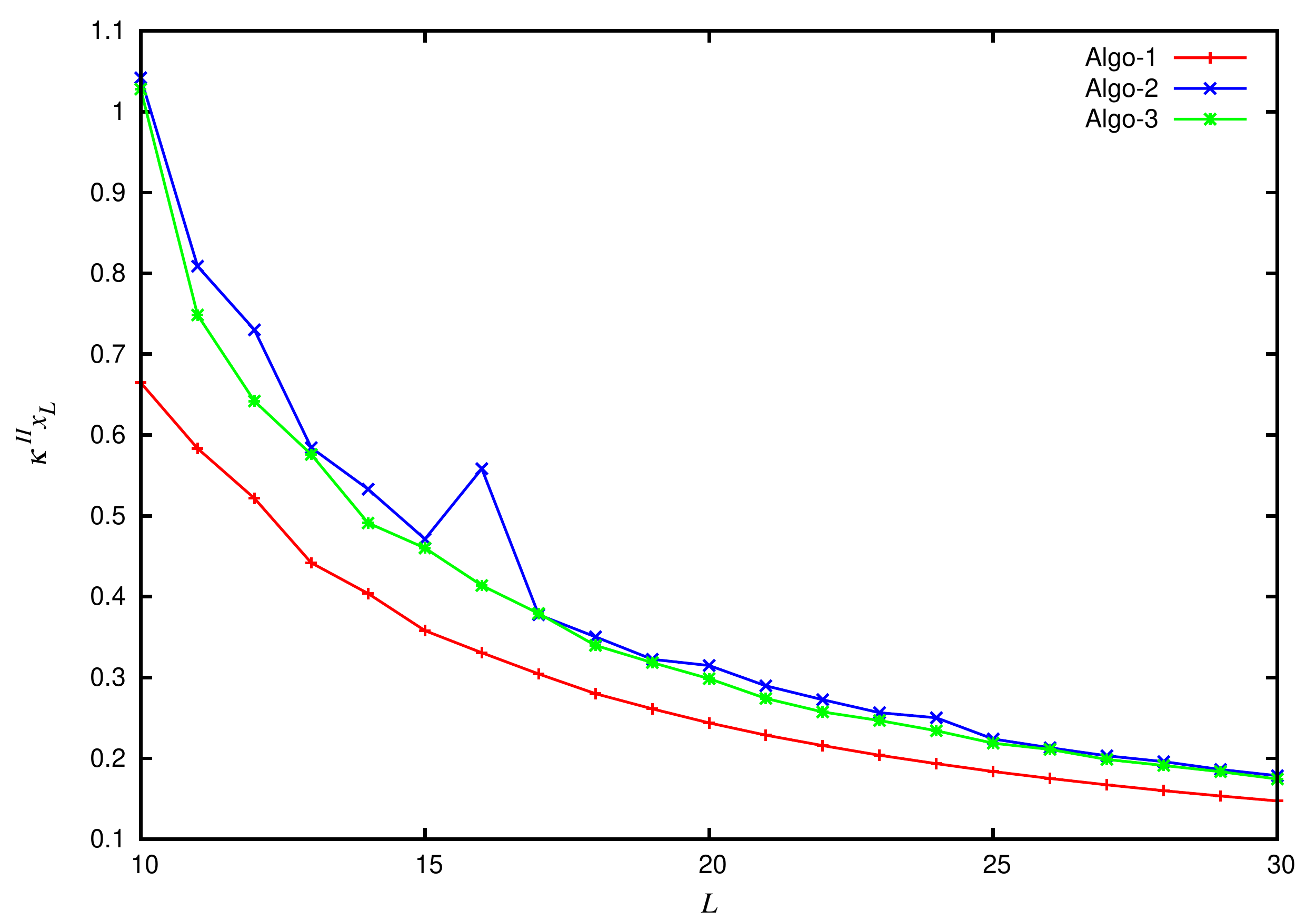}}\\
        \subfloat[]{\label{kappa2-xL-relationship-b}\includegraphics[height=4.75in,angle=-90,trim=0.5cm 0.5cm 0cm 0.3cm,clip=true]{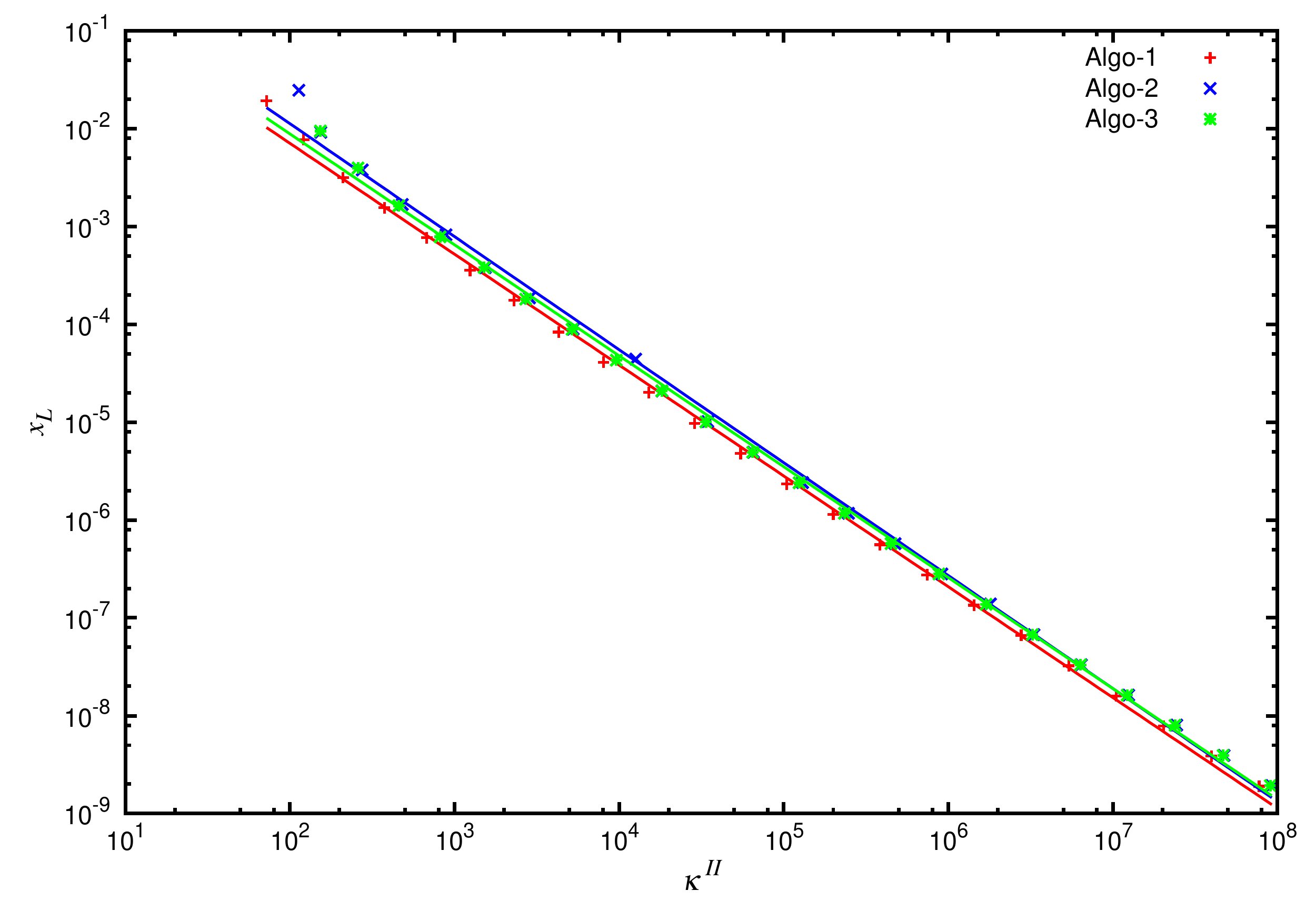}}
    \end{center}
    \caption[The product $\kappa^{II}x_L$ is of order unity.]{{\bf The product $\kappa^{II}x_L$ is of order unity.} For this figure each chemistry was simulated at a value of $\kappa$ equal to the initiation threshold $\kappa^{II}$ corresponding to that chemistry, and the steady state concentration $x_L$ of the catalyst was determined in the low fixed point (starting from the standard initial condition). The parameters values are the same as in Fig. \ref{kappa1-2-vs-L}. {\bf (a)} The product of $\kappa^{II}$ and $x_L$ as a function of $L$. {\bf (b)} $x_L$ versus $\kappa^{II}$ on a log-log plot. The slopes of the fitted straight lines vary in the range -1.13 to -1.16 for the three algorithms (slope = -1 would have meant that $\kappa^{II}x_2$ is strictly constant). The figure shows that while each individual factor $\kappa^{II}$ and $x_L$ ranges over several orders of magnitude, their product, though not constant, is of order unity.}
    \label{kappa2-xL-relationship}
\end{figure}

The contribution of a catalyst to the rate of the reaction it catalyzes appears through the factor $1 + \kappa x$, where $\kappa$ is the catalytic strength of the catalyst and $x$ its concentration. The term unity in the above factor is the relative contribution of the spontaneous (uncatalyzed) reaction rate. If the catalyst is to play a significant role in the reaction, the catalytic contribution to the reaction rate should be at least comparable to the spontaneous rate, \ie, $\kappa x$ should be at least comparable to unity. As we have seen earlier the concentration of large molecules is typically damped exponentially with their size. Therefore the compensating factor $\kappa$ needs to increase exponentially in order for the catalyzed reaction rate to be comparable to the spontaneous reaction rate. For concreteness consider the extremal ACSs of length $L$ and consider the steady state population $x_L$ of the catalyst $\mathbf{A}(L)$ in the low fixed point as $\kappa$ is increased. In the spirit of this rough argument one expects that at the initiation threshold the term $\kappa^{II} x_L$ should be of order unity. In Fig. \ref{kappa2-xL-relationship} we display this product for different values of $L$. Though there is a secular decreasing trend with $L$, this product remains of order unity (Fig. \ref{kappa2-xL-relationship-a}) even as the individual factors change over several orders of magnitude (Fig. \ref{kappa2-xL-relationship-b}). This lends numerical support to the above explanation for the exponential dependence of $\kappa^{II}$ on $L$.

\section{A problem for primordial ACSs to produce large molecules}
A natural initial condition for the origin of life scenario is one where only the food set molecules, and perhaps a few other not very large molecules (dimers, trimers, etc.) have nonzero concentrations, while the large molecules have zero concentrations. It is from such an initial condition that we would like to see the emergence of large molecules through the dynamics.

The exponential increase of the initiation threshold, $\kappa^{II}$, with $L$, quantifies the difficulty in using ACSs to generate large molecules in the primordial scenario of the type modeled above. This means that one needs large molecules with unreasonably high catalytic strengths to exist in the chemistry in order to get them to appear with appreciable concentrations starting from physically reasonable primordial initial conditions.

\thispagestyle{plain}
\cleardoublepage
\chapter{\label{ChapterNestedACSs}Nested ACSs: A structure to promote large molecules}

\lettrine[lines=2, lhang=0, loversize=0.0, lraise=0.0]{W}{e} now discuss a mechanism that may overcome the barrier of large catalytic strengths, and may enable large molecules to arise from primordial initial conditions without exponentially increasing catalytic strengths. This mechanism relies on the existence of multiple ACSs of different sizes in the catalyzed chemistry, in a topology such that the smaller ACSs reinforce the larger ones, thereby enabling large molecules to appear with significant concentrations without exponentially increasing their catalytic strength.

\section{Using a small ACS to reinforce a larger one}
\begin{figure}[ht]
    \begin{center}
        \subfloat[]{\label{nested-3-8-a-network}\includegraphics[width=4.25in]{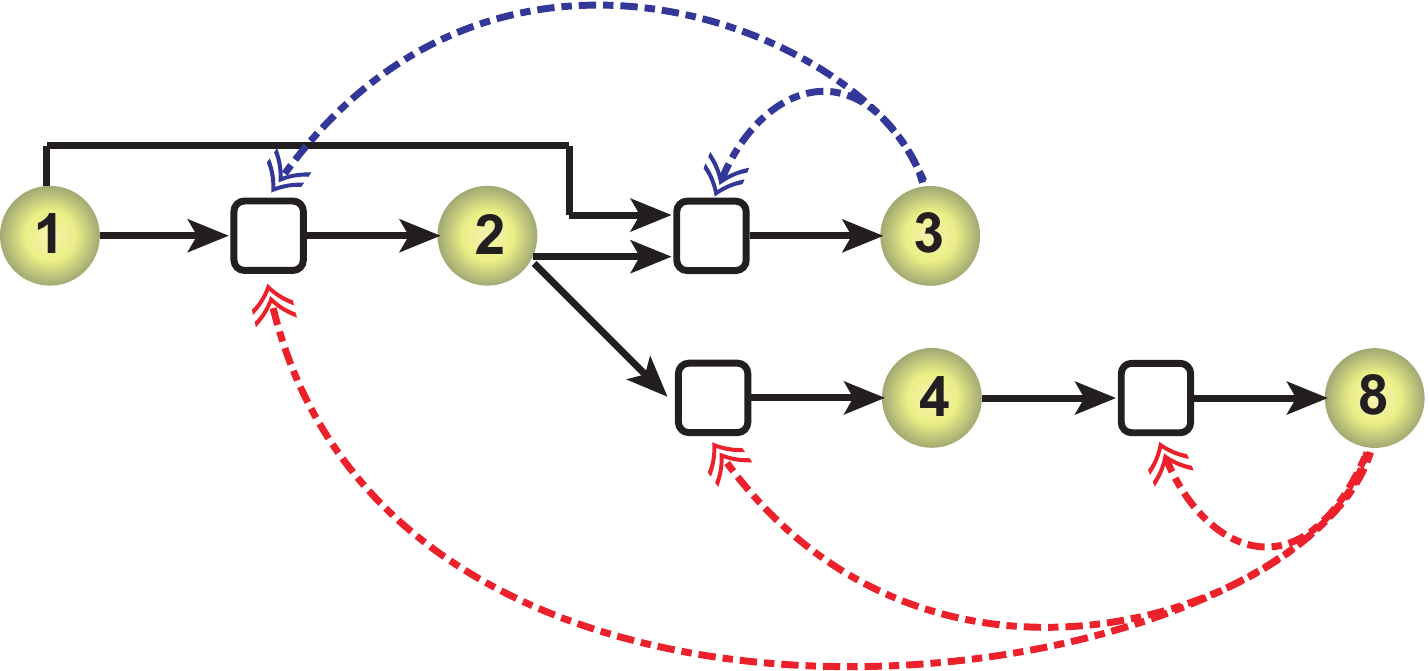}}\\
        \subfloat[]{\label{nested-3-8-b-network}\includegraphics[width=5.60in]{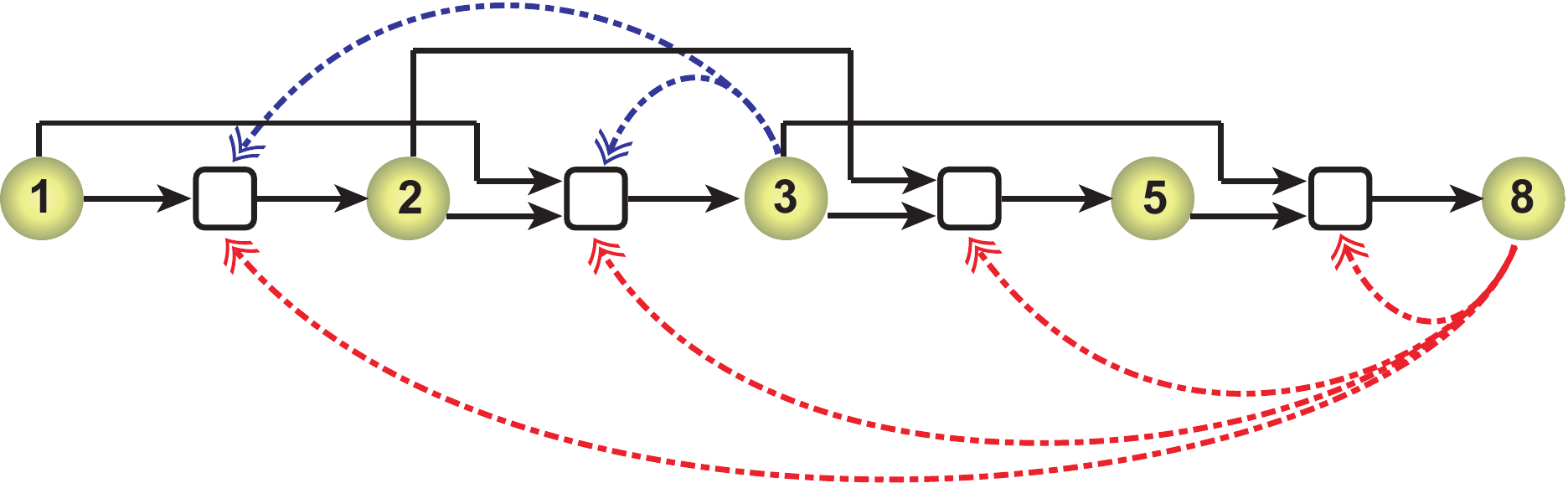}}\\
    \end{center}
    \caption[Pictorial representation of nested ACSs.]{{\bf Pictorial representation of nested ACSs.} {\bf (a)} ACS3+8, defined by Eqs. (\ref{acs3-definition}) and (\ref{acs8-definition}). {\bf (b)} ACS3+8', defined by Eqs. (\ref{acs3-definition}) and (\ref{acs8'-definition}). The notation is the same as for Fig. \ref{acs-eg-network}. The dashed arrows (catalytic links) are given in two colours, blue and red, to distinguish the two ACSs whose catalysts are molecules A(3) and A(8), respectively. Reactions having two catalysts are given by two distinct equations in the text (e.g. (\ref{acs3-rct1}) and (\ref{acs8-rct1})), but in the figure are represented by a single reaction node with two incoming catalytic links (the reaction node is not duplicated to avoid visual clutter).}
    \label{nested-3-8-network}
\end{figure}

\begin{figure}
    \begin{center}
        \subfloat[]{\label{nested-3-8-a}\includegraphics[height=4.75in,angle=-90,trim=1.5cm 2cm 1.5cm 2cm,clip]{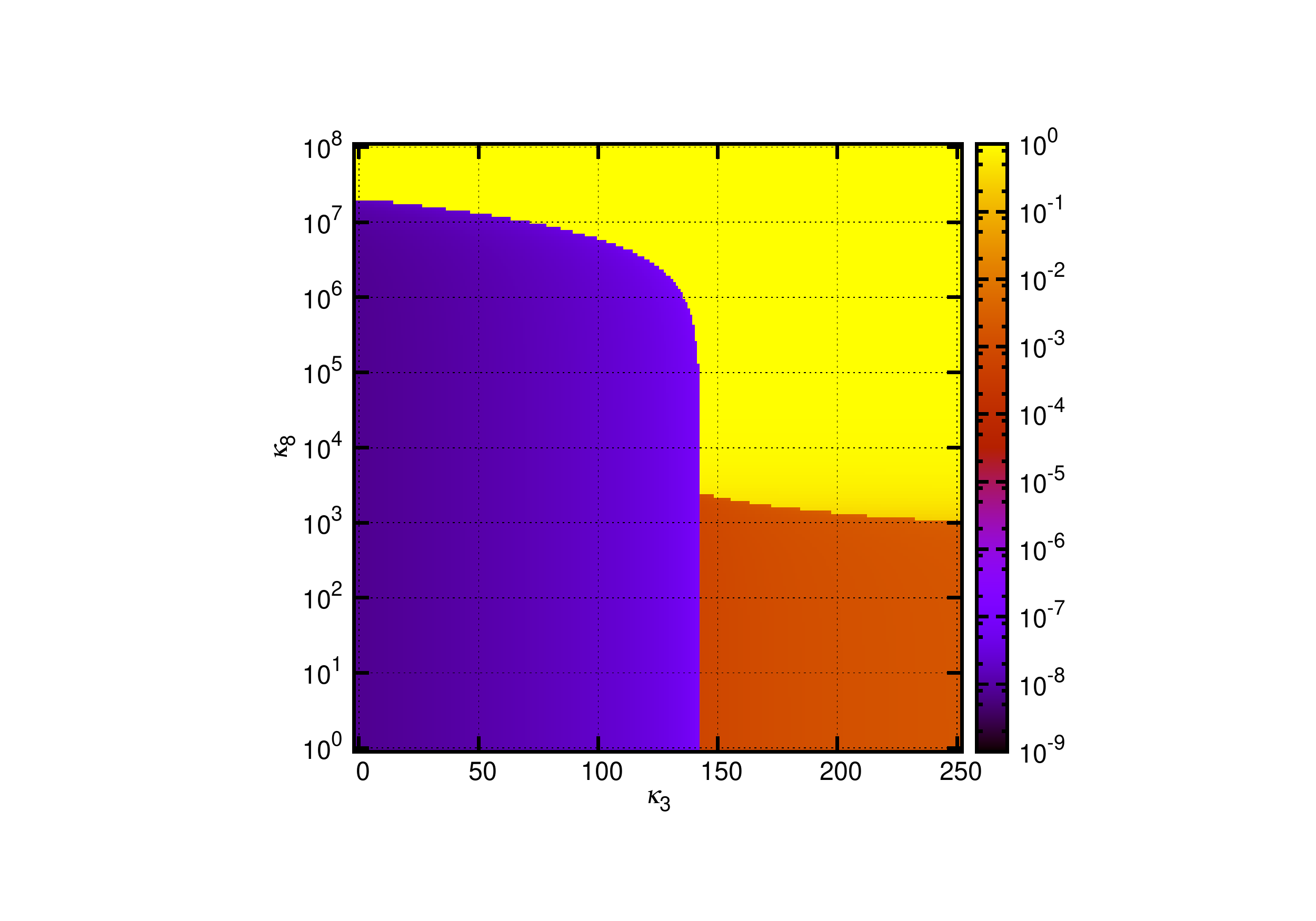}}\\
        \subfloat[]{\label{nested-3-8-b}\includegraphics[height=4.75in,angle=-90,trim=1.5cm 2cm 1.5cm 2cm,clip]{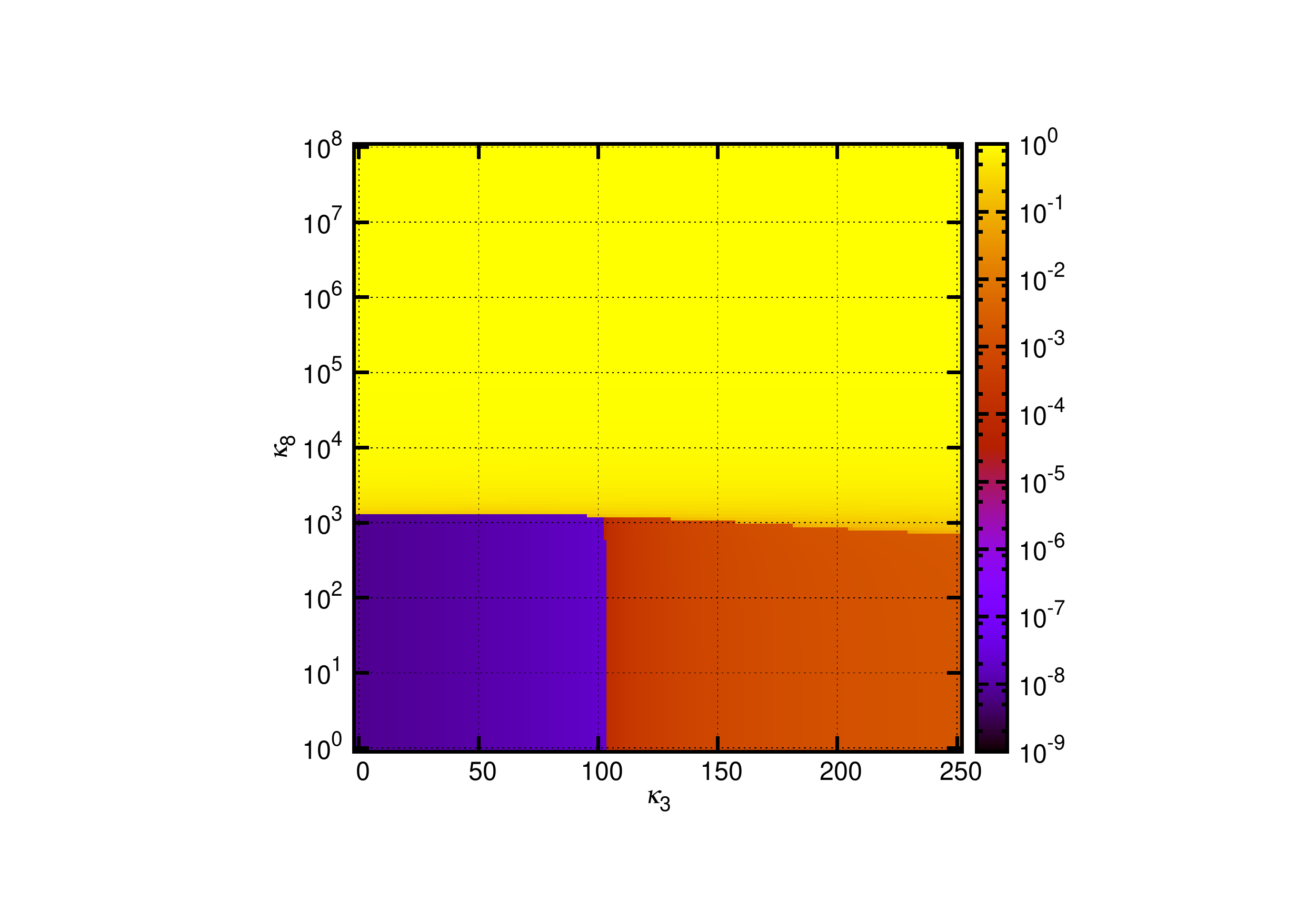}}
    \end{center}
    \caption[Reinforcement of a larger ACS by a smaller one: The case of ACS3+8.]{{\bf Reinforcement of a larger ACS by a smaller one: The case of ACS3+8.} The figure shows the steady state concentration $x_8$ (in colour coding as indicated) for two different initial conditions as a function of $\kappa_3$ and $\kappa_8$, the catalytic strengths of $\mathbf{A}(3)$ and $\mathbf{A}(8)$ respectively. All simulations were done for $k_f = k_r = A = 1, \phi = 20, N=100$. The two figures (a) and (b) differ in the initial condition of the dynamics. {\bf (a)} The standard initial condition, {\bf (b)} initial condition $x_n = 1$ for all $n=2,3,\ldots,N$.}
    \label{nested-3-8}
\end{figure}

To illustrate the basic idea we consider the following simple example where the catalyzed chemistry contains only two ACSs, one of length three and the other of length eight (which we refer to as ACS3 and ACS8, respectively), each generated by the Algorithm 2 mentioned in the previous chapter. All reactions of the former are catalyzed by $\mathbf{A}(3)$ with a catalytic strength $\kappa_3$, and of the latter by $\mathbf{A}(8)$ with the catalytic strength $\kappa_8$. Thus the two ACSs are:
\begin{subequations}
  \label{acs3-definition}
  \begin{eqnarray}
    \label {acs3-rct1} \mathrm{\mathbf A}(1) + \mathrm{\mathbf A}(1) & \reactionrevarrow{\ensuremath{\mathrm{\mathbf A}(3)}}{} & \mathrm{\mathbf A}(2) \\
    \label {acs3-rct2} \mathrm{\mathbf A}(1) + \mathrm{\mathbf A}(2) & \reactionrevarrow{\ensuremath{\mathrm{\mathbf A}(3)}}{} & \mathrm{\mathbf A}(3),
  \end{eqnarray}
\end{subequations}
and
\begin{subequations}
  \label{acs8-definition}
  \begin{eqnarray}
    \label {acs8-rct1} \mathrm{\mathbf A}(1) + \mathrm{\mathbf A}(1) & \reactionrevarrow{\ensuremath{\mathrm{\mathbf A}(8)}}{} & \mathrm{\mathbf A}(2) \\
    \label {acs8-rct2} \mathrm{\mathbf A}(2) + \mathrm{\mathbf A}(2) & \reactionrevarrow{\ensuremath{\mathrm{\mathbf A}(8)}}{} & \mathrm{\mathbf A}(4) \\
    \label {acs8-rct3} \mathrm{\mathbf A}(4) + \mathrm{\mathbf A}(4) & \reactionrevarrow{\ensuremath{\mathrm{\mathbf A}(8)}}{} & \mathrm{\mathbf A}(8).
  \end{eqnarray}
\end{subequations}
The catalyzed chemistry consists of the above five catalyzed reaction pairs (we will refer to this catalyzed chemistry as ACS3+8). The network topology is pictorially depicted in Fig. \ref{nested-3-8-a-network}. The system also exhibits bistability, and the concentration of $\mathbf{A}(8)$ in the two fixed point attractors is exhibited in Fig. \ref{nested-3-8} as a function of $\kappa_3$ and $\kappa_8$.

\begin{figure}
    \begin{center}
        \subfloat[]{\label{nested-3-8-time}\includegraphics[height=5.25in,angle=-90]{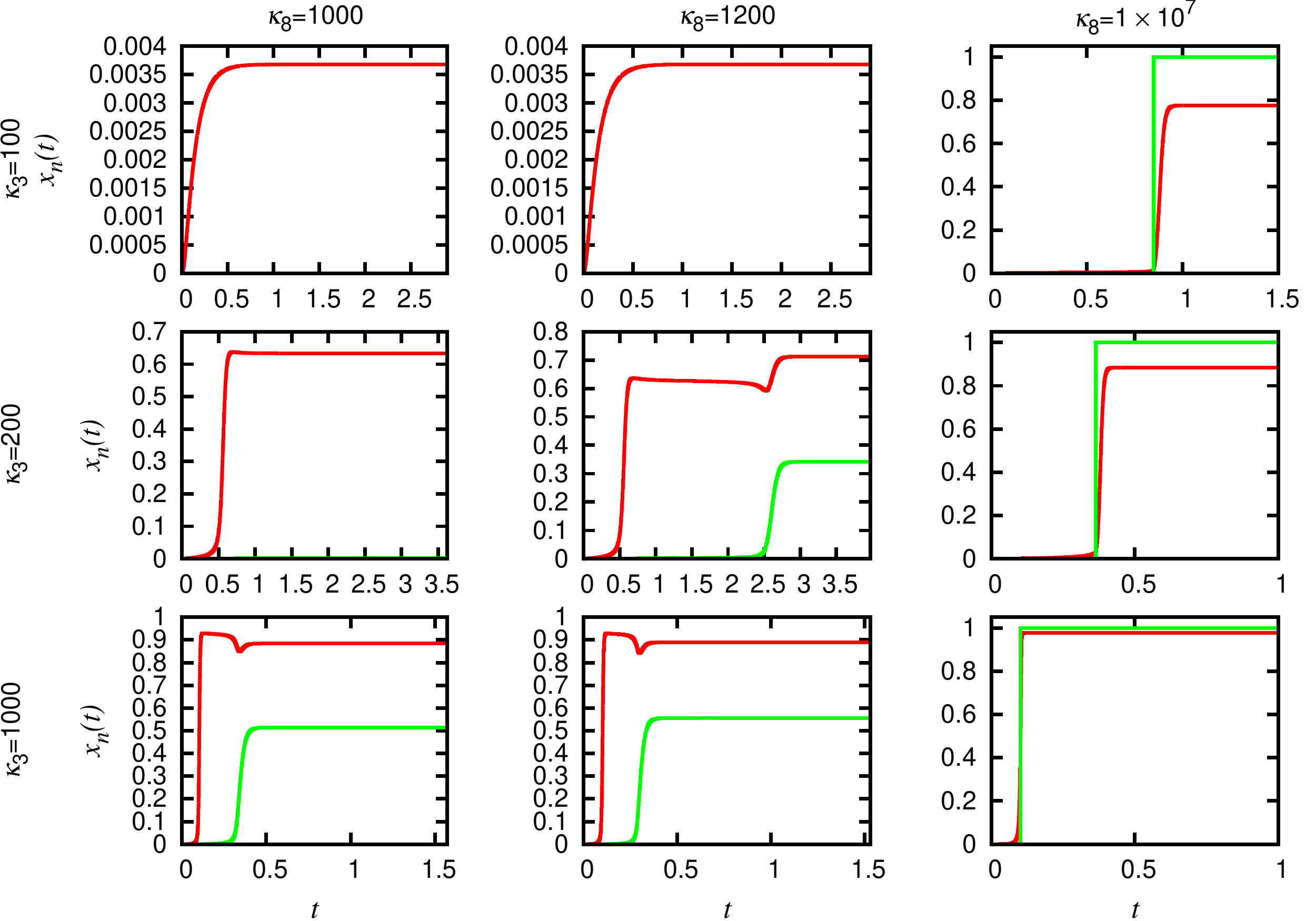}}\\
        \subfloat[]{\label{nested-3-8-ssc}\includegraphics[height=5.25in,angle=-90]{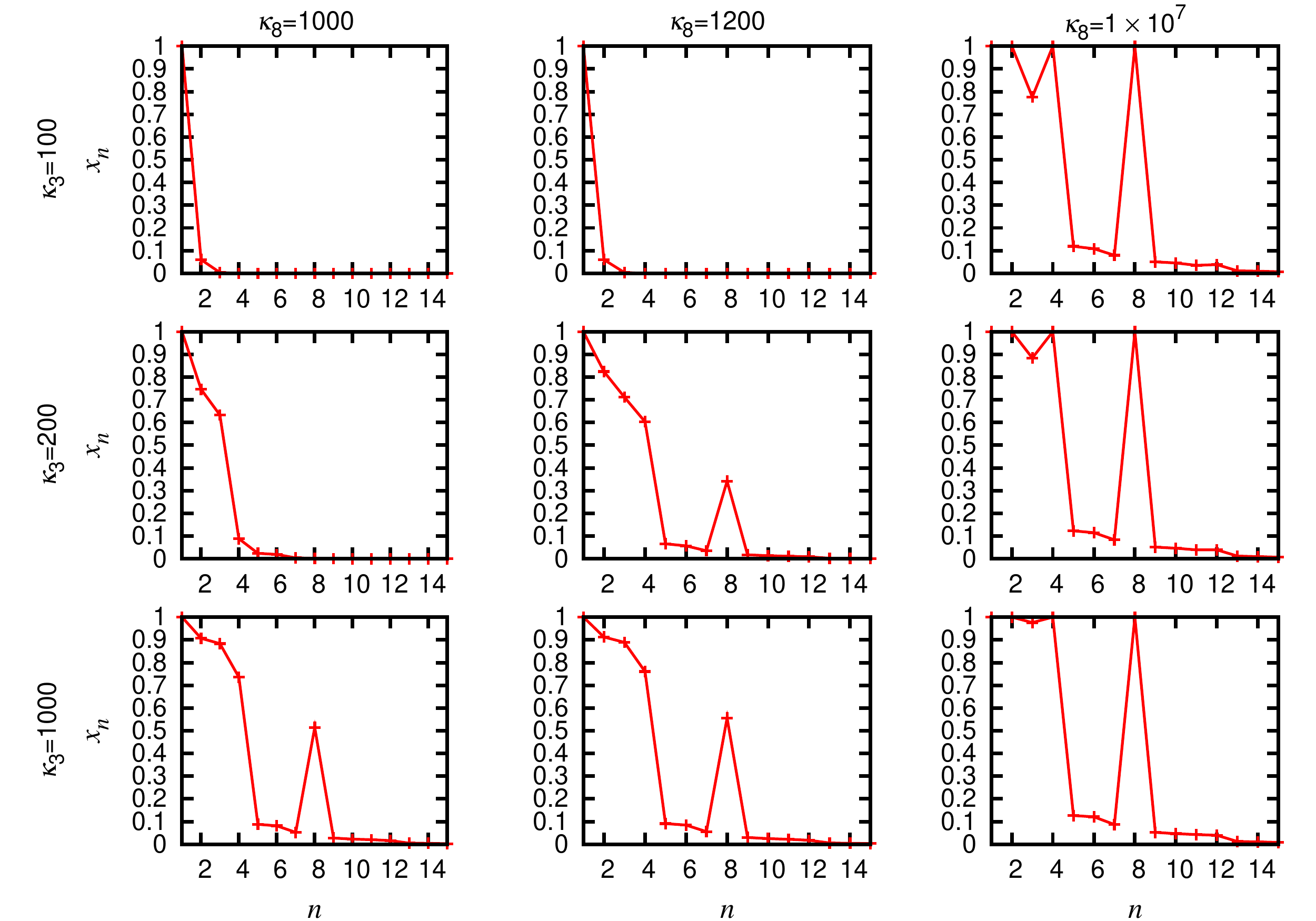}}
    \end{center}
    \caption[Concentrations in ACS3+8.]{{\bf Concentrations in ACS3+8.} {\bf (a)} Time evolution of concentrations of $\mathbf A(3)$ (shown in red curve) and $\mathbf A(8)$ (green curve). {\bf (b)} Steady state concentrations profiles (shown up to $n=15$). The concentrations are shown for three values of $\kappa_3$ and $\kappa_8$ (nine pairs in total). Other parameters take same values as in Fig. \ref{nested-3-8}.}
    \label{nested-3-8-time-ssc}
\end{figure}

When $\kappa_3$ is small the two pictures in Fig. \ref{nested-3-8} show the usual bistability of ACS8 along the $\kappa_8$ axis. The initiation and maintenance thresholds are $\kappa_8^{II} = 1.78 \times 10^7$ and $\kappa_8^{I} = 1145$ given by the location of the boundary between the low concentration region (blue, $x_8 \sim 10^{-7}$) and the high concentration region (yellow $x_8 \sim 1$) along the $\kappa_8$ axis in Figs. \ref{nested-3-8-a} and \ref{nested-3-8-b} respectively. As $\kappa_3$ increases, the initiation threshold of ACS8 decreases slowly for a while (in Fig. \ref{nested-3-8-time-ssc} for $\kappa_3$ = 100, ACS8 dominates at $\kappa_8 = 1 \times 10^7$), then drops sharply near $\kappa_3 = 141$. This value of $\kappa_3$ is the initiation threshold of ACS3 when $\kappa_8=0$. When $\kappa_3$ exceeds this value, the steady state value of $x_8$ is either high (yellow, $x_8 \sim 1$) or intermediate (orange, $x_8 \sim 10^{-3}$), depending upon the value of $\kappa_8$ (see panels corresponding to $\kappa_3 = 200$ and $1000$ in Fig. \ref{nested-3-8-time-ssc}).

The key point is that the initiation threshold of the larger catalyst depends on the catalytic strength of the smaller catalyst. The former plummets sharply when the latter approaches the initiation threshold of the smaller catalyst, dropping to a much lower value than before (compare the lower limit of the yellow region in Fig. \ref{nested-3-8-a} to the left and right of $\kappa_3 = \kappa_3^{II} = 141$; the value of $\kappa_8^{II}$ plunges several orders of magnitude from $1.78 \times 10^7$ at $\kappa_3 = 0$ down to $2178$ at $\kappa_3 = 141$). Starting from the standard initial condition, thus, the larger catalyst can acquire a significant concentration at a much lower value of its catalytic strength in the presence of a smaller ACS operating above its initiation threshold than in its absence.

\subsection{\label{nested-acs-and-background}Why a small ACS reinforces a larger one: The role of the background and spontaneous reactions}
When $\kappa$ exceeds the initiation threshold for a catalyzed chemistry containing an ACS, not only do the steady state concentrations of the ACS product molecules rise by several orders of magnitude, but also those of the background molecules rise. As an example compare the two steady state profiles of ACS65 in Fig. \ref{acs-eg-ssc}, which correspond to values of $\kappa$ below and above the initiation threshold. As one goes from the lower to the upper curve, the concentration of the ACS members of course increases dramatically (as shown by the sharp peaks), but note that the concentrations of other molecules not produced by catalyzed reactions also goes up significantly. Thus in the chemistry containing two ACSs (ACS3+8) as one moves along the $\kappa_3$ axis in Fig. \ref{nested-3-8-a} and crosses the initiation threshold of ACS3 (\ie, $\kappa_3$ exceeds $\kappa_3^{II} = 141$), the concentration of $\mathbf{A}(8)$ (a molecule belonging to the background of ACS3 as its production is not catalysed by ACS3) increases from $\sim 10^{-7}$ (blue region) to $\sim 10^{-3}$ (orange region). This increase in the concentration of $\mathbf{A}(8)$ by a factor of $\sim 10^4$ makes it easier for ACS8 to function and its initiation threshold drops by a corresponding factor of about $10^4$ (from $\sim 10^7$ to $\sim 10^3$).

This fact highlights the role of spontaneous reactions in the overall dynamics. The background molecules are connected to the ACS through spontaneous reactions, and if it were not for the latter, an ACS would not be able to push up the concentrations of its nearby background. We shall refer to a structure such as the one described above containing ACSs of different sizes with the smaller ACS feeding into the larger one through the spontaneous reactions as a `nested ACS' structure.

We remark that the ``distance'' between the catalysts of the two ACSs (as calculated on the network of spontaneous reactions linking the two molecules) is pertinent to the point under discussion in this section. If the two ACSs are very distant, e.g., instead of two catalysts of size 3 and 8 in the above example we had studied catalysts of size 3 and 20, then even though the smaller ACS raises the concentration of the larger catalyst, the rise may not be sufficient to allow the larger ACS to dominate at reasonable catalytic strengths.

\section{The role of `overlapping' catalyzed pathways in nested ACSs}
The above example also serves to highlight some other features of catalyzed chemistries containing multiple ACSs. Note that the production pathway of $\mathbf{A}(8)$ in ACS8 (Eqs. \ref{acs8-definition} and Fig. \ref{nested-3-8-a-network}) contains one reaction pair in common with ACS3, namely the reaction pair $\mathbf{A}(1) + \mathbf{A}(1) \rightleftharpoons \mathbf{A}(2)$. One can consider a situation wherein the overlap is greater. E.g., consider the ACS8$'$ defined by
\begin{subequations}
  \label{acs8'-definition}
  \begin{eqnarray}
    \label {acs8'-rct1} \mathrm{\mathbf A}(1) + \mathrm{\mathbf A}(1) & \reactionrevarrow{\ensuremath{\mathrm{\mathbf A}(8)}}{} & \mathrm{\mathbf A}(2) \\
    \label {acs8'-rct2} \mathrm{\mathbf A}(1) + \mathrm{\mathbf A}(2) & \reactionrevarrow{\ensuremath{\mathrm{\mathbf A}(8)}}{} & \mathrm{\mathbf A}(3) \\
    \label {acs8'-rct3} \mathrm{\mathbf A}(2) + \mathrm{\mathbf A}(3) & \reactionrevarrow{\ensuremath{\mathrm{\mathbf A}(8)}}{} & \mathrm{\mathbf A}(5) \\
    \label {acs8'-rct4} \mathrm{\mathbf A}(3) + \mathrm{\mathbf A}(5) & \reactionrevarrow{\ensuremath{\mathrm{\mathbf A}(8)}}{} & \mathrm{\mathbf A}(8).
  \end{eqnarray}
\end{subequations}
Now the set of reactions in ACS3 is a subset of ACS8$'$ (ignoring the catalyst, which is different in the two cases). The degree of overlap of the catalyzed reaction sets  between a pair of nested ACSs makes a difference in the dynamics. Consider, for example, the catalyzed chemistry consisting of ACS3 and ACS8$'$, \ie, the set of catalyzed reactions given by Eqs. (\ref{acs3-definition}) and (\ref{acs8'-definition}), which we refer to as ACS3+8$'$. This is picturized in Fig. \ref{nested-3-8-b-network}. Like ACS3+8, this chemistry also shows a reduction of $\kappa_8^{II}$, when $\kappa_3$ exceeds its initiation threshold. We find that while at $\kappa_3=0$ the value of $\kappa_8^{II}$ for the two chemistries is not too different ($1.6 \times 10^7$ for ACS3+8$'$ versus $1.8 \times 10^7$ for ACS3+8), at $\kappa_3 = 141$, $\kappa_8^{II}$ reduces to a value 920 in ACS3+8$'$, which is less than half of the value 2178 that it reduces to in ACS3+8. Thus a larger degree of overlap between the catalyzed reaction sets of nested ACSs causes more effective reinforcement.

\begin{figure}
    \begin{center}
        \subfloat[]{\includegraphics[height=1.9in,angle=-90,trim=0.cm 2cm 0.5cm 2cm,clip]{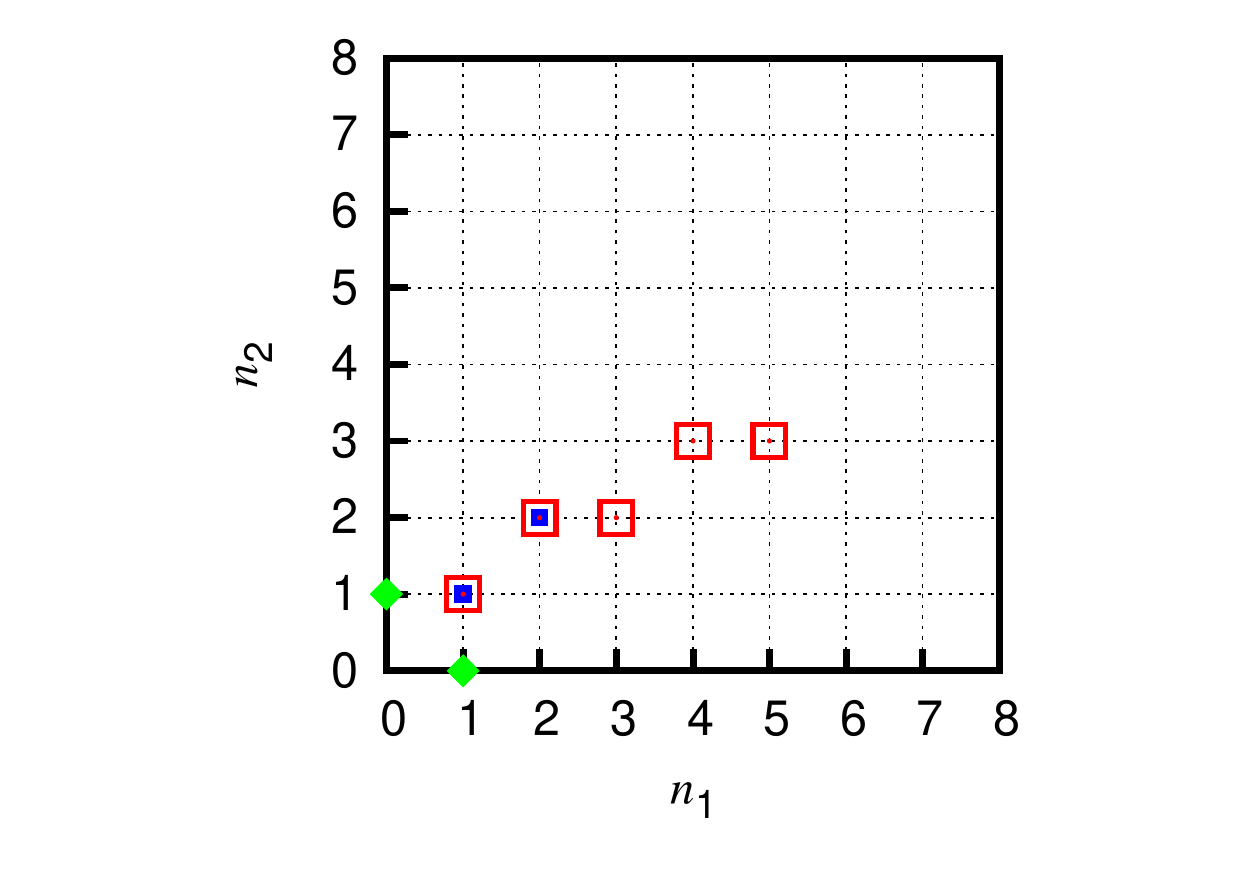}}
        \subfloat[]{\includegraphics[height=1.9in,angle=-90,trim=0.cm 2cm 0.5cm 2cm,clip]{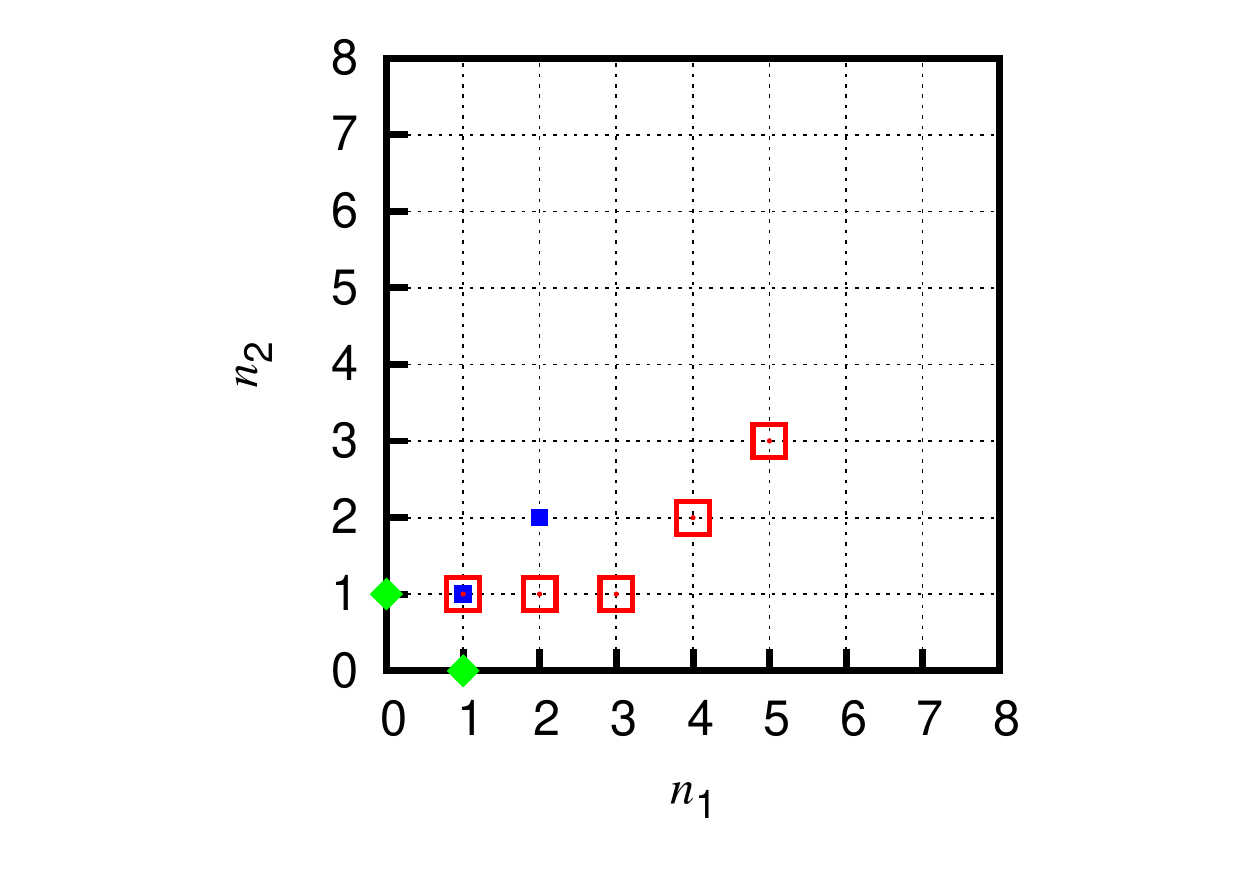}}
        \subfloat[]{\includegraphics[height=1.9in,angle=-90,trim=0.cm 2cm 0.5cm 2cm,clip]{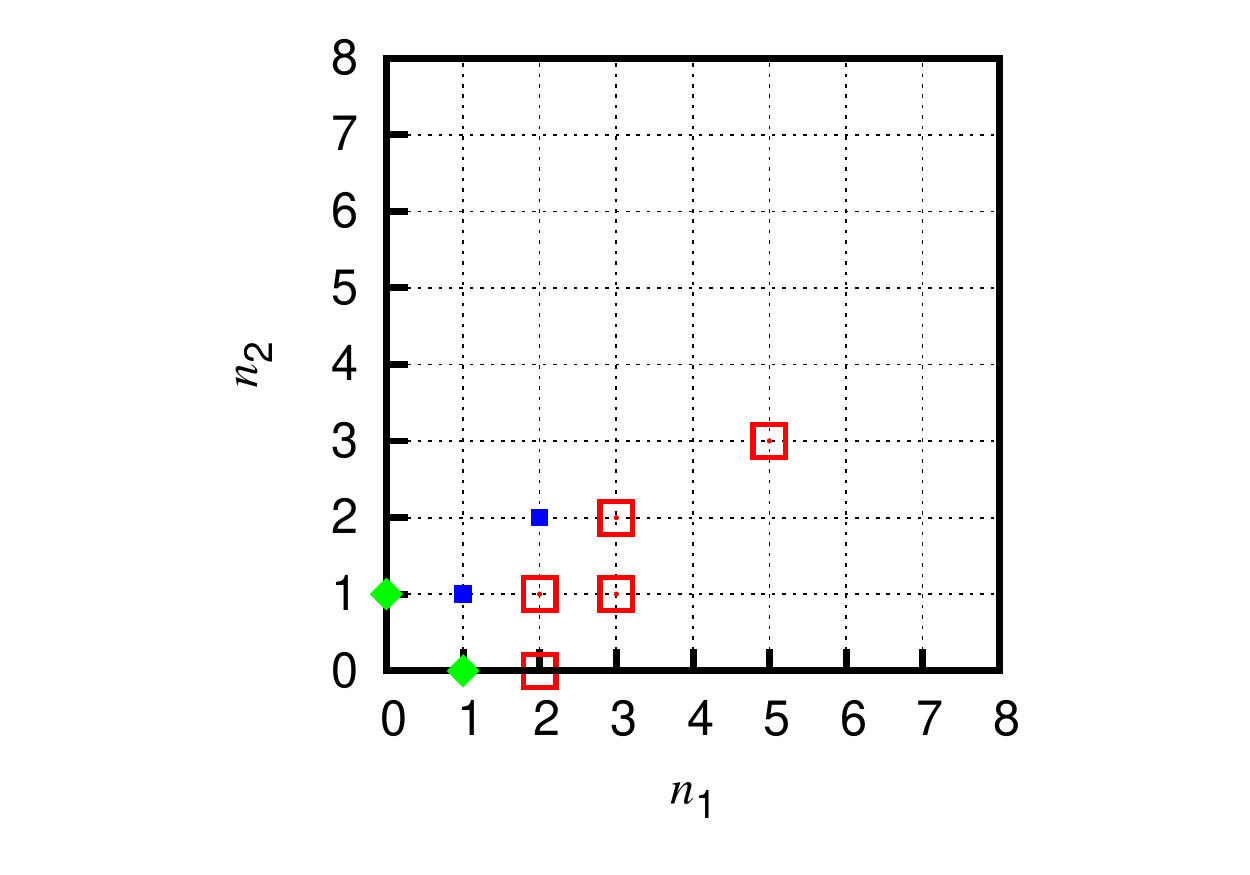}}
    \end{center}
    \caption[Examples of nested ACS pairs with different degrees of overlap for $f=2$.]{{\bf Examples of nested ACS pairs with different degrees of overlap for $f=2$.} In the three cases the reaction sets have (A) maximal overlap, (B) partial overlap, (C) no overlap. The blue and red squares marking the grid points indicate the identity of molecules produced in the two ACSs; blue filled squares correspond to the products of the smaller ACS, red unfilled squares to those of the larger ACS. The $x$ and $y$ axes denote the number of monomers of type (1,0) and (0,1), respectively, in the molecules. The green rhombuses represent the two monomers.}
    \label{acs-2d-nested-4-8}
\end{figure}

Another example with this behaviour for $f=2$ is described in Fig. \ref{acs-2d-nested-4-8}. In each of the three ACS pairs shown in the figure, the smaller ACS, of length 4, is the same, (it will be referred to as ACS(2,2)) and is defined by the reactions (each catalyzed by (2,2))
\begin{subequations}
  \label{ACS(2,2)-definition}
  \begin{eqnarray}
    (0,1) + (1,0) & \reactionrevarrow{(2,2)}{} & (1,1) \\
    (1,1) + (1,1) & \reactionrevarrow{(2,2)}{} & (2,2).
  \end{eqnarray}
\end{subequations}

\noindent The three larger ACSs, called ACS(5,3)(a), ACS(5,3)(b) and ACS(5,3)(c), respectively, can essentially be determined from the figure. For example, ACS(5,3)(a) consists of the two reaction pairs given by Eqs. (\ref{ACS(2,2)-definition}), both catalyzed by (5,3) as well as the three reactions
\begin{subequations}
  \label{ACS(5,3)(a)-definition}
  \begin{eqnarray}
    (1,0) + (2,2) & \reactionrevarrow{(5,3)}{} & (3,2) \\
    (1,1) + (3,2) & \reactionrevarrow{(5,3)}{} & (4,3) \\
    (1,0) + (4,3) & \reactionrevarrow{(5,3)}{} & (5,3).
  \end{eqnarray}
\end{subequations}

\noindent ACS(5,3)(b) consists of the single reaction pair given by the first of Eqs. (\ref{ACS(2,2)-definition}), catalyzed by (5,3), as well as the four reactions
\begin{subequations}
  \label{ACS(5,3)(b)-definition}
  \begin{eqnarray}
    (1,0) + (1,1) & \reactionrevarrow{(5,3)}{} & (2,1) \\
    (1,0) + (2,1) & \reactionrevarrow{(5,3)}{} & (3,1) \\
    (1,1) + (3,1) & \reactionrevarrow{(5,3)}{} & (4,2) \\
    (1,1) + (4,3) & \reactionrevarrow{(5,3)}{} & (5,3),
  \end{eqnarray}
\end{subequations}
and ACS(5,3)(c) consists of the five reaction pairs
\begin{subequations}
  \label{ACS(5,3)(c)-definition}
  \begin{eqnarray}
    (1,0) + (1,0) & \reactionrevarrow{(5,3)}{} & (2,0) \\
    (0,1) + (2,0) & \reactionrevarrow{(5,3)}{} & (2,1) \\
    (1,0) + (2,1) & \reactionrevarrow{(5,3)}{} & (3,1) \\
    (0,1) + (3,1) & \reactionrevarrow{(5,3)}{} & (3,2) \\
    (2,1) + (3,2) & \reactionrevarrow{(5,3)}{} & (5,3).
  \end{eqnarray}
\end{subequations}

The network topologies for three catalyzed chemistries are depicted pictorially in Fig. \ref{nested-2d-4-8-network}.
\begin{figure}
    \begin{center}
        \subfloat[]{\label{nested-2d-4-8-a-network}\includegraphics[width=5.6in]{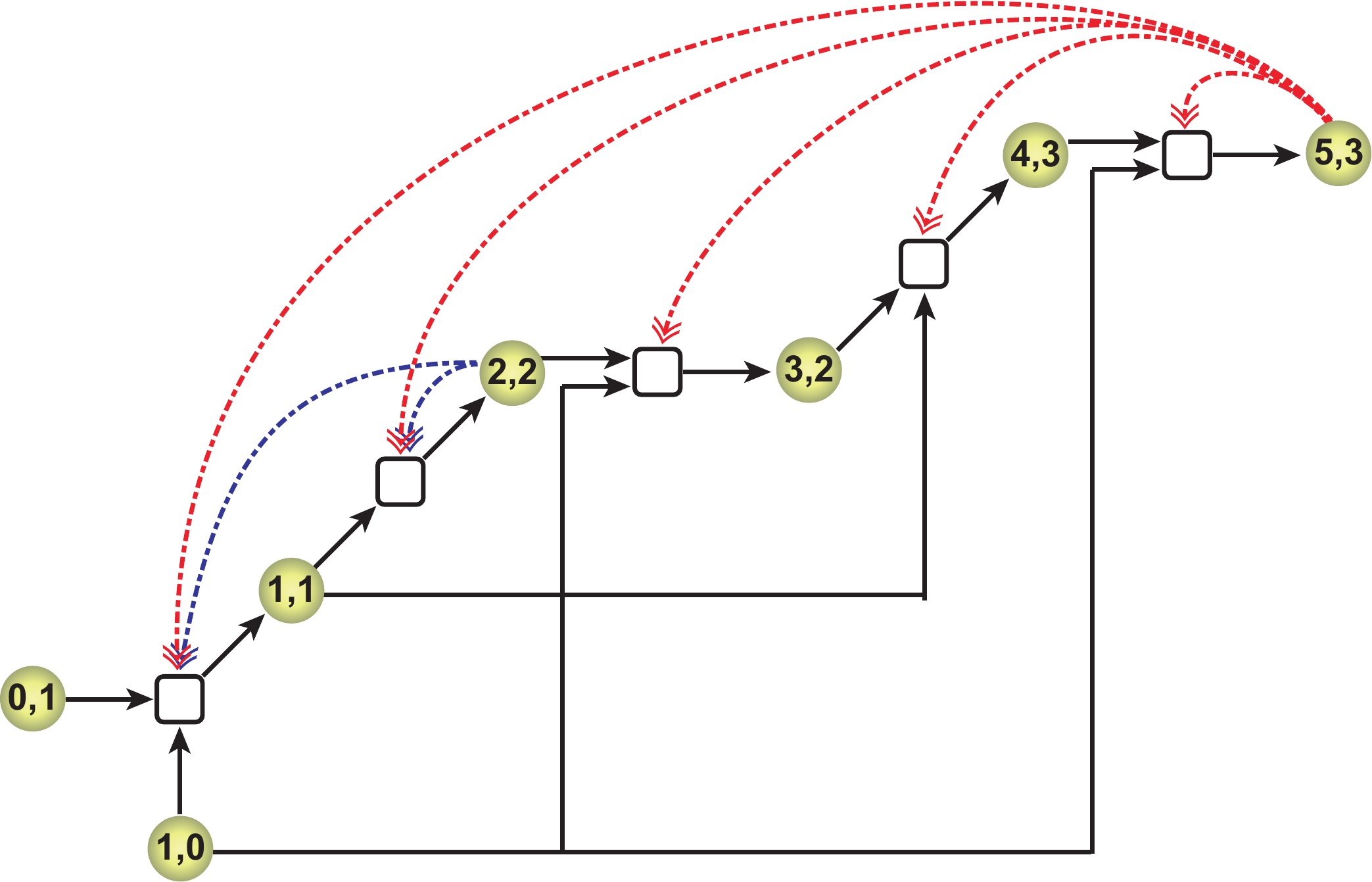}}\\
        \subfloat[]{\label{nested-2d-4-8-b-network}\includegraphics[width=5.75in]{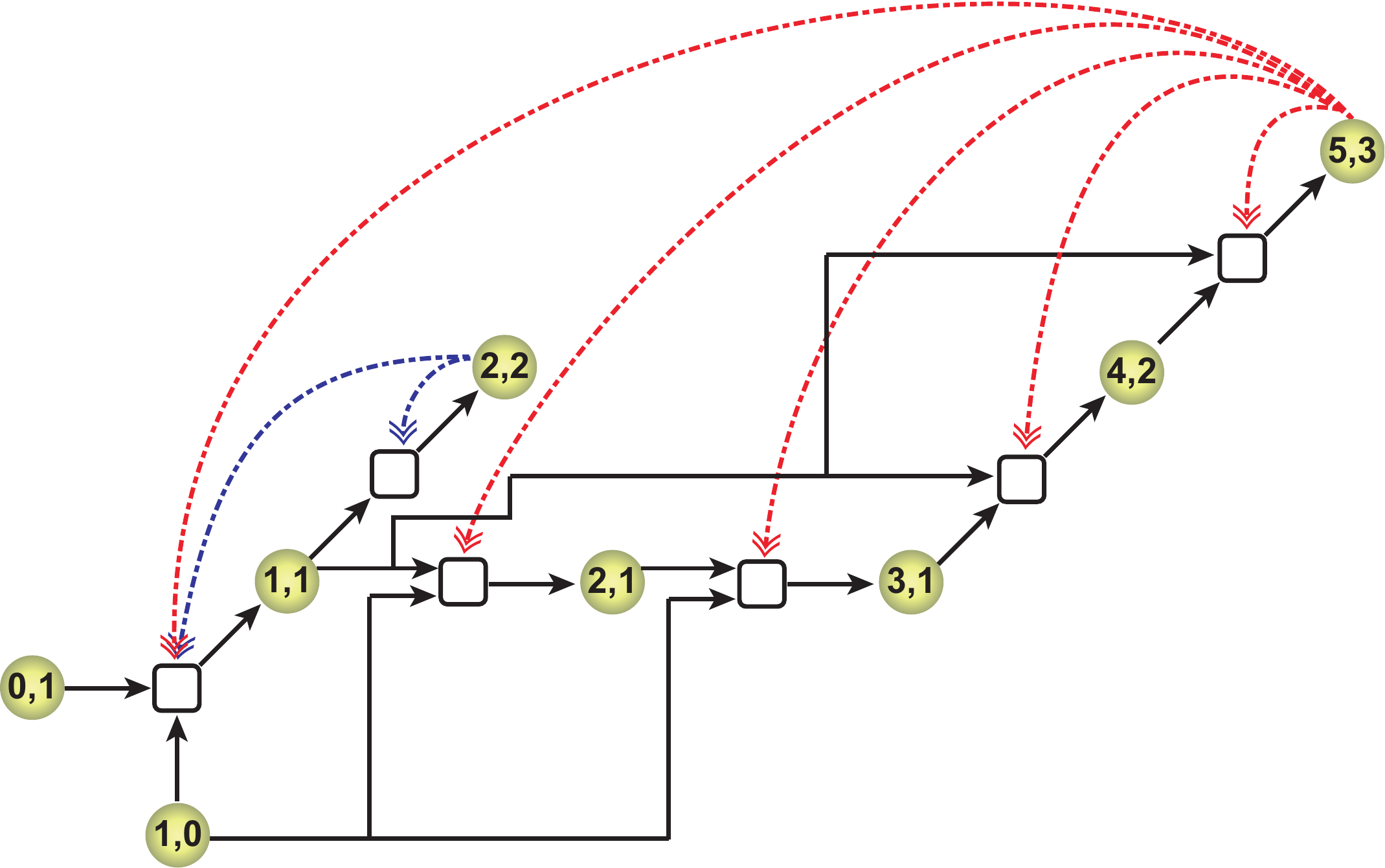}}
    \end{center}
    \caption[Pictorial representation of nested ACSs with $f=2$.]{{\bf Pictorial representation of nested ACSs with $f=2$.} {\it Continued.}}
\end{figure}

\captionsetup{list=no}
\begin{figure}
    \ContinuedFloat
    \begin{center}
        \subfloat[]{\label{nested-2d-4-8-c-network}\includegraphics[width=4.3in]{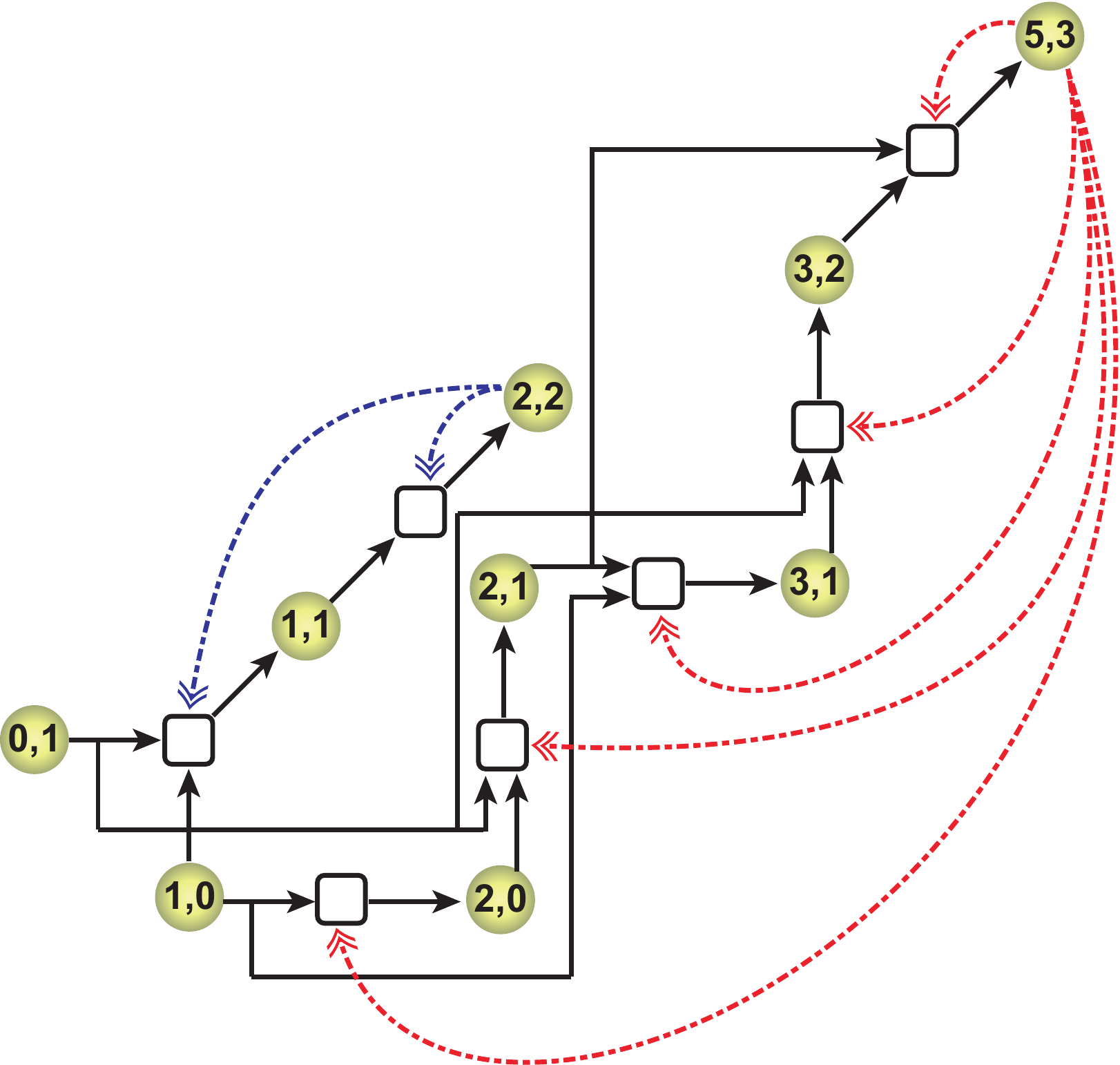}}
    \end{center}
    \caption[]{{\bf Pictorial representation of nested ACSs with $f=2$.} Each of the three figures shows two nested ACSs. The smaller ACS in each figure is ACS(2,2), defined by Eqs. (\ref{ACS(2,2)-definition}). The larger ACS is different and is given by {\bf (a)} ACS(5,3)(a), {\bf (b)} ACS(5,3)(b), and, {\bf (c)} ACS(5,3)(c), defined below Eqs. (\ref{ACS(2,2)-definition}). The notation is the same as in Figs. \ref{acs-eg-network} and \ref{nested-3-8-network}.}
    \label{nested-2d-4-8-network}
\end{figure}
\captionsetup{list=yes}

We consider the population dynamics of chemistries in which the spontaneous part includes all possible ligation and cleavage reactions involving molecules with upto $N=15$ monomers with homogeneous rate constants $k_f = k_r = 1$, $\phi = 15$, and the catalyzed part containing one or more of the above mentioned ACSs. When ACS(2,2) is the only ACS present, the system shows bistability with the initiation threshold being $\kappa_{(2,2)}^{II} = 551$. When ACS(5,3)(a), (b) or (c) are the only ACSs present, the initiation thresholds for them are 1125197, 1031082, and 1000112, respectively. When ACS(2,2) and one of ACS(5,3) (a), (b) or (c) are both present, and the catalytic strength of (2,2) is 552, the initiation thresholds of the
three larger ACSs reduce to 941, 1256, and 2482, respectively. Again, it is seen that the larger the degree of overlap of the two nested ACSs, the more effective is the reinforcement.

\section{Summary}
In this chapter we have considered nested ACSs in a simple setting, wherein the catalyzed chemistry contains just two extremal ACSs, one larger than the other. The catalyst of the larger ACS is in the background of the smaller ACS, \ie, it can be produced from the product molecules of the smaller ACS via pathways of spontaneous reactions. We have shown that this reduces the catalytic strength required for the larger ACS to dominate by a substantial amount (by several order of magnitude in the examples considered). We have given an explanation for this, namely, that when the smaller ACS operates above its own initiation threshold, it boosts the concentrations of its background molecules including that of the larger catalyst by a large factor resulting in a corresponding decrease of the initiation threshold of the larger ACS.

\thispagestyle{plain}
\cleardoublepage
\chapter{\label{ChapterHierarchyNestedACSs}Hierarchy of nested ACSs: A possible route for the appearance of large molecules}

\lettrine[lines=2, lhang=0, loversize=0.0, lraise=0.0]{T}{he} process of nesting discussed in the previous chapter for two ACSs can be extended to multiple levels of ACSs connected to each other. Here we discuss sequences of ACSs of increasing size, with the catalyzed reaction set of each ACS in the sequence partially or completely contained within the next one, and the catalytic strength of molecules increasing with size in a controlled manner. We construct examples of such sequences in which large catalyst molecules containing several hundred monomers can acquire significant concentrations starting from the standard initial condition, even though all catalysts have moderate catalytic strengths.

We construct a hierarchy of nested ACSs using two algorithms. In order to construct a cascade of nested ACSs in which reaction sets of smaller ACSs are completely contained in the larger ones (maximal overlap), we used Algorithm 4 (described below). This algorithm produces a cascade of ACSs with $g$ steps (generations), with the $k^\mathrm{th}$ generation ACS containing $n_k$ new reactions. We studied several catalyzed chemistries containing a cascade of nested ACSs for $f=1$ and 2. One example of each type is presented below; other examples gave qualitatively similar results. To construct ACSs with a variable degree of overlap, we used Algorithm 5. At each step $k$, an extremal ACS of length $L_k$ is added, generated using Algorithm 3. In Algorithm 5 the sequence of lengths, $L_1, L_2, \ldots, L_g$, is specified before running the algorithm.

\section{\label{Algo-4-5}Generating a cascade of nested ACSs}
\begin{description}
  \item[Algorithm 4:] {\bf Incremental, random construction of a sequence of reaction sets with maximal overlap.} We construct successive sets, or `generations', $g$ in number, $P_1, P_2, \ldots P_g$, of product molecules, starting from the food set $\mathcal {F} (\equiv P_0)$. Each generation $P_k$ $(k=1,2,\ldots g)$ has a pre-specifed number of molecules, $n_k$ ($n_1$, $n_2$, $\ldots n_g$ need to be specified before running the algorithm). At step $k$ of the cascade an ACS $S_k$ is constructed from the previous ACS $S_{k-1}$ by adding reactions between molecules belonging to a reactant set $R_k$ consisting of all the products of the previous generations and the food set, $R_k = P_0 \cup P_1 \cup P_2 \ldots P_{k-1}$. Let $L_k$ denote the size of the largest molecule in $R_k$. At the beginning of the $k^\mathrm{th}$ step, $P_k$ is empty and the set of reactions in $S_k$ is the same as in $S_{k-1}$, except that the catalysts of the reactions in $S_k$ are not yet assigned. ($S_0$ is the empty set.) To construct $P_k$ and $S_k$, pick a molecule $\mathbf X$ at random from the previous generation of products $P_{k-1}$ and another molecule $\mathbf Y$ at random from $R_k$, and determine the product $\mathbf Z$ formed if they were to be ligated ($\mathbf X$ being the same as $\mathbf Y$ is allowed). If the size of this product, $L(\mathbf Z)$, is $\leq L_k$, discard the pair and choose another pair. If $L(\mathbf Z) > L_k$, add the molecule $\mathbf Z$ to $P_k$ and the ligation reaction $\mathbf X + \mathbf Y \rightarrow \mathbf Z$ to $S_k$. Repeat this procedure until $n_k$ molecules are added to $P_k$ and $n_k$ reactions are added to $S_k$. Assign catalysts to each reaction in $S_k$ (which includes the reactions inherited from $S_{k-1}$ and the new $n_k$ reactions) randomly from $P_k$. This completes the $k^\mathrm{th}$ step. To get the full cascade with $g$ generations this process is carried out for $k=1,2, \ldots g$. By construction the set of reactions (ignoring the catalyst) of each $S_k$ is fully contained in that of $S_{k+1}$ but the catalysts are different, being drawn from $P_k$ for $S_k$ and $P_{k+1}$ for $S_{k+1}$. The union of the $S_k$'s constitutes the set of reactions in the catalyzed chemistry. Note that in this catalyzed chemistry reactions have multiple catalysts, the multiplicity declining for reactions producing higher generation molecules.

      The size of the food set constrains how large $n_k$ can be. For $f=1$, $n_1 = 1$ and $P_1= \{ \mathbf A(2) \} $ as the only product one can make from a reaction in the food set is $\mathbf {A}(1) + \mathbf {A}(1) \rightarrow \mathbf {A}(2)$, and $n_2$ can only have values 1 and 2 as the only products one can produce in the second generation (from reactants in $R_2 = \{ \mathbf {A}(1),\mathbf {A}(2) \})$ are $\mathbf A(3)$ and $\mathbf A(4)$, etc. Similarly for $f=2$, $n_1$ can be only 1,2 or 3, as reactions among the two food set molecules $(1,0)$ and $(0,1)$ can only produce three molecules $(2,0)$, $(1,1)$ and $(0,2)$.

  \item [Algorithm 5:] {\bf Incremental, random construction of a sequence of reaction sets with partial overlap.} In this algorithm we decide on a sequence of increasing lengths, $L_1, L_2, \ldots, L_g$, and generate an extremal ACS, denoted $S_i$, of length $L_i$ using Algorithm 3 for each $i =1,2,\ldots,g$. The union of the $S_i$'s constitutes the set of reactions in the catalyzed chemistry.

      For each $i$ a different random number seed is used to initialize Algorithm 3. Changing the random number seed results in a different set of reactions defining the ACS.
\end{description}

\section{\label{section-ACS441}Dominance of an ACS of length 441 (ACS441)}
For $f=1$ we describe a cascade with $g=15$ and $n_1=1, n_2 = n_3 = \dots = n_{15} =2$. This catalyzed chemistry had 29 product molecules, the largest of which was $\mathbf A(441)$ having 441 monomers. The list of molecules and reactions is given in Appendix \ref{Appendix-ACS441-reaction-list}. The catalytic strength $\kappa$ of each molecule was chosen by an explicit length dependent rule
\begin{equation}
  \label{cascade-kappa-L-dependence1}
  \kappa(L) = K L^{\beta},
\end{equation}
where $K$ and $\beta$ are constants. We describe a simulation with $K=500$ and $\beta = 1.5$. This particular rule was chosen to contrast with Eq. (\ref{kappa2-exponential-L}) which characterizes the initiation threshold of an extremal ACS of length $L$. For a value of $L$ such as 441, the exponential function in Eq. (\ref{kappa2-exponential-L}) would have given an astronomically large catalytic strength\footnote{Extrapolating Eq. (\ref{kappa2-exponential-L}) with $\rho=0.64$ as given below that equation would give $\kappa = e^{0.64\times441} = 3.76 \times 10^{122}$. As mentioned in Section \ref{Exponential-Kappa}, $\rho$ depends upon other parameters, including $\phi$. The value $\rho=0.64$ given in Section \ref{Exponential-Kappa} is obtained for $\phi=1$ (see Fig. \ref{kappa1-2-vs-L}). In this section we discuss simulations with larger values of $\phi$ for which $\rho$ is large. Thus the corresponding values of $\kappa$ from Eq. (\ref{kappa2-exponential-L}) (for extremal ACSs) would actually be significantly larger than $3.76 \times 10^{122}$.}, whereas the much slower growing power law in Eq. (\ref{cascade-kappa-L-dependence1}) gives $\kappa(441) = 4.6 \times 10^6$ for the above mentioned values of the constants. Starting from the standard initial condition, the steady state concentration profile of this catalyzed chemistry embedded in a fully connected spontaneous chemistry with $N=800$ is shown in Fig. \ref{1d-algo4-a} and for sparsely connected spontaneous chemistry with $k=20$ and $2$ in Fig. \ref{1d-algo4-b} and \ref{1d-algo4-c}, respectively. As also seen for chemistries with a single ACS (Fig. \ref{acs-eg-sparse}, Section \ref{ACS-Sparse-chem}) the increase of sparseness causes the background concentrations to decline, while ACS molecules dominate more strongly.

\begin{figure}
    \begin{center}
        \subfloat[]{\label{1d-algo4-a}\includegraphics[height=4.75in,angle=-90,trim=0.5cm 0.5cm 0cm 0.3cm,clip=true]{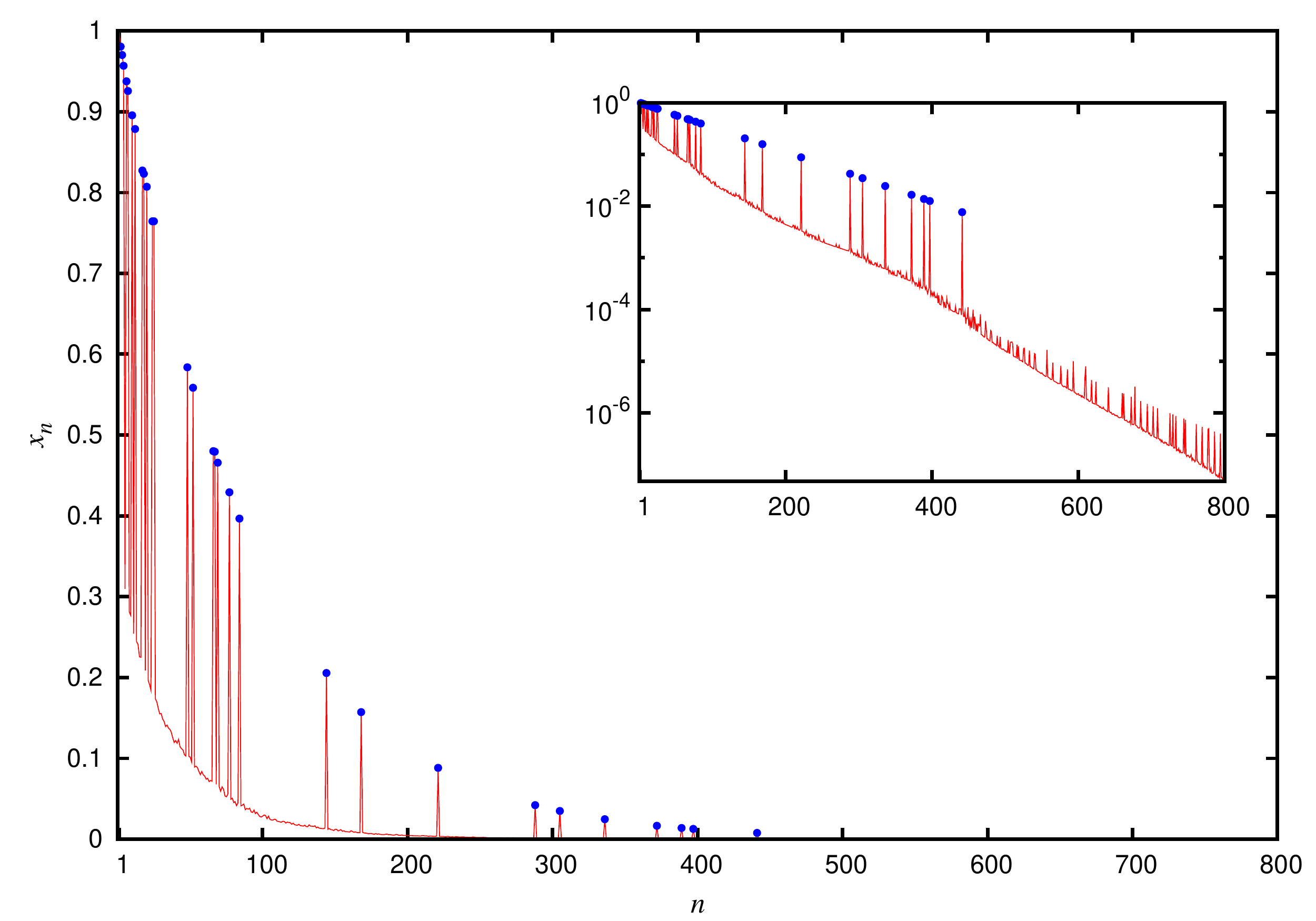}} \\
        \subfloat[]{\label{1d-algo4-b}\includegraphics[height=4.75in,angle=-90,trim=0.5cm 0.5cm 0cm 0.3cm,clip=true]{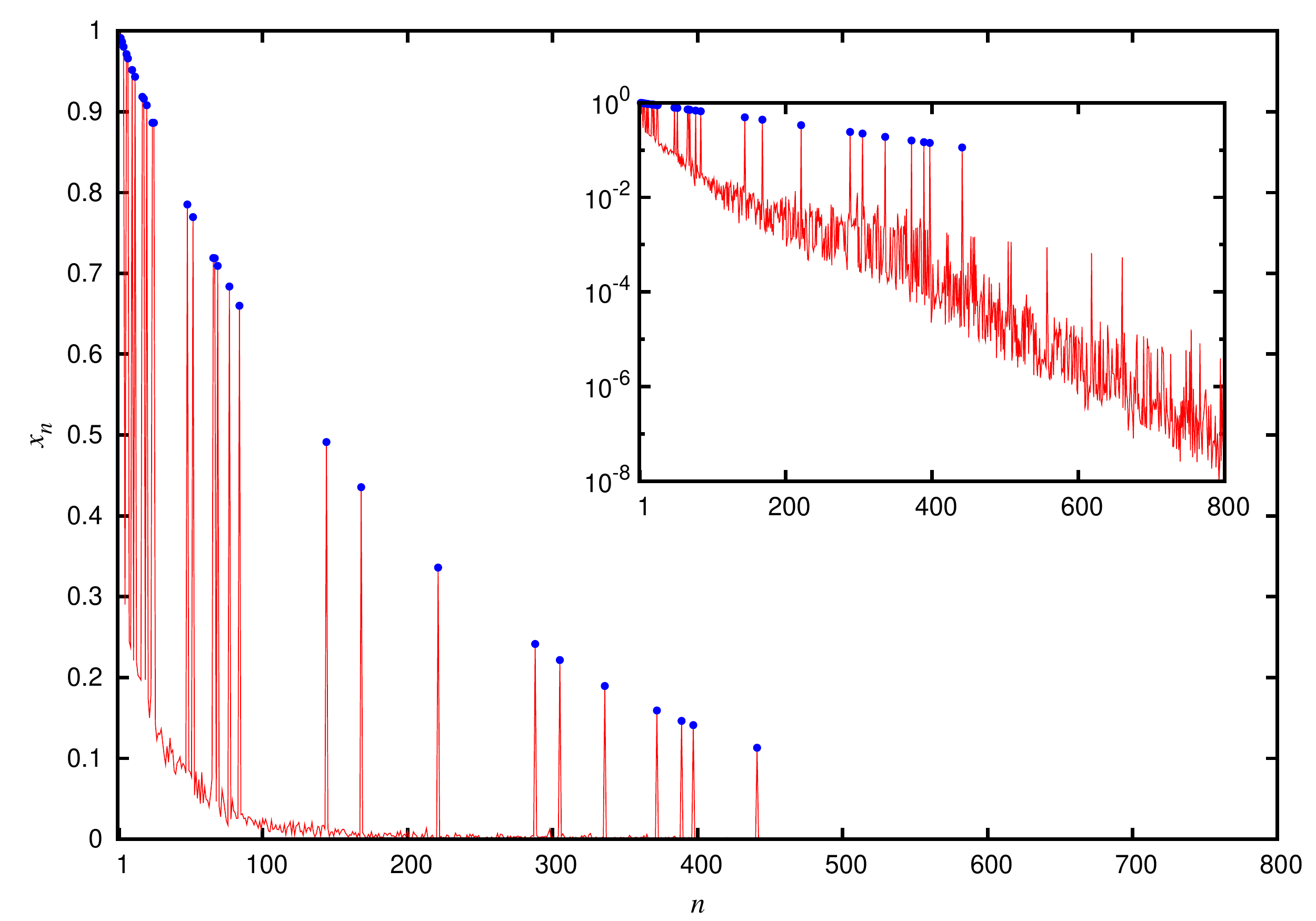}}
    \end{center}
    \caption[The dominance of a cascade of nested ACSs with a molecule of size 441 (ACS441).]{{\bf The dominance of a cascade of nested ACSs with a molecule of size 441 (ACS441).} {\it Continued.}}
\end{figure}

\captionsetup{list=no}
\begin{figure}
    \ContinuedFloat
    \begin{center}
        \subfloat[]{\label{1d-algo4-c}\includegraphics[height=4.75in,angle=-90,trim=0.5cm 0.5cm 0cm 0.3cm,clip=true]{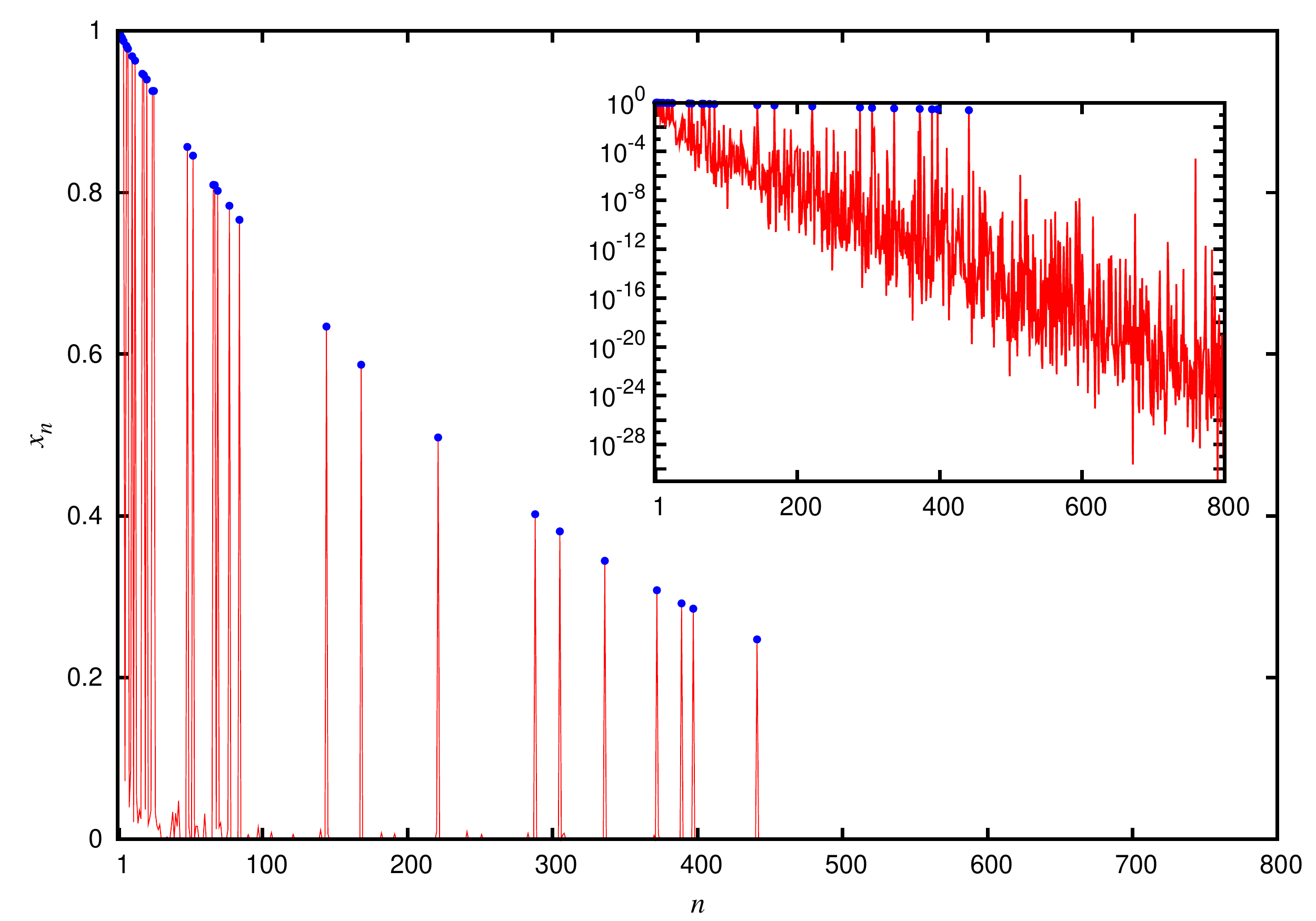}}
    \end{center}
    \caption[]{{\bf The dominance of a cascade of nested ACSs with a molecule of size 441 (ACS441).} The molecules and reactions of this ACS are listed in Appendix \ref{Appendix-ACS441-reaction-list}. The red curves show the steady state concentration $x_n$ of all the molecules as a function of their size $n$, starting from the standard initial condition; blue dots show the concentrations of the ACS molecules. Insets show the same on a semi-log plot. It is evident that the large ACS molecules acquire a significant concentration. The catalytic strengths of the ACS molecules depend upon their size $n$ according to $\kappa(n) = 500 \times n^{1.5}$, and for all cases $k_f = k_r = 1$, $\phi = 50$, $N = 800$. The three figures differ in the level of sparseness of the spontaneous chemistry in which the ACS is embedded. The spontaneous chemistry in {\bf (a)} is fully connected, in {\bf (b)} has degree 20, and in {\bf (c)} has degree 2.}
    \label{1d-algo4}
\end{figure}
\captionsetup{list=yes}

This example shows that with the nested ACS structure in the catalyzed chemistry, large catalyst molecules can acquire significant concentrations starting from an initial condition containing only the monomers, even when catalytic strengths grow quite slowly with the length of the catalyst. It is worth noting that product of the catalytic strength of $\mathbf A(441)$ and its steady state concentration ($x_{441} = 0.0077$) is about 36000, and this is the factor by which it speeds up the reactions it catalyzes over the spontaneous rate. In view of the fact that enzymes containing a few hundred amino acids speed up the reactions they catalyze within cells by factors of about $10^5$ and greater, the catalytic efficiency demanded of $\mathbf A(441)$ does not seem unreasonably high.

\begin{figure}
    \begin{center}
        \includegraphics[height=4.75in,angle=-90,trim=0.5cm 0.5cm 0cm 0.3cm,clip=true]{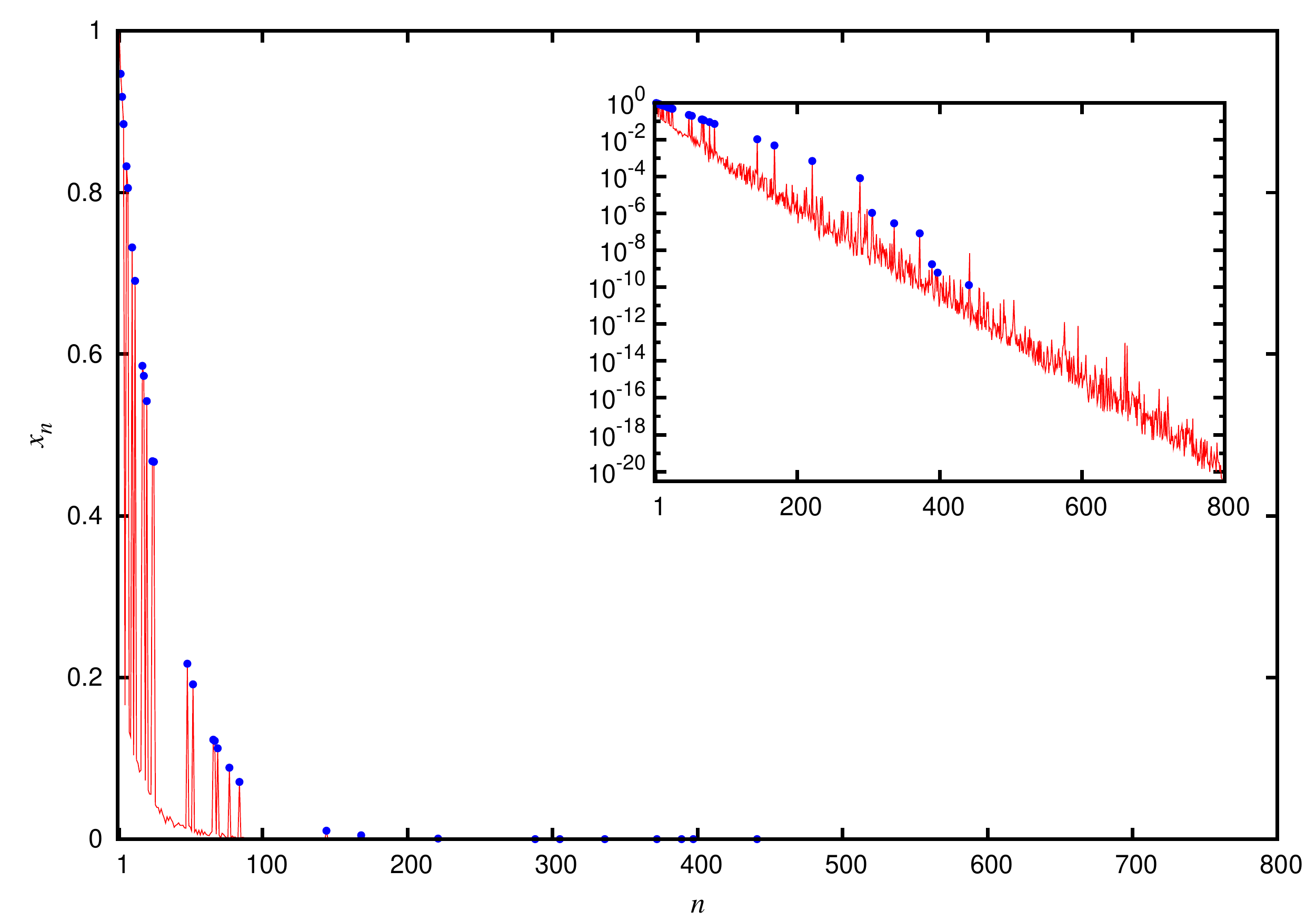}
    \end{center}
    \caption[Dominance of larger molecules in ACS441 trails off for $\beta=1.2$.]{{\bf Dominance of larger molecules in ACS441 trails off for $\beta=1.2$.} This figure is for the same parameter values as Fig. \ref{1d-algo4-b} but with $\kappa(n) = 500 \times n^{1.2}$. It can be seen that ACS molecules upto $\mathbf A(288)$ are able to dominate over the background (albeit with smaller concentrations than in Fig. \ref{1d-algo4-b}) while larger ones trail off and merge into background.}
    \label{1d-algo4-power1.2}
\end{figure}

Eq. (\ref{cascade-kappa-L-dependence1}), which gives a particular functional form for $\kappa(L)$, is ad-hoc, and, at this stage, merely an example given to quantify the level of catalytic strengths that is sufficient for large molecules to arise in appreciable concentrations in the prebiotic scenario under consideration if chemistry has the nested ACS structure of the kind discussed. One may ask if an even weaker requirement would suffice. We have considered smaller values of $\beta$ (1.2 and 1.0) keeping $K$ fixed, and found that ACS molecules upto a particular size (depending on $\beta$) do well but that the concentration of larger ACS molecules trails off and merges with the background (Fig. \ref{1d-algo4-power1.2}). The size range of ACS molecules that do well can be increased by increasing the coefficient $K$. Since the results depend upon several factors, including the topology of the ACS and the uncatalyzed chemistry, a detailed investigation has not been carried out.

\section{\label{section-cascade-f2}Cascading nested ACSs with $f=2$}
\begin{figure}
    \begin{center}
        \subfloat[]{\label{2d-algo4-a}\includegraphics[height=4.75in,angle=-90,trim=1.5cm 0.5cm 1cm 0.3cm,clip=true]{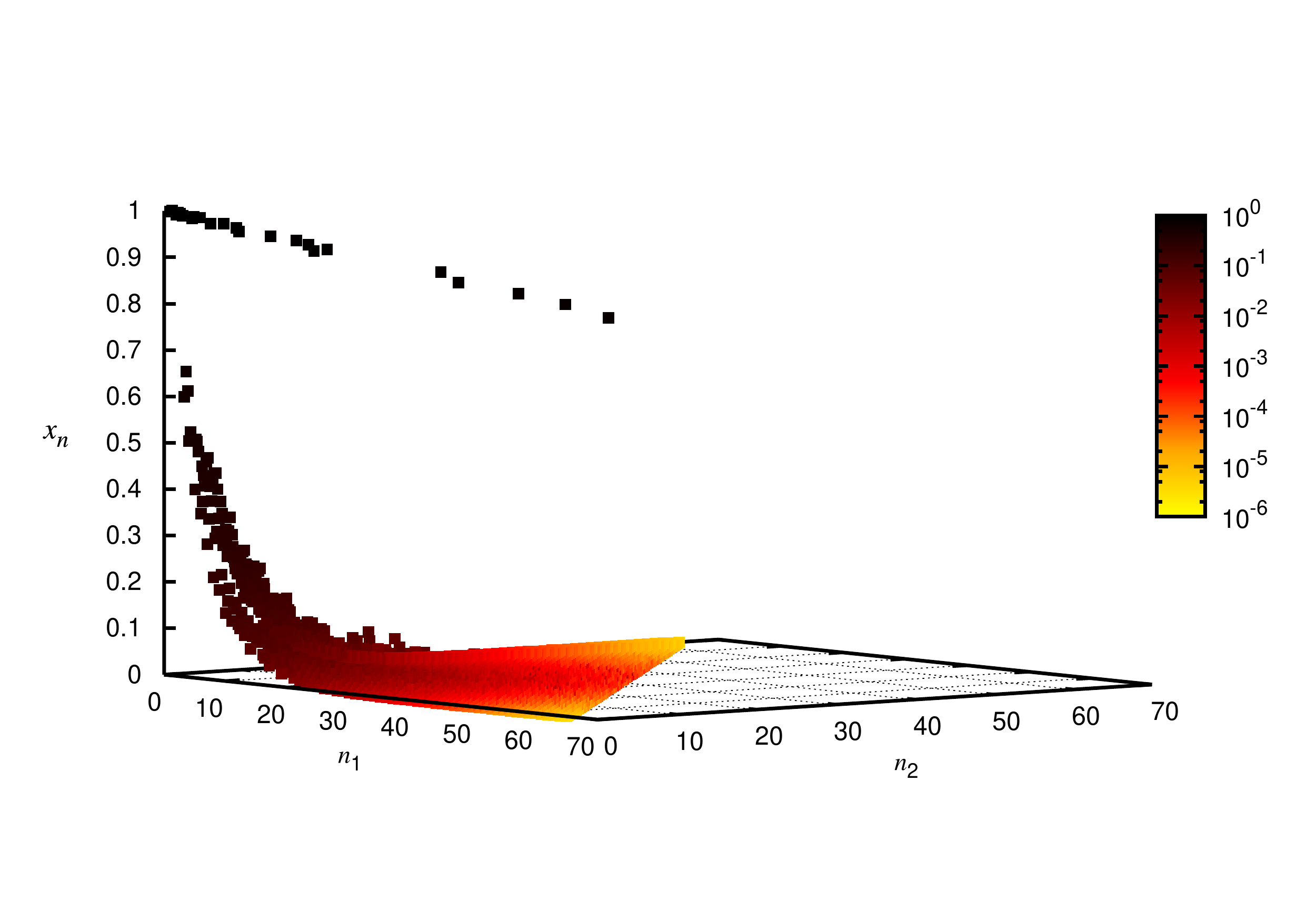}} \\
        \subfloat[]{\label{2d-algo4-b}\includegraphics[height=4.75in,angle=-90,trim=1cm 0.5cm 1cm 0.3cm,clip=true]{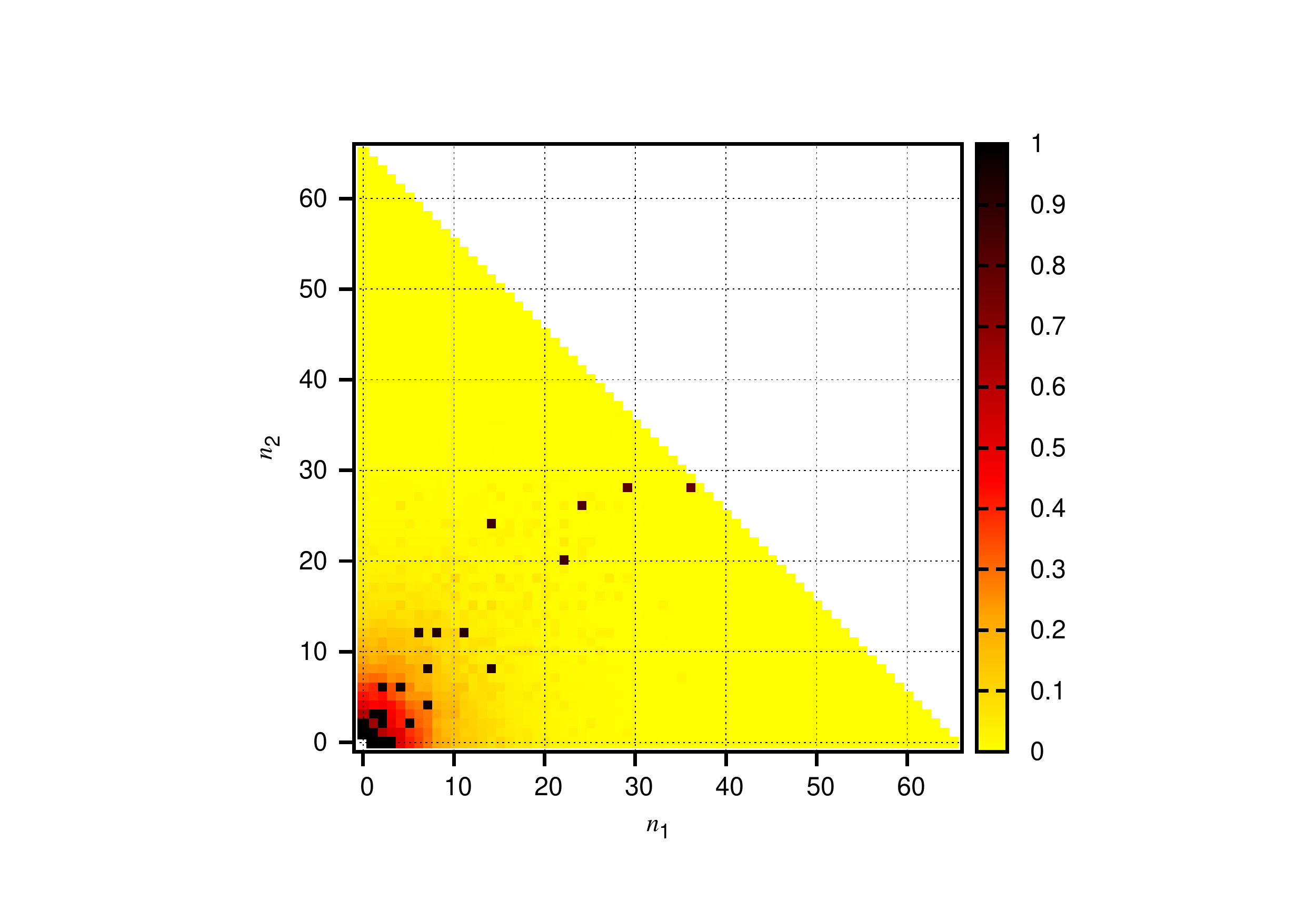}}
    \end{center}
    \caption[Dominance of a cascade of nested ACSs with length 64 (ACS(36,28)) in a $f=2$ chemistry.]{{\bf Dominance of a cascade of nested ACSs with length 64 (ACS(36,28)) in a $f=2$ chemistry.} {\bf (a)} 3D plot showing the steady state concentration $x_n$ of the molecule $n=(n_1,n_2)$ as a function of $n_1$ and $n_2$, starting from the standard initial condition. The colour coding is on a logarithmic scale of the concentration. {\bf (b)} A `top view' of the same so that the ACS molecules and background are more clearly distinguished. The colour coding here is on a linear scale of concentration. The food set and ACS molecules have the highest concentrations and stand out as black dots. The catalytic strengths of the ACS molecules depend upon their size $L \equiv n_1 + n_2$ according to $\kappa(L) = 500 \times L^{1.5}$, and $k_f = k_r = 1$, $\phi = 10$, $N = 65$. The spontaneous chemistry has degree 20.}
    \label{2d-algo4}
\end{figure}

An example of a nested ACS with two food sources, ACS(36,28), is presented in Fig. \ref{2d-algo4}. This is also generated by Algorithm 4 and has 7 generations with $n_k =3$ for each generation, the largest molecule being (36,28) (the full list of molecules and reactions is given in Appendix \ref{Appendix-ACS36-28-reaction-list}). Again starting from the standard initial condition the larger ACS molecules acquire appreciable concentrations with a moderate demand on their catalytic strengths.

As a final example we present in Fig. \ref{2d-partial-nest} a cascade of ACSs, named ACS(18,27) after its largest molecule, in which smaller ACSs have only a partial overlap with longer ones. This is generated using Algorithm 5 and consists of a series of 10 ACSs of increasing length. The detailed list of molecules, reactions and catalytic strengths is given in Appendix \ref{Appendix-ACS18-27-reaction-list}. At each step, we use the random number seed to control the degree of overlap produced between successive generations of ACSs. At each step $i$, the random number seed was so chosen that the catalytic strength required for ACS $S_i$ to dominate (given all the ACSs of previous generations) is not too large compared to the one for ACS $S_{i-1}$ and, at the same time, each successive ACS in the cascade has only a few reactions in common with other ACSs. Unlike in the examples discussed above, generated by Algorithm 4, in the present case molecules (except the small molecules) produced in the catalyzed chemistry have typically only one or two catalyzed ligation reactions producing them. The chemistry also contains a number of catalyzed `side reactions', which produce molecules that are neither catalysts nor reactants in any pathway leading to the largest molecule. In fact there is a `side pathway' consisting of several reactions that may be viewed as `draining the resources' of the main ACS. ACS dominance at moderate catalytic strengths occurs for this chemistry also. The largest $\kappa$ is 50000 for the molecule (18,27), and at a steady state population of 0.26 enhances the rate of a reaction by a factor of 13000 over the spontaneous rate. This shows that the mechanism outlined by us is not restricted to maximally overlapping nested ACSs but is more generic.

\begin{figure}
    \begin{center}
        \subfloat[]{\label{2d-algo3-a}\includegraphics[height=4.75in,angle=-90,trim=1.5cm 0.5cm 1cm 0.3cm,clip=true]{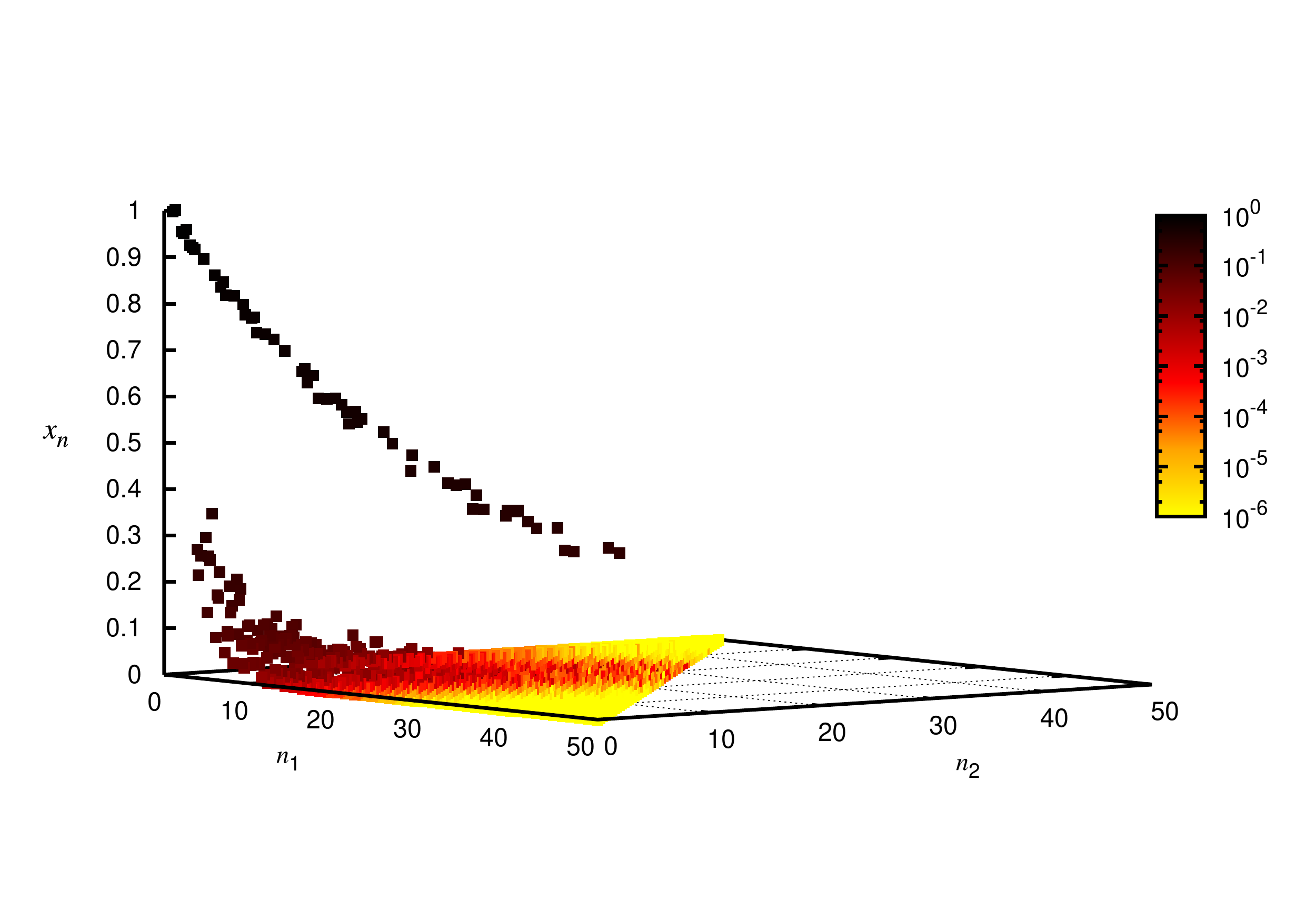}} \\
        \subfloat[]{\label{2d-algo3-b}\includegraphics[height=4.75in,angle=-90,trim=1cm 0.5cm 1cm 0.3cm,clip=true]{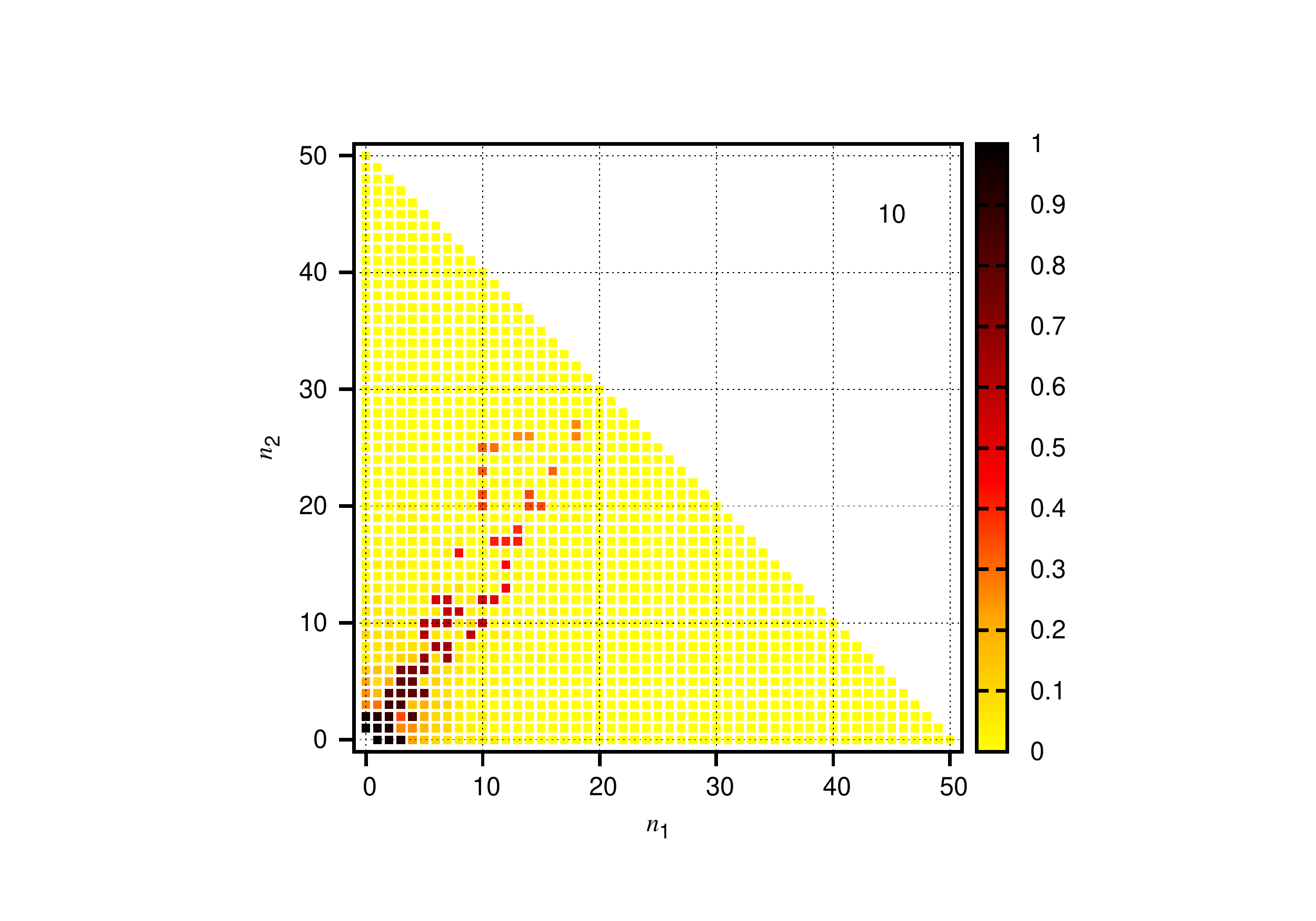}}
    \end{center}
    \caption[Dominance of a cascade of partially overlapping nested ACSs in a $f=2$ chemistry (ACS(18,27)).]{{\bf Dominance of a cascade of partially overlapping nested ACSs (ACS(18,27)).} {\it Continued.}}
\end{figure}

\captionsetup{list=no}
\begin{figure}
    \ContinuedFloat
    \begin{center}
        \subfloat[]{\label{2d-algo3-c}\includegraphics[height=6in,angle=-90,trim=1.0cm 0.0cm 0.0cm 4cm,clip=true]{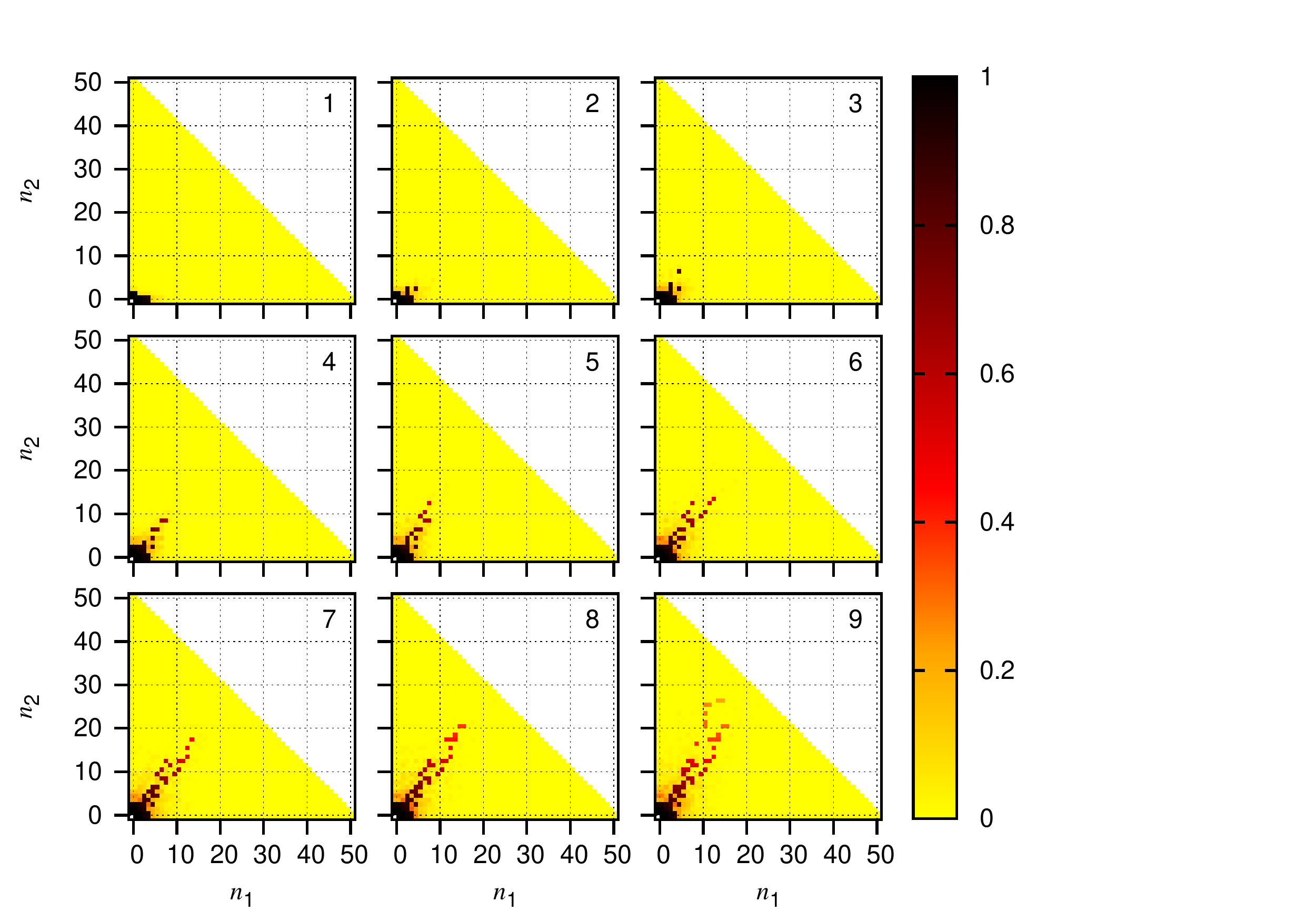}}
    \end{center}
    \caption{{\bf Dominance of a cascade of partially overlapping nested ACSs (ACS(18,27)).} {\bf (a)} Steady state concentration profile starting from the standard initial condition. {\bf (b)} Top view of the same. {\bf (c)} Sequence of steady state concentration profiles as each successive ACS is added to the chemistry. The legend for (a) is the same as for Fig. \ref{2d-algo4-a} and for (b) and (c) the same as Fig. \ref{2d-algo4-b}. $k_f = k_r = 1$, $\phi = 15$, $N = 50$, and the spontaneous chemistry has degree 5.}
    \label{2d-partial-nest}
\end{figure}
\captionsetup{list=yes}

\thispagestyle{plain}
\cleardoublepage
\chapter{\label{ChapterDiscussion}Discussion and future outlook}

\section{Summary}
To summarize, we have considered the population dynamics in a well stirred chemical reactor of an artificial chemistry in which large molecules can in principle be produced by successive ligations of pairs of smaller molecules, and where the populations of the smallest molecules, the `food set' molecules or monomers, are buffered. The chemistry contains a large number of spontaneous reactions of which a small subset could be catalyzed by molecules produced in the chemistry. We have particularly considered the case where the set of catalyzed reactions contains autocatalytic sets. The main questions this thesis has been concerned with are: ({\it a}) under what circumstances do molecules belonging to the ACSs dominate in concentration over the rest of molecules in the chemistry (the background), and ({\it b}) starting from an initial condition that does not contain good catalysts, does the dynamics of such a chemical network permit a sparse set of large molecules (containing several tens or a few hundred monomers) that are good catalysts to arise and be maintained in the reactor at concentrations significantly above the background?

Under ({\it a}) we confirm earlier results (Bagley and Farmer \cite{Bagley1989, Bagley1991}) that the mere existence of ACSs in the chemistry at the level of network topology is not sufficient to guarantee their domination at the population level, and extend them to quantitatively characterize the interplay between ACS topology and rate constants including catalytic strengths, dissipation rate, etc., that results in ACS domination. In the absence of catalyzed reactions, and even the presence of an ACS if the catalytic strength $\kappa$ of catalysts is not large enough, the steady state concentration profile in such a reactor shows an exponential decline with the length of the molecule (see, e.g., Figs. \ref{noacs}, \ref{noacs-2d}, \ref{acs-eg-ssc} and Eqs. (\ref{ss1}), (\ref{sszerophi})). In order for ACS molecules to shine above the background, the catalytic strength needs to be sufficiently large. We show that a high dissipation rate $\phi$ together with a large catalytic strength contributes to ACS domination, the former contributing to keeping the background low and the latter to preferentially enhancing the ACS molecule concentrations (Fig. \ref{acs-eg} and the discussion below Eqs. \ref{kappainftyrateeqns}). We show that ACSs often produce a bistability in the dynamics (within a range of catalytic strength $\kappa^I \leq \kappa \leq \kappa^{II}$) with ACS molecule concentrations being appreciable and much higher than the background in one of the two stable fixed points (the `high' fixed point), and being low and comparable to the background in the `low' fixed point (Figs. \ref{acs-eg}, \ref{acs-eg-bis}, \ref{acs4}). Bistability provides a crisp criterion for ACS domination: If the reactor initially contains only the monomers -- the relevant condition for the origin of life problem under consideration (or more generally, if the initial conditions are in the basin of attraction of the low fixed point), ACS molecules will dominate in the steady state only if $\kappa > \kappa^{II}$. If the initial conditions are in the basin of attraction of the high fixed point (and this means that the concentration of the catalysts is already high), ACS molecules will dominate when $\kappa \geq \kappa^I$. We also show that for extremal ACSs (in which the largest molecule is the catalyst of all the catalyzed reactions) the `maintenance threshold' $\kappa^I$ grows as a power of the size of the catalyst, while the `initiation threshold' $\kappa^{II}$ grows exponentially with the size of the catalyst (Fig. \ref{kappa1-2-vs-L}, Eq. (\ref{kappa2-exponential-L})). This extends the results of Ohtsuki and Nowak \cite{Ohtsuki2009} to a much larger class of models. Since $\kappa^{II}$ characterizes the lower limit of catalytic strength needed to get to the `high' fixed point starting from an initial condition containing only small molecules, the last result quantifies the difficulty for large molecules to dominate in a prebiotic scenario via the ACS mechanism, namely, the requirement of an unreasonably large catalytic strength.

Under ({\it b}) we present a possible resolution to this problem via nested ACSs. We show that if an ACS catalyzed by large molecules contains within it (or partially overlaps with) a smaller ACS catalyzed by smaller molecules (we refer to this as a `nested ACS' structure), the catalytic strength required for the large ACS to dominate comes down drastically (Fig. \ref{nested-3-8} and the associated discussion). Effectively the small ACS reinforces the larger one. We argue that this reinforcement (as measured by a reduced $\kappa^{II}$ of the large ACS) has its roots in the population upliftment that the small ACS provides to its background which includes the large catalyst. One can construct a cascade of nested ACSs with each successive and larger ACS in the cascade lying in the background of the smaller ACSs, thereby receiving reinforcement from them and in turn providing reinforcement to the next larger ACS in the cascade. We use this mechanism to construct several examples of cascaded nested ACSs with moderate catalytic strengths in which large molecules (containing upto several hundred monomers) are found to dominate significantly over the background starting from only monomers in the initial state(Figs. \ref{1d-algo4}, \ref{2d-algo4}, \ref{2d-partial-nest}. We show that even a $\kappa$ growing as slowly as a power of the catalyst size (Eq. (\ref{cascade-kappa-L-dependence1})) is enough to get large molecule domination through the mechanism of nested ACSs. This is a significant improvement over the exponentially growing requirement for $\kappa$ in part ({\it a}). We construct several examples including one in which a large catalyst (with close to 450 monomers) arises with a dominant population in the steady state starting from initial conditions containing only the monomers. This catalyst has a reasonable catalytic strength compared to enzyme catalysts (in the steady state it enhances the rates of reactions that it catalyzes by a factor of less than $10^5$ over the spontaneous rate).

Thus our work presents a possible resolution to the chicken and egg problem in chemistry posited by the existence of large molecules. In the context of the present model the appearance of large molecules is a natural dynamical consequence of chemistry possessing the structure described above -- a cascading nested ACS structure (with a not very demanding set of catalytic strengths) embedded in a spontaneous chemistry -- together with the buffered presence of the food set molecules in a well stirred region of space. The mechanism is an incremental one: to begin we require the small molecules that are built by spontaneous reactions from the food set to have weak catalytic properties, but after that we only require molecular properties (\ie, catalytic strength) that are an extension of existing molecular properties; at no point do we require a molecular property that is a radical departure from what already exists. Through the mechanism of sequential population support to successively larger (and stronger) catalysts from smaller (and weaker) ones, and the nonlinearity and feedback inherent in autocatalytic systems, the large catalysts effectively bootstrap themselves into existence.

\section{Possible applications of the model to prebiotic scenarios}
The kind of mathematical model we have studied, inspired by the work of Bagley and Farmer, is quite abstract; its virtue is the economy of assumptions that go into its structure. The main ingredients are that objects can combine with each other in processes or `reactions' to form other objects and certain objects can facilitate certain processes, \ie, `catalyze' certain reactions. The population dynamics implements a simple scheme for how the abundances of the objects would change with time assuming that the probability of objects combining is proportional to their abundances. Such a generic scheme while it applies in detail to no particular situation allows us to imagine mechanisms at a conceptual level. It is significant that in this scheme an ACS can direct the flows towards itself and cause a certain sparse subset of objects, including some specific large composite ones, to capture a large fraction of the chemical resources.

At this level of abstraction the model (or a variant with qualitatively similar features) could apply to the peptide chemistry as well as an RNA chemistry and to a prebiotic metabolism, as already noted by Bagley and Farmer. Indeed it would be equally applicable if a prebiotic environment actually had a mixture of ingredients from all these classes of chemistries, a possibility that has been advocated in, e.g., \cite{Copley2007, Powner2011}. Copley, Smith and Morowitz \cite{Copley2007} have proposed a scenario which seeks to explain how the RNA world might have originated through a series of incremental steps starting from a primitive metabolism. The food sources for this supposed metabolism are $\mathrm{CO}_2$, $\mathrm{H}_2$, $\mathrm{H}_2\mathrm{S}$, $\mathrm{NH}_3$, etc., in a hydrothermal vent. Their scenario envisages multiple stages of increasing complexity which they refer to as ({\it i}) the monomer stage, in which metabolism, possibly powered by an autocatalytic set such as the reverse TCA cycle, produces nucleotides and simple amino acids, ({\it ii}) the multimer stage, which produces dimers and small cofactors, ({\it iii}) the micro-RNA stage, producing of oligonucleotides of length 3-10, ({\it iv}) the mini-RNA stage, with 11-40mers, followed by ({\it v}) the macro-RNA stage, or the RNA world. In their scenario each successive stage produced better catalysts that collectively catalyzed not only their own production from the molecules of the previous stage, but also the reactions of the lower stages. This structure is very similar to the cascade of nested ACSs that we have discussed. A suitably modified version of our model could be constructed to explore the dynamics of this scenario in more detail. At a general level, in the fact that the dynamics of our model results in the stable domination, or concentrating, of the large catalysts, our mathematical work perhaps lends support to the workability of such a scenario.

There is another level at which the present model (or its variants) might talk to prebiotic chemistry. Morowitz \cite{Morowitz1999} has suggested that the metabolic network itself has a shell like structure which can convert simple molecules like $\mathrm{CO}_2$, $\mathrm{H}_2$, $\mathrm{NH}_3$, etc., through ``a hierarchy of nested reaction networks involving increasing complexity" into purines, pyrimidines, complex cofactors, etc. Reaction sets created by our Algorithms 4 and 5 are reminiscent of this picture. Missing from Morowitz's picture is a catalyst assignment for each reaction from among the molecules in the various shells or from among other catalysts accessible prebiotically (e.g., surfaces in hydrothermal vents). It might be worthwhile to attempt to add in that information for a more complete scenario and for potential modeling.

\section{Caveats and future directions}
\begin{enumerate}
  \item We have studied the properties of autocatalytic systems, by choosing specific examples of ACS topology and special algorithms for constructing them. This has allowed us to systematically investigate the dynamical properties that ACSs offer. We believe that similar dynamical properties would hold for more general topological structures than we have considered. Nevertheless the question arises as to whether all these structures are very special structures and whether or not they are likely to arise within real chemistry and `generic' artificial chemistries. ACSs have been shown to be quite generic in a large class of randomly constructed artificial chemistries \cite{Steel2000, Hordijk2011}, and a similar analysis could be extended to cascading nested ACSs. This would help parametrize or characterize chemistries that would contain such structures and those that would not. In this context it would useful to go beyond the simple case we have considered in which a molecule is defined by the number of monomers of each type and not their sequence. It may also be interesting to look for structures similar to nested ACSs in real metabolic networks using methods similar to those in \cite{Kun2008}.
  \item An important related question is one of side reactions (discussed in \cite{Orgel2000, Orgel2008}) which might destroy the efficacy of ACSs. In the real chemistry one expects that even if cascading nested ACSs exist, there would also exist other catalyzed reactions channelizing the ACS products into pathways leading in other directions. Whether substantial populations of large molecules in the nested ACSs can be maintained in the presence of such diversion is a question that remains to be systematically investigated. Our last example of cascading nested ACSs in fact has several side reactions and it may be noted that large ACSs still dominated in that case. We remark that while side reactions can drain resources from ACSs, they also help the system to explore new directions in chemical space in an evolutionary scenario.
  \item We have considered deterministic dynamics in this paper. Stochastic fluctuations are important when molecular populations are small and can have interesting effects on ACS dynamics \cite{Togashi2003, Togashi2005, Awazu2007, Filisetti2011}. For a chemistry containing multiple ACSs, Bagley, Farmer and Fontana \cite{Bagley1991a} used stochasticity to produce examples of trajectories that differed from each other in the sequence of ACSs that came to dominate the reactor. It would be interesting to explore such effects in the context of the present model.
  \item Another simplification we have made is that of homogeneous chemistries, wherein the rate constants of all spontaneous reactions have been taken to be the same, and even catalytic strengths, where variable, have been taken to be smooth functions of the length. We have also shown that introducing a small amount of heterogeneity or randomness in the rate constants does not change the qualitative behaviour significantly. However, the effect of cranking up the heterogeneity has not been studied. From studies of disordered systems in statistical mechanics and condensed matter physics it has become clear that such heterogeneity can lead to rugged landscapes, multiple attractors and timescales, and paths that are difficult to locate \cite{Wales2004}. The dynamics of such systems when they are driven by a non-equilibrium flux or buffering of food set molecules is an open question. It is possible that the constraints placed by the ruggedness of the landscape will reduce the number of accessible paths. The populating of molecules at different levels in a nested hierarchy of ACSs is likely to happen in fits and starts on multiple timescales when heterogeneity is accounted for. It is perhaps in such a scenario that one should look for answers to the questions raised under (1), (2) and (3) above.
  \item The present model describes a population dynamics in a fixed environment. However the environment can change over a longer time scale leading to a complex evolutionary dynamics \cite{Jain1998, Jain2001}. We remark that in the context of the present model, the exogenous supply or depletion of certain molecules may affect the dynamics and allow the system to explore other steady states. For example, the system in the lower attractor of the bistable region may get elevated to the higher attractor if the exogenously supplied molecules can move the system from the basin of the lower attractor to that of the higher attractor. A sudden flux out of the system, may deplete the system of some molecules; this may cause a transition from the higher to the lower attractor. It may be interesting to study the effect of exogenously supplied molecules as well as of the sudden removal of molecules from the system and eventually incorporate an evolutionary scenario in which a more realistic setting for the origin of life could be investigated.
  \item The model we have discussed has no spatial structures, \ie, the dynamics of the system happen in a homogeneous well stirred region of a prebiotic niche. One of the important steps in the origin of life was the appearance of enclosures. One can consider a scenario in which one or more of the molecules produced in our chemistry can spontaneously aggregate to form vesicles thereby providing an enclosure to the ACS dynamics. Enclosures such as vesicles can selectively allow exchange of molecules with the environment, thereby allowing the system to access new steady states. An important body of work in this respect is the one that deals with the growth and evolution of protocells \cite{Schwegler1985, Segre2000, Furusawa2006, Shenhav2007, Macia2007, Szathmary2007, Sole2009, Zhu2009}. In our model we have also excluded self replicating molecules. One can also think of scenarios in which one of the molecules produced in the chemistry has the property of template replication. While systems of replicators including those where the replicator is catalyzed have been extensively studied \cite{Stadler1991a, Stadler2000, Szathmary2006, Cornish-Bowden2008, Ohtsuki2009}. The dynamics of systems with replication reactions as well as catalyzed non-replication reactions may be interesting in its own right. Further these systems may exist in the presence of enclosures. It will be interesting to integrate these different approaches.
\end{enumerate}

\appendix
\renewcommand\chaptername{Appendix}
\thispagestyle{plain}
\cleardoublepage
\chapter{\label{Appendix-N-independence}$N$-independence of results at large $N$}
While the chemistry considered in this model is an infinite one, numerical exploration requires us to work with a finite set of molecules that participate in the chemistry. This introduces an additional parameter $N$, the size of the largest molecule produced in the chemistry, in the model. Here we present evidence showing that certain important properties of the model become $N$-independent at large enough $N$. Specifically we show that the steady state profile of concentrations become independent for $N$ for both uncatalyzed chemistries (discussed in Chapter \ref{ChapterSpontaneousChemistry}) and chemistries with ACS (discussed in Chapter \ref{ChapterACS}).

\section{Uncatalyzed chemistry}

\begin{figure}[ht]
    \begin{center}
        \includegraphics[width=3.5in,angle=-90,trim=0.1cm 0.5cm 0cm 0cm,clip=true]{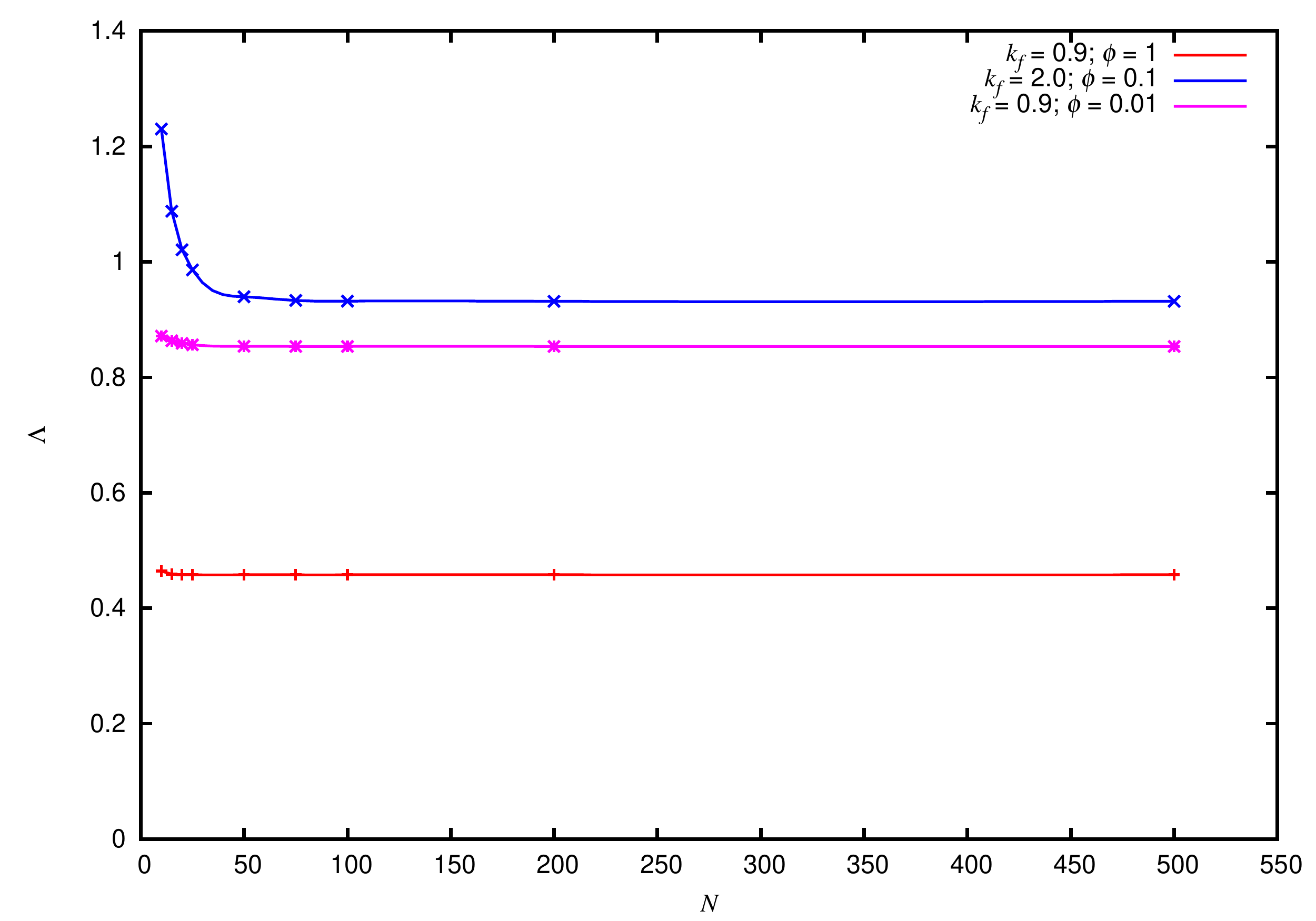}
    \end{center}
    \caption[Dependence of $\Lambda$ on $N$ for uncatalyzed chemistries.]{{\bf Dependence of $\Lambda$ on $N$ for uncatalyzed chemistries.} $A=k_r=1$. We determine $\Lambda$ at different values of $N$ by fitting the steady state profiles to an exponential, $x_n = ce^{-\gamma n}$ (excluding the concentrations $x_1$ to $x_4$). We see a dependence of $\Lambda$ on $N$ for values of $N<50$, but $\Lambda$ becomes essentially independent of $N$ for $N>100$.}
    \label{Nvslamda}
\end{figure}

The steady state concentrations for an uncatalyzed chemistry can be approximated by an exponential function, $x_n = ce^{-\gamma n} = c\Lambda^n$, where $c$ and $\Lambda=e^{-\gamma}$ are constants (see Eq. (\ref{ss1})). We calculate $\Lambda$ from the slope of the straight line fit on the plot of the log of the steady state concentrations, $x_n$, versus length, $n$. We found that for sufficiently large $N$, $\Lambda$ becomes independent of $N$; see Fig. \ref{Nvslamda}.

When $\Lambda<1$, the concentrations of the large molecules are small. It is evident that if the concentrations of the large molecules produced in the chemistry are so small that their contributions to the dynamics of the (relatively) smaller molecules is negligible, then, including molecules larger than those already considered in such a chemistry would cause no difference to the results.

\section{Catalyzed chemistry}

\begin{figure}[ht]
    \begin{center}
        \includegraphics[width=3.5in,angle=-90,trim=0.1cm 0.5cm 0cm 0cm,clip=true]{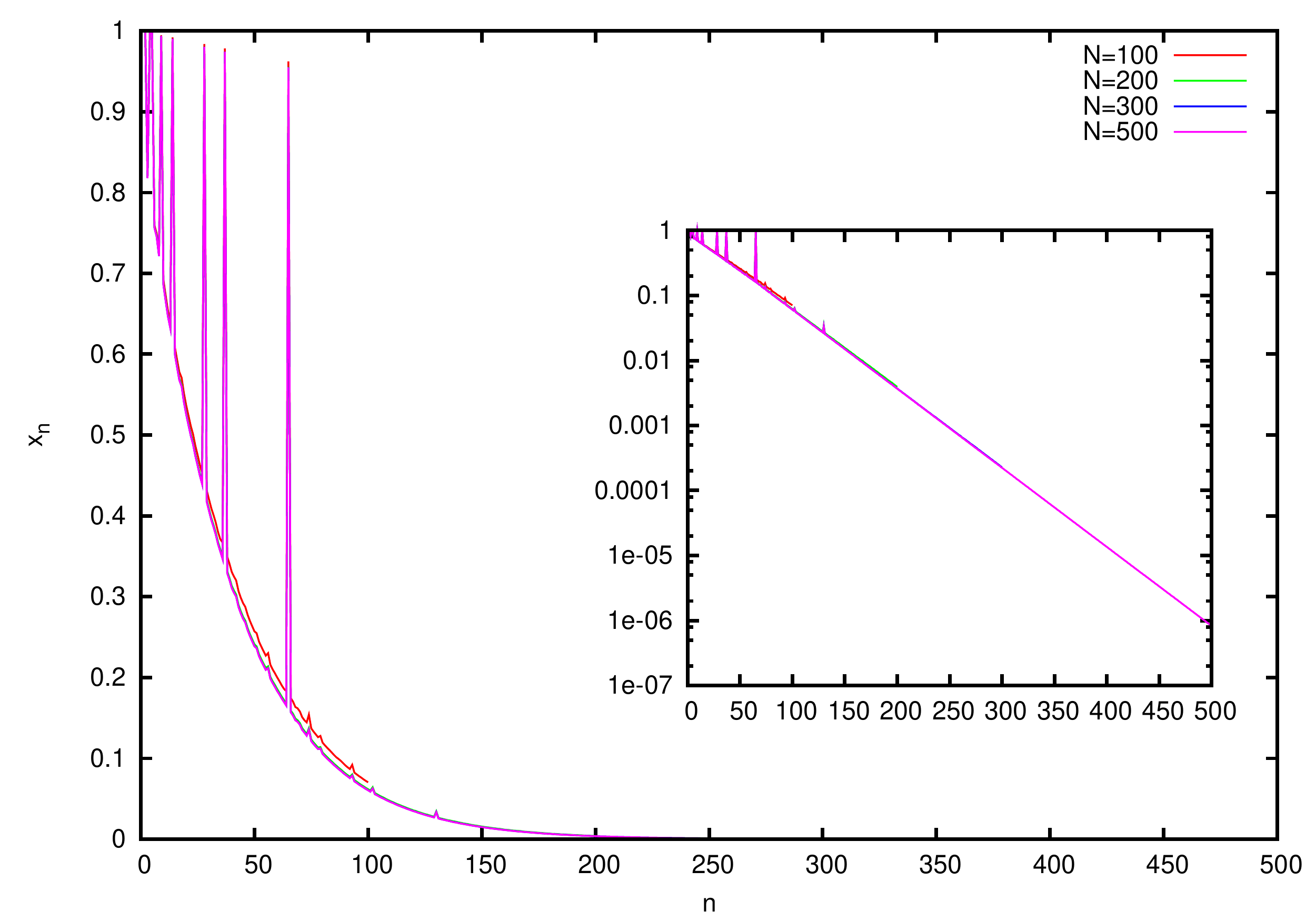}
    \end{center}
    \caption[Dependence of steady state concentrations on $N$ for a chemistry that includes an ACS.]{{\bf Dependence of steady state concentrations on $N$ for a chemistry that includes an ACS.} The figure shows the steady state concentrations for the chemistry that includes ACS65 (Eq. (\ref{acs65-definition})) for $N=100, 200, 300, 500$. In all cases $A = k_f = k_r = 1, \phi=5$. The steady state concentrations show an $N$-dependence upto $N=100$, but for $N\ge200$ the profile becomes $N$-independent.}
    \label{N-depend-acs}
\end{figure}

For chemistries with catalyzed reactions, including ACSs, we find that for sufficiently large $N$, the numerical results are independent of $N$. In Fig. \ref{N-depend-acs} we show how the steady state concentrations in the chemistry that includes ACS65 (defined by Eq. (\ref{acs65-definition})) depend upon $N$.

The results for the chemistry with $f=2$ were also found to be $N$-independent (for sufficiently large values of $N$). We find that for $f=2$ the $N$-independence is reached at much smaller values of $N$ than for $f=1$.

\thispagestyle{plain}
\cleardoublepage
\chapter{\label{Appendix-Bagley}Model run at fixed input rate of monomer instead of buffered concentration}

In Section \ref{Comparison-Bagley} we have discussed the differences between our model and that studied by Bagley and Farmer \cite{Farmer1986, Bagley1989, Bagley1991}. One of the differences is that of the boundary condition. In our model, we have considered the food set molecules to be buffered, \ie, their concentrations remain constant in time. Whereas, Bagley and Farmer consider a constant influx of the monomers; they consider the system as a chemical reactor in a which a fixed supply of monomers enter per unit time and a dilution flux takes away the products of the system. In this Appendix we discuss our model with a fixed input rate of monomers instead of buffered concentrations.

Consider the case of a fully connected and homogenous spontaneous chemistry with $f=1$. The rate equations for $n=2,3,\ldots,N$ (for $n>N/2$ the third term is absent) are given by (see Eq. (\ref{rateequation-N-homogeneous})),
\begin{align}
    \label{bagley-system}
    \nonumber
    \dot x_n = & \sum_{i \leq j,i+j=n}\left(k_fx_ix_j - k_rx_n\right) - \sum_{i=1,i \neq n}^{(N-n)} \left(k_fx_ix_n - k_rx_{i+n}\right) \\
              &- 2\left(k_fx_n^2 - k_rx_{2n}\right) - \phi x_n.
\end{align}
And for the monomer,
\begin{align}
    \label{bagley-system-monomer}
    \dot x_1 = \delta - \sum_{i=2}^{(N-1)} \left(k_fx_ix_1 - k_rx_{i+1}\right) - 2\left(k_fx_1^2 - k_rx_{2}\right) - \phi x_1,
\end{align}
where, $\delta$, is the fixed rate at which the monomer enters the system.

If $m_1$ is the molecular weight in atomic mass units (amu) of the monomer molecules and $x_n$ the concentration of $\mathbf A(n)$ in moles per unit volume, the total mass of the system is given by $M = \sum_{n=1}^N m_1nx_n$. All the terms in Eqs. (\ref{bagley-system}) and (\ref{bagley-system-monomer}) conserve mass except $\delta$, that corresponds to addition of new monomers to the system, and terms with $\phi$, which models the loss of monomers (as monomers as well as larger molecules) from the system. Therefore, the rate of change of total mass is given by,
\begin{align}
    \dot M = \delta m_1 - \phi M.
\end{align}
At steady state, the total mass of the system goes to a fixed point, $M = \frac{\delta m_1}{\phi}$. Thus fixing $\delta$ and $\phi$ fixes the steady state value of $M$.

\begin{figure}
    \begin{center}
        \includegraphics[height=4.75in,angle=-90,trim=0.25cm 0.5cm 0cm 0.3cm,clip=true]{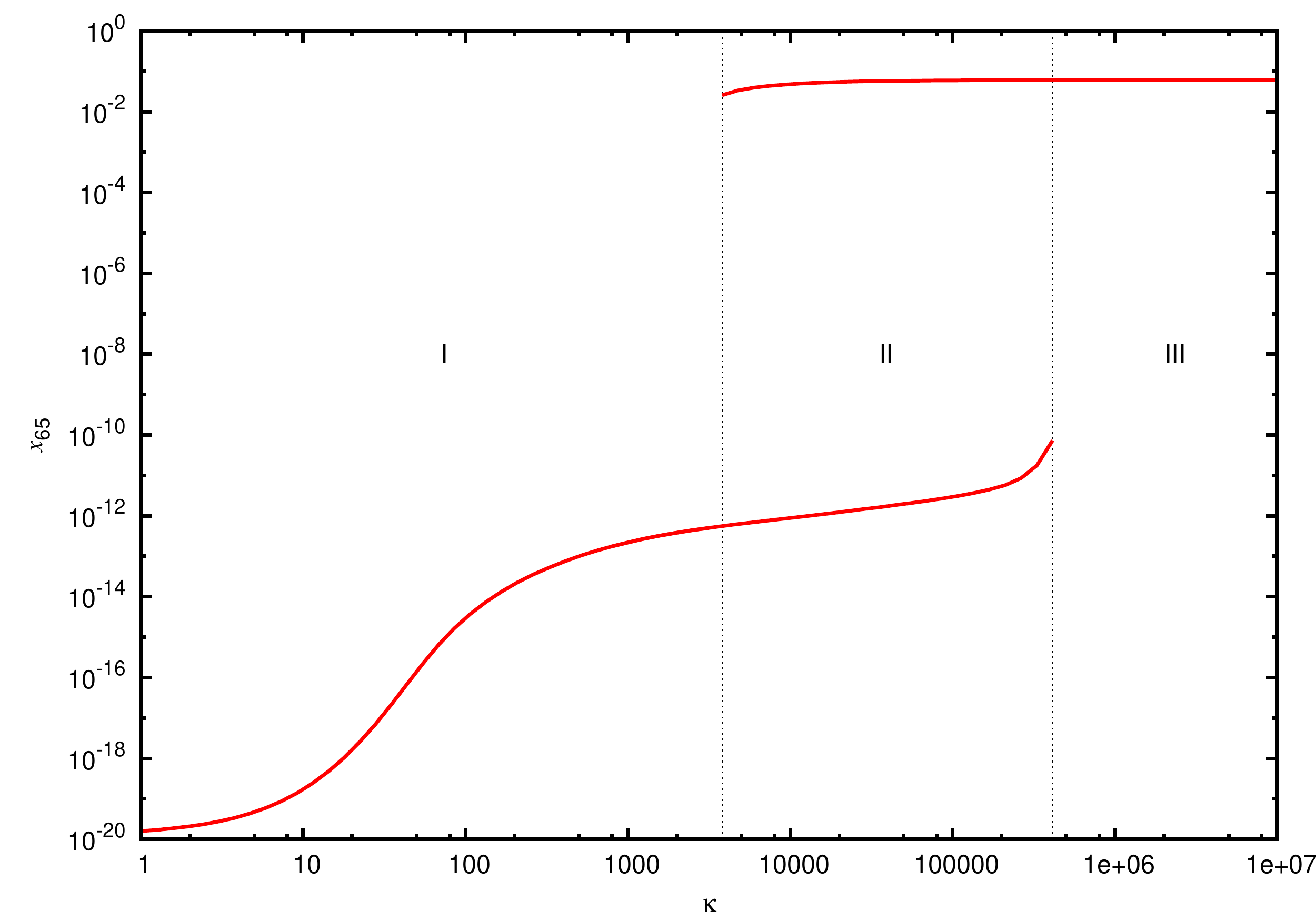}
    \end{center}
    \caption[Bistability in the dynamics of ACS65 for a fixed input rate of monomers.]{{\bf Bistability in the dynamics of ACS65 for a fixed input rate of monomers.} The steady state concentration of $\mathbf A(65)$ versus $\kappa$ for $k_f = k_r = 1, \delta = 5000, \phi=100$. The steady state mass in the system is $M=50$. This figure was generated in same way as Fig. \ref{acs-eg-bis} starting from two different initial conditions, but with a fixed input rate instead of buffered concentration of the monomer (\ie, using Eq. (\ref{bagley-system-monomer}) for the monomer concentration instead of $x_1=1$. In region I, both initial conditions lead to the same attractor in which the mass is largely accumulated in small molecules. Also in region III, the system has a single attractor in which the ACS molecules dominate over the background. The system exhibits bistability, \ie, it has two possible attractors, in region II. See Fig. \ref{ssc-fixedinput} for the steady state concentration profiles of the system in the two attractors.}
    \label{bistab-fixedinput}
\end{figure}

\begin{figure}
    \begin{center}
        \includegraphics[height=4.75in,angle=-90,trim=0.25cm 0.5cm 0cm 0.3cm,clip=true]{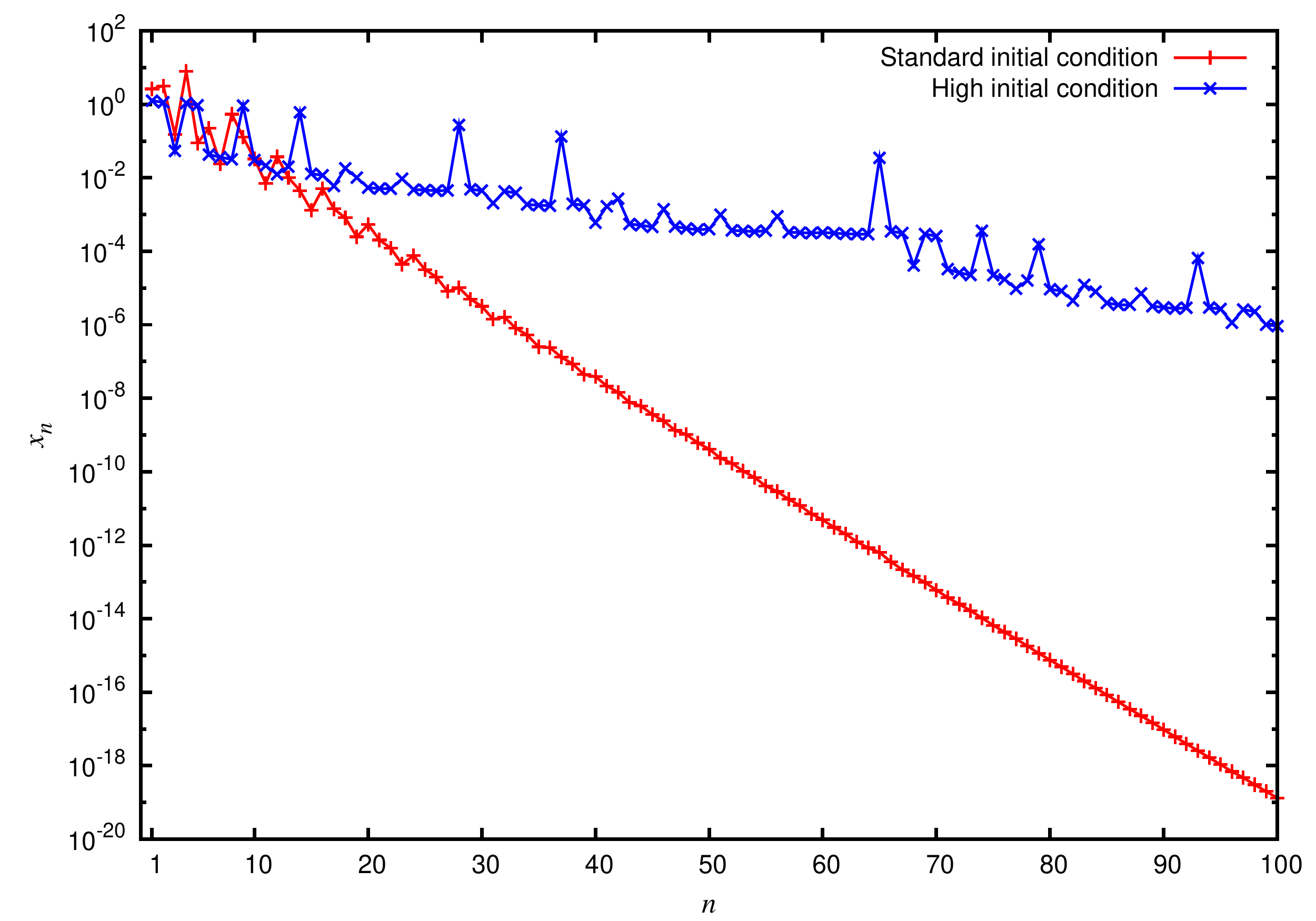}
    \end{center}
    \caption[Steady state concentration profile for the system with fixed input rate of monomers.]{{\bf Steady state concentration profile for the system with fixed input rate of monomers.} The steady state concentration $x_n$ versus the length of molecule $n$ is plotted for the system with fixed input rate of monomers, starting from two initial conditions. The red curve plots the steady state concentration profile for the attractor achieved from the standard initial conditions ($x_n = 0\ \forall\ n=1,2,\ldots,N$); blue curve plots the same for the attractor achieved from the high initial condition ($x_n = 1\ \forall\ n$).}
    \label{ssc-fixedinput}
\end{figure}

In this system we are able to reproduce the result of ACS domination, which has been discussed for the buffered food set case in Chapter \ref{ChapterACS}. For ACS65 (defined by Eqs. (\ref{acs65-definition})) we are able to see that the molecules produced in ACS are able to dominate well above the background as was the case in the buffered input system. We even find bistability in the dynamics of the system as $\kappa$ changes (see Fig. \ref{bistab-fixedinput}). In the lower attractor, most of the mass is accumulated in the small molecules, whereas in the higher attractor the ACS molecules are able to dominate over the background (see Fig. \ref{ssc-fixedinput}).

\thispagestyle{plain}
\cleardoublepage
\chapter{\label{Appendix-Nowak}Comparison with the model of Ohtsuki and Nowak}

Ohtsuki and Nowak discussed a model of prelife catalysts \cite{Ohtsuki2009}. They consider two type of activated monomers: $0^*$ and $1^*$, produced spontaneously and abundantly in prebiotic chemistry. These monomers can deactivate to produce inactivated monomers, 0 and 1. Activated monomers can combine with inactivated molecules to produce larger molecules, $i + 0^* \rightarrow i0$ and $i + 1^* \rightarrow i1$. It is assumed that the molecules only grow from one side by the addition of activated monomers. Therefore, each string $i$ has a unique precursor, $i'$, and two followers, denoted $i0$ and $i1$ (\ie, if $i = 010, i0 = 0100$). The dynamics of the system are given by,
\begin{align}
    \label{nowak-eq}
    \dot y_i = a_iy_{i'} - (d + a_{i0} + a_{i1})y_i, i = 0, 1, 00, 01, \ldots,
\end{align}
where $i$ represent all binary strings, $x_i$ is the concentration of sequence $i$ and $a_i$ is the rate constants for the reaction in which $i$ is produced form $i'$. $a_{i0}$ is the rate constant for the reaction $i + 0^* \rightarrow i0$ and $a_{i1}$ the rate constant for the reaction $i + 1^* \rightarrow i1$. It is assumed that the concentration of activated monomers is buffered, and the rate constants subsume their concentrations in Eq. (\ref{nowak-eq}). For production of deactivated monomers, 0 and 1, via reactions $0^* \rightarrow 0$ and $1^* \rightarrow 1$, the predecessors $i'$ are $0^*$ and $1^*$ respectively. $d$ is a loss rate at which the molecules are removed from the system. Each sequence of strings, obtained by repeated addition of 0s or 1s, is termed as a lineage, e.g., $0, 00, 000, \ldots$, or $1, 10, 101, 1010, \ldots$, etc. For the case of `fully symmetric' prelife, assume that $a_0 = a_1 = \lambda/2$ and $a_i = a$ for all other sequences, $i$. In this case, all sequences of length $n$ have the same steady state concentration and one may focus on a particular lineage, which for convenience may be chosen to be $0, 00, 000, \ldots$. Between them the reactions are
\begin{subequations}
  \begin{align}
    \label{rct-1-ONmodel}0^* & \rightarrow 0 \\
    0 + 0^* & \rightarrow 00 \\
    00 + 0^* & \rightarrow 000 \\
    000 + 0^* & \rightarrow 0000 \\
    \nonumber \vdots
  \end{align}
\end{subequations}
and the rate equations are given by,
\begin{subequations}
  \begin{align}
    \dot y_0 = & \frac{\lambda}{2}y_{0*} - (d + 2a)y_0, \\
    \dot y_{00} = & a - (d + 2a)y_{00}, \\
    \dot y_{000} = & a - (d + 2a)y_{000}, \\
    \nonumber \vdots &
  \end{align}
\end{subequations}
$y_{0^*}$ is the concentration of $0^*$. Let a sequence of length $k$ be denoted as $0^k, k=1,2,\ldots$, then one may write the equations as
\begin{subequations}
  \label{nowak-eq-0seq}
  \begin{align}
    \dot y_1 = & \frac{\lambda}{2}y_{0^*} - (d + 2a)y_1, \\
    \dot y_k = & ay_{k-1} + (d + 2a)y_k, k = 2, 3, \ldots
  \end{align}
\end{subequations}

In their catalyzed prelife model, Ohtsuki and Nowak assume that some strings can catalyze certain reactions. For example, if sequence $j$ is a catalyst for the reaction $i + 0^* \rightarrow i0$, it enhances the rate of the reaction from $a_{i0}$ $(a_{i0} + cx_j)$, where $a_{i0}$ is the spontaneous reaction rate and $c$ is the strength at which $j$ catalyzes the reaction. They call a `perfect prelife catalyst' as one which catalyzes all the ligation reaction in its own lineage. In fully symmetric prelife with a perfect catalyst of length 5 we can write following reactions
\begin{subequations}
  \label{Nowak-Catalyzed}
  \begin{eqnarray}
    0^* & \reactionarrow{}{} & 0 \\
    0 + 0^* & \reactionarrow{00000}{} & 00 \\
    00 + 0^* & \reactionarrow{00000}{} & 000 \\
    000 + 0^* & \reactionarrow{00000}{} & 0000 \\
    0000 + 0^* & \reactionarrow{00000}{} & 00000 \\
    00000 + 0^* & \reactionarrow{}{} & 000000 \\
    000000 + 0^* & \reactionarrow{}{} & 0000000 \\
    \vdots
  \end{eqnarray}
\end{subequations}
The rate equations for the system described by Eqs. (\ref{Nowak-Catalyzed}) are given by,
\begin{subequations}
  \label{nowak-eq-0seq-catalyzed}
  \begin{align}
    \dot y_1 = & \frac{\lambda}{2}y_{0^*} - (d + a + (a + cy_5))y_1, \\
    \dot y_2 = & (a + cy_5)y_1 + (d + a + (a + cy_5))y_2, \\
    \dot y_3 = & (a + cy_5)y_2 + (d + a + (a + cy_5))y_3, \\
    \dot y_4 = & (a + cy_5)y_3 + (d + a + (a + cy_5))y_4, \\
    \dot y_5 = & (a + cy_5)y_4 + (d + 2a)y_5, \\
    \dot y_6 = & ay_5 + (d + 2a)y_6, \\
    \dot y_7 = & ay_6 + (d + 2a)y_7, \\
    \vdots &
  \end{align}
\end{subequations}

\section{Reducing our $f=1$ model to Ohtsuki and Nowak model}
Consider the following forward reactions in our model,
\begin{subequations}
  \begin{eqnarray}
    \label{rct-1-ourmodel}{\bf A}(1) + {\bf A}(1) & \rightarrow & {\bf A}(2) \\
    {\bf A}(1) + {\bf A}(2) & \rightarrow & {\bf A}(3) \\
    {\bf A}(1) + {\bf A}(3) & \rightarrow & {\bf A}(4) \\
    {\bf A}(1) + {\bf A}(4) & \rightarrow & {\bf A}(5) \\
    \nonumber \vdots
  \end{eqnarray}
\end{subequations}
All other reactions are omitted from the chemistry by setting their respective reaction rate constants equal to zero, $k^F_{ij} = 0$ for all $i \neq 1, j$, $k^R_{ij} = 0$ for all $i,j$ pairs. Thus from Eq. (\ref{rateequationF1}) we get,
\begin{align}
    \label{reduced-eq}
    \dot x_n =& k^F_{1(n-1)}x_1x_{n-1} - k^F_{1n}x_1x_n - \phi_n x_n, n = 2,3,\ldots
\end{align}

We now make following relabeling of species in Ohtsuki and Nowak model to compare it with our model, $0^k$ is now labeled as $\mathbf A(k+1)$ and $0^*$, the buffered activated monomer in the system, is now labeled $\mathbf A(1)$. Explicitly,
\begin{subequations}
    \begin{align}
        \nonumber 0^* & \leftrightarrow \mathbf A(1), \\
        \nonumber 0 & \leftrightarrow \mathbf A(2), \\
        \nonumber 00 & \leftrightarrow \mathbf A(3), \\
        \nonumber 000 & \leftrightarrow \mathbf A(4), \\
        \nonumber \vdots &
    \end{align}
\end{subequations}

Comparing Eqs. (\ref{reduced-eq}) and (\ref{nowak-eq-0seq}) we get, $k^F_{11}x_1 = \lambda/2, k^F_{1n} = a$ for all $n \geq 2$, and $\phi_n = (d + a)$ for all $n$. Note, reaction (\ref{rct-1-ourmodel}) is a bimolecular reaction, where as reaction (\ref{rct-1-ONmodel}) is a unimolecular reaction, but as $x_1 (=A)$ is a constant it can be absorbed in the rate constant. For a catalysed reaction, $\kappa k^F = c$. (Note that Ohtsuki and Nowak only consider spontaneous transformation of activated monomers into deactivated ones, thus reaction (\ref{rct-1-ourmodel}) is not catalysed.)

\begin{figure}
    \begin{center}
        \includegraphics[height=4.75in,angle=-90,trim=0.25cm 0.5cm 0cm 0.3cm,clip=true]{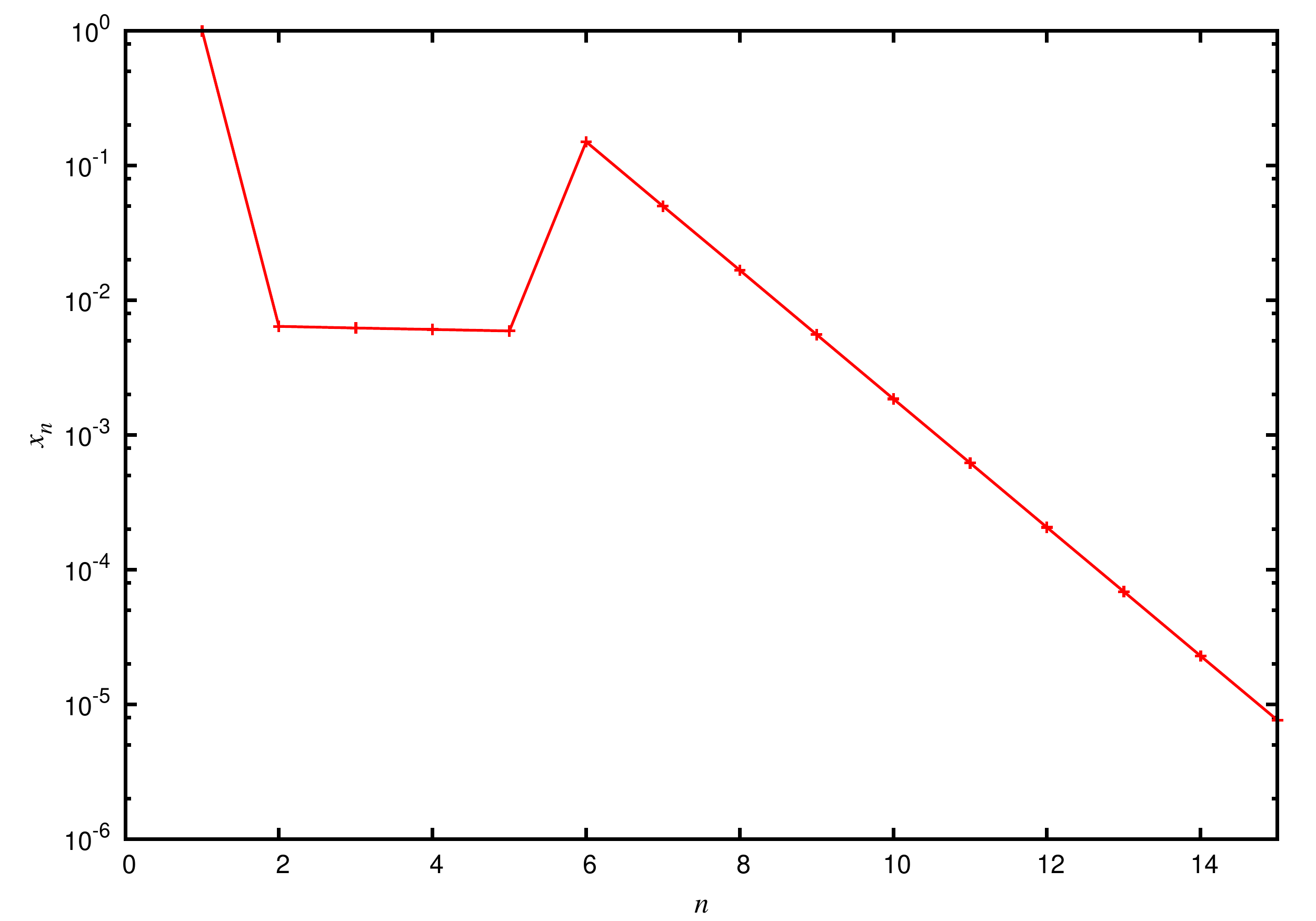}
    \end{center}
    \caption[Steady state concentration profile for reduced model.]{{\bf Steady state concentration profile for reduced model.} The plot shows the steady state concentrations $x_n$ versus length of molecule $n$ for the catalyzed chemistry defined by Eqs. (\ref{nowak-acs-def}). $\mathbf A(1)$ or the activated monomer $0^*$ has a buffered concentration fixed at 1. $x_n, n \geq 2$ gives the population of molecule $\mathbf A(n)$ that models the member of the lineage with length $n-1$. $\mathbf A(6)$, which is the largest molecule produced in the catalysed set dominated the population. Parameters take following values $A = 1, k^F_{11} (= \lambda/2) = 1/2, k^F_{1j} = 1, \phi = 2, \kappa = 500$.}
    \label{nowak-ssc}
\end{figure}

With these transformations our model reduces exactly to the fully symmetric prelife model with catalysis given by Ohtsuki and Nowak. All their analytical results also apply to this (reduced) model. Numerically we have been able to reproduce all their results. An example of steady state concentration profile for a chemistry containing the following set of catalyzed reactions is shown in Fig. \ref{nowak-ssc}.
\begin{subequations}
  \label{nowak-acs-def}
  \begin{eqnarray}
    {\bf A}(1) + {\bf A}(2) & \reactionarrow{{\bf A}(6)}{} & {\bf A}(3) \\
    {\bf A}(1) + {\bf A}(3) & \reactionarrow{{\bf A}(6)}{} & {\bf A}(4) \\
    {\bf A}(1) + {\bf A}(4) & \reactionarrow{{\bf A}(6)}{} & {\bf A}(5) \\
    {\bf A}(1) + {\bf A}(5) & \reactionarrow{{\bf A}(6)}{} & {\bf A}(6)
  \end{eqnarray}
\end{subequations}
In this reduced form of the model only the largest molecule produced in catalyzed reactions dominates the population.

\thispagestyle{plain}
\cleardoublepage
\chapter{\label{Appendix-Catalyzed-Chemistries-wo-ACS}The origin of multistability in ACS65}

In this appendix we attempt to give an intuitive understanding of why ACS65 has a multistability in its dynamics, in particular the existence of the third attractor with intermediate concentrations shown in Fig. \ref{acs-eg-multi}, The reason rests on the substructure of ACS65. Notice that any catalyzed chemistries defined by a subset of the reactions given in Eqs. \ref{acs65-definition} (that define ACS65) is not an ACS. Thus our discussion naturally leads to non-ACS catalyzed chemistries.

\section{Catalyzed chemistries not containing an ACS}
A particularly interesting case of catalyzed chemistries that do not contain an ACS arises when the set of catalyzed reactions form unbroken pathways from the food set to higher molecules, but the catalysts are not drawn from the product set of these reactions (\ie, the catalyzed chemistry satisfies condition (b) for an ACS but not the condition (a); see Section \ref{Definition-ACS}). In this case we again observe the domination of the molecules produced by the catalyzed pathways, but this occurs at even higher catalytic strengths than for the ACSs that include production for these catalysts. The reasons are similar to those discussed above in Section \ref{kappa-infty} in the context of the $\kappa \rightarrow \infty$ limit for ACSs (note that in that limit the identity of the catalyst is irrelevant for the argument as long as the catalyst has a nonzero concentration). We have also observed bistability in these chemistries (an example is discussed below) as well as a dependence on $\phi$ similar to the ACS case. In general, catalyzed chemistries (with and without ACSs) seem to have a rich and complex phase structure but more comprehensive investigations than we have done are needed to nail down the possible range of interesting behaviours.

Consider the following set of catalyzed reactions along with full spontaneous chemistry. Notice that there is a continuous path from the food set to $\mathbf A(4)$. And the catalysts, namely, $\mathbf A(9)$ and $\mathbf A(5)$, are not produced in the catalyzed reactions themselves. This chemistry is in fact a subset of ACS65, containing the first two catalyzed reactions of the set.
\begin{subequations}
  \label{catalyzed-chem-definition}
  \begin{eqnarray}
    \mathrm{\mathbf A}(1) + \mathrm{\mathbf A}(1) & \reactionrevarrow{\ensuremath{\mathrm{\mathbf A}(9)}}{} & \mathrm{\mathbf A}(2) \\
    \mathrm{\mathbf A}(2) + \mathrm{\mathbf A}(2) & \reactionrevarrow{\ensuremath{\mathrm{\mathbf A}(5)}}{} & \mathrm{\mathbf A}(4)
  \end{eqnarray}
\end{subequations}

\begin{figure}
    \begin{center}
        \subfloat[]{\label{cat-chem-a}\includegraphics[height=4.75in,angle=-90,trim=0.5cm 0.5cm 0cm 0.3cm,clip=true]{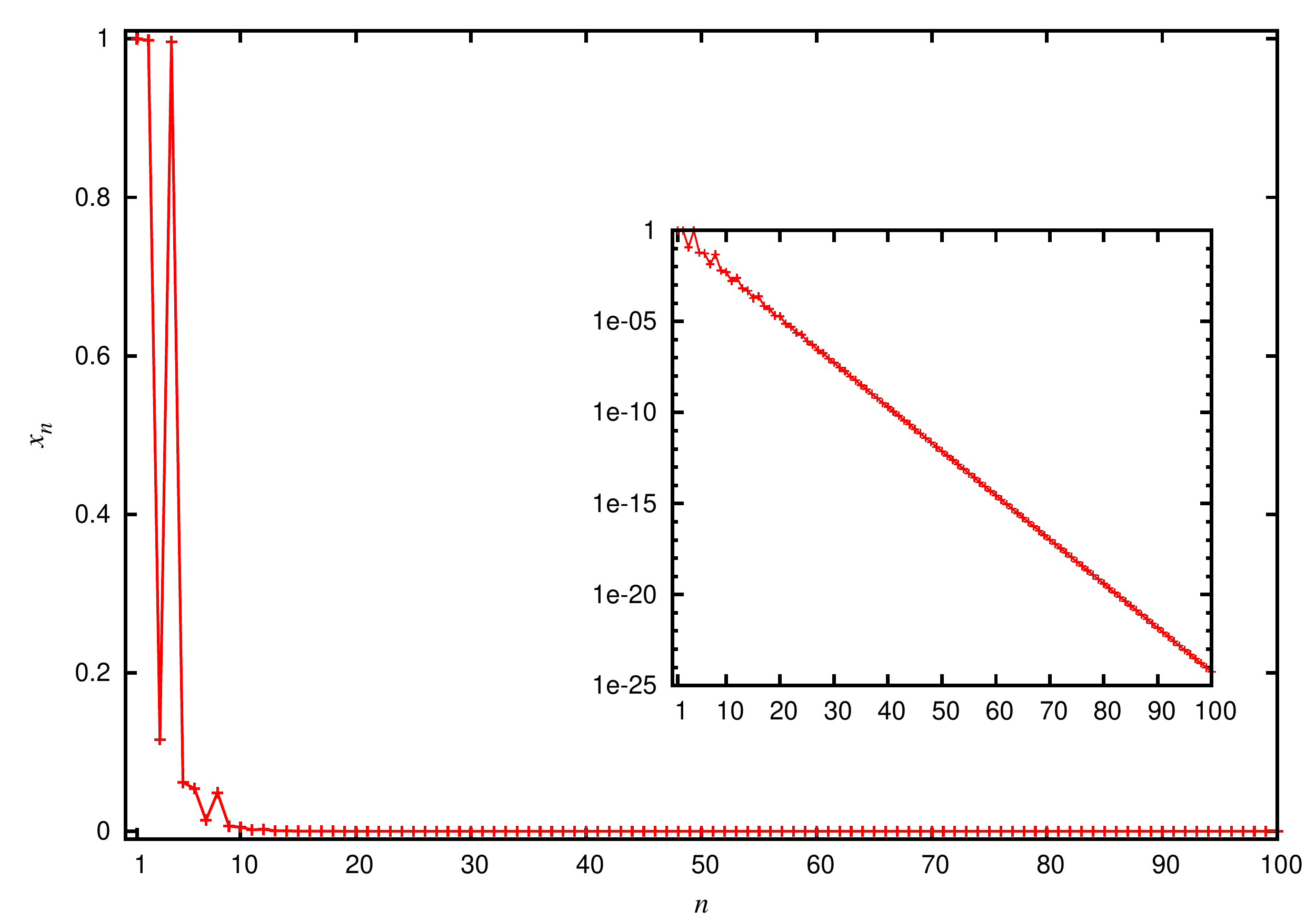}} \\
        \subfloat[]{\label{cat-chem-b}\includegraphics[height=4.75in,angle=-90,trim=0.5cm 0.5cm 0cm 0.3cm,clip=true]{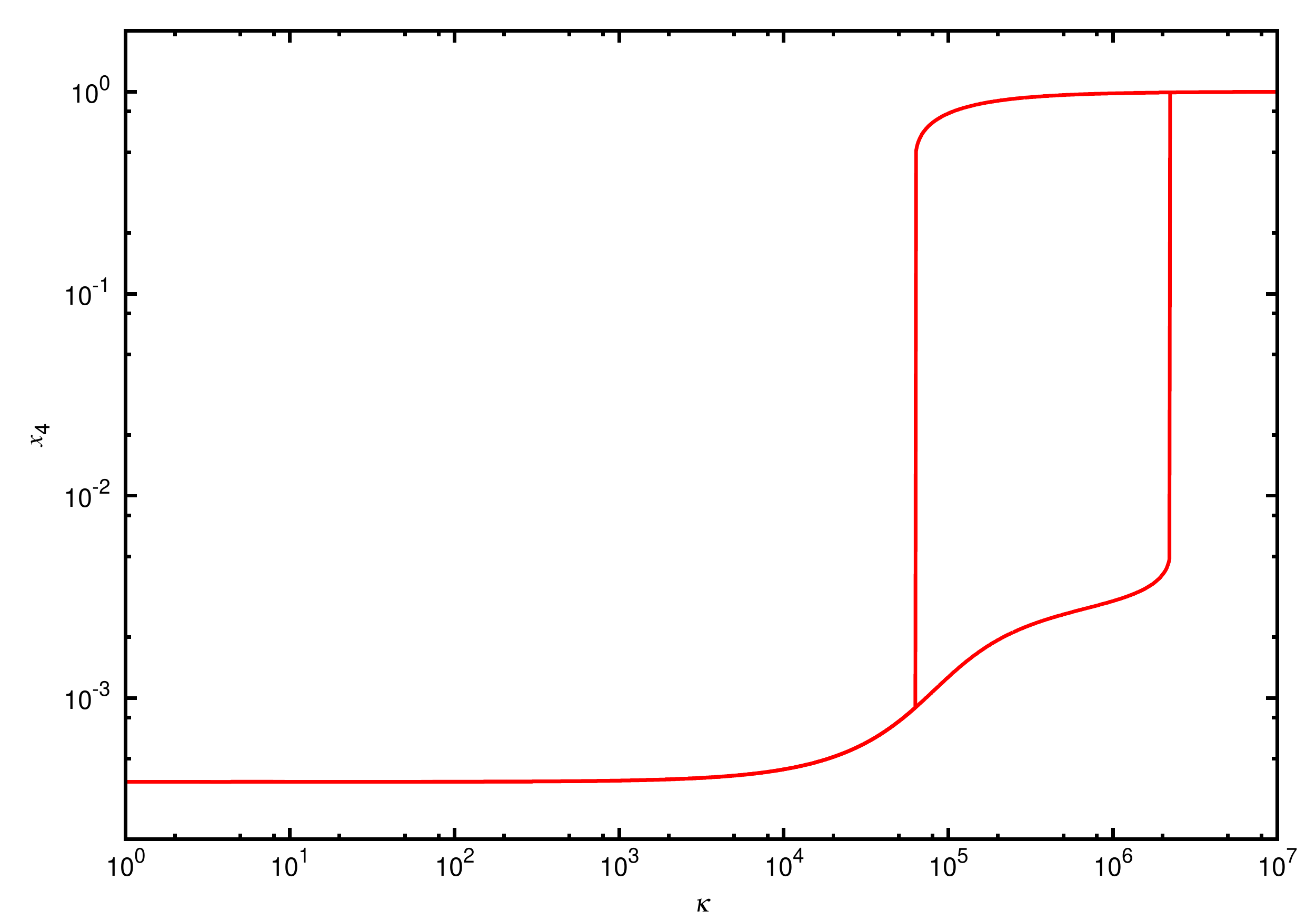}}
    \end{center}
    \caption[Steady state concentrations and bistability of a catalyzed chemistry.]{{\bf Steady state concentrations and bistability of a catalyzed chemistry.} {\bf(a)} The steady state concentration profile for the catalyzed chemistry (\ref{catalyzed-chem-definition}). $A = k_f = k_r = 1, \phi = 15, \kappa = 5 \times 10^6$. $\mathbf A(2)$ and $\mathbf A(4)$ have higher concentrations compared to other molecules in the background. {\bf(b)} The hysteresis curve showing bistability in the dynamics of this catalyzed chemistry without an ACS. The parameters take same values as in (a), except $\kappa$, which is varied.}
    \label{cat-chem}
\end{figure}

The steady state concentration profile for this chemistry which we refer to as C9 (as $\mathbf A(9)$ is the largest catalyst in this chemistry) is shown in Fig. \ref{cat-chem-a}. As for an ACS, the molecules produced in the catalyzed reactions, viz., $\mathbf A(2)$ and $\mathbf A(4)$ have much larger concentrations as compared to other species in the chemistry. Fig. \ref{cat-chem-b} shows bistability in the steady state concentrations as $\kappa$ is varied. Note that the initiation threshold happens to be the same as for ACS65 namely $\kappa^{II} = 2226392$, but the maintenance threshold is now considerably higher.

\section{\label{multistability-acs65}Reasons for multistability in dynamics of ACS65}
\begin{figure}
    \begin{center}
        \subfloat[]{\label{acs-eg-delrct-a}\includegraphics[height=4.75in,angle=-90,trim=0.5cm 0.5cm 0cm 0.3cm,clip=true]{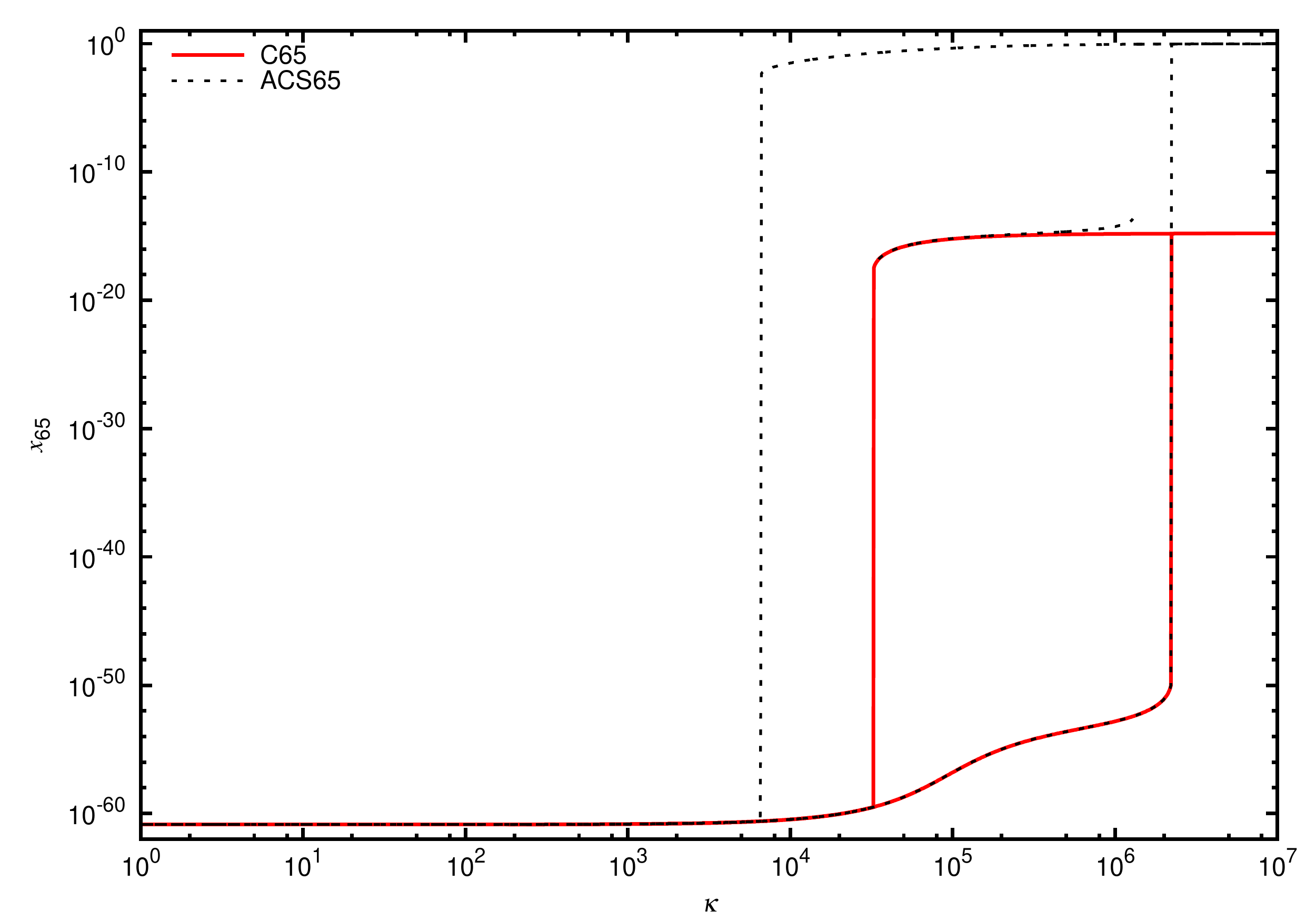}} \\
        \subfloat[]{\label{acs-eg-delrct-b}\includegraphics[height=4.75in,angle=-90,trim=0.5cm 0.5cm 0cm 0.3cm,clip=true]{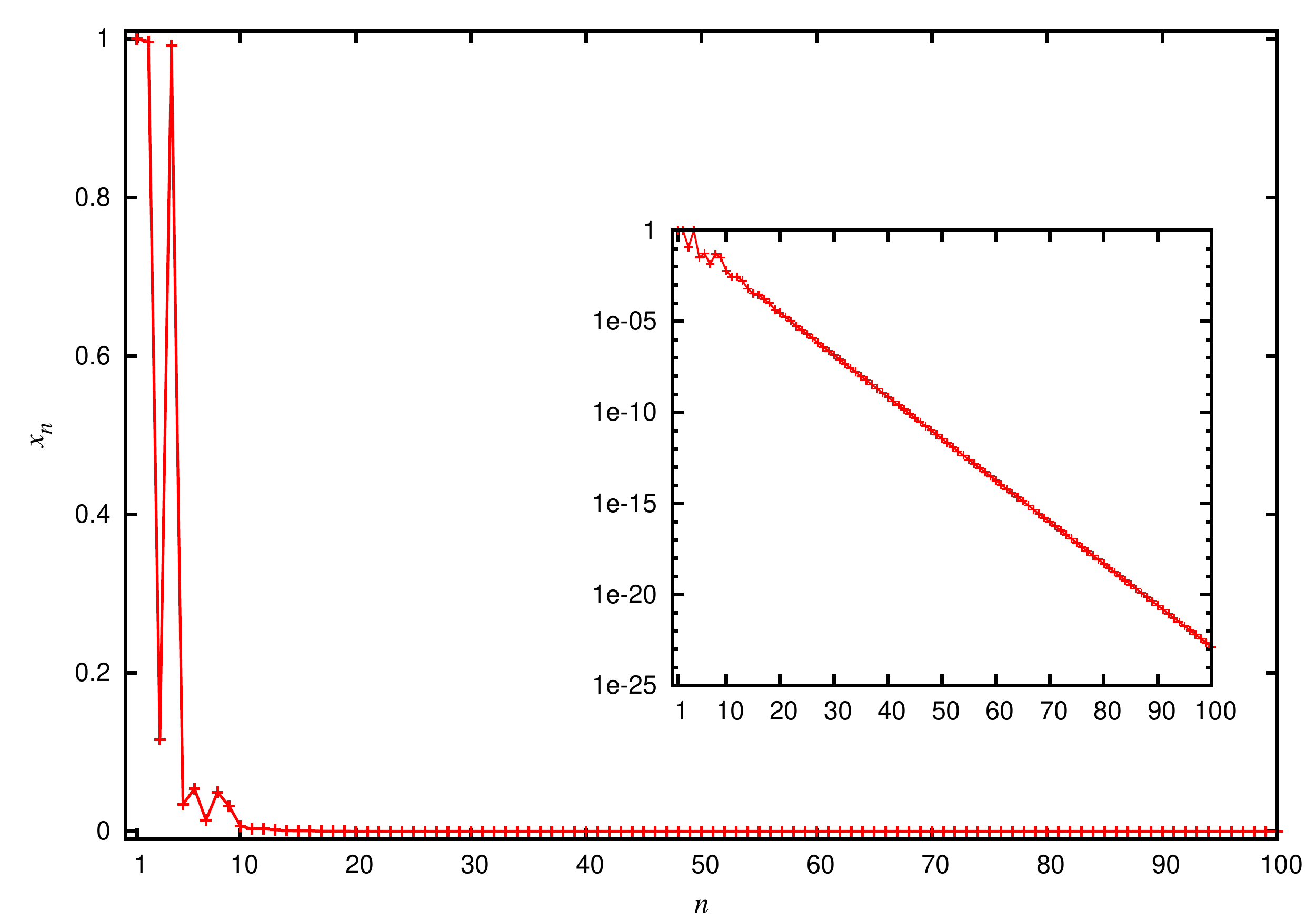}}
    \end{center}
    \caption[Bistability in the dynamics of the catalyzed chem C65, which is the same as ACS65, but with the catalyzed reaction (\ref{acs65-rct3}) deleted.]{{\bf Bistability in the dynamics of the catalyzed chem C65, which is the same as ACS65, but with the catalyzed reaction (\ref{acs65-rct3}) deleted.} {\bf(a)} The hysteresis curve for the catalyzed chemistry C65. For comparison the phase portrait for ACS65 is also shown in dotted line. The intermediate steady state in ACS65 corresponds to the upper steady state of C65. The parameters are $k_f = k_r = A = 1, \phi = 15, N=100$  {\bf(b)} The steady state concentration profile in upper steady state. $\kappa = 5 \times 10^5$ other parameters take same values as in (a). Note that this profile is essentially the same as that for ACS65 shown in Fig. \ref{acs-eg-multi-b}.}
    \label{acs-eg-delrct}
\end{figure}

The above example shows that bistability can arise even in a simple catalyzed chemistry that has continuous catalyzed pathways originating from the food source. This is in fact the origin of multistability in dynamics of ACS65. Consider the case of ACS65 defined in Eqs. \ref{acs65-definition}, but with reaction (\ref{acs65-rct3}) deleted. Note that the resulting set of 7 catalyzed reaction pairs is no more an ACS, but merely a catalyzed chemistry which we refer to as C65. The continuous catalyzed pathways end at $\mathbf A(4)$ just as in C9 above. The catalyzed chemistry C65 with 7 catalyzed reaction pairs also shows a bistability (Fig. \ref{acs-eg-delrct-a}). The steady state concentration profile for the system in the upper attractor is shown in Fig. \ref{acs-eg-delrct-b}.

The main point we wish to draw attention to is that the three catalyzed chemistries C9, C65 and ACS65, all have been found empirically to have bistability with the same value of $\kappa^{II} = 2226342$. Since C9 is the smallest catalyzed chemistry that is common to all three (note that as sets of reactions, $\mathrm {C9} \subset \mathrm {C65} \subset \mathrm {ACS65}$), it is reasonable to suppose that it is the structure of C9 that is determining the value of $\kappa^{II}$ for all three. That is, the additional catalyzed reactions in C65 and ACS65 do not seem to matter as far as the value of $\kappa^{II}$ is concerned.

Notice that the catalyst for reaction (\ref{acs65-rct3}) in ACS65 is $\mathbf A(28)$. As $\mathbf A(28)$ a is considerably large molecule, its concentration in the lower attractor is very small; the product $\kappa \times x_{28}$ is thus much smaller than 1. That is, the rate of the catalyzed reaction (\ref{acs65-rct3}) is much less than the corresponding spontaneous reaction. For this reason, it can be assumed that the catalyzed reaction (\ref{acs65-rct3}) is absent in the range where $\kappa \times x_{28}$ remains much less than 1 (see Section \ref{kappa-infty} and also \ref{exponential-initaition-threshold}). Therefore ACS65 in this range also exhibits the same behaviour as C65.

Now C65 has an upper attractor given by the upper red line in Fig. \ref{acs-eg-delrct-a}. Therefore ACS65 also acquires this steady state. This adds an additional stable state, inducing multistability, in the dynamics of ACS65. The multistability curve for ACS65 is also shown in Fig. \ref{acs-eg-delrct-a}, where the intermediate steady state of ACS65, overlaps with higher steady state of C65. The steady state profiles of C65 in its higher attractor and ACS65 in its intermediate attractor, shown in Figs. \ref{acs-eg-delrct-b} and \ref{acs-eg-multi-b}, are essentially indistinguishable.

\thispagestyle{plain}
\cleardoublepage
\chapter{\label{Appendix-ACS-topology-and-Bistability}Effect of ACS topology on bistability}

In Section \ref{acs-bistability} we have discussed how bistability arises in simple ACSs. We extend the discussion to include how the assignment of catalysts to the reactions of an ACS affects the dynamics, specifically the existence and extent of bistability. The assignment of the catalysts to the reactions of the ACS is an aspect of the topology of the ACS. ACSs with the same set of reactions (and hence the same product set $P(S)$) may differ from each other in the assignment of catalysts to the reactions. For example, consider following set of reactions comprising an ACS,
\begin{subequations}
 \label{acs-x-definition}
 \begin{eqnarray}
    \label {acs-x-rct1} \mathrm{\mathbf A}(1) + \mathrm{\mathbf A}(1) & \reactionrevarrow{\ensuremath{\mathrm{\mathbf A}(x)}}{} & \mathrm{\mathbf A}(2) \\
    \label {acs-x-rct2} \mathrm{\mathbf A}(2) + \mathrm{\mathbf A}(2) & \reactionrevarrow{\ensuremath{\mathrm{\mathbf A}(y)}}{} & \mathrm{\mathbf A}(4), \\
    \label {acs-x-rct3} \mathrm{\mathbf A}(2) + \mathrm{\mathbf A}(2) & \reactionrevarrow{\ensuremath{\mathrm{\mathbf A}(z)}}{} & \mathrm{\mathbf A}(8),
\end{eqnarray}
\end{subequations}
where, $x,y,z \in \{2,4,8\}$. This ACS has 27 possible topologies of catalyzed reactions, enumerated by the values of $x,y,z$. The list of topologies and the $\kappa^{I}$ and $\kappa^{II}$ values for each topology assuming a homogeneous catalyzed chemistry (\ie, all catalysts have the same catalytic strengths) are listed in Table \ref{topology-kappa1-kappa2}. For some cases, such as whenever $x=2$, \ie, $\mathbf A(2)$ is the catalyst of the reaction (\ref{acs-x-rct1}), we do not find bistability in the dynamics of the ACS (irrespective of the values $y$ and $z$ take) for the value of $\phi$ studied; recall from the discussion in the Section \ref{acs-bistability} that there exists a critical length and a $\bar\phi$ below which we do not see bistability. The bistability may exist for some topologies for a higher value of $\phi$, e.g., $x=2,y=8,z=8$ has a bistability at $\phi=25$ with $\kappa^{I}= 481.5$ and $\kappa^{II} = 2807.4$. In general, the parameter region where bistability exists in the dynamics depends upon $x$, $y$ and $z$. We illustrate this in the table below for a homogeneous chemistry.

\setlength{\tabcolsep}{15pt}
\renewcommand{\arraystretch}{1.5}
\setlength{\LTcapwidth}{5in}
\begin{center}
  \begin{longtable}{|c|c|c|c|c|}
    \caption[List of all the possible topologies of ACS given by Eq. (\ref{acs-x-definition}) and respective $\kappa^{I}$ and $\kappa^{II}$ values.]{\label{topology-kappa1-kappa2}{\bf List of all the possible topologies of ACS given by Eq. (\ref{acs-x-definition}) and respective $\kappa^{I}$ and $\kappa^{II}$ values.} $\mathbf A(x), \mathbf A(y), \mathbf A(z)$ are the catalysts for the reactions (\ref{acs-x-rct1}), (\ref{acs-x-rct2}), (\ref{acs-x-rct3}) respectively. The parameters used to calculate $\kappa^{I}$ and $\kappa^{II}$ are $A=k_f=k_r=1$, $\phi=5$, $N=15$.} \\
      \hline
      $x$ & $y$ & $z$ & $\kappa^{I}$ & $\kappa^{II}$ \\
      \hline \hline
      \endfirsthead
      \multicolumn{3}{l}{\tablename\ \thetable{}} & \multicolumn{2}{r}{{\it $\ldots$ continued from previous page}} \\   \hline
      $x$ & $y$ & $z$ & $\kappa^I$ & $\kappa^{II}$ \\
      \hline \hline
      \endhead
          \multicolumn{5}{l}{{\it continued on next page $\ldots$}} \\
      \endfoot
      \endlastfoot
      2 & - & - & \multicolumn{2}{c|}{\multirow{4}[8]{*}{Bistability does not exists for $\phi=5$}} \\ \cline{1-3}
      4 & 2 & 2 & \multicolumn{2}{c|}{}  \\ \cline{1-3}
      4 & 2 & 4 & \multicolumn{2}{c|}{} \\ \cline{1-3}
      4 & 2 & 8 & \multicolumn{2}{c|}{}  \\ \hline
      4 & 4 & 2 & 6.558503e+01 & 6.677801e+01 \\ \hline
      4 & 4 & 4 & 6.332721e+01 & 6.554611e+01 \\ \hline
      4 & 4 & 8 & 6.101660e+01 & 6.520190e+01 \\ \hline
      4 & 8 & 2 & 1.449191e+02 & 1.504237e+02 \\ \hline
      4 & 8 & 4 & 1.498893e+02 & 1.594422e+02 \\ \hline
      4 & 8 & 8 & 1.713392e+02 & 2.072384e+02 \\ \hline
      8 & 2 & 2 & 2.795060e+02 & 8.610626e+02 \\ \hline
      8 & 2 & 4 & 2.862641e+02 & 1.026008e+03 \\ \hline
      8 & 2 & 8 & 3.140126e+02 & 3.388361e+03 \\ \hline
      8 & 4 & 2 & 2.927453e+02 & 1.053617e+03 \\ \hline
      8 & 4 & 4 & 2.991615e+02 & 1.223987e+03 \\ \hline
      8 & 4 & 8 & 3.251521e+02 & 3.564402e+03 \\ \hline
      8 & 8 & 2 & 3.487341e+02 & 5.303906e+03 \\ \hline
      8 & 8 & 4 & 3.553727e+02 & 5.791569e+03 \\ \hline
      8 & 8 & 8 & 3.786843e+02 & 1.194718e+04 \\ \hline
  \end{longtable}
\end{center} 

\thispagestyle{plain}
\cleardoublepage
\chapter{\label{Appendix-ACS441-reaction-list}List of reactions and their catalysts in ACS441}

The table lists all the reactions with their respective catalysts in the example of a catalyzed chemistry, discussed in Section \ref{section-ACS441}, containing a cascade of nested ACSs for $f=1$ generated using Algorithm 4. The steady state concentrations for this chemistry are displayed in Fig. \ref{1d-algo4}. This chemistry was generated with $g=15$ and $n_1=1, n_2 = n_3 = \ldots = n_{15} = 2$.

Note that a general molecule for the $f=1$ case, represented by A($n$) in the main text, has been represented here by $n$ for brevity.

The molecules in various generations are as follows:
$P_0 = \{1\}$, $P_1 = \{2\}$, $P_2 = \{3,4\}$, $P_3 = \{6,7\}$, $P_4 = \{10,12\}$, $P_5 = \{17,18\}$, $P_6 = \{20,24\}$, $P_7 = \{25,48\}$, $P_8 = \{52,66\}$, $P_9 = \{67,69\}$, $P_{10} = \{77,84\}$, $P_{11} = \{144,168\}$, $P_{12} = \{221,288\}$, $P_{13} = \{305,336\}$, $P_{14} = \{372,389\}$, $P_{15} = \{397,441\}$.

The catalyst for a reaction listed under generation $P_k$ is added at step $k$ of algorithm. It is apparent from the reaction table that the ACSs are maximally overlapping, \ie, any ACS of generation $k$ contains all the reactions of generation $k-1$.

\setlength{\tabcolsep}{5pt}
\setlength{\LTcapwidth}{8in}
\begin{landscape}
\renewcommand{\arraystretch}{1.5}
\begin{center}
\begin{longtable}{|c|c|c|c|c|c|c|c|c|c|c|c|c|c|c|c|}
  \caption[List of reactions and their catalysts in ACS441.]{{\bf List of reactions and their catalysts in ACS441.}} \\
  \hline
  \multirow{2}{*}{\hspace{1cm}{\bf Reaction}\hspace{1cm}} & \multicolumn{15}{|c|}{\bf Catalyst added in generation} \\ \cline{2-16}
  & \hspace{0.2cm}{\bf $P_1$}\hspace{0.2cm}  & \hspace{0.2cm}{\bf $P_2$}\hspace{0.2cm} & \hspace{0.2cm}{\bf $P_3$}\hspace{0.2cm} & \hspace{0.2cm}{\bf $P_4$}\hspace{0.2cm} & \hspace{0.2cm}{\bf $P_5$}\hspace{0.2cm} & \hspace{0.2cm}{\bf $P_6$}\hspace{0.2cm} & \hspace{0.2cm}{\bf $P_7$}\hspace{0.2cm} & \hspace{0.2cm}{\bf $P_8$}\hspace{0.2cm} & \hspace{0.2cm}{\bf $P_9$}\hspace{0.2cm} & \hspace{0.1cm}{\bf $P_{10}$}\hspace{0.1cm} & \hspace{0.1cm}{\bf $P_{11}$}\hspace{0.1cm} & \hspace{0.1cm}{\bf $P_{12}$}\hspace{0.1cm} & \hspace{0.1cm}{\bf $P_{13}$}\hspace{0.1cm} & \hspace{0.1cm}{\bf $P_{14}$}\hspace{0.1cm} & \hspace{0.1cm}{\bf $P_{15}$}\hspace{0.1cm} \\ \hline
  \endfirsthead
  \multicolumn{1}{l}{\tablename\ \thetable{}} & \multicolumn{15}{r}{{\it $\ldots$ continued from previous page}} \\  \hline
  \multirow{2}{*}{{\bf Reaction}} & \multicolumn{15}{|c|}{\bf Catalyst in generation} \\ \cline{2-16}
  & {\bf $P_1$}  & {\bf $P_2$} & {\bf $P_3$} & {\bf $P_4$} & {\bf $P_5$} & {\bf $P_6$} & {\bf $P_7$} & {\bf $P_8$} & {\bf $P_9$} & {\bf $P_{10}$} & {\bf $P_{11}$} & {\bf $P_{12}$} & {\bf $P_{13}$} & {\bf $P_{14}$} & {\bf $P_{15}$} \\ \hline
  \endhead
      \multicolumn{16}{l}{{\it continued on next page $\ldots$}} \\
  \endfoot
  \endlastfoot

  $1 + 1 \rightleftharpoons 2$ & 2 & 3 & 6 & 10 & 17 & 24 & 48 & 52 & 69 & 84 & 144 & 288 & 336 & 372 & 397\\ \hline
  $1 + 2 \rightleftharpoons 3$ & & 3 & 7 & 10 & 18 & 20 & 25 & 52 & 67 & 77 & 168 & 221 & 305 & 372 & 397  \\ \hline
  $2 + 2 \rightleftharpoons 4$ & & 4 & 7 & 10 & 18 & 20 & 25 & 66 & 67 & 77 & 168 & 221 & 336 & 372 & 397 \\ \hline
  $3 + 3 \rightleftharpoons 6$ & & & 6 & 12 & 17 & 24 & 25 & 66 & 67 & 84 & 144 & 221 & 336 & 372 & 441 \\ \hline
  $3 + 4 \rightleftharpoons 7$ & & & 7 & 12 & 18 & 24 & 48 & 52 & 67 & 84 & 144 & 221 & 336 & 372 & 397 \\ \hline
  $4 + 6 \rightleftharpoons 10$ & & & & 10 & 18 & 24 & 48 & 52 & 69 & 77 & 168 & 288 & 336 & 389 & 397 \\ \hline
  $6 + 6 \rightleftharpoons 12$ & & & & 12 & 17 & 20 & 48 & 66 & 69 & 77 & 144 & 221 & 305 & 372 & 441 \\ \hline
  $7 + 10 \rightleftharpoons 17$ & & & & & 17 & 24 & 25 & 52 & 69 & 84 & 168 & 221 & 336 & 389 & 441 \\ \hline
  $6 + 12 \rightleftharpoons 18$ & & & & & 18 & 20 & 48 & 66 & 69 & 77 & 168 & 288 & 336 & 372 & 441 \\ \hline
  $2 + 18 \rightleftharpoons 20$ & & & & & & 20 & 48 & 52 & 69 & 84 & 144 & 221 & 336 & 389 & 441 \\ \hline
  $7 + 17 \rightleftharpoons 24$ & & & & & & 20 & 25 & 66 & 69 & 77 & 168 & 221 & 336 & 372 & 397 \\ \hline
  $1 + 24 \rightleftharpoons 25$ & & & & & & & 48 & 52 & 69 & 84 & 144 & 288 & 305 & 372 & 441 \\ \hline
  $24 + 24 \rightleftharpoons 48$ & & & & & & & 48 & 66 & 67 & 77 & 144 & 221 & 305 & 389 & 441 \\ \hline
  $4 + 48 \rightleftharpoons 52$ & & & & & & & & 66 & 69 & 84 & 168 & 221 & 336 & 372 & 397 \\ \hline
  $18 + 48 \rightleftharpoons 66$ & & & & & & & & 66 & 69 & 84 & 144 & 288 & 336 & 389 & 441 \\ \hline
  $1 + 66 \rightleftharpoons 67$ & & & & & & & & & 67 & 77 & 168 & 221 & 336 & 389 & 441 \\ \hline
  $3 + 66 \rightleftharpoons 69$ & & & & & & & & & 67 & 84 & 168 & 221 & 305 & 372 & 441 \\ \hline
  $10 + 67 \rightleftharpoons 77$ & & & & & & & & & & 77 & 168 & 288 & 305 & 372 & 441 \\ \hline
  $17 + 67 \rightleftharpoons 84$ & & & & & & & & & &  77 & 144 & 221 & 305 & 389 & 397 \\ \hline
  $67 + 77 \rightleftharpoons 144$ & & & & & & & & & & & 144 & 288 & 305 & 389 & 397 \\ \hline
  $84 + 84 \rightleftharpoons 168$ & & & & & & & & & & & 144 & 221 & 336 & 389 & 441 \\ \hline
  $77 + 144 \rightleftharpoons 221$ & & & & & & & & & & & & 221 & 305 & 372 & 397 \\ \hline
  $144 + 144 \rightleftharpoons 288$ & & & & & & & & & & & & 221 & 336 & 372 & 441 \\ \hline
  $17 + 288 \rightleftharpoons 305$ & & & & & & & & & & & & & 305 & 372 & 441 \\ \hline
  $48 + 288 \rightleftharpoons 336$ & & & & & & & & & & & & & 336 & 389 & 397 \\ \hline
  $67 + 305 \rightleftharpoons 372$ & & & & & & & & & & & & & & 389 & 397 \\ \hline
  $84 + 305 \rightleftharpoons 389$ & & & & & & & & & & & & & & 389 & 397 \\ \hline
  $25 + 372 \rightleftharpoons 397$ & & & & & & & & & & & & & & & 441 \\ \hline
  $69 + 372 \rightleftharpoons 441$ & & & & & & & & & & & & & & & 397 \\ \hline
\end{longtable}
\end{center}
\end{landscape}

\thispagestyle{plain}
\cleardoublepage
\chapter{\label{Appendix-ACS36-28-reaction-list}List of reactions and their catalysts in ACS(36,28)}

The table lists all the reactions with their respective catalysts in the example of a catalyzed chemistry, discussed in Section \ref{section-cascade-f2}, containing a cascade of nested ACSs for $f=2$ generated using Algorithm 4. The steady state concentrations for this chemistry are displayed in Fig. \ref{2d-algo4}. This chemistry was generated with $g=7$ and $n_k=3$.

The molecules in various generations are as follows:
$P_0 = \{(1,0),(0,1)\}$, $P_1 = \{(1,1),(0,2),(2,0)\}$, $P_2 = \{(1,3),(2,2),(3,0)\}$, $P_3 = \{(2,6),(2,3),(5,2)\}$, $P_4 = \{(4,6),(7,4),(7,8)\}$, $P_5 = \{(8,12),(6,12),(14,8)\}$, $P_6 = \{(14,24),(22,20),(11,12)\}$, $P_7 = \{(29,28),(36,28), (24,26)\}$.

The catalyst for a reaction listed under generation $P_k$ is added at step $k$ of algorithm. It is apparent from the reaction table that the ACSs are maximally overlapping, \ie, any ACS of generation $k$ contains all the reactions of generation $k-1$.

\begin{landscape}
\setlength{\tabcolsep}{10pt}
\renewcommand{\arraystretch}{1.5}
\begin{center}
\begin{longtable}{|c|c|c|c|c|c|c|c|}
  \caption[List of reactions and their catalysts in ACS(36,28).]{{\bf List of reactions and their catalysts in ACS(36,28).}} \\
  \hline
  \multirow{2}{*}{\hspace{1.75cm}{\bf Reaction}\hspace{1.75cm}} & \multicolumn{7}{|c|}{\bf Catalyst added in generation} \\ \cline{2-8}
  & \hspace{0.5cm}{\bf $P_1$}\hspace{0.5cm}  & \hspace{0.5cm}{\bf $P_2$}\hspace{0.5cm} & \hspace{0.5cm}{\bf $P_3$}\hspace{0.5cm} & \hspace{0.5cm}{\bf $P_4$}\hspace{0.5cm} & \hspace{0.5cm}{\bf $P_5$}\hspace{0.5cm} & \hspace{0.5cm}{\bf $P_6$}\hspace{0.5cm} & \hspace{0.5cm}{\bf $P_7$}\hspace{0.5cm} \\ \hline
  \endfirsthead
  \multicolumn{1}{l}{\tablename\ \thetable{}} & \multicolumn{7}{r}{{\it $\ldots$ continued from previous page}} \\  \hline
  \multirow{2}{*}{{\bf Reaction}} & \multicolumn{7}{|c|}{\bf Catalyst in generation} \\ \cline{2-8}
  & {\bf $P_1$}  & {\bf $P_2$} & {\bf $P_3$} & {\bf $P_4$} & {\bf $P_5$} & {\bf $P_6$} & {\bf $P_7$} \\ \hline
  \endhead
      \multicolumn{8}{l}{{\it continued on next page $\ldots$}} \\
  \endfoot
  \endlastfoot

  $(0,1) + (0,1) \rightleftharpoons  (0,2)$ & $(1,1)$ &  $(1,3)$ &  $(2,3)$ &  $(7,4)$ &  $(8,12)$ &  $(11,12)$ &  $(29,28)$ \\ \hline
  $(0,1) + (1,0) \rightleftharpoons  (1,1)$ & $(0,2)$ &  $(1,3)$ &  $(2,6)$ &  $(4,6)$ &  $(14,8)$ &  $(22,20)$ &  $(36,28)$ \\ \hline
  $(1,0) + (1,0) \rightleftharpoons  (2,0)$ & $(1,1)$ &  $(3,0)$ &  $(5,2)$ &  $(7,4)$ &  $(14,8)$ &  $(14,24)$ &  $(29,28)$ \\ \hline
  $(1,0) + (2,0) \rightleftharpoons  (3,0)$ & & $(2,2)$ &  $(5,2)$ &  $(4,6)$ &  $(14,8)$ &  $(14,24)$ &  $(24,26)$ \\ \hline
  $(1,1) + (0,2) \rightleftharpoons  (1,3)$ & & $(1,3)$ &  $(2,6)$ &  $(7,8)$ &  $(8,12)$ &  $(14,24)$ &  $(24,26)$ \\ \hline
  $(0,2) + (2,0) \rightleftharpoons  (2,2)$ & & $(1,3)$ &  $(2,6)$ &  $(7,8)$ &  $(6,12)$ &  $(22,20)$ &  $(29,28)$ \\ \hline
  $(0,1) + (2,2) \rightleftharpoons  (2,3)$ & & & $(2,6)$ &  $(7,4)$ &  $(14,8)$ &  $(22,20)$ &  $(24,26)$ \\ \hline
  $(3,0) + (2,2) \rightleftharpoons  (5,2)$ & & & $(2,3)$ &  $(7,8)$ &  $(6,12)$ &  $(11,12)$ &  $(29,28)$ \\ \hline
  $(1,3) + (1,3) \rightleftharpoons  (2,6)$ & & & $(5,2)$ &  $(7,8)$ &  $(14,8)$ &  $(22,20)$ &  $(24,26)$ \\ \hline
  $(2,0) + (2,6) \rightleftharpoons  (4,6)$ & & & & $(4,6)$ &  $(14,8)$ &  $(22,20)$ &  $(36,28)$ \\ \hline
  $(2,2) + (5,2) \rightleftharpoons  (7,4)$ & & & & $(7,4)$ &  $(6,12)$ &  $(14,24)$ &  $(36,28)$ \\ \hline
  $(5,2) + (2,6) \rightleftharpoons  (7,8)$ & & & & $(7,4)$ &  $(14,8)$ &  $(14,24)$ &  $(36,28)$ \\ \hline
  $(2,6) + (4,6) \rightleftharpoons  (6,12)$ & & & & & $(14,8)$ &  $(11,12)$ &  $(29,28)$ \\ \hline
  $(4,6) + (4,6) \rightleftharpoons  (8,12)$ & & & & & $(14,8)$ &  $(22,20)$ &  $(36,28)$ \\ \hline
  $(7,4) + (7,4) \rightleftharpoons  (14,8)$ & & & & & $(8,12)$ &  $(11,12)$ &  $(36,28)$ \\ \hline
  $(3,0) + (8,12) \rightleftharpoons  (11,12)$ & & & & & & $(22,20)$ &  $(36,28)$ \\ \hline
  $(8,12) + (6,12) \rightleftharpoons  (14,24)$ & & & & & & $(22,20)$ &  $(24,26)$ \\ \hline
  $(14,8) + (8,12) \rightleftharpoons  (22,20)$ & & & & & & $(11,12)$ &  $(29,28)$ \\ \hline
  $(2,6) + (22,20) \rightleftharpoons  (24,26)$ & & & & & & & $(29,28)$ \\ \hline
  $(7,8) + (22,20) \rightleftharpoons  (29,28)$ & & & & & & & $(29,28)$ \\ \hline
  $(14,8) + (22,20) \rightleftharpoons  (36,28)$ & & & & & & & $(29,28)$ \\ \hline
\end{longtable}
\end{center}
\end{landscape}

\thispagestyle{plain}
\cleardoublepage
\chapter{\label{Appendix-ACS18-27-reaction-list}List of reactions and their catalysts in ACS(18,27)}

The table lists all the reactions with their respective catalysts and catalytic strengths in the example of a catalyzed chemistry, discussed in Section \ref{section-cascade-f2}, containing a cascade of partially overlapping ACSs for $f=2$ generated using Algorithm 5. The steady state concentrations for this chemistry are displayed in Fig. \ref{2d-partial-nest}.

The catalyzed chemistry contains 10 generations of ACSs of lengths 3, 6, 10, 15, 19, 25, 30, 35, 40, and 45.

\renewcommand{\arraystretch}{1.5}
\setlength{\LTcapwidth}{5in}
\begin{center}
\begin{longtable}{|c|c|}
  \caption[List of all the nested ACSs in ACS(18,27).]{{\bf List of reactions in the ACSs of different length and their catalysts.} ACSs of increasing length (using Algorithm 5) are added in the chemistry. The length of the ACS and the catalytic strength of the catalyst are mentioned in the table.} \\
  \hline
  \hspace{2.5cm}{\bf Reaction}\hspace{2.5cm} & \hspace{1cm}{\bf Catalyst}\hspace{1cm} \\ \hline
  \endfirsthead
  \multicolumn{1}{l}{\tablename\ \thetable{}} & \multicolumn{1}{r}{{\it $\ldots$ continued from previous page}} \\   \hline
  \hspace{2cm}{\bf Reaction}\hspace{2cm} & \hspace{1cm}{\bf Catalyst}\hspace{1cm} \\ \hline
  \endhead
    \multicolumn{2}{l}{{\it continued on next page $\ldots$}} \\
  \endfoot
  \endlastfoot

  \multicolumn{2}{|c|}{\bf Generation 1 of length 3 ($\kappa=1000$)} \\ \hline
  $(1,0) + (1,0) \rightleftharpoons (2,0)$ & $(3,0)$ \\ \hline
  $(2,0) + (1,0) \rightleftharpoons (3,0)$ & $(3,0)$ \\ \hline

  \multicolumn{2}{|c|}{\bf Generation 2 of length 6 ($\kappa=2000$)} \\ \hline
  $(1,0) + (1,0) \rightleftharpoons (2,0)$ & $(4,2)$ \\ \hline
  $(0,1) + (2,0) \rightleftharpoons (2,1)$ & $(4,2)$ \\ \hline
  $(2,1) + (0,1) \rightleftharpoons (2,2)$ & $(4,2)$ \\ \hline
  $(2,1) + (2,1) \rightleftharpoons (4,2)$ & $(4,2)$ \\ \hline

  \multicolumn{2}{|c|}{\bf Generation 3 of length 10 ($\kappa=4000$)} \\ \hline
  $(1,0) + (0,1) \rightleftharpoons (1,1)$ & $(4,6)$ \\ \hline
  $(1,1) + (1,0) \rightleftharpoons (2,1)$ & $(4,6)$ \\ \hline
  $(1,1) + (1,1) \rightleftharpoons (2,2)$ & $(4,6)$ \\ \hline
  $(0,1) + (2,2) \rightleftharpoons (2,3)$ & $(4,6)$ \\ \hline
  $(2,3) + (2,3) \rightleftharpoons (4,6)$ & $(4,6)$ \\ \hline

  \multicolumn{2}{|c|}{\bf Generation 4 of length 15 ($\kappa=7000$)} \\ \hline
  $(0,1) + (0,1) \rightleftharpoons (0,2)$ & $(7,8)$ \\ \hline
  $(0,2) + (1,0) \rightleftharpoons (1,2)$ & $(7,8)$ \\ \hline
  $(1,2) + (1,0) \rightleftharpoons (2,2)$ & $(7,8)$ \\ \hline
  $(2,2) + (2,2) \rightleftharpoons (4,4)$ & $(7,8)$ \\ \hline
  $(1,2) + (4,4) \rightleftharpoons (5,6)$ & $(7,8)$ \\ \hline
  $(5,6) + (1,2) \rightleftharpoons (6,8)$ & $(7,8)$ \\ \hline
  $(5,6) + (2,2) \rightleftharpoons (7,8)$ & $(7,8)$ \\ \hline

  \multicolumn{2}{|c|}{\bf Generation 5 of length 19 ($\kappa=10000$)} \\ \hline
  $(0,1) + (1,0) \rightleftharpoons (1,1)$ & $(7,12)$ \\ \hline
  $(1,1) + (0,1) \rightleftharpoons (1,2)$ & $(7,12)$ \\ \hline
  $(1,2) + (1,2) \rightleftharpoons (2,4)$ & $(7,12)$ \\ \hline
  $(1,1) + (2,4) \rightleftharpoons (3,5)$ & $(7,12)$ \\ \hline
  $(2,4) + (3,5) \rightleftharpoons (5,9)$ & $(7,12)$ \\ \hline
  $(1,1) + (5,9) \rightleftharpoons (6,10)$ & $(7,12)$ \\ \hline
  $(1,2) + (6,10) \rightleftharpoons (7,12)$ & $(7,12)$ \\ \hline

  \multicolumn{2}{|c|}{\bf Generation 6 of length 25 ($\kappa=15000$)} \\ \hline
  $(0,1) + (1,0) \rightleftharpoons (1,1)$ & $(12,13)$ \\ \hline
  $(1,1) + (1,0) \rightleftharpoons (2,1)$ & $(12,13)$ \\ \hline
  $(2,1) + (0,1) \rightleftharpoons (2,2)$ & $(12,13)$ \\ \hline
  $(2,2) + (0,1) \rightleftharpoons (2,3)$ & $(12,13)$ \\ \hline
  $(1,0) + (2,3) \rightleftharpoons (3,3)$ & $(12,13)$ \\ \hline
  $(2,1) + (3,3) \rightleftharpoons (5,4)$ & $(12,13)$ \\ \hline
  $(2,3) + (5,4) \rightleftharpoons (7,7)$ & $(12,13)$ \\ \hline
  $(7,7) + (2,2) \rightleftharpoons (9,9)$ & $(12,13)$ \\ \hline
  $(7,7) + (3,3) \rightleftharpoons (10,10)$ & $(12,13)$ \\ \hline
  $(9,9) + (2,3) \rightleftharpoons (11,12)$ & $(12,13)$ \\ \hline
  $(11,12) + (1,1) \rightleftharpoons (12,13)$ & $(12,13)$ \\ \hline

  \multicolumn{2}{|c|}{\bf Generation 7 of length 30 ($\kappa=20000$)} \\ \hline
  $(0,1) + (1,0) \rightleftharpoons (1,1)$ & $(13,17)$ \\ \hline
  $(1,1) + (0,1) \rightleftharpoons (1,2)$ & $(13,17)$ \\ \hline
  $(1,2) + (1,1) \rightleftharpoons (2,3)$ & $(13,17)$ \\ \hline
  $(1,0) + (2,3) \rightleftharpoons (3,3)$ & $(13,17)$ \\ \hline
  $(3,3) + (1,1) \rightleftharpoons (4,4)$ & $(13,17)$ \\ \hline
  $(3,3) + (1,2) \rightleftharpoons (4,5)$ & $(13,17)$ \\ \hline
  $(2,3) + (4,5) \rightleftharpoons (6,8)$ & $(13,17)$ \\ \hline
  $(2,3) + (6,8) \rightleftharpoons (8,11)$ & $(13,17)$ \\ \hline
  $(6,8) + (4,4) \rightleftharpoons (10,12)$ & $(13,17)$ \\ \hline
  $(10,12) + (2,3) \rightleftharpoons (12,15)$ & $(13,17)$ \\ \hline
  $(12,15) + (1,2) \rightleftharpoons (13,17)$ & $(13,17)$ \\ \hline

  \multicolumn{2}{|c|}{\bf Generation 8 of length 35 ($\kappa=27000$)} \\ \hline
  $(1,0) + (0,1) \rightleftharpoons (1,1)$ & $(15,20)$ \\ \hline
  $(0,1) + (1,1) \rightleftharpoons (1,2)$ & $(15,20)$ \\ \hline
  $(1,1) + (1,2) \rightleftharpoons (2,3)$ & $(15,20)$ \\ \hline
  $(1,2) + (2,3) \rightleftharpoons (3,5)$ & $(15,20)$ \\ \hline
  $(3,5) + (1,1) \rightleftharpoons (4,6)$ & $(15,20)$ \\ \hline
  $(1,1) + (4,6) \rightleftharpoons (5,7)$ & $(15,20)$ \\ \hline
  $(3,5) + (4,6) \rightleftharpoons (7,11)$ & $(15,20)$ \\ \hline
  $(7,11) + (4,6) \rightleftharpoons (11,17)$ & $(15,20)$ \\ \hline
  $(1,0) + (11,17) \rightleftharpoons (12,17)$ & $(15,20)$ \\ \hline
  $(1,1) + (12,17) \rightleftharpoons (13,18)$ & $(15,20)$ \\ \hline
  $(13,18) + (1,2) \rightleftharpoons (14,20)$ & $(15,20)$ \\ \hline
  $(1,0) + (14,20) \rightleftharpoons (15,20)$ & $(15,20)$ \\ \hline

  \multicolumn{2}{|c|}{\bf Generation 9 of length 40 ($\kappa=35000$)} \\ \hline
  $(0,1) + (0,1) \rightleftharpoons (0,2)$ & $(14,26)$ \\ \hline
  $(1,0) + (0,2) \rightleftharpoons (1,2)$ & $(14,26)$ \\ \hline
  $(1,2) + (1,2) \rightleftharpoons (2,4)$ & $(14,26)$ \\ \hline
  $(1,2) + (2,4) \rightleftharpoons (3,6)$ & $(14,26)$ \\ \hline
  $(3,6) + (1,0) \rightleftharpoons (4,6)$ & $(14,26)$ \\ \hline
  $(3,6) + (2,4) \rightleftharpoons (5,10)$ & $(14,26)$ \\ \hline
  $(5,10) + (1,2) \rightleftharpoons (6,12)$ & $(14,26)$ \\ \hline
  $(6,12) + (2,4) \rightleftharpoons (8,16)$ & $(14,26)$ \\ \hline
  $(2,4) + (8,16) \rightleftharpoons (10,20)$ & $(14,26)$ \\ \hline
  $(0,1) + (10,20) \rightleftharpoons (10,21)$ & $(14,26)$ \\ \hline
  $(10,21) + (0,2) \rightleftharpoons (10,23)$ & $(14,26)$ \\ \hline
  $(0,2) + (10,23) \rightleftharpoons (10,25)$ & $(14,26)$ \\ \hline
  $(10,23) + (1,2) \rightleftharpoons (11,25)$ & $(14,26)$ \\ \hline
  $(8,16) + (5,10) \rightleftharpoons (13,26)$ & $(14,26)$ \\ \hline
  $(4,6) + (10,20) \rightleftharpoons (14,26)$ & $(14,26)$ \\ \hline

  \multicolumn{2}{|c|}{\bf Generation 10 of length 45 ($\kappa=50000$)} \\ \hline
  $(0,1) + (1,0) \rightleftharpoons (1,1)$ & $(18,27)$ \\ \hline
  $(0,1) + (1,1) \rightleftharpoons (1,2)$ & $(18,27)$ \\ \hline
  $(1,2) + (1,0) \rightleftharpoons (2,2)$ & $(18,27)$ \\ \hline
  $(1,2) + (1,1) \rightleftharpoons (2,3)$ & $(18,27)$ \\ \hline
  $(1,2) + (2,2) \rightleftharpoons (3,4)$ & $(18,27)$ \\ \hline
  $(2,3) + (2,3) \rightleftharpoons (4,6)$ & $(18,27)$ \\ \hline
  $(4,6) + (1,1) \rightleftharpoons (5,7)$ & $(18,27)$ \\ \hline
  $(4,6) + (3,4) \rightleftharpoons (7,10)$ & $(18,27)$ \\ \hline
  $(7,10) + (7,10) \rightleftharpoons (14,20)$ & $(18,27)$ \\ \hline
  $(0,1) + (14,20) \rightleftharpoons (14,21)$ & $(18,27)$ \\ \hline
  $(14,21) + (2,2) \rightleftharpoons (16,23)$ & $(18,27)$ \\ \hline
  $(2,3) + (16,23) \rightleftharpoons (18,26)$ & $(18,27)$ \\ \hline
  $(14,21) + (4,6) \rightleftharpoons (18,27)$ & $(18,27)$ \\ \hline
\end{longtable}
\end{center}

\setlength{\LTcapwidth}{8in}
\begin{landscape}
\begin{center}
\begin{longtable}{|c|c|c|c|c|c|c|c|c|c|c|}
  \caption [List of reactions and their catalysts in ACS(18,27).]{{\bf List of reactions in the catalyzed chemistry and their catalysts.} The table lists all the reactions that are part of the catalyzed chemistry with all its catalysts. The catalysts that belong to different generations ($G_1$ to $G_{10}$) have been separated in different columns. It is easy to see from this table the amount of overlap between any two nested ACSs. For example, between the ACSs $G_5$ and $G_6$ which contain, respectively, 7 and 11 reactions, only one is common.} \\
  \hline
     \multirow{2}{*}{\hspace{1.7cm}{\bf Reaction}\hspace{1.7cm}} & \multicolumn{10}{|c|}{\bf Catalyst} \\ \cline{2-11} & \hspace{0.35cm}$G_1$\hspace{0.35cm} & \hspace{0.35cm}$G_2$\hspace{0.35cm} & \hspace{0.35cm}$G_3$\hspace{0.35cm} & \hspace{0.35cm}$G_4$\hspace{0.35cm} & \hspace{0.35cm}$G_5$\hspace{0.35cm} & \hspace{0.35cm}$G_6$\hspace{0.35cm} & \hspace{0.35cm}$G_7$\hspace{0.35cm} & \hspace{0.35cm}$G_8$\hspace{0.35cm} & \hspace{0.35cm}$G_9$\hspace{0.35cm} & $G_{10}$ \\ \hline
  \endfirsthead
     \multicolumn{1}{l}{\tablename\ \thetable{}} & \multicolumn{10}{r}{{\it $\ldots$ continued from previous page}} \\   \hline
     \multirow{2}{*}{{\bf Reaction}} & \multicolumn{10}{|c|}{\bf Catalyst} \\ \cline{2-11} & $G_1$ & $G_2$ & $G_3$ & $G_4$ & $G_5$ & $G_6$ & $G_7$ & $G_8$ & $G_9$ & $G_{10}$ \\ \hline
  \endhead
    \multicolumn{11}{l}{{\it continued on next page $\ldots$}} \\
  \endfoot
  \endlastfoot
  $(0,1) + (0,1) \rightleftharpoons  (0,2)$ & & & & (7,8) & & & & & (14,26) & \\ \hline
  $(1,0) + (0,1) \rightleftharpoons  (1,1)$ & & & (4,6) & & (7,12) & (12,13) & (13,17) & (15,20) & & (18,27) \\ \hline
  $(1,0) + (1,0) \rightleftharpoons  (2,0)$ & (3,0) & (4,2) & & & & & & & & \\ \hline
  $(0,2) + (1,0) \rightleftharpoons  (1,2)$ & & & & (7,8) & & & & & (14,26) & \\ \hline
  $(1,1) + (0,1) \rightleftharpoons  (1,2)$ & & & & & (7,12) & & (13,17) & (15,20) & & (18,27) \\ \hline
  $(0,1) + (2,0) \rightleftharpoons  (2,1)$ & & (4,2) & & & & & & & & \\ \hline
  $(1,1) + (1,0) \rightleftharpoons  (2,1)$ & & & (4,6) & & & (12,13) & & & & \\ \hline
  $(2,0) + (1,0) \rightleftharpoons  (3,0)$ & (3,0) & & & & & & & & & \\ \hline
  $(2,1) + (0,1) \rightleftharpoons  (2,2)$ & & (4,2) & & & & (12,13) & & & & \\ \hline
  $(1,1) + (1,1) \rightleftharpoons  (2,2)$ & & & (4,6) & & & & & & & \\ \hline
  $(1,2) + (1,0) \rightleftharpoons  (2,2)$ & & & & (7,8) & & & & & & (18,27) \\ \hline
  $(0,1) + (2,2) \rightleftharpoons  (2,3)$ & & & (4,6) & & & (12,13) & & & & \\ \hline
  $(1,2) + (1,1) \rightleftharpoons  (2,3)$ & & & & & & & (13,17) & (15,20) & & (18,27) \\ \hline
  $(1,2) + (1,2) \rightleftharpoons  (2,4)$ & & & & & (7,12) & & & & (14,26) & \\ \hline
  $(1,0) + (2,3) \rightleftharpoons  (3,3)$ & & & & & & (12,13) & (13,17) & & & \\ \hline
  $(2,1) + (2,1) \rightleftharpoons  (4,2)$ & & (4,2) & & & & & & & & \\ \hline
  $(1,2) + (2,2) \rightleftharpoons  (3,4)$ & & & & & & & & & & (18,27) \\ \hline
  $(1,1) + (2,4) \rightleftharpoons  (3,5)$ & & & & & (7,12) & & & & & \\ \hline
  $(1,2) + (2,3) \rightleftharpoons  (3,5)$ & & & & & & & & (15,20) & & \\ \hline
  $(2,2) + (2,2) \rightleftharpoons  (4,4)$ & & & & (7,8) & & & & & & \\ \hline
  $(3,3) + (1,1) \rightleftharpoons  (4,4)$ & & & & & & & (13,17) & & & \\ \hline
  $(1,2) + (2,4) \rightleftharpoons  (3,6)$ & & & & & & & & & (14,26) & \\ \hline
  $(3,3) + (1,2) \rightleftharpoons  (4,5)$ & & & & & & & (13,17) & & & \\ \hline
  $(2,1) + (3,3) \rightleftharpoons  (5,4)$ & & & & & & (12,13) & & & & \\ \hline
  $(2,3) + (2,3) \rightleftharpoons  (4,6)$ & & & (4,6) & & & & & & & (18,27) \\ \hline
  $(3,5) + (1,1) \rightleftharpoons  (4,6)$ & & & & & & & & (15,20) & & \\ \hline
  $(3,6) + (1,0) \rightleftharpoons  (4,6)$ & & & & & & & & & (14,26) & \\ \hline
  $(1,2) + (4,4) \rightleftharpoons  (5,6)$ & & & & (7,8) & & & & & & \\ \hline
  $(1,1) + (4,6) \rightleftharpoons  (5,7)$ & & & & & & & & (15,20) & & (18,27) \\ \hline
  $(2,4) + (3,5) \rightleftharpoons  (5,9)$ & & & & & (7,12) & & & & & \\ \hline
  $(5,6) + (1,2) \rightleftharpoons  (6,8)$ & & & & (7,8) & & & & & & \\ \hline
  $(2,3) + (4,5) \rightleftharpoons  (6,8)$ & & & & & & & (13,17) & & & \\ \hline
  $(2,3) + (5,4) \rightleftharpoons  (7,7)$ & & & & & & (12,13) & & & & \\ \hline
  $(3,6) + (2,4) \rightleftharpoons  (5,10)$ & & & & & & & & & (14,26) & \\ \hline
  $(5,6) + (2,2) \rightleftharpoons  (7,8)$ & & & & (7,8) & & & & & & \\ \hline
  $(1,1) + (5,9) \rightleftharpoons  (6,10)$ & & & & & (7,12) & & & & & \\ \hline
  $(4,6) + (3,4) \rightleftharpoons  (7,10)$ & & & & & & & & & & (18,27) \\ \hline
  $(5,10) + (1,2) \rightleftharpoons  (6,12)$ & & & & & & & & & (14,26) & \\ \hline
  $(3,5) + (4,6) \rightleftharpoons  (7,11)$ & & & & & & & & (15,20) & & \\ \hline
  $(7,7) + (2,2) \rightleftharpoons  (9,9)$ & & & & & & (12,13) & & & & \\ \hline
  $(1,2) + (6,10) \rightleftharpoons  (7,12)$ & & & & & (7,12) & & & & & \\ \hline
  $(2,3) + (6,8) \rightleftharpoons  (8,11)$ & & & & & & & (13,17) & & & \\ \hline
  $(7,7) + (3,3) \rightleftharpoons  (10,10)$ & & & & & & (12,13) & & & & \\ \hline
  $(6,8) + (4,4) \rightleftharpoons  (10,12)$ & & & & & & & (13,17) & & & \\ \hline
  $(9,9) + (2,3) \rightleftharpoons  (11,12)$ & & & & & & (12,13) & & & & \\ \hline
  $(6,12) + (2,4) \rightleftharpoons  (8,16)$ & & & & & & & & & (14,26) & \\ \hline
  $(11,12) + (1,1) \rightleftharpoons  (12,13)$ & & & & & & (12,13) & & & & \\ \hline
  $(10,12) + (2,3) \rightleftharpoons  (12,15)$ & & & & & & & (13,17) & & & \\ \hline
  $(7,11) + (4,6) \rightleftharpoons  (11,17)$ & & & & & & & & (15,20) & & \\ \hline
  $(1,0) + (11,17) \rightleftharpoons  (12,17)$ & & & & & & & & (15,20) & & \\ \hline
  $(2,4) + (8,16) \rightleftharpoons  (10,20)$ & & & & & & & & & (14,26) & \\ \hline
  $(12,15) + (1,2) \rightleftharpoons  (13,17)$ & & & & & & & (13,17) & & & \\ \hline
  $(0,1) + (10,20) \rightleftharpoons  (10,21)$ & & & & & & & & & (14,26) & \\ \hline
  $(1,1) + (12,17) \rightleftharpoons  (13,18)$ & & & & & & & & (15,20) & & \\ \hline
  $(10,21) + (0,2) \rightleftharpoons  (10,23)$ & & & & & & & & & (14,26) & \\ \hline
  $(13,18) + (1,2) \rightleftharpoons  (14,20)$ & & & & & & & & (15,20) & & \\ \hline
  $(7,10) + (7,10) \rightleftharpoons  (14,20)$ & & & & & & & & & & (18,27) \\ \hline
  $(0,2) + (10,23) \rightleftharpoons  (10,25)$ & & & & & & & & & (14,26) & \\ \hline
  $(0,1) + (14,20) \rightleftharpoons  (14,21)$ & & & & & & & & & & (18,27) \\ \hline
  $(1,0) + (14,20) \rightleftharpoons  (15,20)$ & & & & & & & & (15,20) & & \\ \hline
  $(10,23) + (1,2) \rightleftharpoons  (11,25)$ & & & & & & & & & (14,26) & \\ \hline
  $(8,16) + (5,10) \rightleftharpoons  (13,26)$ & & & & & & & & & (14,26) & \\ \hline
  $(14,21) + (2,2) \rightleftharpoons  (16,23)$ & & & & & & & & & & (18,27) \\ \hline
  $(4,6) + (10,20) \rightleftharpoons  (14,26)$ & & & & & & & & & (14,26) & \\ \hline
  $(2,3) + (16,23) \rightleftharpoons  (18,26)$ & & & & & & & & & & (18,27) \\ \hline
  $(14,21) + (4,6) \rightleftharpoons  (18,27)$ & & & & & & & & & & (18,27) \\ \hline
\end{longtable}
\end{center}
\end{landscape}

\thispagestyle{plain}
\cleardoublepage
\chapter{\label{Programs}Description of the programs used to generate the results in the thesis.}

This appendix discusses the programs used to generate various numerical results in the thesis. The code source files can be downloaded from the website:\\\url{http://sites.google.com/site/varungiri/research/phd-thesis}.

The numerical work reported in this thesis was largely done using C codes. For the numerical integration of differential equations these codes use CVODE solver library of the SUNDIALS package v.2.3.0 \cite{SUNDIALS} which can be downloaded from \url{https://computation.llnl.gov/casc/sundials/download/download.html} (November 2006 release version). The list of programs is as follows:

\subsubsection*{Chapter 3}
\begin{itemize}
  \item \verb"spontaneous-f1.c, spontaneous-f2.c" : Constructs a fully connected spontaneous chemistry for $f=1, 2$ (respectively) and integrates the system of differential equations to find steady state concentrations. This program produces the time courses and the steady state concentration values for a class of user specified initial conditions.
  \item \verb"spontaneous-sparse-f1.c, spontaneous-sparse-f2.c" : Constructs a sparse spontaneous chemistry with degree $k$ specified as a parameter to the program for $f=1, 2$ (respectively) and produces the steady state concentration values.
  \item \verb"spontaneous-hetro-f1.c" : Constructs a fully connected spontaneous chemistry. It then assigns the rate constants for each reaction as described in Section \ref{section-spont-hetero} (for heterogeneous chemistries) and computes the steady state concentration values.
\end{itemize}

\subsubsection*{Chapter 4}
\begin{itemize}
  \item \verb"acs-f1.c" : Constructs the fully connected spontaneous chemistry for $f=1$ with catalyzed reactions defined by ACS65 (Eqs. (\ref{acs65-definition})) and computes the steady state concentrations.
  \item \verb"acs-f2.c" : Constructs the fully connected spontaneous chemistry for $f=2$ with catalyzed reactions defined by ACS(8,10) (Eqs. (\ref{acs8-10-definition})) and computes the steady state concentrations.
  \item \verb"ACS-2-4.ode" : Input file for XPPAUT for ACS4 (Eqs. (\ref{acs4-definition})).
  \item \verb"acs-sparse-f1.c" : Constructs a sparse spontaneous chemistry for $f=1$ with catalyzed reactions defined by ACS65 (Eqs. (\ref{acs65-definition})) and computes the steady state concentrations.
  \item \verb"acs-hetro-f1.c" : Constructs a fully connected spontaneous chemistry and assign rate constants for each reaction as described in Section \ref{section-spont-hetero}. It catalyzes the reactions as defined by ACS65 (Eqs. (\ref{acs65-definition})) and computes steady state concentration values.
\end{itemize}
These programs can be appropriately changed to study other ACSs. The instructions to do so are included with the code.

\subsubsection*{Chapter 5}
\begin{itemize}
  \item \verb"algo1.c, algo2.c, algo3.c" : Constructs a fully connected spontaneous chemistry for $f=1$ with an ACS of length $L$ constructed using Algorithm 1, 2, or 3 (respectively) described in Section \ref{Algos-1-2-3}.
\end{itemize}

\subsubsection*{Chapter 6}
\begin{itemize}
  \item \verb"acs-f1-ACS3_8.c" : Constructs a fully spontaneous chemistry for $f=1$ with a pair of nested ACSs defined by Eqs. (\ref{acs3-definition}) and (\ref{acs8-definition}) and computes the steady state concentration values.
  \item \verb"acs-f2-ACS4_8.c" : Constructs a fully spontaneous chemistry for $f=2$ with a pair of nested ACSs defined by Eqs. (\ref{ACS(2,2)-definition}) and (\ref{ACS(5,3)(c)-definition}) and computes the steady state concentration values. The ACSs defined in this program can be modified to study the dynamics of the other ACSs mentioned in the text below Eq. (\ref{ACS(2,2)-definition}).
\end{itemize}

\subsubsection*{Chapter 7}
\begin{itemize}
  \item \verb"algo4-1f.c, algo4-2f.c" : Constructs a sparse spontaneous chemistry for $f=1,2$ (respectively) with nested ACSs constructed using Algorithm 4 described in Section \ref{Algo-4-5}, and computes steady state concentration values.
  \item \verb"algo5-2f.c" : Constructs a sparse spontaneous chemistry for $f=2$ with nested ACSs constructed using Algorithm 5 described in Section \ref{Algo-4-5}, and computes steady state concentration values.
\end{itemize} 

\thispagestyle{plain}
\cleardoublepage
\phantomsection
\addcontentsline{toc}{chapter}{Bibliography}
\bibliography{thesis}

\pagestyle{plain}
\thispagestyle{plain}
\cleardoublepage
\phantomsection
\addcontentsline{toc}{chapter}{List of Figures}
\listoffigures

\thispagestyle{plain}
\cleardoublepage
\phantomsection
\addcontentsline{toc}{chapter}{List of Tables}
\listoftables

\thispagestyle{plain}
\cleardoublepage
\renewcommand{\glossarypreamble}{The definitions given here are for quick reference. For precise definitions see text.}
\printglossary


\end{document}